%% file: main.tex
\documentclass[a4paper,11pt]{article}
\pdfoutput=1

\usepackage{jheppub}
\usepackage[pass]{geometry}

\usepackage[T1]{fontenc}
\usepackage{caption}
\hypersetup{bookmarksnumbered}
\usepackage{mathtools,slashed}
\usepackage[table,dvipsnames]{xcolor}
\definecolor{linkblue}{rgb}{0.07,0.28,0.70}
\hypersetup{allcolors=linkblue}
\usepackage{dsfont}
\usepackage{amssymb}

\usepackage{titlesec}

\usepackage{tikz}
\usetikzlibrary{arrows.meta,decorations.markings,decorations.pathmorphing,calc,intersections}

\titleformat{\paragraph}
{\normalfont\normalsize\bfseries}{\theparagraph}{1em}{}
\titlespacing*{\paragraph}
{0pt}{3.25ex plus 1ex minus .2ex}{1.5ex plus .2ex}

\begin{document}

\newgeometry{textwidth=6.5in,textheight=9in,top=2in,left=1.885in,headheight=0pt,headsep=0pt}
\thispagestyle{empty}
\begingroup
\renewcommand{\baselinestretch}{1}\setlength{\parskip}{5pt plus 1pt}
\fontsize{12}{14.5}\selectfont\setlength{\leftmargini}{2.5em}

\vspace*{-0.4in}

\noindent\hfill
\begin{tabular}{@{}l@{}}
    PUPT-2668 \\
    CALT-TH 036
\end{tabular}

\begin{center}
$$$$
{\fontsize{17.28}{22}\selectfont\textbf{\mathversion{bold}
The OPE of lines beyond CFT
}\par}

\vspace{1.0cm}

\textrm{Jan Albert,$^{\mathcal D}$ Matthew Forslund$^{\mathcal D}$ and Alexandre Homrich$^{{S_\mathrm{eff}}}$}
\\ \vspace{1cm}
{\fontsize{10}{12}\selectfont\textit{$^{\mathcal D}$Princeton Center for Theoretical Science, Princeton University, Princeton, NJ 08544, USA}}
\\ \vspace{2mm}
{\fontsize{10}{12}\selectfont\textit{$^{{S_\mathrm{eff}}}$Walter Burke Institute for Theoretical Physics, Caltech, Pasadena, CA 91125, USA}}
\end{center}

\vspace{2mm}
{\fontsize{10.95}{13.6}\selectfont
\begin{center}{\bfseries Abstract\vspace{-.5em}}\end{center}
\begin{quotation}
We discuss the fusion of parallel line defects in theories defined by relevant deformations of a conformal fixed point.
We propose an operator product expansion to describe this limit in terms of a Wilsonian effective action whose coefficients are computable in conformal perturbation theory around the UV.
This allows one to fix, order by order, the fusion limit of correlation functions involving the defects, up to non-perturbative coefficients associated to correlation functions (condensates) on the fused defect.
It thus generalizes the familiar OPE of local operators in QFT.
As a non-perturbative test of our proposal, we consider the  short-distance potential between two magnetic impurities in the $2d$ Ising model. We find excellent agreement between the OPE framework 
and \texttt{DMRG} extractions of the potential and condensates (the latter also computed exactly with integrability). In addition, we consider a number of solvable models to illustrate different features of the framework.
\end{quotation}}

\endgroup
\restoregeometry
\makeatletter
\setcounter{page}{0}\newpage
\beforetochook\hrule
\setcounter{tocdepth}{2}
\tableofcontents
\afterTocSpace
\hrule
\afterTocRuleSpace
\setcounter{footnote}{0}
\clearpage
\pagestyle{myplain}\pagenumbering{arabic}
\makeatother

\section{Introduction and summary}

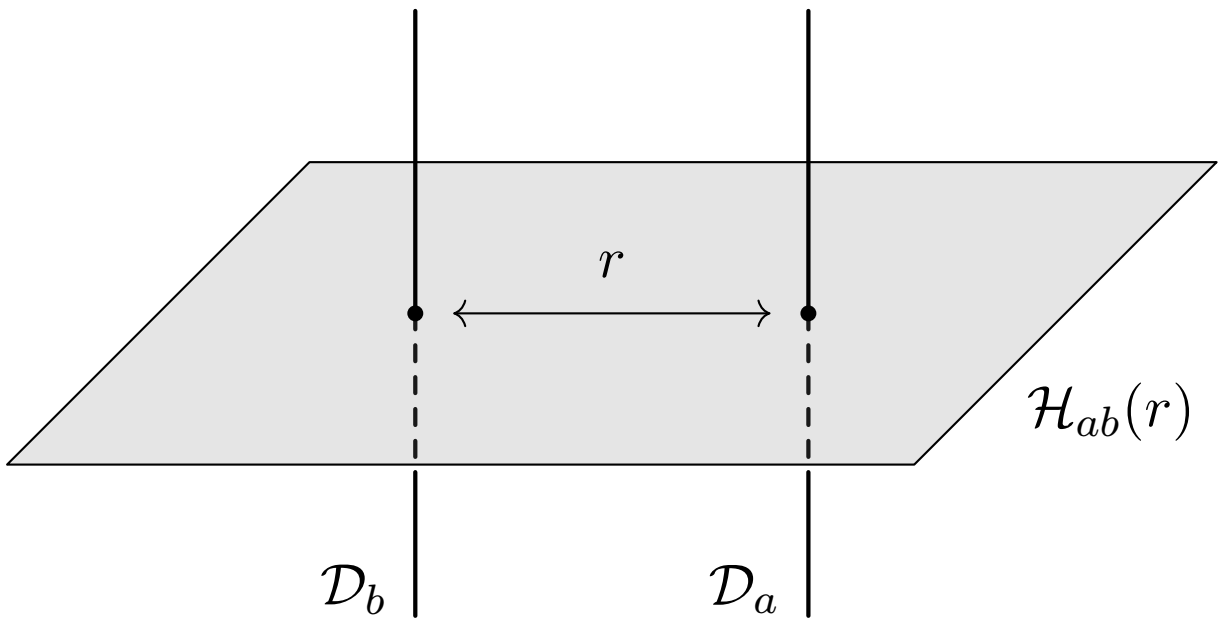
\begin{figure}[h]
    \centering
    \input{Figures/Hab.tikz}
    \captionsetup{singlelinecheck=false}
    \caption{
The potential $V_{ab}(r)$ is the ground-state energy of the Hilbert space $\mathcal H_{ab}(r)$ on a time slice pierced by the line defects $\mathcal{D}_a$ and $\mathcal{D}_b$ describing the worldlines of the probes.
Equivalently,
    \begin{equation}\label{eq:V-def}
        V_{ab}(r) \equiv -\lim_{L\to \infty} \frac{1}{L}\log \langle \mathcal D_a[r]\, \mathcal D_b[0]\rangle\,,
    \end{equation}
with $L$ the length of the defects. It is defined up to an
$r$-independent constant,  which can be shifted by a mass counterterm $\int ds\, m$ on each defect. 
}
    \label{fig:Hab}
\end{figure}

The potential between two static probes is among the simplest observables of a quantum field theory. 
It depends only on the separation $r$ between the defects, see figure \ref{fig:Hab}, and yet by dialing this parameter we can explore the theory at every energy scale. 
At long distances it can serve as a diagnostic for rich phenomena such as superfluidity, 
superconductivity and confinement.
In this paper we study the opposite regime, $r\to 0$, which probes the UV of the theory. We will show that the potential in this limit has a universal structure, largely fixed by the UV fixed point.

We will focus on QFTs defined as relevant deformations of a CFT in the UV,
\begin{equation}\label{eq:g-def}
    S = S_\text{CFT} + g\int d^d x\, \varPhi \qquad (\Delta_\varPhi <d)\,,
\end{equation}
and consider line defects obtained as relevant deformations of conformal (or topological) line defects,
\begin{equation}\label{eq:h-def}
    \mathcal D = \mathcal D_{\text{CFT}}\, e^{h\int ds\, \varphi} \qquad (\Delta_\varphi <1)\,.
\end{equation}
Here $\varphi(s)$ is a local operator living on the conformal defect $\mathcal D_{\text{CFT}}$. 

\begin{figure}[t]
    \centering
    \includegraphics[width=0.99\linewidth]{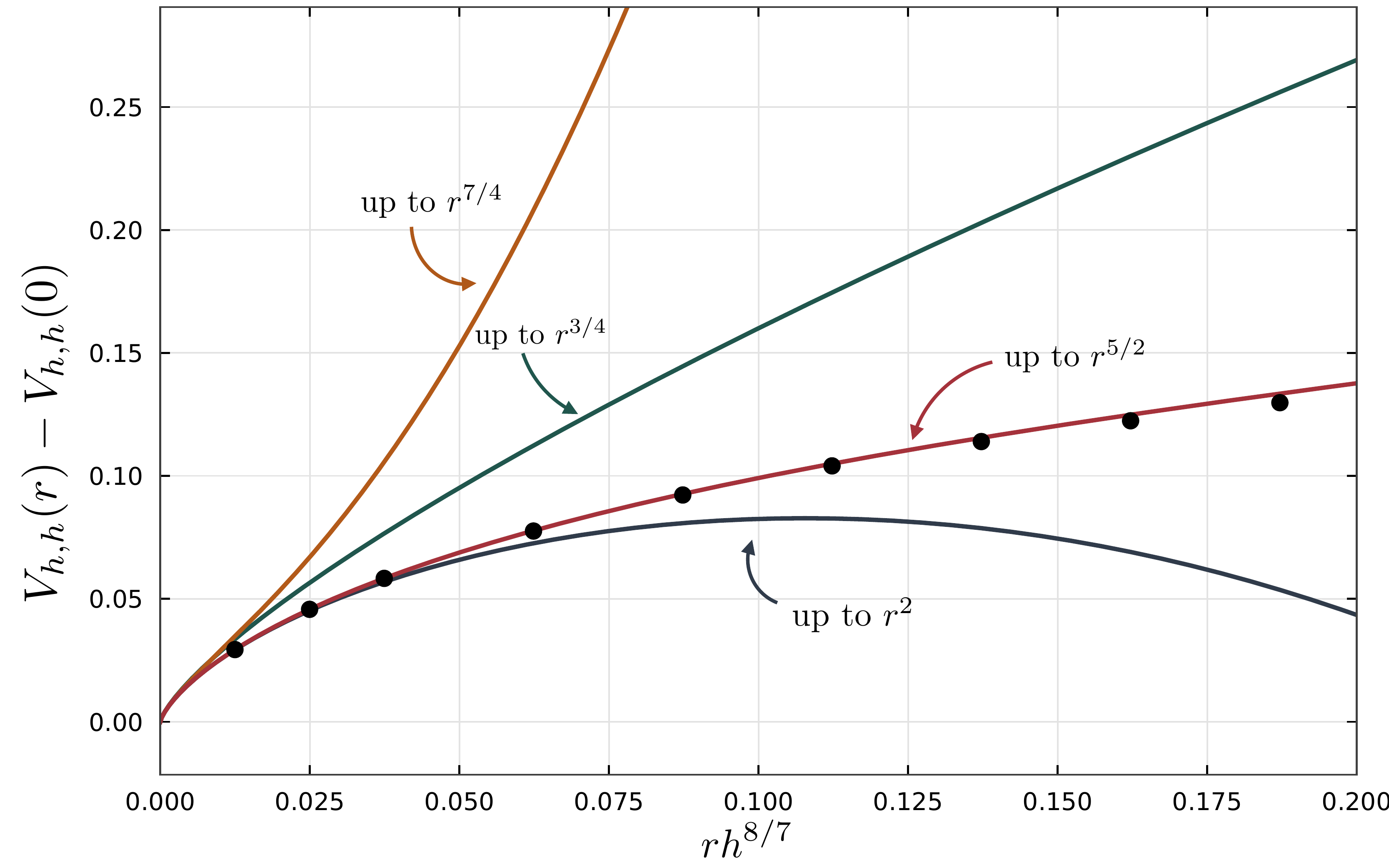}
  \caption{
The potential $V_{h,h}(r)-V_{h,h}(0)$ in units $h=1$
(black points), obtained with \texttt{DMRG},
compared with truncations of
\eqref{eq:VIsing-intro} through orders $r^{3/4}$ (green),
$r^{7/4}$ (orange), $r^2$ (blue), and $r^{5/2}$ (red).
All coefficients are fixed by \eqref{eq:Ising-constants-intro} and the exact condensates \eqref{eq:Ising-condensates-intro}. The constant $V_0=V_{h,h}(0)$ is estimated by fitting to the two smallest separations.
}
    \label{fig:Ising-intro}
 \end{figure}

When $\mathcal D_\text{CFT}$ is the trivial line, $\varphi$ is a bulk operator restricted to the line, and $\mathcal D$ is known as a (generalized) pinning field defect \cite{ParisenToldin:2016zxc,Cuomo:2021kfm,Popov:2025cha}.
Such defects are usually studied at the
endpoint of their RG flow, as conformal defects.
Here, instead, we keep the
scales set by $g$ and $h$ finite, and study the potential at separations much shorter
than these scales.
This setup includes magnetic impurities in the Ising model and, with minor modifications,
Wilson lines in pure Yang-Mills.
We will discuss these examples shortly, but the reader might want to anticipate the final result for $2d$ Ising depicted in figure \ref{fig:Ising-intro}.

In the limit $r\to 0$ the two defects can no longer be told apart, and they
fuse. Our strategy is to describe this fusion with a defect operator product expansion, generalizing the effective theory of fusion for conformal defects, recently introduced in
\cite{Diatlyk:2024zkk,Diatlyk:2024qpr,Kravchuk:2024qoh,Cuomo:2024psk}, to theories with mass scales.
We propose an expansion of the following form,
\begin{equation}\label{eq:DOPE-intro}
    \mathcal D_a[r]\mathcal D_b[0] \simeq \mathcal D_c^\text{CFT}[0] e^{-S_\text{eff}} \qquad (\text{as } r\to 0)\,,
\end{equation}
where $\mathcal D_c^\text{CFT}$ is a new conformal (or topological) defect of the UV CFT, and
$S_\text{eff}$ an effective action of operators on said defect,
\begin{equation}\label{eq:Seff-intro}
    S_\text{eff} = \int ds\, \left(-\frac{a_0(r)}{r} + \sum_{i}a_i(r) r^{\Delta_i -1} \mathcal{O}_i(s)\right)\,.
\end{equation}
We note that the defect $\mathcal{D}_c^{\text{CFT}}$ might not be simple, in which case the action contains defect-changing operators. We discuss this point further in section \ref{sec:non-simple}.

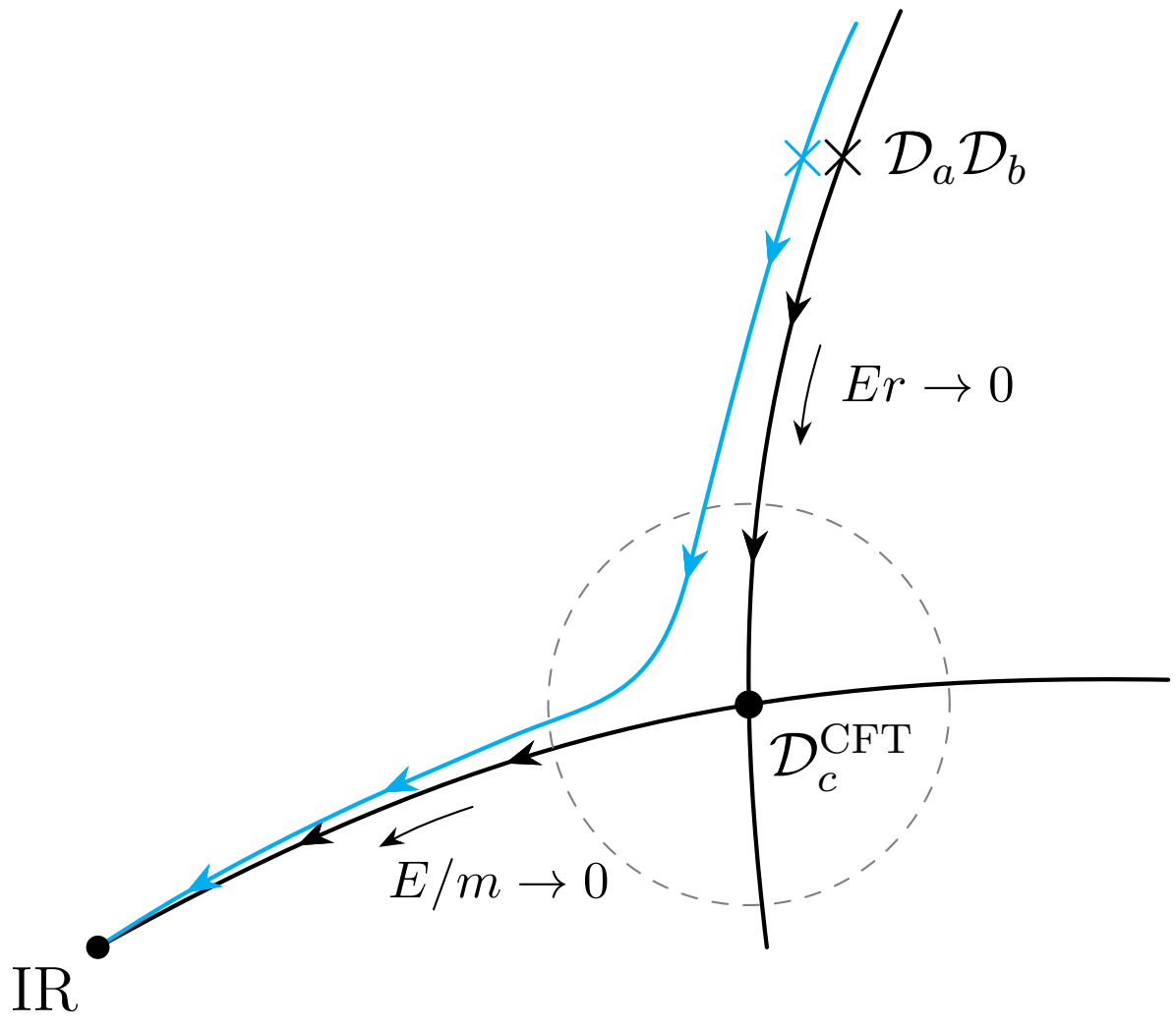
\begin{figure}[t]
    \centering
    \input{Figures/RG.tikz}
    \caption{Schematic representation of defect fusion as an RG flow. In a CFT, the defects
    $\mathcal D_a\mathcal D_b$ flow to the single conformal defect $\mathcal D_c^\text{CFT}$ along
    the vertical black line. In the massive theory, a relevant perturbation is turned on, and
    the flow (depicted in blue) continues on to a true IR fixed point. Since $mr\ll 1$, the
    massive flow passes arbitrarily close to the fixed point $\mathcal D_c^\text{CFT}$, where
    it can be described with an effective action (dashed circle).}
    \label{fig:RG}
\end{figure}

The logic behind this proposal is Wilsonian. After integrating out the modes with energy higher
than $1/r$, we can no longer resolve the two defects, and they merge into a single one with an
effective action. Since $r\to 0$, the cutoff is arbitrarily high, and this step does not affect
the bulk theory. The OPE thus becomes an operator statement within the original theory.
The resulting picture is sketched in figure~\ref{fig:RG}. In a CFT, the fusion can be thought of as an RG flow starting in
the UV with the separated defects $\mathcal D_a,\mathcal D_b$ and flowing
down to a single conformal defect (the black dot). In the massive theory, instead, the
flow continues (blue line). The limit $mr\ll 1$, with $m$ the mass scale
set by $g,h$, ensures that the relevant deformations triggered by $g,h$ remain small at the scale $1/r$, and so
the blue curve passes near the $\mathcal{D}_c^\text{CFT}$ fixed point before continuing towards the IR. Accordingly, our action
differs from the conformal one in two aspects:
\begin{enumerate}
    \item Due to the loss of scale invariance, the fusion coefficients $a_i(r)$ are promoted to dimensionless functions of $r$ and the
    scales set by the relevant couplings $g,h$.
    We will further argue that these coefficients can be computed in conformal perturbation theory (CPT) as an expansion in $r$.

    \item The sum over operators is not restricted to irrelevant $(\Delta_i>1)$ defect operators. This is due to the fact that we have not flowed the fused defect all the way to the IR; we have kept modes at the scales set
by $g,h$. The mass deformations can produce relevant operators on the fused defect that will
trigger subsequent RG flows.
\end{enumerate}

A crucial property of the OPE (\ref{eq:DOPE-intro}) is locality.
Indeed, the same effective action should be valid when fusing only small portions of the defects, irrespective of how they are continued far away, or the presence of distant insertions.
 In particular, the fusion coefficients $a_i(r)$ of the local action cannot depend 
 on global properties of the defects, such as the length $L$
 or their global geometry.

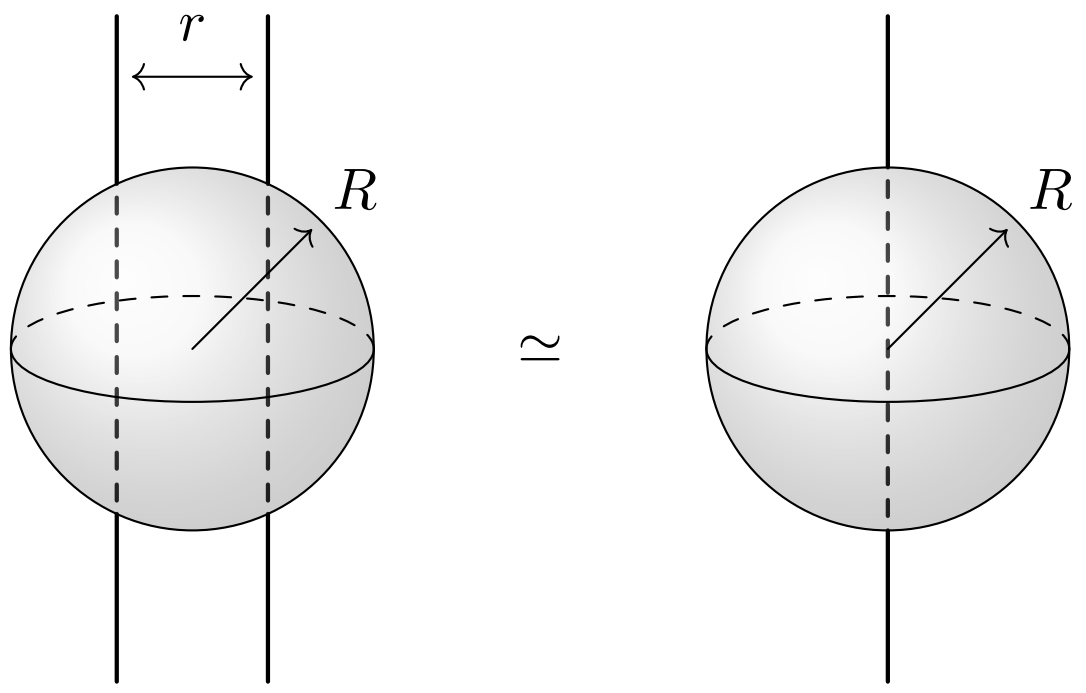
\begin{figure}[t]
    \centering
    \input{Figures/CPTball.tikz}
 \caption{To compute the (local) fusion coefficients $a_i(r)$, we consider a
configuration where we only turn on the relevant deformations \eqref{eq:g-def} and
\eqref{eq:h-def} inside a small ball of radius $R$ intersecting the defects, with
$r\ll R\ll 1/m$, so that CPT is reliable.
The cutoff $R$ regulates the infamous IR divergences of naive conformal perturbation theory (see discussions in sections \ref{sec:prescription_local_ops} and \ref{sec:prescription_lines}).
With this prescription, we can insert any  bulk operator $\mathcal{O}_k$ away from the defects,
and \textit{match} the computation of the correlators  
$\langle \mathcal O_k(x) \mathcal D_a[r] \mathcal D_b[0]\rangle$ and $\langle \mathcal O_k(x) \mathcal D_c^\text{CFT}[0] e^{-S_\text{eff}}\rangle$. Selecting many $\mathcal{O}_k$ produces enough equations so that we can
solve for the coefficients $a_i(r)$ to any given order in $r$.
This is the defect version of the strategy developed in \cite{Zamolodchikov:1990bk} for the case of local operators.
In agreement with locality, we find coefficients that are independent of the size of the ball, order by order, in all of our examples. This provides a highly nontrivial check of the framework.}
\label{fig:CPTball}
\end{figure}

Locality motivates our proposal that the fusion coefficients are computable in CPT.
They can be determined by matching both sides of \eqref{eq:DOPE-intro}
at distances large compared
with~$r$, but still much shorter than the scales set by $g,h$, where CPT applies.
We thus conjecture that the $a_i$ admit an asymptotic expansion polynomial in the relevant couplings,
\begin{equation}
    a_i(r) \simeq a_i^{(0,0)} + a_i^{(1,0)}g r^{d-\Delta_\varPhi} + a_i^{(0,1)}h r^{1-\Delta_\varphi} + a_i^{(1,1)}g h r^{d+1-\Delta_\varPhi-\Delta_\varphi} + \cdots\,,
\end{equation}
up to logarithms of $r$ due to operator mixing.
This expansion is analogous to that of structure functions in the local-operator OPE, fleshed out by Zamolodchikov in \cite{Zamolodchikov:1990bk} and reviewed in section~\ref{sec:localOps}.
The leading constant $a_i^{(0,0)}$ is the fusion coefficient in the conformal case \cite{Diatlyk:2024zkk,Diatlyk:2024qpr,Kravchuk:2024qoh,Cuomo:2024psk}.
It vanishes for relevant defect operators $\mathcal{O}_i$.  We briefly sketch the method for computing the coefficients $a_i$ in figure \ref{fig:CPTball}, and spell it out in detail in section \ref{sec:Lines}.

With this ansatz for the defect OPE, we can go ahead and make predictions for the UV behavior of the defect correlator. We first organize the effective action in increasing powers of $r$, and then bring down from $e^{-S_\text{eff}}$ the terms with positive powers of $r$ as operator insertions,
\begin{equation}\label{eq:DDprediction-intro}
    \langle \mathcal D_a[r] \mathcal D_b[0]\rangle \simeq \langle \mathcal D_c^\text{CFT}[0] e^{-S_\text{eff}}\rangle = e^{\frac{a_0(r)}{r}L}\left(\langle\mathcal D_c\rangle - \sum_i a_i(r)r^{\Delta_i-1} \int ds\langle \mathcal D_c \mathcal O_i(s)\rangle + \cdots\right)\,.
\end{equation}
The terms of $S_\text{eff}$ that do not vanish as $r\to 0$ cannot be expanded, and we keep them
in the exponent. Apart from the identity term, whose contribution we wrote explicitly,
they can only come from relevant defect operators. Together with $\mathcal D_c^\text{CFT}$,
they define the fused defect $\mathcal D_c$, which in general ceases to be conformal.

As a consequence, the operators $\mathcal O_i$ on the fused defect acquire expectation values and nontrivial higher-point functions. These condensates are non-perturbative and must be computed in the full deformed theory. We note that, due to the locality of the $a_i$, the dependence of the defect correlator (in the $r\rightarrow 0$ limit) on $L$ and the shape of the defects enters only through these condensates, and through explicit volume integrals. 

The prediction for
the potential then follows from \eqref{eq:V-def}:
\begin{equation}\label{eq:Vprediction-intro}
    V_{ab}(r) \simeq -\frac{a_0(r)}{r} +\sum_i a_i(r) r^{\Delta_i -1}\langle \mathcal{O}_i\rangle_c +\cdots\qquad (\text{as } r\rightarrow0),
\end{equation}
up to an additive constant, where $\langle \mathcal O\rangle_c\equiv \langle \mathcal D_c \mathcal{O}\rangle/\langle \mathcal D_c \rangle$.
 Similarly to the local operator OPE \cite{Wilson:1969zs,Shifman:1978bx,Shifman:1978by,Zamolodchikov:1990bk}, CPT fixes the asymptotic functional form of the potential as $r\to 0$ through the coefficients $a_i(r)$. They multiply non-perturbative constants associated to defect $n$-point functions $\langle\mathcal D_c\rangle$, $\langle\mathcal D_c\mathcal O_i\rangle$, $\langle\mathcal D_c\int\mathcal O_i\mathcal O_j\rangle$, etc. Only the contribution from the identity, known as the Casimir,\footnote{This is a bit of a misnomer in the massive theory, because the potential responsible for the Casimir effect between probes is the full potential $V_{ab}(r)$ rather than $-a_0(r)/r$. In the massless theory, all non-perturbative condensates vanish in the $L\to \infty$ limit, and the two agree. We will notwithstanding keep the terminology from the massless case.} is fully perturbative.

As a non-perturbative check of the framework, we considered the potential between magnetic impurities $\mathcal D_{h} \equiv \exp h\int ds\, \sigma$ in the $2d$ Ising model (see section \ref{sec:IsingModel}).
For simplicity, we keep the bulk at criticality, so that the only relevant couplings are the magnetic fields $h$ on the defects. It would be straightforward to extend the CPT and numerical analysis to include thermal or magnetic deformations in the bulk. 
For two impurities with equal magnetic fields $h$, which fuse into $\mathcal D_{2h}$, we obtain the following short-distance asymptotic expansion,
\begin{align}\label{eq:VIsing-intro}
{V_{h,h}(r) \simeq}& \ {V_0 - h^2 C_0\, r^{3/4}+2h^3\left(C_\epsilon \log r \mu + C_{\sigma} \right)\langle \sigma\rangle_{2h}\, r^{7/4} -h^2  C_\epsilon\langle[\epsilon ]_\mu \rangle_{2h}\, r^{7/4}}\\
    & \ { - \frac{h}{4} \langle \partial^2_x\sigma \rangle_{2h}\, r^2 + h^4 C_0^{(4)}\, r^{5/2} + O(r^{11/4})\,,}\nonumber
\end{align}
where $\langle   \mathcal O\rangle_{2h} = \langle\mathcal D_{2h} \mathcal O \rangle/\langle\mathcal D_{2h}\rangle$, and $V_0=V_{h,h}(0)$ is a scheme-dependent constant not determined by CPT.
The coefficients are given by
\begin{equation}\label{eq:Ising-constants-intro}
{
    C_0 = \frac{\sqrt{\pi }\, \Gamma \left(-\frac{3}{8}\right)}{\Gamma \left(\frac{1}{8}\right)} \, , \qquad C_\epsilon = \frac{\sqrt{\pi }\, \Gamma \left(-\frac{7}{8}\right)}{2\,\Gamma \left(-\frac{3}{8}\right)}\, , \qquad C_\sigma \approx 5.60020\, , \qquad C_0^{(4)} \approx 5.2678\,,}
\end{equation}
the last two being convergent integrals that we evaluate numerically. These are computed in CPT in section \ref{sec:IsingModel}.

Let us make a few remarks about this result. First, the $r$-dependence of the identity coefficient $a_0(r)$ overwhelms the conformal scaling $\sim 1/r$, suppressing the short-distance blowup in the Casimir term. This occurs when fusing pinning field defects generated by operators of dimension $\Delta_\varphi<1/2$.
Next, we note the appearance of the condensates  $\langle\sigma\rangle_{2h}$ and $\langle[\epsilon]_\mu \rangle_{2h}$ on the 
fused defect $\mathcal D_{2h}$.  This confirms that all defect operators can be generated in the effective action
 \eqref{eq:Seff-intro}, including relevant and marginal ones. 
 
 Finally, when fixing the logarithmic piece of $a_\sigma(r)$, we were forced to introduce a 
 reference scale $\mu$. This scale is correlated with the choice of scheme in the 
 definition of the renormalized operator $[\epsilon]_{\mu}$. This is a manifestation of 
 the additive mixing between the defect operator $[\epsilon]_{\mu}$ and $h\sigma$. Combined, the $\sigma$ and $[\epsilon]_\mu$ 
 terms make the $r^{7/4}$ coefficient $\mu$-independent, as it should be. However, one does need to be careful to match schemes when measuring the condensate $\langle [\epsilon]_\mu \rangle_{2h}$, e.g.\ in lattice simulations.

The three condensates $\langle \sigma\rangle_{2h},\langle [\epsilon]_\mu\rangle_{2h}$ and $\langle \partial_x^2\sigma\rangle_{2h}$ are non-perturbative IR data. A special fact, which we discuss in section \ref{sec:IsingCondensates}, is that they can be computed exactly thanks to integrability. The result is
\begin{equation}\label{eq:Ising-condensates-intro}
    \langle \sigma\rangle_{2h} = \frac{8}{7}\, C\, (2h)^{1/7},\quad
    \langle [\epsilon]_{\mu}\rangle_{2h} = (2h)^{8/7}\left[-C' + \frac{8}{7}\, C \log \frac{2\mu}{(2h)^{8/7}}\right],\quad
    \langle \partial_x^2\sigma\rangle_{2h} = A\, (2h)^{17/7},
\end{equation}
with
\begin{equation}
    C = \frac{\Gamma\left(\frac{3}{7}\right)\Gamma\left(\frac{7}{8}\right)^{8/7}}{\sqrt{\pi}\,\Gamma\left(\frac{13}{14}\right)}
    \,,\qquad
    A = \frac{64\,\Gamma\left(\frac{2}{7}\right)\Gamma\left(\frac{7}{8}\right)^{24/7}}{5\sqrt{\pi}\,\Gamma\left(\frac{11}{14}\right)}
    \,,\qquad
    C' \approx 2.6040\,,
\end{equation}
where $C'$ is also known in closed form. Inserting them into \eqref{eq:VIsing-intro} provides a completely fixed expression for the asymptotic expansion of the potential $V_{h,h}(r)$ as $r \to 0$, to order $r^{5/2}$, up to the constant $V_0$.

To the best of our knowledge, there exists no exact solution for the potential 
$V_{h,h}(r)$. To test our prediction {\eqref{eq:VIsing-intro}}, we have therefore resorted to numerical simulations.
 We consider the realization of the $2d$ critical Ising model as a transverse-field 
 Ising spin chain, and turn on magnetic impurities as Pauli-$z$ deformations. We compute the ground-state energy of the deformed Hamiltonian 
 with \texttt{DMRG} \cite{White:1992zz,White:1993zza} for varying values of $r$ to reconstruct 
 the potential, and extrapolate to the continuum limit. The comparison is shown in figure~\ref{fig:Ising-intro}. The agreement is manifest.
 
 \texttt{DMRG} is also used to measure the condensates of $\sigma$ and $\epsilon$. We find agreement with \eqref{eq:Ising-condensates-intro}. The same condensates can be extracted by fitting the potential, providing a three-way check.
 
\bigskip

Besides the Ising stress test, we studied a number of examples that can be solved completely
with elementary methods in section \ref{sec:examples}. In these, the defect correlators and the condensates are
available in closed form, so every element of the proposal can be checked explicitly, including
aspects of the framework that the Ising model study did not probe, such as relevant deformations in the bulk and defect curvature effects.

First, we applied the proposal to the fusion of pinning field defects in free massive scalar
theory in $d=4$ (see section \ref{sec:massive_scalar}). Since the theory is Gaussian, the full defect correlator is
solvable using the massive propagator, and so are the defect
condensates. Combined with the fusion coefficients computed in CPT in $m^2$, they
reproduce the $r\sim 0$ expansion of the exact result. One can also see
explicitly how the IR divergences of a naive perturbative expansion are absorbed into the
condensates.

Next, we considered the massless Schwinger model, i.e.\ $2d$ QED with $N_f$
massless fermions (see section \ref{sec:Schwinger}). This theory is interacting, but it is
again solvable, as we review in appendix \ref{app:Schwinger}. Here the relevant deformation is the gauge coupling, making this a simple toy
model for QCD-like theories. Once more, the perturbative predictions, combined with the
explicit condensates, match the $r\sim 0$ expansion of the exact result. 

These analytic and numerical explorations provide substantial evidence in favor of the validity
of the defect OPE \eqref{eq:DOPE-intro} which, we should stress, we have not proven in general.
Accepting its validity, we point out that the OPE has applications beyond potentials between
probe particles. It can be used in more complicated correlators involving parallel defects,
e.g.\ $\langle \mathcal O_1(x_1)\cdots\mathcal O_n(x_n)\mathcal D_a[r]\mathcal D_b[0]\rangle$,
in the limit where the defects fuse. The non-perturbative data are then the correlators in the
presence of the fused defect. It can also be used with finite-length defects wrapping a
nontrivial contour $\gamma$, although when $\gamma$ is not a straight line, the effective action
\eqref{eq:Seff-intro} needs to be modified to include curvature corrections, as discussed in
section \ref{sec:curvature}. In particular, taking $\gamma$ to be a non-contractible loop
wrapping a thermal cycle of length $\beta$ allows us to probe the finite-temperature potential,
defined as the free energy
$V_{ab}^\beta(r) \equiv -\frac{1}{\beta}\log \langle \mathcal D_a[r] \mathcal D_b[0]\rangle_\beta$.
Remarkably, as explained above, the OPE predicts the same short-distance asymptotic structure as
\eqref{eq:Vprediction-intro}, with the $\beta$-dependence entering only through the condensates
$\langle \mathcal D_c \rangle_\beta$, $\langle \mathcal D_c \mathcal O_i\rangle_\beta$, etc.

The rest of the paper is organized as follows. In section~\ref{sec:localOps} we review how to compute the OPE of local operators in CPT and we apply it to the free massive scalar in $d=4$. Section~\ref{sec:Lines} contains our proposal for the OPE of line defects, the prescription to compute fusion coefficients, and a discussion on curvature corrections, non-simple defects, translation invariance, and the bulk-to-defect OPE. In section \ref{sec:examples} we test the defect OPE in solvable examples. We consider massless and massive scalars, Maxwell theory, and the Schwinger model. Then in section~\ref{sec:IsingModel} we graduate to a more nontrivial case: the Ising model. We derive our prediction for the short-distance potential between magnetic impurities, describe the exact computation of the condensates and compare the asymptotic prediction against numerical simulations. We conclude in section~\ref{sec:conclusions}, where we also comment on the implications of the defect OPE for the quark-antiquark potential in pure Yang-Mills. Since CPT computations are rather involved, we relegate technical details to a series of appendices: 
appendix~\ref{app:curvature} for curvature corrections,
appendix~\ref{app:FreeCFTs} for the free massless theories, appendix~\ref{app:MassScalar} for the massive scalar, appendix~\ref{app:Schwinger} for the Schwinger model and appendix~\ref{app:Ising} for the Ising model. Appendix \ref{app:TBA} complements the integrability subsection.

\section{OPE of local operators}\label{sec:localOps}
We start with a review of the OPE of local operators in CPT. The OPE is originally due to Wilson \cite{Wilson:1969zs}, and it has been used extensively in QCD, following \cite{Shifman:1978bx,Shifman:1978by}, where it describes high-energy experiments like the scattering of $e^+e^- \to \text{hadrons}$. In CPT, it was formulated by Al.~B.~Zamolodchikov, who successfully tested his proposal in the massive $2d$ Lee-Yang model \cite{Zamolodchikov:1990bk}, and it has since been applied to many $2d$ models, e.g.\ \cite{Zamolodchikov:1991vx,Delfino:1995zk,Belavin:2003pu}.

In this section, we discuss this framework in preparation for the defect case in section~\ref{sec:Lines}. We describe the general logic in section~\ref{sec:LocalGenLogic}, give the prescription to compute structure functions in section~\ref{sec:prescription_local_ops}, and work out the free massive scalar in $d=4$ as an example in section~\ref{sec:OPE_massive_scalar}. The reader already familiar with the subject might want to skip this section.

\subsection{General logic}\label{sec:LocalGenLogic}
The simplest configuration in which the OPE features is in studying the high-energy limit $|x|\to 0$ of a two-point function of local scalar operators in a QFT in $\mathbb R^d$,
\begin{equation}\label{eq:2pt}
	\langle \mathcal O_1(x) \mathcal O_2(0)\rangle_\text{QFT}\,.
\end{equation}
In a CFT, two-point functions are trivially fixed by symmetry considerations. In a theory with a mass scale $m$, they are nontrivial functions $f(m|x|)$ of a dimensionless ratio encoding information of the spectrum and couplings of the theory. Throughout, we will have in mind a QFT defined by the relevant deformation of a UV fixed point,\footnote{Similar conclusions are expected to apply when $\varPhi$ is only \textit{marginally relevant}, but the vanishing classical mass dimension of the coupling obscures the discussion. We leave a detailed study of the marginally relevant case (in the defect setup) for the future.} as in \eqref{eq:g-def}, where the mass scale is fixed by the relevant coupling as $m\sim g^{1/(d-\Delta_\varPhi)}$. The short-distance limit $m|x|\to 0$ corresponds to the relevant coupling being small, $g|x|^{d-\Delta_\varPhi}\ll 1$, so we expect the two-point function to be described by the vicinity of the UV CFT.

The OPE makes this precise. The basic idea is to approximate the product of colliding operators as an infinite sum of single-operator insertions,\footnote{
In QFT this is in general only an asymptotic expansion as $x\to 0$. In a CFT, the situation is qualitatively better thanks to the state-operator correspondence, which lets one prove its convergence (see e.g.\ \cite{Pappadopulo:2012jk} and references therein). In a theory with a mass scale, the state-operator correspondence is missing and, with it, any general statement about the convergence of the OPE.
We use the symbol $\simeq$ for asymptotic expansions rather than the conventional $\sim$ in the math literature, to reserve the latter for the standard physicist interpretation meaning ``scaling up to proportionality constants''.}
\begin{equation}\label{eq:OPE1}
	\mathcal O_1(x) \mathcal O_2(0) \simeq \sum_i C_{12 i}(x) \mathcal O_i(0) \qquad (\text{as }x\to0)\,,
\end{equation}
where the sum runs over a complete basis of local operators in the QFT. This is done at an operator level, and the expectation value is only taken afterwards, producing
\begin{equation}\label{eq:OPE2}
		\langle \mathcal O_1(x) \mathcal O_2(0)\rangle_\text{QFT} \simeq \sum_i C_{12 i}(x) \langle \mathcal O_i\rangle_\text{QFT}\,.
\end{equation}
The OPE thus achieves a short-distance expansion of the correlator in terms of vacuum expectation values (VEVs), or condensates $\langle \mathcal O_i\rangle_\text{QFT}$, which need not vanish in a general QFT. Let us comment on various aspects of this expansion.

\subsubsection*{Locality of structure functions}

The expansion \eqref{eq:OPE1} is an operator statement.
It holds inside any correlation function $\langle \mathcal O_1(x) \mathcal O_2(0) \cdots \rangle$ with additional insertions placed a typical distance $L\gg |x|$ away from them. This can be motivated from a Wilsonian RG picture, as in the introduction: seen from that distance, the resolution is too poor to distinguish the two operators, and they become a single insertion. By locality, the expansion coefficients $C_{12i}(x)$, known as structure functions, cannot depend on the distant insertions. They are fixed by the dynamics in an arbitrarily small neighborhood of the two points. This is the property that lets us compute them in section~\ref{sec:prescription_local_ops}, and the one we will carry over to line defects.

\subsubsection*{Basis of operators}

When writing any operator expansion like \eqref{eq:OPE1}, it is necessary to fix our basis of operators. In a CFT, we exploit conformal symmetry to arrange them in conformal families, and then we typically pick a basis where primaries are orthonormal, $\langle \mathcal O_i(x)\mathcal O_j(0)\rangle=\delta_{ij}/|x|^{2\Delta_i}$. OPE coefficients are usually reported in this basis. In a massive theory, there is no such canonical choice of basis, and so one needs to be diligent to specify the basis in which OPE coefficients are being computed. Different choices of basis are known as different \textit{schemes}. While the coefficients $C_{12i}(x)$ are scheme dependent, the full combination $\sum_i C_{12i}(x) \mathcal O_i(0)$ is not, making the expansion \eqref{eq:OPE1} meaningful.

Operators in the QFT are in one-to-one correspondence with those in the UV CFT, but they may mix into linear combinations involving the mass parameter. It is convenient to label our operators by their scaling dimension in the UV limit. This can be done by requiring that the leading short-distance divergence in their two-point functions reproduces the CFT result, i.e.
\begin{equation}\label{eq:leading-div}
    \lim_{x\to 0} |x|^{\Delta_i+\Delta_j}\langle \mathcal O_i(x)\mathcal O_j(0)\rangle_\text{QFT} = \delta_{ij}\,.
\end{equation}
This introduces a hierarchy among operators by their scaling dimension: it forbids an operator $\mathcal O_i$ from mixing with operators with $\Delta>\Delta_i$, as they would lead to stronger divergences. The condition \eqref{eq:leading-div} also fixes the normalization of our operators. Unless otherwise noted, we will throughout be working in this scheme, which only leaves open the possibility that operators mix with operators of lower dimension. The operator basis can be fully specified by fixing subleading terms in \eqref{eq:leading-div}, see e.g.\ section~\ref{sec:OPE_massive_scalar}.

\subsubsection*{Perturbative control}

Let us now look back at equation \eqref{eq:OPE2} and note that the OPE achieved a separation between UV and IR data. On the UV side, we have the structure functions $C_{12i}(x)$, which are local; on the IR side, we have the expectation values $\langle \mathcal O_i\rangle_\text{QFT}$ on the long-distance vacuum. The latter, being determined from the end of the RG flow, are naturally non-perturbative with respect to CPT. Indeed, on dimensional grounds, they must scale as $\langle \mathcal O_i\rangle_\text{QFT} \sim g^{\Delta_i/(d-\Delta_\varPhi)}$, which is in general non-analytic in the coupling. The former, on the other hand, being encoded at the beginning of the flow, are expected to have a perturbative expansion in $g$~\cite{Zamolodchikov:1990bk}. Dimensional analysis then fixes their form to
\begin{equation}\label{eq:C12i}
	C_{12i}(x) = \frac{c_{12i}\left(g|x|^{d-\Delta_\varPhi}\right)}{|x|^{\Delta_1+\Delta_2-\Delta_i}}\,, \quad \text{where}\quad c_{12i}\big(t\big) \simeq \sum_n c_{12i}^{(n)} t^n\,,
\end{equation}
up to logs of $|x|$ due to additive operator mixing, as we will see.

In the coming subsection, we will discuss the steps to compute the coefficients $c_{12i}^{(n)}$, order by order in CPT. When plugged into \eqref{eq:OPE2}, they provide a prediction for the asymptotic $|x|$-dependence of the two-point function, only up to the non-perturbative VEVs. This is how the OPE lets us organize non-perturbative corrections to the asymptotic UV limit of correlation functions. We should stress, however, that non-perturbative corrections that are subleading to any power, such as $e^{-1/g|x|^{d-\Delta_\varPhi}}$, are beyond its reach. 

The predictive uses of the OPE are twofold. One can use it as a fitting function in measurements of correlation functions to determine the low-energy VEVs. Or, if the VEVs have already been measured by other means, one can plug them in to produce a prediction for the high-energy behavior of correlation functions. We will use the defect OPE in both ways in section~\ref{sec:IsingModel}.

\subsection{Prescription to extract coefficients}\label{sec:prescription_local_ops}

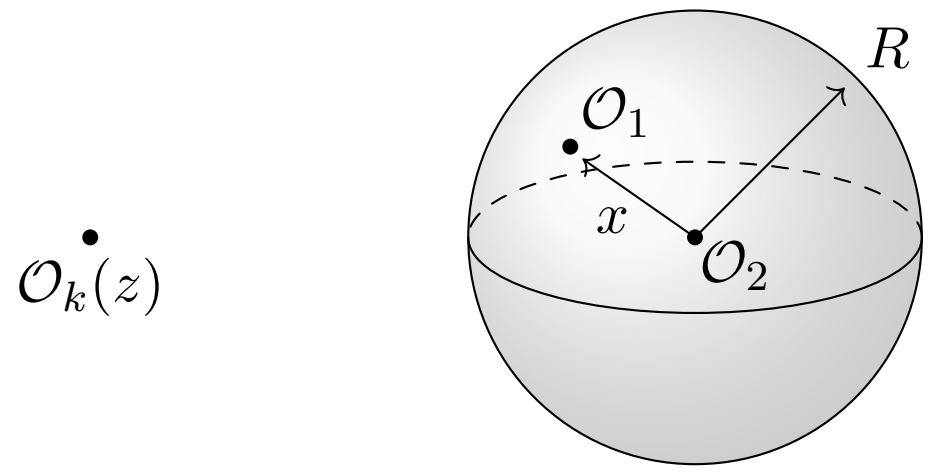
\begin{figure}[h]
    \centering
    \input{Figures/CPTballLocal.tikz}
    \caption{Pictorial representation of $\langle \mathcal O_k(z) \mathcal O_1(x)\mathcal O_2(0)\rangle_\text{CPT}$.}
    \label{fig:CPTballLocal}
\end{figure}

To compute the structure functions $C_{12i}(x)$ appearing in \eqref{eq:OPE1}, all we need to do is compare correlators before and after replacing the OPE. Since the coefficients are assumed to be perturbative, we can do so in CPT. In order to keep CPT reliable, though, we only turn on the relevant deformation from \eqref{eq:g-def} inside a ball of radius $R\gg |x|$, with $R\ll 1/m$, containing the pair of operators but no other insertion, see figure~\ref{fig:CPTballLocal}. By locality, we can use the OPE inside the ball, and the OPE coefficients should remain independent of global details like the radius $R$. We can therefore consider equations of the form
\begin{equation}\label{eq:3=2CPT}
    \langle \mathcal O_k(z) \mathcal O_1(x)\mathcal O_2(0)\rangle_\text{CPT} \simeq \sum_i \frac{c_{12i}\left(g|x|^{d-\Delta_\varPhi}\right)}{|x|^{\Delta_1+\Delta_2-\Delta_i}} \langle \mathcal O_k(z) \mathcal O_i(0)\rangle_\text{CPT}\,,
\end{equation}
where $\mathcal O_k(z)$ is some spectator insertion outside the ball, and the correlators on either side are expansions of the form
\begin{equation}\label{eq:CPTnaive}
	\langle \mathcal O_1(x) \mathcal O_2(0)\rangle_\text{CPT} = \sum_n \frac{(-g)^n}{n!}\int d^dy_1\cdots d^dy_n\,\langle \mathcal O_1(x) \mathcal O_2(0) \varPhi(y_1)\ldots \varPhi(y_n)\rangle_\text{CFT}\,,
\end{equation}
with integrals restricted to $|y_i|<R$.

The first step is then to evaluate two- and three-point functions in CPT. We start by considering, for the insertions $\mathcal O_1,\mathcal O_2,\mathcal O_i,\mathcal O_k$ and the relevant deformation $\varPhi(y)$, operators as defined in the UV CFT. They can be regarded as ``bare'' operators, in perturbation theory parlance. The corresponding CPT integrals will typically have divergences from collisions of the operators $\varPhi(y)$ with $\mathcal O_1,\mathcal O_2,\mathcal O_i$ and among themselves. These UV divergences are resolved with renormalization, which involves regulating divergences (e.g.\ with a UV cutoff~$|y-x|>\delta$) and removing them by adding suitable counterterms to the action. Adding counterterms amounts to redefining operators through linear combinations, and introducing contact terms supported where multiple operators collide. We will walk through some examples below.

For now, we note that power-like divergences $\sim 1/\delta^{\alpha}$ ($\alpha>0$) can be removed without introducing any scale, whereas to remove logarithmic divergences $\sim \log \delta$, one needs to introduce a reference scale $\mu$. This arbitrary scale is a manifestation of the scheme dependence discussed above. Renormalized operators $[\mathcal O(x)]_\text{r}$ and OPE coefficients depend on it, but their combination does not. Thus, a more faithful depiction of the OPE looks like
\begin{equation}\label{eq:OPE3}
	[\mathcal O_1(x)]_\text{r} [\mathcal O_2(0)]_\text{r} \simeq \sum_i \frac{c_{12i}\left(g|x|^{d-\Delta_\varPhi}, \mu\right)}{|x|^{\Delta_1+\Delta_2-\Delta_i}} [\mathcal O_i(0)]_\text{r}\,.
\end{equation}
This is an operator statement between well-defined operators in a QFT. It can be written in any basis of our liking.

Once UV divergences are accounted for, the correlation functions $\langle \mathcal O_k(z) [\mathcal O_i(0)]_\text{r}\rangle_\text{CPT}$ and $\langle \mathcal O_k(z) [\mathcal O_1(x)]_\text{r} [\mathcal O_2(0)]_\text{r}\rangle_\text{CPT}$ become meaningful, and can be evaluated to any order in CPT. Listing the renormalized version of \eqref{eq:3=2CPT} for various operators $\mathcal O_k(z)$ provides a system of equations that can be used to solve for OPE coefficients. Since operators $[\mathcal O_i(0)]_\text{r}$ enter the OPE in a hierarchical way, we can truncate both the OPE and CPT expansions. Indeed, to obtain an asymptotic short-distance expansion up to order $O(|x|^{\alpha})$, it is enough to consider operators of dimension $\Delta_i\leq \alpha + \Delta_1 + \Delta_2$ and CPT expansions to order $n\leq (\alpha + \Delta_1+\Delta_2)/(d-\Delta_\varPhi)$. If we are concerned with applications to two-point functions as in \eqref{eq:OPE2}, we can moreover restrict to the only operators that can acquire a VEV: scalar primaries of the UV CFT that do not transform under unbroken symmetries. In more general correlators, spinning operators and descendants may also contribute. This leaves us with a finite number of unknowns, $c_{12i}$, and enough equations to solve the system.

If the OPE is a meaningful concept with the structure discussed above, the resulting coefficients $c_{12i}$ should be independent of the radius $R$ of the CPT ball. In particular, they should be IR finite, i.e.\ survive the limit of $R\to \infty$. Zamolodchikov showed that this was indeed the case in an example to first order in CPT~\cite{Zamolodchikov:1990bk}. Guida and Magnoli later proved it in general to any order in CPT \cite{Guida:1995kc}. That is, the IR divergences as $R\to\infty$ in the three-point function on the left of \eqref{eq:3=2CPT} are always precisely reproduced by the IR divergences in the two-point functions on the right, with $R$-independent proportionality constants as shown. This cancellation of IR divergences serves as a very nontrivial check of the structure of the OPE and the perturbative nature of its coefficients: they are local UV data, independent of any long-distance structure.

This is far from obvious.
Indeed, a naive attempt to describe the short-distance limit of the two-point function $\langle [\mathcal O_1(x)]_\text{r}[\mathcal O_2(0)]_\text{r}\rangle_\text{QFT}$ by evaluating it
in perturbation theory would yield an expansion of the form \eqref{eq:CPTnaive} with the integrals over all of space. In infinite volume, these integrals are plagued with IR divergences that prevent us from assigning any meaning to it.
After the discussion above,
we can identify the reason behind this issue: these divergences are the result of perturbation theory attempting to reproduce the non-perturbative IR condensates $\langle[\mathcal O_i]_\text{r}\rangle_\text{QFT}$. We would only recover these condensates upon resumming perturbation theory to all orders, as we will do in section~\ref{sec:OPE_massive_scalar}, but this is in general beyond our capabilities. It is therefore much more fruitful to separate perturbative from non-perturbative data using the OPE, and only compute the former in CPT.

\subsection{Example: Free massive scalar}\label{sec:OPE_massive_scalar}
To illustrate the above discussion, in this section we work out explicitly the OPE of $\phi\times\phi$ in four-dimensional free massive scalar theory,
\begin{equation}\label{eq:MassScalar}
    S = \int d^4x \left( \frac{1}{2}\partial_\mu \phi \partial^\mu \phi  + \frac{1}{2}m^2 \phi^2\right)\,.
\end{equation}
This simple-looking theory becomes very instructive when regarded as a deformation of the massless CFT by the relevant operator $\varPhi = \phi^2$ with coupling $g= m^2/2$. It being a free theory, the two-point function is known exactly,
\begin{equation}
    \langle \phi(x) \phi(0)\rangle_{\text{QFT}} = \int\frac{d^4 p}{(2\pi)^4}\frac{e^{ix\cdot p}}{p^2+m^2} = \frac{m}{4\pi^2|x|} K_{1}(m|x|)\,,
\end{equation}
where $K_\nu(z)$ is the modified Bessel function of the second kind.
We are interested in the short-distance limit, $x\to 0$, of this function,
\begin{align}\label{eq:MassScalarUVasym}
    \langle \phi(x) \phi(0)\rangle_\text{QFT} =&\, \frac{1}{4\pi^2 |x|^2} + \left(-1+2\gamma + 2\log\tfrac{m|x|}{2}\right)\frac{m^2}{16\pi^2} \\
    &+ \left(-5+4\gamma + 4\log\tfrac{m|x|}{2}\right)\frac{m^4|x|^2}{256\pi^2} + O\left(|x|^4\right)\,, \nonumber
\end{align}
where $\gamma$ is the Euler–Mascheroni constant.

Intuitively, this UV limit should be described by the vicinity of the massless theory, but, as anticipated, a naive application of conformal perturbation theory as in \eqref{eq:CPTnaive} cannot do the job. That would lead to an expansion in integer powers of the relevant coupling $m^2/2$, and fall short of reproducing the logarithms of $m$. We will confirm that this non-analytic dependence is encoded in the VEVs $\langle \mathcal O_i\rangle$, entering the two-point function through the OPE as in \eqref{eq:OPE2}. Since we aim for an asymptotic expansion up to order $\sim |x|^2$, it is enough to keep operators with $\Delta_i\leq 4$ and work at second order in CPT. The lowest operators that can acquire a VEV are $\mathds 1$, $\phi^2$ and $\phi^4$, so we are aiming for an OPE of the form
\begin{equation}
    \phi(x)\phi(0) = \frac{c_{\phi\phi}^{\mathds 1}(m^2|x|^2)}{|x|^2} + c_{\phi\phi}^{\phi^2}(m^2|x|^2)\phi^2(0) + c_{\phi\phi}^{\phi^4}(m^2|x|^2)|x|^2\phi^4(0) +\cdots\,.
\end{equation}

\subsubsection{Three-point functions in CPT}
The steps to extract the OPE coefficients will be the following: (1) compute three- and two-point functions in CPT, (2) renormalize operators to handle UV divergences, and (3) solve the system of equations of the form \eqref{eq:3=2CPT}. We will verify that the resulting OPE coefficients are independent of the size of the CPT ball. We begin with three-point functions of the form $\lim_{z\to \infty}|z|^{2\Delta_k}\langle \mathcal O_k(z) \phi(x)\phi(0)\rangle_\text{CPT}$, where we place the spectator insertion at infinity to simplify expressions. We have three unknowns, so we consider three equations; for $\mathcal O_k = \mathds 1$, $\phi^2$ and $\phi^4$. For the identity, we are to compute
\begin{equation}\label{eq:phiphiCPT}
    \langle \phi(x)\phi(0)\rangle_\text{CPT} = \sum_n \frac{1}{n!}\left(-\frac{m^2}{2}\right)^n\int_{|y_1|,\ldots,|y_n|<R}\hspace{-50pt} d^4y_1\cdots d^4y_n\,\langle \phi^2(y_1)\cdots \phi^2(y_n) \phi(x)\phi(0) \rangle_\text{CFT}\,.
\end{equation}
This is what we called ``naive CPT'' in \eqref{eq:CPTnaive}, but with integrals now contained inside a finite ball. The correlators on the right-hand side should be evaluated in the massless scalar theory. Since it is a Gaussian theory, they can be computed through Wick contractions using the massless propagator,
\begin{equation}
    \langle \phi(x) \phi(0)\rangle_{\text{CFT}} = \frac{1}{4\pi^2|x|^2}\,.
\end{equation}

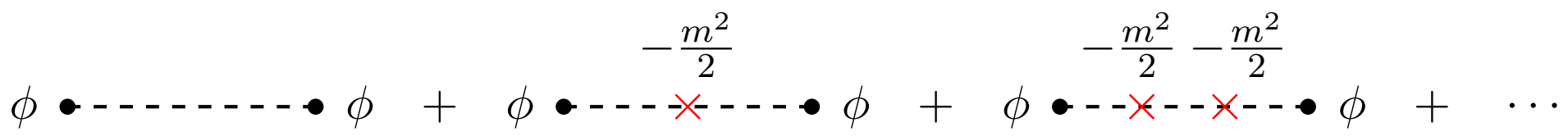
\begin{figure}[t]
    \centering
    \input{Figures/phiphiCPT.tikz}
    \caption{Diagrams contributing to $\langle \phi(x)\phi(0) \rangle_\text{CPT} $ to second order in CPT. Red crosses indicate insertions of the relevant deformation $-\frac{m^2}{2}\int d^4 y\, \phi^2(y)$.}
    \label{fig:phiphiCPT}
\end{figure}

The diagrams contributing to \eqref{eq:phiphiCPT} are shown in figure \ref{fig:phiphiCPT}.\footnote{In this section, we implicitly divide CPT correlators $\langle \cdots\rangle_\text{CPT}$ by the vacuum partition function $\langle 1\rangle_\text{CPT}$. As usual, this amounts to discarding diagrams with isolated vacuum bubbles. As a word of caution, we note that we do not use the same convention throughout all examples. We will discuss it case by case.} They give
\begingroup
\allowdisplaybreaks
\begin{align}\label{eq:3pt1CPT}
    \langle \phi(x)\phi(0)\rangle_\text{CPT} =&\, \frac{1}{4\pi^2|x|^2} -\frac{m^2}{(4\pi^2)^2} \int_{|y|<R}\hspace{-15pt}d^4y\hspace{5pt}\, \frac{1}{|x-y|^2|y|^2} \\
    &+ \frac{m^4}{(4\pi^2)^3}\int_{|y_1|,|y_2|<R}\hspace{-35pt}d^4y_1d^4y_2 \hspace{7pt}\frac{1}{|x-y_2|^2|y_2-y_1|^2|y_1|^2} + O(m^6)\nonumber
    \\
    =&\, \frac{1}{4\pi^2|x|^2} - m^2\frac{1+2\log\frac{R}{|x|}}{16\pi^2} + m^4 \frac{8R^2 - |x|^2\left(5+4\log\frac{R}{|x|}\right)}{256\pi^2} + O(m^6)\,, \nonumber
\end{align}
\endgroup
where the integrals can be easily evaluated e.g.\ by passing to radial coordinates.
This result makes manifest the pitfalls of naive CPT: the correlator is badly divergent in the limit of $R\to \infty$, and there is no trace of the non-analytic $\log m$ from \eqref{eq:MassScalarUVasym}. We will see how the OPE addresses both issues.

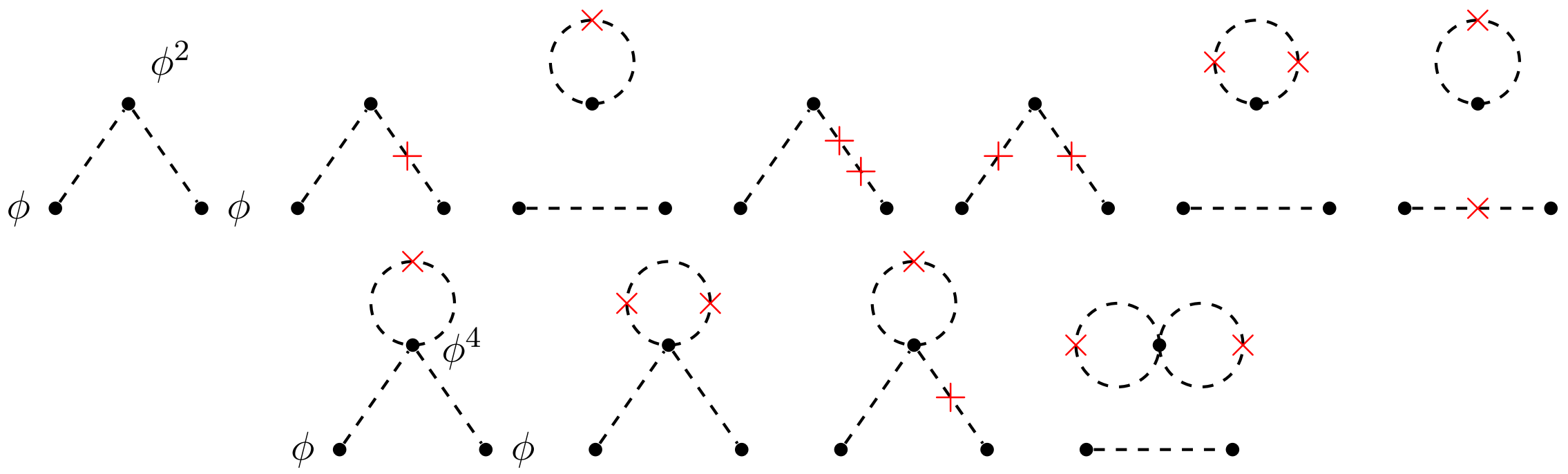
\begin{figure}[h]
    \centering
    \input{Figures/phi2phiphiCPT.tikz}
    \input{Figures/phi4phiphiCPT.tikz}
    \caption{Diagrams contributing to $\langle \phi^2(\infty)\phi(x)\phi(0) \rangle_\text{CPT} $ and $\langle \phi^4(\infty)\phi(x)\phi(0) \rangle_\text{CPT}$ to second order in CPT.}
    \label{fig:phi24phiphiCPT}
\end{figure}

For now, we proceed to evaluate the remaining three-point functions in CPT. The relevant diagrams are shown in figure \ref{fig:phi24phiphiCPT}, and they evaluate to
\begingroup
\allowdisplaybreaks
\begin{align}\label{eq:3pt2CPT}
    \langle \phi^2(\infty)\phi(x)\phi(0)\rangle_\text{CPT} 
    &= \frac{1}{8\pi^4} -m^2\frac{R^4+8R^2|x|^2-2|x|^4}{128\pi^4|x|^2} \\
    &+m^4 \frac{2 R^6 + 3R^4|x|^2\left(11+2\log\frac{R}{|x|}\right) - 12R^2|x|^4 +|x|^6}{1536\pi^4|x|^2} + O(m^6)\,,\nonumber\\
    \langle \phi^4(\infty)\phi(x)\phi(0)\rangle_\text{CPT} 
    &= -m^2\frac{3R^4}{128\pi^6} +m^4\frac{3R^8 + 64R^6|x|^2 - 12R^4|x|^4}{4096\pi^6|x|^2} + O(m^6)\,.
\end{align}
\endgroup
We note that pushing the spectator insertion to $z\to \infty$ (after multiplying by $|z|^{2\Delta_k}$) effectively freezes the propagators attached to it, simplifying computations significantly. We also note that by the CFT operators $\phi^2$, $\phi^4$, we mean the normal-ordered operators $:\phi^2:$, $:\phi^4:$, so that self-contractions with the massless propagator vanish. These results are not yet very illuminating, just another set of awfully IR-divergent expressions.

\subsubsection{Two-point functions in CPT}
Having computed the perturbative three-point functions, we now move on to the two-point functions $\langle \mathcal O_k(\infty)\mathcal O_i(0)\rangle_\text{CPT}$. Start with $\mathcal O_k = \mathds 1$ and $\mathcal O_i = \phi^2$. We find
\begin{align}\label{eq:phi2(0)CPT}
    \langle \phi^2(0)\rangle_\text{CPT} &= -\frac{m^2}{(4\pi^2)^2} \int_{\delta<|y|<R}\hspace{-22pt}d^4y\hspace{5pt}\, \frac{1}{|y|^4} + \frac{m^4}{(4\pi^2)^3} \int_{|y_1|,|y_2|<R}\hspace{-35pt}d^4y_1 d^4y_2\, \frac{1}{|y_1|^2|y_2|^2|y_1-y_2|^2} + O(m^6)\nonumber\\
    &= -\frac{m^2}{8\pi^2}\log\frac{R}{\delta} + m^4 \frac{R^2}{32\pi^2} + O(m^6)\,.
\end{align}
Here, in the first integral, we encounter a new feature; a UV divergence from the collision of the relevant operator with the insertion $\phi^2(0)$. This indicates that our $\phi^2$ is not a well-defined operator in the massive QFT. This can be fixed with renormalization. Regulate the divergence with a UV cutoff $|y|>\delta$ (as we did in the integral above), and remove it by defining the renormalized operator
\begin{equation}\label{eq:phi2Ren}
    [\phi^2(0)]_\text{r} \equiv \phi^2(0) - \frac{m^2}{8\pi^2}\log (\mu\delta)\mathds 1\,.
\end{equation}
Since this particular UV divergence is logarithmic, we are forced to choose a reference scale~$\mu$. This scale is arbitrary; different choices of $\mu$ correspond to different \textit{schemes}, which simply amount to different bases in the space $\{\mathds 1, [\phi^2]_\text{r}\}$, e.g.\
\begin{equation}
    [\phi^2(0)]_\text{r}' = [\phi^2(0)]_\text{r} - \frac{m^2}{8\pi^2}\log \frac{\mu'}{\mu}\mathds 1\,.
\end{equation}

We find a similar issue with $\phi^4$,
\begin{align}
    \langle \phi^4(0)\rangle_\text{CPT} &= \frac{3m^4}{(4\pi^2)^4} \int_{\delta<|y_1|,|y_2|<R}\hspace{-45pt}d^4y_1 d^4y_2\hspace{3pt}\, \frac{1}{|y_1|^4|y_2|^4} + O(m^6) = m^4 \frac{3}{64\pi^4}\log^2 \frac{R}{\delta} + O(m^6)\,.
\end{align}
This divergence is removed by considering the renormalized operator
\begin{equation}
    [\phi^4(0)]_\text{r} \equiv \phi^4(0) - \frac{3m^2}{4\pi^2} \log(\mu \delta) [\phi^2(0)]_\text{r} - m^4 \frac{3}{64\pi^4}\log^2(\mu \delta)\mathds 1\,.
\end{equation}
Again, changing $\mu$ amounts to a change of basis in the space of well-defined operators. With these new proper notions of operators in the QFT, the overlaps with the identity are finite,
\begin{align}
    \langle [\phi^2(0)]_\text{r}\rangle_\text{CPT} &= -\frac{m^2}{8\pi^2}\log(R\mu) + m^4 \frac{R^2}{32\pi^2} + O(m^6)\,, \\
    \langle [\phi^4(0)]_\text{r}\rangle_\text{CPT} &= m^4 \frac{3}{64\pi^4}\log^2(R\mu) + O(m^6)\,.
\end{align}

Continuing with the two-point functions $\langle \mathcal O_k(\infty)\mathcal O_i(0)\rangle_\text{CPT}$, consider now $\mathcal O_k = \phi^2,\phi^4$ and $\mathcal O_i =\mathds 1$. In this case, we find
\begin{align}
    \langle \phi^2(\infty)\rangle_\text{CPT} &= -m^2\frac{R^4}{32\pi^2} + m^4 \frac{R^6}{192\pi^2} + O(m^6)\,, \\
    \langle \phi^4(\infty)\rangle_\text{CPT} &= m^4 \frac{3R^8}{1024\pi^4} + O(m^6)\,.
\end{align}
It may seem puzzling at first that this result differs from the one above, where the insertions were placed at the origin, but one should remember that our prescription for the CPT computations is to restrict the relevant deformation only within a ball of radius $R$ around the origin. This IR regulator breaks translation invariance, and so one-point functions at different positions differ. Note also that, outside the ball, the theory remains at the UV fixed point, so we do not need renormalization for the spectator operators.

We continue by evaluating the remaining two-point functions. We obtain
\begingroup
\allowdisplaybreaks
\begin{align}
    \langle \phi^2(\infty) [\phi^2(0)]_\text{r}\rangle_\text{CPT} &= \frac{1}{8\pi^4} - m^2\frac{R^2}{16\pi^4} +m^4\frac{R^4(5 + \log R\mu)}{256\pi^4} + O(m^6)\,,\\
    \langle \phi^4(\infty) [\phi^2(0)]_\text{r}\rangle_\text{CPT} &= - m^2\frac{3R^4}{128\pi^6} +m^4\frac{R^6}{64\pi^6} + O(m^6)\,,\\
    \langle \phi^2(\infty) [\phi^4(0)]_\text{r}\rangle_\text{CPT} &= - m^2\frac{3}{32\pi^6}\log R\mu +m^4\frac{3R^2(1+2\log R\mu)}{128\pi^6} + O(m^6)\,,\\
    \langle \phi^4(\infty) [\phi^4(0)]_\text{r}\rangle_\text{CPT} &= \frac{3}{32\pi^8} - m^2\frac{3R^2}{32\pi^8} +m^4\frac{9R^4(3+\log R\mu)}{512\pi^8} + O(m^6)\,.
\end{align}
\endgroup
Again, when using ``bare operators'', several diagrams diverge due to the collisions of relevant operators with the insertion at the origin. Using the renormalized operators as defined above cancels the UV divergences, giving finite perturbative two-point functions.

\subsubsection{OPE coefficients}
To extract the OPE coefficients, or structure functions, now we just need to build the system of equations of the form \eqref{eq:3=2CPT}, plug the perturbative results on either side, and solve it order by order in $m^2$. We obtain the following result:
\begin{subequations}
\begin{align}
    c^{\mathds 1}_{\phi\phi}(m^2|x|^2,\mu) &= \frac{1}{4\pi^2} - \frac{\left(1-2\log\mu|x|\right)}{16\pi^2}m^2|x|^2 - \frac{\left(5-4\log \mu|x|\right)}{256\pi^2}m^4|x|^4 + O(m^6)\,,\\
    c^{\phi^2}_{\phi\phi}(m^2|x|^2,\mu) &= 1 + \frac{1}{8}m^2|x|^2 + \frac{1}{192}m^4|x|^4 + O(m^6)\,,\\
    c^{\phi^4}_{\phi\phi}(m^2|x|^2,\mu) &= 0 + O(m^6)\,.
\end{align}
\end{subequations}
Note that all $R$-dependence has disappeared. We performed the OPE inside a finite ball containing the QFT, and found that its structure is independent of the global details of the ball. In particular, the coefficients clearly survive the $R\to \infty$ limit, despite the IR divergences that plagued the intermediate CPT computations. The OPE rearranged the IR divergences of the three-point functions $\langle \mathcal O_k(\infty) \phi(x)\phi(0)\rangle_\text{CPT}$ into the IR divergences of the two-point functions $\langle \mathcal O_k(\infty)[\mathcal O_i(0)]_\text{r}\rangle_\text{CPT}$, with finite coefficients. This serves as a rather nontrivial check of the expectations presented in section~\ref{sec:LocalGenLogic}: the OPE is a local operator statement with coefficients computable in CPT.

Now that we have established its $R$-independence, we can go on and use the OPE in the full QFT, where the relevant deformation \eqref{eq:g-def} is turned on in all of spacetime. It gives the following prediction for the short-distance limit of the two-point function,
\begin{equation}\label{eq:MassScalarPrediction}
    \boxed{
    \begin{aligned}
    \langle \phi(x) \phi(0)\rangle_\text{QFT} =&\, \frac{1}{4\pi^2|x|^2} +\left(\frac{m^2\left(-1+2\log\mu|x|\right)}{16\pi^2} + \langle [\phi^2]_\text{r}\rangle_\text{QFT}\right) \\
    &+\left(\frac{m^2\left(-5+4\log \mu|x|\right)}{32\pi^2}+ \langle [\phi^2]_\text{r}\rangle_\text{QFT}\right)\frac{m^2|x|^2}{8} + O(|x|^4)\,.
    \end{aligned}}
\end{equation}
We see that the OPE correctly predicts the functional $x$-dependence of the exact short-distance expansion \eqref{eq:MassScalarUVasym}! Moreover, the coefficients are written in terms of vacuum expectation values of the operators entering the OPE. While the VEV of the identity is normalized to $1$, that of $ [\phi^2]_\text{r}$ remains undetermined, as it lies beyond the reach of CPT.
By writing unknown coefficients in terms of VEVs, the OPE grants us \textit{universality}; the same quantities can be measured in entirely different experiments.
One just needs to be careful to match renormalization schemes.

\subsubsection{Condensates}
In a genuinely non-perturbative QFT, this is as far as we would be able to push the OPE. The condensates would have to be determined with independent non-perturbative methods. In this simple free theory example, however, we can go further and evaluate $\langle [\phi^2]_\text{r}\rangle_\text{QFT}$ explicitly to verify that \eqref{eq:MassScalarPrediction} reproduces the UV expansion \eqref{eq:MassScalarUVasym} down to coefficients.
This can be done by resumming the CPT contributions to all orders. Starting with the bare operator, we write
\begin{equation}\label{eq:MassScalarPhi2QFT}
    \langle \phi^2(0)\rangle_\text{QFT} = \frac{1}{4\pi^2}\sum_{n=1}^\infty \left(-\frac{m^2}{4\pi^2}\right)^n \int d^4y_1\cdots d^4y_n\,\frac{1}{|0-y_1|^2}\frac{1}{|y_1-y_2|^2}\cdots\frac{1}{|y_n-0|^2}\,.
\end{equation}
To evaluate these convolution integrals, it is convenient to pass to momentum space, where they factorize. The massless propagator becomes $\int d^4x\, e^{-i p\cdot x}(4\pi^2|x|^2)^{-1} = 1/p^2$, and we can resum the geometric series to
\begin{equation}
    \langle \phi^2(0)\rangle_\text{QFT} = \sum_{n=1}^\infty \left(-m^2\right)^n \int \frac{d^4p}{(2\pi)^4} \left(\frac{1}{p^2}\right)^{n+1} = \int \frac{d^4p}{(2\pi)^4}\left(\frac{1}{p^2+m^2} - \frac{1}{p^2}\right) \,.
\end{equation}
This is simply a loop of the massive propagator, as expected.\footnote{The shift by $-1/p^2$ is due to our definition of the CFT operator as the normal-ordered product $:\phi^2:$.} Thanks to the finite mass $m$, the resummed integral is no longer IR divergent. It is however UV divergent. We can regulate it with a hard-momentum cutoff $|p|<\Lambda$ to obtain
\begin{equation}
    \langle \phi^2(0)\rangle_\text{QFT} = \frac{m^2}{8\pi^2}\log\frac{m}{\Lambda} + O(\Lambda^{-2})\,.
\end{equation}

The next step is to cancel divergences with renormalization as in \eqref{eq:phi2Ren}. However, since we are using a UV regulator different from $\delta$, we need to match schemes before we can use the resulting VEV in \eqref{eq:MassScalarPrediction}. One way to do so is by recomputing the leading-order CPT one-point function \eqref{eq:phi2(0)CPT} with a hard-momentum UV cutoff and comparing results. We must keep the IR cutoff the same, so we use the regulated momentum-space propagator
\begin{equation}
    G_R(p) \equiv \int_{|x|<R}\hspace{-18pt}d^{4}x \,  \frac{e^{-ip\cdot x}}{4\pi^2 |x|^2} = \frac{1-J_0(|p| R)}{p^2}\,,
\end{equation}
where $J_\nu(z)$ is a Bessel function of the first kind. The first order CPT result is then
\begin{align}
    \langle \phi^2(0)\rangle_\text{CPT} &= -m^2 \int_{|p|<\Lambda}\hspace{-18pt}d^4 p\hspace{3pt} \frac{G_R(p)^2}{(2\pi)^4} +O(m^4) = -\frac{m^2}{8\pi^2}\left(\gamma + \log \frac{R\Lambda}{2}\right) + O(m^4)\,,
\end{align}
where we only kept terms up to order $O(\Lambda^{-1})$. Comparing results with \eqref{eq:phi2(0)CPT}, we conclude that $\Lambda = 2e^{-\gamma}/\delta $. Using then the renormalized operator as defined in \eqref{eq:phi2Ren}, we obtain
\begin{equation}
    \langle [\phi^2]_\text{r}\rangle_\text{QFT} = \frac{m^2}{8\pi^2}\left(\gamma+\log\frac{m}{2\mu}\right)\,,
\end{equation}
which, when inserted into \eqref{eq:MassScalarPrediction}, cancels the scheme dependence and reproduces the original expansion \eqref{eq:MassScalarUVasym} exactly.

\section{OPE of line defects}\label{sec:Lines}

Having reviewed the OPE of local operators, we now graduate to the OPE (or fusion) of extended operators. 
Such operators, or defects, have received a lot of attention in recent years.
See \cite{Cuomo:2021rkm,Cuomo:2022xgw,Giombi:2022vnz,Casini:2022bsu,Giombi:2023dqs,Cuomo:2023qvp,Wu:2023ezm,Hu:2023ghk,Aharony:2023amq,Gabai:2023lax,Cai:2024ppo,Choi:2024ktc,Krishnan:2024wrc,Toldin:2024pqi,Diatlyk:2024ngd,Kristjansen:2024map,Gabai:2025zcs,Barkeshli:2025cjs,Kim:2025tvu,Chalabi:2025nbg,Ambrosino:2025myh,Meineri:2025qwz,Gomis:2025gzb,Komargodski:2025jbu,Kravchuk:2025evf,Ambrosino:2025pjj,Giombi:2025evu,Drukker:2025dfm,qin2026revealingaltermagneticfermisurfaces,Komargodski:2026vwn,Abdalla:2026wdx,Giombi:2026kdz,Sarma:2026lte,Diatlyk:2026gab,Zhu:2026cls,Bianchi:2026sax,Chernikov:2026lcv,Ambrosino:2026ovo,Cuomo:2026mop,Kristjansen:2026qlx,Chandra:2026wnf,Bianchi:2026dhj,Lanzetta:2026pvr,Ambrosino:2026zsi,Feng:2026rrk,Barrat:2026ozj,Cuomo:2026kuo,Barrat:2026pip}
for a necessarily incomplete list of recent studies.
For simplicity, here we restrict our attention to \textit{line defects}, although we expect our methods will generalize to higher-dimensional defects. We use the terms line operator and line defect interchangeably, but we always have in mind a relativistic Euclidean line object which can extend along space, where it acts as a fixed-time operator on the Hilbert space, or along time, where it is seen as a point impurity or defect on the time slices leading to a modified Hilbert space. Throughout, we also assume that our defects are compatible with locality, in the sense that path integrals cut across the defects factorize according to the usual rules.

\subsection{General logic}
Consider a correlation function of a QFT in $\mathbb R^d$ ($d\geq 2$) with two parallel line defects stretching along a contour $\gamma$ separated by a distance $r$,\footnote{In this paper, we normalize correlators so that the partition function with no insertions (defects or local operators) is one. Unless otherwise specified, we do not divide $\langle...\rangle$ by any defect correlator. We use square brackets $\mathcal D[...]$ to specify the support and/or location of the defects, as should be clear from context. Here $\gamma$ is the contour that the defects span and $r,0$ the positions along the transverse direction.}
\begin{equation}\label{eq:2line}
    \langle \mathcal D_a[\gamma,r]\mathcal D_b[\gamma,0]\cdots\rangle_\text{QFT}\,,
\end{equation}
where the ellipsis represents any other insertions away from the defects. We are interested in the asymptotic behavior of this correlation function as $r\to 0$, while we keep the intrinsic scale of the theory, $m$, the length of the contours, $L$, and any other scale pertinent to the external insertions fixed. We will argue that, much like with local operators, this limit is controlled by an OPE with coefficients that are under perturbative control.

\subsubsection*{Conformal defects}
Before discussing the general case, let us focus on a massless theory and review the OPE of \textit{conformal} line defects. This featured in early works in $d=2$ \cite{Bachas:2007td,Bachas:2013ora,Konechny:2015qla}, but it has only been fleshed out recently for higher dimensions \cite{Diatlyk:2024zkk,Diatlyk:2024qpr,Kravchuk:2024qoh,Cuomo:2024psk}. When the theory has no intrinsic mass scale, the correlator \eqref{eq:2line} depends only on $r/L$ and ratios of $r$ with the separation of other insertions (which we will henceforth leave implicit). Therefore, the short-distance limit that we are interested in is equivalent to the IR limit $L\to \infty$, with $r$ fixed, where we look at the defects from far away. This can be described as a defect RG flow starting in the UV with the separated defects $\mathcal D_a,\mathcal D_b$, and flowing down to a single conformal defect $\mathcal D_c$ of the same shape $\gamma$ in the IR.

The tail of this RG flow is described by a low-energy effective action on the worldline of the IR defect, so we write
\begin{equation}\label{eq:dOPE}
    \mathcal D_a[\gamma,r]\mathcal D_b[\gamma,0] \simeq \mathcal D_c[\gamma, 0] e^{-S_\text{eff}[\gamma]} \qquad (\text{as } r\to 0)\,,
\end{equation}
where $S_\text{eff}$ consists of an expansion in irrelevant operators, suppressed by powers of the EFT cutoff $\Lambda\sim 1/r$,\footnote{
For brevity, in this discussion we leave curvature corrections implicit. We address these terms, which are necessary when $\gamma$ is not a straight line, in section~\ref{sec:curvature}.}
\begin{equation}\label{eq:Seff}
    S_\text{eff}[\gamma] = \int_\gamma ds\, \left(-\frac{a_0}{r} + \sum_{i\in\text{irr}}a_i r^{\Delta_i -1} \mathcal O_i(s)\right)\,.
\end{equation}
The cosmological constant of this action, $-a_0/r$, is the contribution from the identity operator $\mathds 1$, and it describes a Casimir energy for the fusion of the probes $\mathcal D_a,\mathcal D_b$ into $\mathcal D_c$. The other $a_i$ are further (dimensionless) Wilson coefficients for the irrelevant operators ($\Delta_i>1$) living on the defect $\mathcal D_c$. These constants $a_i$, dubbed fusion coefficients, are the direct analogues of the structure constants in the OPE of local operators. Note that we do not include relevant operators in \eqref{eq:Seff} because we assume that $\mathcal D_c$ is the endpoint of the flow. Such deformations would trigger subsequent flows, which one should follow all the way to the true endpoint. Similarly, (exactly) marginal operators are not included; they are viewed as part of the definition of the conformal defect~$\mathcal D_c$ (i.e.\ specifying a particular point on a defect conformal manifold, if there is one). 

The form of the defect OPE \eqref{eq:dOPE} is very constrained by locality. Naively, one might have attempted to write down a sum of conformal defects with real $r$-dependent coefficients, $\sum_k C_{abk}(r)\mathcal D_k$, mimicking the OPE of local operators. But such an expansion would spoil the factorization properties of the path integral. The exponential of a local action is essential for partition functions to multiply under cuts of the path integral across the defect.  The OPE of lines \textit{can} involve a sum over conformal defects, but it must be a direct sum ${\mathcal D_c = \bigoplus_k\mathcal D_k}$. In such cases, the defect $\mathcal D_c$ is called non-simple. For now, we assume that $\mathcal D_c$ is simple, and will return to this case in section~\ref{sec:non-simple}.

The $r$-dependence in (\ref{eq:Seff}) enters only through the scale-suppression of the Wilson coefficients in \eqref{eq:Seff}. In retrospect, this structure is actually not that exotic; the OPE of local operators $\mathcal O_1(x)\mathcal O_2(0)$ in CFT can be regarded as an effective action for the fusion of zero-dimensional defects, with terms suppressed by powers of the cutoff $\Lambda \sim 1/|x|$. See~\cite{Kravchuk:2024qoh} for a more detailed comparison of the two OPEs, taking into account constraints from Weyl invariance.

\subsubsection*{Adding a mass scale}
We now turn on a mass scale. Compared with the local operator case, here there is a richer space of massive flows that we can consider. First, we can have a UV CFT, with $\mathcal D_a,\mathcal D_b$ corresponding conformal defects, perturbed by a bulk relevant deformation,
\begin{equation}\label{eq:g-defSec3}
    S = S_\text{CFT} + g\int d^d x\, \varPhi \qquad (\Delta_\varPhi <d)\,.
\end{equation}
In this case the intrinsic mass scale is $m\sim g^{1/(d-\Delta_\varPhi)}$. Second, we can keep a CFT in the bulk, and only turn on a relevant deformation on the defects $\mathcal D_a,\mathcal D_b$, starting from conformal (or topological) defects,
\begin{equation}\label{eq:h-defSec3}
    \mathcal D[\gamma] = \mathcal D_{\text{CFT}}[\gamma]\, e^{h\int_\gamma ds\, \varphi} \qquad (\Delta_\varphi <1)\,.
\end{equation}
The most notable examples of this kind are (generalized) \textit{pinning field defects}, which are deformations on top of the trivial line \cite{ParisenToldin:2016zxc,Cuomo:2021kfm,Popov:2025cha}. Notice that, in contrast with most CFT treatments, we do not flow the defects to the IR; we retain the intrinsic mass scale $m\sim h^{1/(1-\Delta_\varphi)}$. We can also have both deformations $g$ and $h$ at once. 

In the massive case, we still expect a defect OPE of the form \eqref{eq:dOPE} to survive. The logic is again Wilsonian. We introduce a cutoff $\Lambda <1/r$ and integrate out high-energy modes so that the resolution becomes too poor to distinguish the defects, and they fuse into a single one. The situation is very analogous to the one described above for CFTs, but the $r\to 0$ limit is no longer a true IR limit because the length of the defects (and distance to external insertions) remains finite in units of $m$. In other words, in the Hilbert space on the time slice pierced by the two defects, there are states whose energy scales with $\sim 1/r$ and others with $\sim m$. The Wilsonian step only removes the former. So the defect retains an energy scale after the fusion, and it may continue to flow down.

One should view the flow in two steps, as sketched in blue in figure \ref{fig:RG} from the introduction, which we probe at different energy scales $E$ set by the length of the defects and distance to other insertions. For energies $E\sim 1/r$, we are close to the blue cross, and we need the full configuration of the two defects. For energies $m\ll E\ll 1/r$, we can fuse the defects and use an effective description around the UV CFT in the bulk, with a single conformal defect $\mathcal D_c^\text{CFT}$ in it. The limit $mr\ll 1$, which we are assuming throughout, ensures that the relevant deformations triggered by $g,h$ remain small, and so the blue curve passes nearby this fixed point. For energies $E\ll m$, the bulk and/or fused defect theories flow down to the true IR limit, where we could use a new effective description in terms of IR degrees of freedom. We will be concerned with the first step of the flow; i.e.\ the effective description around $\mathcal D_c^\text{CFT}$.

The above discussion leads us to propose that the defect OPE in QFT still reads
\begin{equation}\label{eq:dOPE2}
    \mathcal D_a[\gamma,r]\mathcal D_b[\gamma,0] \simeq \mathcal D_c^\text{CFT}[\gamma, 0] e^{-S_\text{eff}[\gamma]} \qquad (\text{as } r\to 0)\,,
\end{equation}
where $\mathcal D_c^\text{CFT}[\gamma, 0]$ is a conformal (or topological) defect of the UV CFT, i.e.\ before the relevant deformation of scale $m$ is turned on. Due to the presence of a scale, though, the effective action now takes the form
\begin{equation}\label{eq:Seff2}
    S_\text{eff}[\gamma] = \int_\gamma ds\, \left(-\frac{a_0(mr)}{r} + \sum_{\hspace{-5pt}i\in\text{rel}, \text{mar}\hspace{-5pt}}\frac{a_i(mr)}{r^{1-\Delta_i}} \mathcal O_i(s) + \sum_{i\in\text{irr}}a_i(mr) r^{\Delta_i -1} \mathcal O_i(s)\right)\,.
\end{equation}
That is, the fusion coefficients $a_i$ get promoted from constants to functions of $mr$; the only dimensionless parameter locally available on the defect.\footnote{Notice that locality forbids these coefficients from depending on $L$ or any other global information of the curve $\gamma$. The same effective action should be valid when fusing only a small portion of the defects.} We also allow now for relevant operators on the defect ($\Delta_i<1$), which can trigger the second step of the flow in figure~\ref{fig:RG}, and exactly marginal operators ($\Delta_i=1$), which capture how $g,h$ move $\mathcal D_c^\text{CFT}$ around defect conformal manifolds. See sections~\ref{sec:Schwinger} and \ref{sec:IsingModel} for examples where this occurs.

Paralleling the OPE of local operators, the fusion of defects happens locally, i.e.\ in the UV. It is therefore naturally perturbative,
and so we expect the fusion coefficients $a_i(mr)$ to admit power-series expansions in the bulk and/or defect relevant couplings $g,h$,
\begin{equation}\label{eq:ai-exp}
    a_i(mr) \simeq \sum_{n,\ell} a_i^{(n,\ell)}\left(g\, r^{d-\Delta_\varPhi}\right)^n \left(h\, r^{1-\Delta_\varphi}\right)^\ell\,,
\end{equation}
accessible in conformal perturbation theory. The first coefficient, $a_i^{(0,0)}$, is the result from fusion in the CFT. It should vanish for relevant defect operators, as they are only turned on by the massive deformations. It should also vanish for exactly marginal operators, since our convention is to absorb marginal couplings into the definition of $\mathcal D_c^\text{CFT}$ in the CFT limit. Higher coefficients follow from higher orders in bulk and/or defect CPT. We will give a practical prescription to extract these coefficients in the following subsection.

Meanwhile, long-distance data entering in the second step of the flow (e.g.\ down to the energy scale of long defects, or distant external insertions) is non-perturbative. In this way, the defect OPE separates UV and IR data and provides a prediction for the asymptotic functional form of a correlator in the limit where two parallel defects fuse. As an example, for the correlation function of two parallel line defects, it predicts\footnote{This expression is somewhat schematic. $\mathcal D_c$ represents the conformal defect $\mathcal D_c^\text{CFT}$ dressed with the relevant deformations in the effective action \eqref{eq:Seff2} entering with non-positive powers of $r$ (after expanding fusion coefficients as in \eqref{eq:ai-exp}). Terms with positive powers of $r$, from irrelevant operators or from relevant/marginal operators at subleading CPT orders, are brought down from the exponential as shown.}
\begin{align}\label{eq:Prediction}
    \langle \mathcal D_a[\gamma,r]\mathcal D_b[\gamma,0]\rangle_\text{QFT} \simeq&\, e^{\frac{a_0}{r}\int_\gamma ds}\Bigg(\langle \mathcal D_c[\gamma]\rangle_\text{QFT}- \sum_i a_ir^{\Delta_i -1}\int_\gamma \hspace{-3pt}ds\langle \mathcal D_c[\gamma] \mathcal O_i(s)\rangle_\text{QFT}\\
    & \hspace{-3pt}+ \frac{1}{2}\sum_{i,j} a_ia_jr^{\Delta_i + \Delta_j - 2} \int_\gamma\hspace{-3pt} ds_1 \int_\gamma\hspace{-3pt} ds_2\,\langle \mathcal D_c[\gamma] \mathcal O_i(s_1)\mathcal O_j(s_2)\rangle_\text{QFT}+\cdots\Bigg)\nonumber.
\end{align}
While the $r$-dependent fusion coefficients $a_i(mr)$ are fixed order by order in CPT, integrated $n$-point functions on the fused defect are non-perturbative condensates that need to be determined by other means. They generalize the VEVs that entered the OPE of local operators, cf.\ \eqref{eq:OPE2}. Note that the dependence on the geometry $\gamma$ of the defects, and in particular their length, enters only through these condensates (and explicit integrals).

Like with local operators in massive QFTs, in general we expect the expansion \eqref{eq:Prediction} to be only asymptotic. In fact, in the case of defects, it is not even known whether the OPE converges already in CFT \cite{Kravchuk:2024qoh}, as there is no state-operator argument for it. Nevertheless, it remains a very useful tool to study short-distance limits of correlators. It is worth re-emphasizing that the existence of a defect OPE, as well as the perturbative nature of its fusion coefficients, is conjectural and has not yet been proven. In sections \ref{sec:examples} and \ref{sec:IsingModel}, we verify the prediction \eqref{eq:Prediction} in a variety of examples of increasing complexity. They all pass the test, providing very nontrivial support for its validity.

\subsection{Prescription to extract coefficients}\label{sec:prescription_lines}
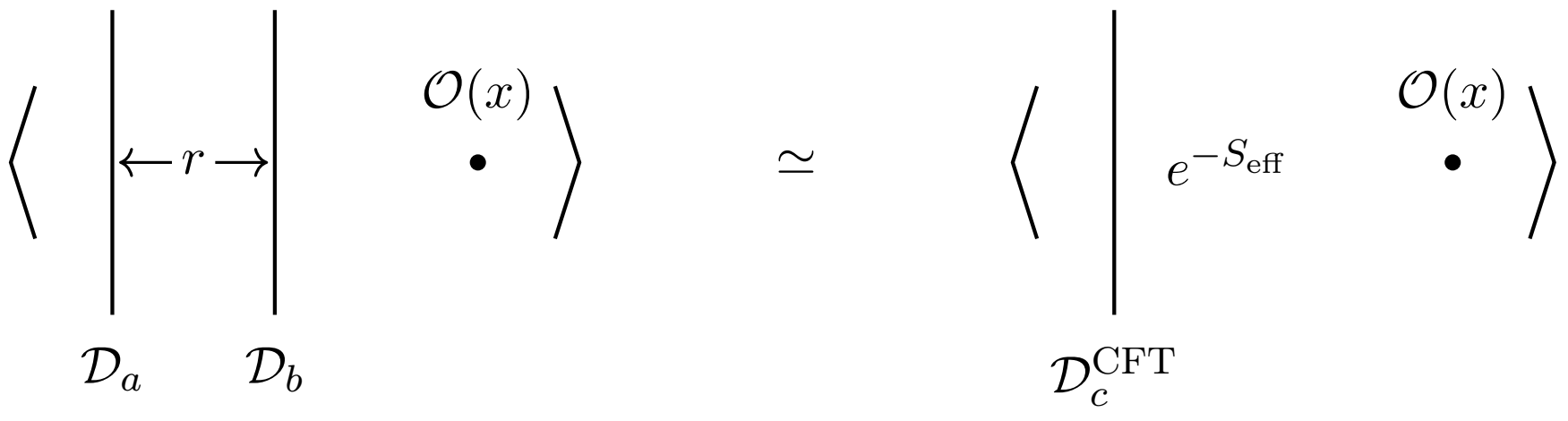
\begin{figure}[t]
    \centering
    \input{Figures/Defect_Prescription.tikz}
    \caption{Our prescription to extract the coefficients in the effective action $S_{\text{eff}}$. We equate correlation functions with an operator $\mathcal O(x) $ inserted at $|x| \gg r$ before and after the fusion, order by order in $r$ using CPT.}
    \label{fig:Defect_prescription}
\end{figure}
The prescription to extract the fusion coefficients $a_i(mr)$ is straightforward, following a similar logic to section~\ref{sec:prescription_local_ops} for local operators.
We simply need to compare correlators with and without our defect OPE replacement, as in figure~\ref{fig:Defect_prescription}, in a configuration where both sides are computable in CPT.
We can achieve this by turning on the relevant deformations \eqref{eq:g-defSec3} and \eqref{eq:h-defSec3} in a finite region of size $R$ (with $r\ll R\ll 1/m$) intersecting the defects, as we sketched in figure \ref{fig:CPTball}.
By locality, the fusion coefficients should be independent of global details like $R$ or the length of the defects, so the same coefficients will be valid when fusing the defects in the full QFT, where the relevant deformations are turned on everywhere.

We consider correlators with an operator $\mathcal O_k(x)$ inserted at a position $|x| \gg r$ away from the defects, which gives us the matching\footnote{Throughout the paper, we adopt the notation that operators on the left of the defect in correlation functions, i.e.\ $\langle \mathcal O(x) \mathcal D\rangle $, correspond to bulk insertions away from the defect, while operators to the right of the defect, i.e.\ $\langle \mathcal D \mathcal O(s)\rangle$, are localized to the defect.}
{
\allowdisplaybreaks
\begin{align}\nonumber
    \langle \mathcal O_k(x)&\mathcal D_a[\gamma,r] \mathcal D_b[\gamma,0] \rangle_\text{CPT} \\\nonumber \simeq & \ e^{\frac{a_0(mr)}{r}\int_\gamma ds}\bigg[\langle \mathcal O_k(x) \mathcal D_c[\gamma,0]\rangle_\text{CPT}-\sum_{i} a_i(mr) r^{\Delta_i-1}\int_\gamma ds\langle \mathcal O_k(x) \mathcal D_c[\gamma,0] \mathcal O_i(s) \rangle_\text{CPT}\\\nonumber
    &+\frac{1}{2}\sum_{i,j} a_i(mr)a_j(mr) r^{\Delta_i+\Delta_j-2} \int_\gamma ds_1 \int_\gamma ds_2 \langle \mathcal{O}_k(x) \mathcal D_c[\gamma,0] \mathcal O_i(s_1) \mathcal{O}_j(s_2) \rangle_\text{CPT} \\
    &+O(r^{\Delta_i+\Delta_j+\Delta_k-3})\bigg]\,.\label{eq:prescription_defect_fusion}
\end{align}
}
Again, this equation is somewhat schematic. Here $\mathcal D_c$ represents the conformal defect $\mathcal D_c^\text{CFT}$ dressed by relevant operators in the effective action \eqref{eq:Seff2} with non-positive powers of $r$, while the operators $\mathcal O_i$ that we brought down from the exponential are both irrelevant operators and subleading contributions from relevant/marginal operators with positive powers of $r$.
We have truncated the expansion at three-point functions, though generically all $n$-point functions appear on the RHS and must be included up to the desired order in $r$.

By CPT, in this case we mean order-by-order expansions in both bulk and defect relevant couplings, $g,h$, around the CFT configuration. On the left-hand side, these follow from bringing down insertions of $g\int d^d x\,\varPhi$ from the bulk action \eqref{eq:g-defSec3} and $h\int_\gamma ds \,\varphi$ from the definition \eqref{eq:h-defSec3} of each defect. On the right-hand side, bulk relevant deformations work the same way, while on the fused defect we bring down order by order the relevant operators from the effective action that we kept implicit in the definition of $\mathcal D_c$. All CPT correlators can be evaluated from knowledge of the conformal defect (DCFT) data \cite{Cardy:1989ir,Cardy:1991tv,McAvity:1995zd,Liendo:2012hy,Billo:2013jda,Gaiotto:2013nva,Billo:2016cpy}. 
Once CPT correlation functions are computed, they can be plugged into \eqref{eq:prescription_defect_fusion} to extract fusion coefficients. At a given order in $r$ and the relevant couplings $g,h$, one can determine $\mathcal D_c^\text{CFT}$ and the fusion coefficients $a_i(mr)$ by considering enough distinct operator insertions~$\mathcal O_k$.

Like in the case of local operators, the correlation functions on both sides of~\eqref{eq:prescription_defect_fusion} are generically IR divergent in CPT. 
These IR divergences arise from perturbation theory trying to reproduce long distance non-perturbative data, and they are regulated by the radius $R$ of the ball from figure~\ref{fig:CPTball}.
If the fusion coefficients $a_i$ are indeed local UV data, the CPT results for them should be independent of $R$, and in particular remain finite when $R$ is sent to infinity.
That is, IR divergences in~\eqref{eq:prescription_defect_fusion} should cancel between LHS and RHS with finite coefficients. While we do not have a general all-orders proof for it,\footnote{We suspect a proof is possible following similar reasoning as~\cite{Guida:1995kc}, but this is beyond our present scope.} we have observed this cancellation in all of our examples up to the order that we considered, providing a very nontrivial consistency check of our formalism.
In practical computations, this cancellation allows us to use any regularization method we wish in the extraction of fusion coefficients. See e.g.\ section~\ref{sec:Schwinger}, where we use dimensional regularization (rather than the hard cutoff~$R$) to illustrate this point.

We should emphasize that all information about the defect geometry on the RHS of~\eqref{eq:prescription_defect_fusion} is encoded in the correlation functions and not in the fusion coefficients $a_i$, which only contain local information.
When extracting these coefficients one can therefore choose whatever geometry is most convenient to simplify calculations. 
For this reason, we will often focus on infinite straight lines in our examples.

The CPT correlation functions in~\eqref{eq:prescription_defect_fusion}, computed with DCFT data, are generically UV divergent as well.
In CPT, we regard the conformal defects and local operators living on them as ``bare'' operators, and handle these divergences with renormalization.
The UV divergences that one can encounter in these computations fall into two main classes.
The first class, like in the case of local operators, is from collisions of the relevant deformations $g\int d^d x \, \varPhi$ and $h \int ds\, \varphi$ with the defect, with local operators on the defect, or among themselves.
As before, we can regulate these divergences with a cutoff $\delta$, and remove them by adding counterterms to the (defect) action that amount to operator-mixing of the $\mathcal O_i(s)$ on the defect, addition of contact terms, etc.

The second class is new compared to the local operator case.
They arise in higher point functions in the RHS of \eqref{eq:prescription_defect_fusion} when two operators on the fused defect collide, for example when $s_1 \to s_2$ in $\int_\gamma ds_1\int_\gamma ds_2\langle \mathcal D_c \mathcal{O}_i(s_1)\mathcal{O}_j(s_2) \rangle $.
These UV divergences can be understood by considering the OPE of the defect operators $\mathcal O_i$ and $\mathcal{O}_j$,
\begin{equation}
    \mathcal O_i(s) \mathcal{O}_j(0) \simeq \frac{c_{ijk}(m s)}{|s|^{\Delta_i+\Delta_j-\Delta_k}}\mathcal{O}_k(0) \, .
\end{equation}
Operators with $\Delta_i+\Delta_j-\Delta_k \ge 1$ generate divergences.
Introducing a cutoff regulator $|s|>\delta$, the divergences for $\Delta_i+\Delta_j-\Delta_k > 1$ may be removed by adding to the effective action a term\footnote{For the case $\Delta_i+\Delta_j-\Delta_k = 1$, we must introduce an auxiliary scale $\mu$, adding the counterterm
\begin{equation}
    \delta S_\text{eff} = -\sum_{ijk} a_i a_j r^{\Delta_i +\Delta_j -2} c_{ijk}\log{\mu}{\delta}\int_\gamma ds\, \mathcal O_k(s)\, .
\end{equation}
The $\mu$-dependence cancels when combining the $\langle \mathcal D_c\mathcal O_i \mathcal O_j \rangle$ contributions with this new $\langle \mathcal D_c \mathcal O_k \rangle $ counterterm, yielding a scheme-independent result. 
This is also the only counterterm necessary in dimensional regularization where power divergences vanish.}
\begin{equation}
    \delta S_\text{eff} = \frac{1}{2}\sum_{ijk} a_i a_j r^{\Delta_i +\Delta_j -2} c_{ijk}\frac{2\delta^{1-\Delta_i-\Delta_j+\Delta_k}}{-1+\Delta_i+\Delta_j-\Delta_k}\int_\gamma ds \,\mathcal O_k(s)\, ,
\end{equation}
which can also be thought of as renormalizing the coefficients $a_k$ in the effective action.
Notice that this generally requires adding to the action operators $\mathcal O_k$ of higher mass dimension than the fusing operators. This is, in fact, the familiar renormalization procedure from effective field theories, which are only renormalizable order by order in the cutoff $\Lambda \sim 1/r$. We discuss this further in the example of section \ref{sec:massive_scalar}.

When the dust settles, we obtain the effective action \eqref{eq:Seff2} written in terms of well-defined renormalized defect operators $[\mathcal O_i]_\text{r}$ to a certain order in $r$. 
Like structure functions in the OPE for local operators, the fusion coefficients $a_i(mr)$ will depend on scheme choices, but so will the basis of operators, making $\sum_i a_i [\mathcal O_i]_\text{r}$ meaningful.
One just needs to be careful to match schemes when measuring IR condensates on the fused defect, to ensure that the scheme dependence in the prediction \eqref{eq:Prediction} cancels out.

In the remainder of this section, we discuss generalizations of the defect OPE and provide more details on some aspects omitted above to streamline the discussion. We discuss how to put the defects on curved paths $\gamma$ (subsection~\ref{sec:curvature}), how to handle corrections when $\mathcal D_c$ is non-simple (subsection~\ref{sec:non-simple}), the translation invariance of the defect OPE (subsection~\ref{sec:translation_invariance}), and the OPE of bulk operators approaching a defect (subsection~\ref{sec:bulk-to-defect}).
Since they are somewhat technical discussions, the reader might want to skip ahead to the examples in section~\ref{sec:examples}, and come back to them when needed.

\subsection{Curvature corrections}\label{sec:curvature}
In the discussions above we implicitly focused, for pedagogical purposes, on the case where the defect worldlines $\gamma$ are straight lines. We now discuss how to extend the formalism to more general shapes. 
As long as $r$ remains the shortest length in the problem (in particular shorter than any feature on the curve) the logic for the defect OPE continues to hold virtually verbatim. The only change is in the content of the fusion effective action \eqref{eq:Seff2}. The basic new ingredients to construct local actions on a generic curve $\gamma$ are nontrivial geometric invariants. By the standard EFT logic, these must be incorporated into the fusion effective action, providing so-called curvature corrections.

These corrections are obtained by constructing invariants of the tangent and normal bundles of the curve $\gamma$. We review the necessary background for constructing these invariants for manifolds of general dimension in appendix~\ref{app:curvature}. Here we simply note that, for lines, they are built out of the extrinsic curvature $K^i$ (where $i$ points in the directions transverse to $\gamma$) and the covariant derivative $\nabla$ along the curve. The only curvature corrections to consider in the fusion effective action are thus products of $\nabla^a K_i \nabla^b K^i$.\footnote{In this paper we will only consider configurations where the defects $\mathcal D_a[\gamma,r]\mathcal D_b[\gamma,0]$ are rigidly displaced along a direction transverse to $\gamma$. For more general configurations there can be additional invariants. See~\cite{Kravchuk:2024qoh} for the general formalism (in the case of conformal defects).} Both $K^i$ and $\nabla$ have mass dimension one, so these objects should be understood as (geometric) operators of dimension $\Delta_{\nabla^{\ell}K^{2n}}=\ell+2n$. As such, they enter the effective action \eqref{eq:Seff2} as\footnote{We schematically denote a product with $\ell$ covariant derivatives and $2n$ $K_i$'s by $\nabla^\ell K^{2n}$. Note that there are nontrivial multiplicities.}
\begin{equation}
    S_\text{eff}[\gamma] = \int_\gamma ds \sqrt{h_{ss}}\, \left(-\frac{a_0(mr)}{r} + \sum_{\ell,n} a_{\nabla^{\ell}K^{2n}}(mr) r^{\ell+2n-1}\nabla^{\ell}K^{2n} + \sum_{i}a_i(mr) r^{\Delta_i -1} \mathcal O_i\right)\,.
\end{equation}

Some comments are in order. First, we have written the curvature corrections as new background terms independent of dynamical operators. This is only the first tower, associated to the identity operator $\mathds 1$. Combinations with dynamical operators, such as $K^2 \mathcal O$, ought also to be included. We leave such combinations implicit in the sum $\sum_i$.

Second, in the CFT limit the new fusion coefficients $a_{\nabla^{\ell}K^{2n}}(mr)$, much like the original ones, tend to constants. The terms that we have written out (i.e.\ the ones associated to the identity) then become the Casimir coefficients discussed in \cite{Diatlyk:2024zkk}. In fact, in the CFT limit, there are additional constraints on fusion constants stemming from Weyl invariance. They have been worked out in detail in \cite{Kravchuk:2024qoh}. We do not discuss Weyl constraints here because in general they need not apply to the full functions $a_{\nabla^{\ell}K^{2n}}(mr)$ of the massive theory.

Third, these same curvature corrections classify the additional counterterms available for the worldline action of curved defects, used e.g.\ when performing renormalization. For a long straight defect of length $L$, the only available counterterm is the cosmological constant, contributing a factor of $\exp(c_0 L)$. For a more general curve $\gamma$, we have further counterterms $\exp\left(\int ds \sqrt{h_{ss}}( c_0 + c_1 K^2 + \cdots)\right)$ at our disposal.

Just like with dynamical operators, we expect the coefficients $a_{\nabla^{\ell}K^{2n}}(mr)$ to be perturbative and admit an expansion of the form \eqref{eq:ai-exp}, for them, too, are UV data at the top of the RG flow. To extract them, we simply need to proceed as in the previous section, but with defects in some curved configuration on which curvature corrections do not vanish. Any such configuration will do, as the coefficients are independent of the geometry.

\vspace{5pt}
\subsection{Non-simple defects}\label{sec:non-simple}
The other generalization that we will discuss is to cases where the fused (conformal) defect is non-simple. In this case, it can be written as a nontrivial direct sum of other defects,
\begin{equation}
    \mathcal D_c^\text{CFT} = \bigoplus_k \mathcal D_k^\text{CFT}\,.
\end{equation}
This occurs e.g.\ when fusing defects obtained as relevant deformations of non-invertible topological lines, or when fusing Wilson lines in non-abelian gauge theories.
The Hilbert space on a time slice pierced by a non-simple defect $\mathcal D = \bigoplus_k \mathcal D_k$ decomposes into the direct sum of Hilbert spaces associated to each defect, $\mathcal H_{\mathcal D} = \bigoplus_k \mathcal H_{\mathcal D_k}$.\footnote{This illustrates why direct sums of defects are allowed while general linear combinations over $\mathbb R$ are not: direct sums are an allowed operation for Hilbert spaces, but linear combinations over $\mathbb R$ are meaningless.}

Local operators $\mathcal O_i(s)$ living on $\mathcal D$ act on the states in the various components of $\mathcal H_{\mathcal D}$. They may map a component $\mathcal H_{\mathcal D_k}$ back to itself,\footnote{Here we mean this in a broad sense. We will not be careful about whether states remain normalizable.} or exchange between components. The former are operators living on the simple defects $\mathcal D_k$; we denote them by $\mathcal O_i^{kk}(s)$. The latter are known as \textit{defect-changing} operators, and they connect different defects $\mathcal D_k \to \mathcal D_\ell$; we denote them by $\mathcal O_i^{k\ell}(s)$. One particular class of the latter type are \textit{defect-ending} operators, $\mathcal O_i^{k0}(s)$, connecting a simple defect to the trivial line $\mathcal D_k \to 1$.

How do we write a fusion effective action on a non-simple defect? As with any EFT, we should include all operators that can live on $\mathcal D_c^\text{CFT}$. In particular, this means including defect-changing operators too. This can be done consistently with a path-ordered exponential of a block-matrix acting on the various components $\mathcal H_{\mathcal D_k^\text{CFT}}$. For brevity, let us focus on a case with two simple defects, $\mathcal D_c^\text{CFT} = \mathcal D_1^\text{CFT} \oplus \mathcal D_2^\text{CFT}$. The effective action then reads\footnote{Note that while we suppress arguments to reduce clutter, in the massive theory, fusion coefficients $a_i^{k\ell}$ are functions of $mr$, and the sums $\sum_i$ run over relevant, marginal and irrelevant operators, as discussed above. They should also include curvature corrections if $\gamma$ is not straight.}
\begin{equation}\label{eq:fusion_with_mixing}
    e^{-S_\text{eff}[\gamma]} = \mathcal P\exp \int_\gamma ds \begin{pmatrix}
        \frac{a_0^{11}}{r} - \sum_i a_i^{11} r^{\Delta_i-1} \mathcal O_i^{11}(s) & -\sum_i a_i^{12} r^{\Delta_i-1} \mathcal O_i^{12}(s)\\
        -\sum_i a_i^{21} r^{\Delta_i-1} \mathcal O_i^{21}(s) & \frac{a_0^{22}}{r} - \sum_i a_i^{22} r^{\Delta_i-1} \mathcal O_i^{22}(s)
    \end{pmatrix}\,.
\end{equation}
The diagonal entries contain effective actions of the form \eqref{eq:Seff2} on the simple defects $\mathcal D_1^\text{CFT}$, $\mathcal D_2^\text{CFT}$. Off-diagonal entries contain expansions of defect-changing operators. They do not contain a cosmological constant term, as the identity $\mathds 1$ is not a defect-changing operator. For defects with no orientation, the matrix is symmetric.

Let us explain what effect this has on the prediction \eqref{eq:Prediction} for the short-distance behavior of the defect two-point function on a closed homogeneous curve $\gamma$ of length $L$. The identity contributions amount to traces over the Hilbert spaces $\mathcal H_{\mathcal D_1},\mathcal H_{\mathcal D_2}$ and reduce to the sum\footnote{In writing this, we are implicitly assuming that there are no relevant defect-changing operators (i.e.\ with $\Delta<1$) in \eqref{eq:fusion_with_mixing}, so that we can separately flow the defects $\mathcal D_1, \mathcal D_2$. When the conformal defect $\mathcal D_c^\text{CFT}$ is non-simple but there are relevant defect-changing operators, the initially distinct defects $\mathcal D_1^\text{CFT}, \mathcal D_2^\text{CFT}$ mix along the RG flow and we cannot study their IR expectation values separately. See \cite{Zhou:2023fqu,Nagar:2024mjz,Lanzetta:2025xfw} for recent examples of such flows.}
\begin{align}
    \langle \mathcal D_a[\gamma,r]\mathcal D_b[\gamma,0]\rangle_\text{QFT} \simeq&\, e^{a_0^{11}(mr)L/r}\langle \mathcal D_1[\gamma]\rangle_\text{QFT} + e^{a_0^{22}(mr)L/r}\langle \mathcal D_2[\gamma]\rangle_\text{QFT} + \cdots\,.
\end{align}
If one of the Casimirs $a_0(mr)$ scales with a power of $r$ smaller than 1, the corresponding prefactor blows up exponentially, and the contribution from defects with lower Casimirs can be neglected. Continuing on, bringing down one insertion from the action \eqref{eq:fusion_with_mixing} gives again diagonal contributions, as defect-changing operators do not survive the trace:
\begin{align}
    \cdots &- e^{a_0^{11}(mr)L/r}L\sum_i a_i^{11}(mr)r^{\Delta_i -1}\langle \mathcal D_1[\gamma] \mathcal O_i^{11}\rangle_\text{QFT}\nonumber\\
    &- e^{a_0^{22}(mr)L/r}L\sum_i a_i^{22}(mr)r^{\Delta_i -1}\langle \mathcal D_2[\gamma] \mathcal O_i^{22}\rangle_\text{QFT} +\cdots\nonumber\,.
\end{align}

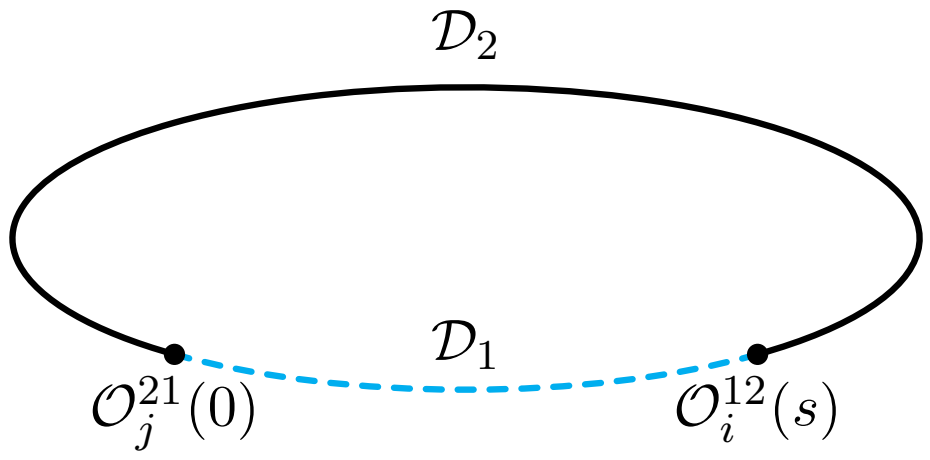
\begin{figure}[h]
    \centering
    \input{Figures/segment.tikz}
    \caption{Graphical representation of the correlator $\langle \mathcal D_1[\gamma_{[0,s)}] \mathcal O_i^{12}(s) \mathcal D_2[\gamma_{[s,L)}] \mathcal O_j^{21}(0)\rangle$.}
    \label{fig:segment}
\end{figure}

For two-point functions the story is qualitatively different. There are diagonal contributions of the usual form
\begin{align}
    \cdots + e^{a_0^{11}(mr)L/r}\frac{L}{2}\sum_{i,j} a_i^{11}(mr)a_j^{11}(mr)r^{\Delta_i + \Delta_j - 2}\int_\gamma ds\,\langle \mathcal D_1[\gamma] \mathcal O_i^{11}(s)\mathcal O_j^{11}(0)\rangle_\text{QFT}+\cdots\nonumber\,,
\end{align}
and similarly for $\mathcal D_2$, but now there are also off-diagonal contributions of the form
\begin{align}
    \cdots + L\sum_{i,j}& a_i^{12}(mr)a_j^{21}(mr)r^{\Delta_i + \Delta_j - 2}\nonumber\\
    &\times\int_\gamma ds\, 
    e^{a_0^{11}(mr)s/r}e^{a_0^{22}(mr)(L-s)/r}
    \langle \mathcal D_1[\gamma_{[0,s)}] \mathcal O_i^{12}(s) \mathcal D_2[\gamma_{[s,L)}] \mathcal O_j^{21}(0)\rangle_\text{QFT}+\cdots\nonumber\,.
\end{align}
This contribution is depicted in figure \ref{fig:segment}. It splits $\gamma$ into two portions of $\mathcal D_1$ and $\mathcal D_2$ connected by defect-changing operators. The expansion continues on to higher orders in a straightforward way. Clearly, the pattern of integrated segments of the two defects features whenever we have an even number of defect-changing operators.

\subsection{Translation invariance}\label{sec:translation_invariance}
Let us now continue with a succinct discussion on the translation invariance of the OPE for line defects. In \eqref{eq:dOPE2}, we fused the defects $\mathcal D_a,\mathcal D_b$, respectively at positions $r$ and $0$, to a defect at position $0$. This choice is of course arbitrary; we could have instead fused them to $r$, or to $r/2$, etc. All choices are physically equivalent, but the fusion coefficients $a_i$ in the effective action change between them, much like the structure functions in the local operator OPE change when we expand the operators around a different location.

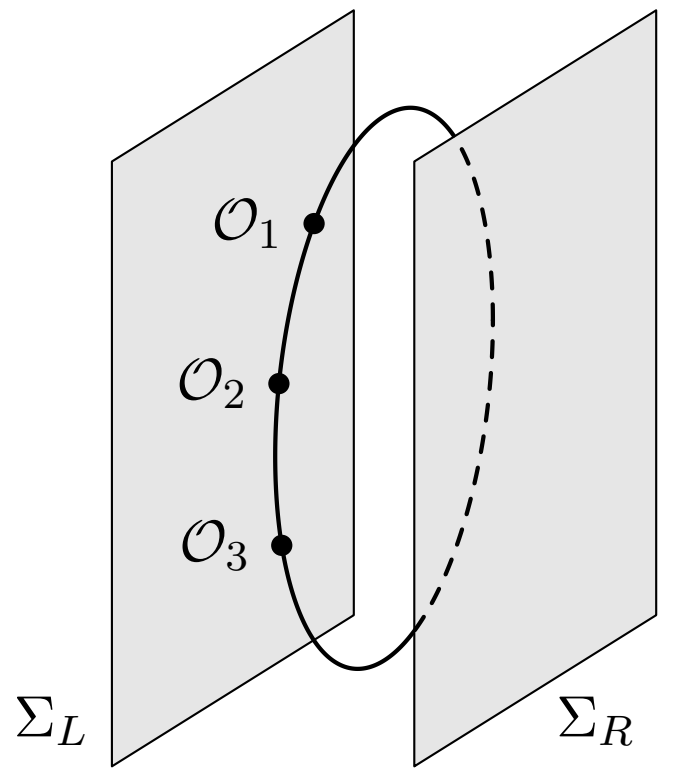
\begin{figure}[h]
    \centering
    \input{Figures/translation.tikz}
    \caption{Graphical representation of the right-hand side of \eqref{eq:ai-Shift}.}
    \label{fig:translation}
\end{figure}

The precise change in the coefficients can be worked out by applying a translation along the transverse direction. Say we fuse the defects at a location $y\in [0,r]$, and we want to relate the new effective action $\mathcal D_a[r]\mathcal D_b[0]\simeq \mathcal D_c^\text{CFT}[y] e^{-\tilde S_\text{eff}}$ to the one in \eqref{eq:dOPE2}. We can do this by expanding the action as in \eqref{eq:Prediction}, and noting that the fused defects (with insertions) are related by
\begin{equation}\label{eq:ai-Shift}
    \mathcal D_c[y] \mathcal O_1(s_1)\ldots \mathcal O_n(s_n) = e^{y \widehat P_r[\Sigma_L]} \mathcal D_c[0] \mathcal O_1(s_1)\ldots \mathcal O_n(s_n) e^{-y \widehat P_r[\Sigma_R]}\,,
\end{equation}
where $\widehat P_r[\Sigma_i]$ is a topological conserved charge, based on the codimension-one surface $\Sigma_i$, implementing translations along the $r$-direction. The configuration is sketched in figure~\ref{fig:translation}. As usual, the momentum charge is given by an integral of the stress tensor,
\begin{equation}\label{eq:Phat}
    \widehat P_\nu[\Sigma_i] \equiv -\int_{\Sigma_i} d^{(d-1)}x\, u^\mu T_{\mu \nu}\,, 
\end{equation}
where $u^\mu$ is a unit vector orthogonal to $\Sigma_i$ (pointing to the left in figure~\ref{fig:translation}).

In the presence of a line defect $\mathcal D[\gamma]$, the stress tensor ceases to be conserved. It satisfies
\begin{equation}\label{eq:Displ}
    \partial^\mu T_{\mu i}(x) = -\delta_\gamma(x) D_i(x_\parallel)\,,
\end{equation}
where the delta function localizes the coordinate to the curve $\gamma$, $x_\parallel$ is a coordinate along it, and $i$ points in the $d-1$ directions orthogonal to it. Here $D_i$ is the displacement operator \cite{McAvity:1993ue,Bianchi:2015liz,Billo:2016cpy}. It implements infinitesimal translations of the defect\footnote{More generally, small deformations of the contour $\gamma$ are realized by localized insertions of the displacement $D_i$, see e.g.~\cite{Billo:2016cpy}.}
\begin{equation}\label{eq:D-transl}
    \delta_{\epsilon}\mathcal D[\gamma,0]= \epsilon^i\widehat P_i[\Sigma_L]\mathcal D[\gamma,0] - \mathcal D[\gamma,0]\epsilon^i\widehat P_i[\Sigma_R] = \epsilon^i\int_\gamma ds\, \mathcal D[\gamma,0] D_i(s)\,,
\end{equation}
as follows from \eqref{eq:Phat} and \eqref{eq:Displ} by applying Stokes' theorem. The matching \eqref{eq:ai-Shift} therefore involves multiple integrals of displacement operator insertions. We will not work out this matching explicitly here. 

As an example, consider a (generalized) pinning field defect $\mathcal D = \exp(h\int ds\, \varphi )$ on a straight line. The displacement operator is given by
\begin{equation}
    D_i = h\partial_i \varphi\,.
\end{equation}
Note that \eqref{eq:Displ} forces $D_i$ to have mass dimension 2. On the other hand, the dimension $\Delta_D$, defined here through the scaling of the two-point function in the short-distance limit, is $\Delta_D = \Delta_\varphi + 1<2$.

For later reference, we point out that in correlation functions of pinning field defects, we can use translation invariance to trade integrated insertions of the displacement operator for transverse derivatives of other insertions. Namely, from $\delta_\epsilon\langle \mathcal D[\gamma,0]\mathcal O_1(s_1)\ldots \mathcal O_n(s_n) \rangle = 0$, it follows that
\begin{equation}\label{eq:Di-Ward-id}
    \int_\gamma ds\langle \mathcal D[\gamma,0]D_i(s) \mathcal O_1(s_1)\ldots \mathcal O_n(s_n)\rangle = -\sum_{k=1}^n\langle \mathcal D[\gamma,0] \mathcal O_1(s_1)\ldots (\partial_i\mathcal O_k)(s_k)\ldots \mathcal O_n(s_n)\rangle \,,
\end{equation}
where transverse derivatives of defect operators are well-defined for pinning field defects because their operators have meaning in the bulk.
Since the displacement operator $D_i$ is itself an operator appearing in the effective action \eqref{eq:Seff2}, the relation above can often be used to simplify correlators in the prediction \eqref{eq:Prediction}. See sections~\ref{sec:massive_scalar} and \ref{sec:IsingModel} for examples.

\subsection{Bulk-to-defect OPE}\label{sec:bulk-to-defect}
\begin{figure}[h]
    \centering
    \input{Figures/bulk-to-defect.tikz}
    \caption{Graphical representation of the bulk-to-defect OPE.}
    \label{fig:bulk-to-defect}
\end{figure}
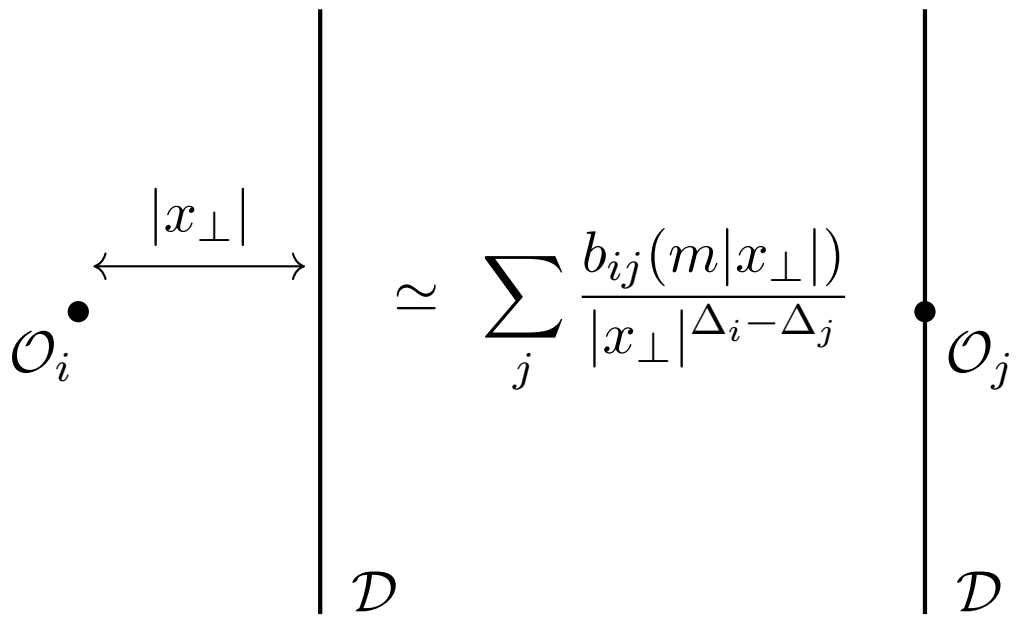

We finish this section by discussing one more immediate generalization of the tools of this section and section~\ref{sec:localOps}: the bulk-to-defect OPE. When a bulk operator approaches a defect, it can be expanded as an infinite sum of local defect operators on the defect at the location of the collision. Namely,
\begin{equation}\label{eq:bulk-to-defect}
    \mathcal O_i(s,x_\perp) \mathcal D\simeq \sum_j b_{ij}(m|x_\perp|) |x_\perp|^{\Delta_j-\Delta_i} \mathcal D \mathcal O_j(s)\,,
\end{equation}
where $s$ is a coordinate along the line defect and $x_\perp$ is perpendicular to it. The configuration is drawn in figure~\ref{fig:bulk-to-defect}.
We recall our conventions that $\mathcal O(x) \mathcal D$ denotes bulk operators in the presence of the defect, while $\mathcal D \mathcal O(s)$ denotes local defect operators on $\mathcal D$.

This expansion follows from the Wilsonian logic advocated in previous sections: integrating out modes with energy higher than $1/|x_\perp|$ makes the resolution too poor to tell apart the bulk operator, and it merges with the defect. In the limit of $|x_\perp|\to 0$ this does not affect the theory, leading to the asymptotic operator expansion \eqref{eq:bulk-to-defect}. In a CFT, the coefficients $b_{ij}$ are constant DCFT data, and \eqref{eq:bulk-to-defect} reduces to the familiar (convergent) bulk-to-defect OPE \cite{Cardy:1989ir,Cardy:1991tv,McAvity:1995zd,Liendo:2012hy,Billo:2013jda,Gaiotto:2013nva,Billo:2016cpy}. In the massive theory, the $b_{ij}$ are regarded as Wilson coefficients and they are promoted to dimensionless functions of $m|x_\perp|$.

Similarly to fusion coefficients and structure functions, we expect the $b_{ij}(m|x_\perp|)$ to be local data independent of global details of the defect and distant insertions. In particular, we should be able to compute them by using \eqref{eq:bulk-to-defect} at distances $R\gg |x_\perp|$ (with $R\ll 1/m$), where CPT is reliable. We therefore expect the coefficients $b_{ij}$ to admit a power-series expansion in the relevant couplings $g,h$ analogous to \eqref{eq:ai-exp}. They can be computed order by order in CPT by building a system of equations of the form \begin{equation}
    \langle \mathcal O_k(y) \mathcal O_i(s,x_\perp) \mathcal D\rangle_\text{CPT} \simeq \sum_j b_{ij}(m|x_\perp|)|x_\perp|^{\Delta_j-\Delta_i}\langle \mathcal O_k(y) \mathcal D \mathcal O_j(s)\rangle_\text{CPT}\,,
\end{equation}
where we restrict the relevant deformations to a ball of radius $R$ as in figures~\ref{fig:CPTball} and \ref{fig:CPTballLocal}. We illustrate this computation in an example in appendix~\ref{app:bulk-to-defect}, which will be used in section \ref{sec:IsingModel}.

\section{Solvable examples}\label{sec:examples}

We now test our proposal for the fusion of line defects in exactly solvable models. Beyond mere tests, which, we anticipate, are all satisfactory, the examples below serve to illustrate various tricks and observations that come in handy when extracting fusion coefficients. We begin with free CFTs, extending the analysis of \cite{Diatlyk:2024zkk,Diatlyk:2024qpr,Kravchuk:2024qoh,Cuomo:2024psk} to determine the $a_i$'s. The free massless scalar (subsection~\ref{sec:FreeCFT-scalar}) provides the opening act, setting the stage for computations of fusion coefficients both for straight and curved defects. It is followed by free Maxwell (subsection~\ref{sec:Maxwell}), where we further discuss simplifications from the $O(d-2)$ symmetry on the plane transverse to the defects, and how to handle spinning operators in the effective~action.

We then proceed to turn on a bulk relevant deformation: a mass term in the $4d$ scalar (subsection~\ref{sec:massive_scalar}).
Treating the deformation in CPT, we extract the fusion coefficients $a_i(m^2r^2)$ order by order in $m^2$ both for straight and curved defects. The results showcase the locality of the $a_i$ and the cancellation of IR divergences. Since the theory is Gaussian, we further compute the exact correlators and compare to the predictions from the defect OPE, leading to the first check of the framework from section~\ref{sec:Lines}. 

We conclude with what is arguably the simplest interacting gauge theory: the massless Schwinger model (subsection~\ref{sec:Schwinger}). Aside from a toy model for future QCD computations, we use this example to illustrate that one can perform computations with any regularization scheme. Choosing dimensional regularization, which kills all power divergences, we show that log IR divergences still cancel in computing the $a_i$'s, and the results reproduce the exact solution for the correlator of Wilson lines.

\subsection{Free CFTs}\label{sec:free_CFTs}

In this section we begin by demonstrating how to extract the fusion coefficients in free CFTs.
Some of these details have been discussed in~\cite{Diatlyk:2024zkk,Diatlyk:2024qpr,Kravchuk:2024qoh,Cuomo:2024psk}, though the subleading coefficients $a_i$ were not extracted explicitly.
The basic procedure is the same in the more complicated examples to follow, so this serves as an instructive warm-up.

\subsubsection{Free massless scalar}\label{sec:FreeCFT-scalar}

We first consider a free massless scalar in $\mathbb{R}^4$ with the usual action $S= \frac{1}{2}\int d^4 x\, \partial_\mu \phi \partial^\mu \phi$.
We consider the fusion of two parallel pinning field defects of the form
\begin{equation}
    \tilde{\mathcal D}_{h} \equiv \exp\left(h \int_\gamma ds\, \phi(s)\right)\, ,
\end{equation}
where the tilde is to differentiate this ``bare'' defect from the proper divergence-free CFT defect ${\mathcal D}_{h}$ to be introduced below.
Since $[\phi]=1$ in $4d$, this is exactly marginal, with the coupling $h$ spanning a defect conformal manifold.
We are interested in the fusion as $r \to 0$ of two parallel defects with couplings $h_1$ and $h_2$ placed at transverse positions $x=r$ and $0$,
\begin{equation}\label{eq:Dh1Dh2-Seff}
    {\mathcal D}_{h_1}[\gamma,r]{\mathcal D}_{h_2}[\gamma,0]  \simeq {\mathcal D}_\text{fused}[\gamma,0]e^{-S_{\mathrm{eff}}[\gamma]}\,.
\end{equation}

\begin{figure}[t]
    \centering
    \input{Figures/DD_MasslessScalar.tikz}
    \caption{Contributions to the $\langle \mathcal D_{h_1}(r) \mathcal D_{h_2}(0) \rangle$ correlator for the free massless scalar. The first diagram gives the Casimir, while the second and third are disconnected contributions.}
    \label{fig:MasslessScalar_DD}
\end{figure}
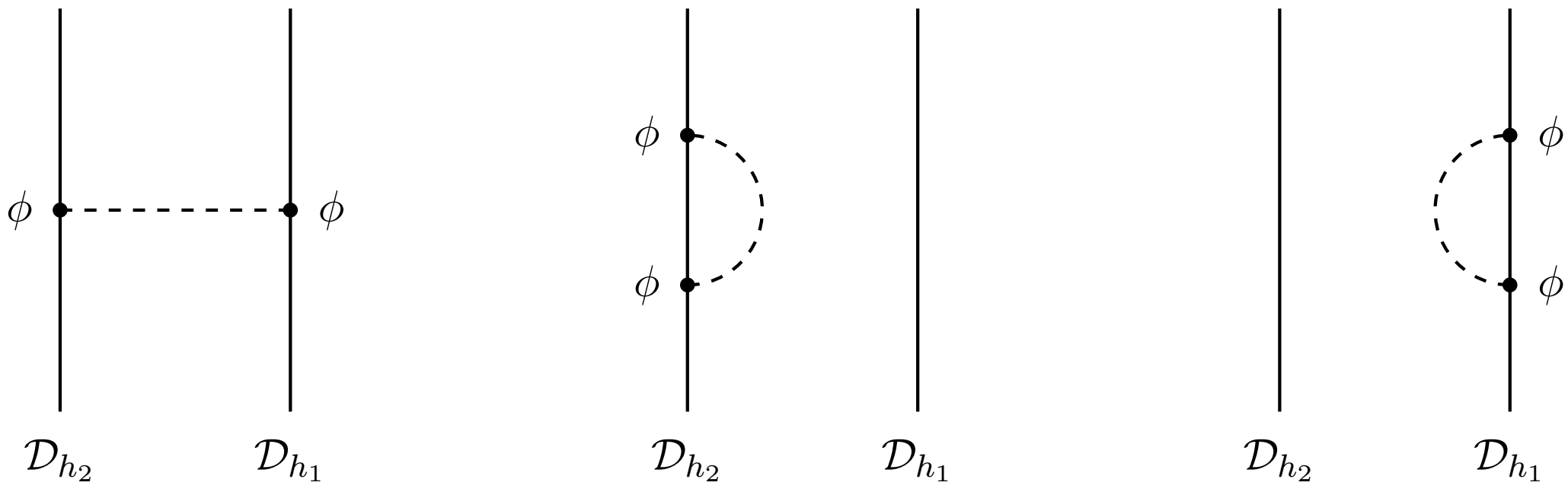

Since the fusion coefficients are independent of the path $\gamma$, we can consider the simplest geometry to extract them.
We will start by considering the two defects to be infinite parallel lines.
The two-point function can be determined by computing the three diagrams in figure~\ref{fig:MasslessScalar_DD}, and it gives
\begin{align}
\begin{split}
    \log\langle \tilde{\mathcal D}_{h_1}[r] \tilde{\mathcal D}_{h_2}[0] \rangle &= \frac{1}{4\pi^2}\int ds_1 \int ds_2 \bigg[\frac{h_1 h_2}{(s_1-s_2)^2+r^2} +\frac{h_1^2+h_2^2}{2(s_1-s_2)^2}\bigg]\\
    &= \frac{h_1 h_2 L}{4\pi r}+\frac{(h_1^2+h_2^2)L}{4\pi^2 \delta}\,,
\end{split}
\end{align}
where the exponentiation of diagrams is immediate since it is a free theory.
Here we have introduced the length of the defect $L$ as an IR regulator, and $|s_1-s_2|>\delta$ as a UV regulator in the second term.
The UV divergence indicates that we have not yet identified the proper action for our defects.
This is done by introducing a cosmological constant counterterm on each defect and defining\footnote{For the free CFT examples there is a natural scheme choice where the entire divergent piece is absorbed, leaving only the Casimir.
Once we turn on relevant deformations, we are generically forced to introduce a scheme-dependent scale $\mu$ to renormalize our defects and operators.}
\begin{equation}\label{eq:MasslessScalarDefectRenorm}
\mathcal{D}_{h_i}\equiv\tilde{\mathcal{D}}_{h_i}\exp\left(-\frac{h_i^2}{4\pi^2 \delta} \int ds\right)\, .\end{equation}
We have now dropped the tilde, as this is the well-defined conformal defect in the CFT. In the following, we will similarly use tildes to indicate ``bare'' defect local operators $\tilde{\mathcal O}(s)$ before identifying their proper CFT definitions $\mathcal O(s)$.

Now we want to reproduce the correlator
\begin{equation}\label{eq:MasslessScalarDefect2Pt}
    \langle \mathcal D_{h_1}[r] \mathcal D_{h_2}[0] \rangle = \exp\left(\frac{h_1 h_2 L}{4\pi r}\right)
\end{equation}
from the defect OPE \eqref{eq:Dh1Dh2-Seff}. The fused defect is known to be another pinning field defect, $\mathcal D_\text{fused} = \mathcal D_{h_1+h_2}$ \cite{Cuomo:2024psk}, we will fix the fusion coefficients $a_{i}$ entering the effective action $S_\mathrm{eff}$. The action should include all irrelevant operators living on the defect.
The first few are
\begin{align}\label{eq:MasslessScalarSeff}
\begin{split}
    S_\mathrm{eff}
    = \int ds \bigg[& -\frac{a_0}{r} + r\left(a_{\partial_i  \phi} \partial_i {\phi} + a_{\phi^2}  \phi^2\right)\\
    &+r^2\left(a_{\partial_i\partial_j  \phi} \partial_i \partial_j  \phi + a_{ \phi^3}  \phi^3 + a_{\partial_i \phi^2}\partial_i ( \phi^2)\right) + O(r^3)\bigg]\,,
\end{split}
\end{align}
where $\partial_i,\partial_j$ are derivatives in the $3$ directions transverse to the defect.\footnote{Derivatives in the longitudinal direction are total derivatives and can be omitted.} Note that while $\partial_i \phi$, $\partial_i\partial_j \phi$, and $\partial_i ( \phi^2)$ are descendants in the bulk, they are primaries on the defect.

\paragraph*{Identifying the CFT operator basis}

Before extracting the coefficients $a_{i}$, we must identify an appropriate basis for the operators in $S_{\text{eff}}$ where UV divergences are removed, just as we did for the defect itself.
Denoting the naive operators appearing in~\eqref{eq:MasslessScalarSeff} with a tilde, we can identify the appropriate definitions by computing their one-point functions on the infinitely-long defect and requiring that they vanish, i.e.\ $\langle \mathcal D_{h_1+h_2} \mathcal O(s)\rangle=0$. For the operators at $O(r)$, we find\footnote{Note that by $\tilde \phi^2$ we mean the normal-ordered product $:\tilde \phi^2:$. There is therefore no self-contraction.}
\begin{subequations}
\begin{align}
    \langle \mathcal D_{{h_{1}+h_2}}   \partial_i \tilde \phi(s) \rangle &= 0\, ,\\
    \langle \mathcal D_{{h_{1}+h_2}} \tilde \phi^2(s) \rangle &= \frac{(h_1+h_2)^2}{4\pi^4 \delta^2}\, .
\end{align}
\end{subequations}
Since $\partial_i \phi$ has vanishing one-point function on the defect, it needs no redefinition.
The other divergence is removed with the change of basis\footnote{We could repeat the exercise for the marginal operator $\phi(s)$ living on the defect, where we find that the necessary redefinition is precisely what we absorbed as a cosmological constant counterterm in \eqref{eq:MasslessScalarDefectRenorm}.}
\begin{align}
    \phi^2(s) &\equiv \left(\tilde{\phi}(s) - \frac{h_1+h_2}{2\pi^2 \delta} \mathds{1}\right)^2 \, .
\end{align}

Carrying out this procedure also for the operators appearing at $O(r^2)$, we find that we also need the redefinitions
\begin{subequations}
\begin{align}
    \partial_i \partial_j \phi(s) &\equiv \partial_i \partial_j \tilde{\phi}(s) +\frac{h_1+h_2}{3\pi^2 \delta^3}\delta_{ij} \mathds 1 ,\\
    \phi^3(s)& \equiv \left(\tilde{\phi}(s) - \frac{h_1+h_2}{2\pi^2 \delta} \mathds{1}\right)^3 \, .
\end{align}
\end{subequations}
The effective action in~\eqref{eq:MasslessScalarSeff} is written in terms of these operators without tildes. 
This procedure is equivalent to throwing out all the diagrams in any correlation function with propagators connecting the defect with the operator $\mathcal{O}(s)$ living on it. 
One can check that this removes all divergences in correlation functions of the form $\int ds \langle \mathcal{O}_i(x) \mathcal D_{{h_1+h_2}} \mathcal{O}_j(s) \rangle $ for operators up to this order.

In addition to the above, we will also encounter UV divergences from the collision of operators on the defect after expanding $S_\text{eff}$.
For example,
\begin{equation}
    \frac{1}{2}a_{\partial_i\phi}^2r^2\int ds_1 ds_2 \langle \partial_i \phi(s_1) \partial_i \phi (s_2) \rangle = \left(\frac{1}{2}a_{\partial_i\phi}^2r^2\right)\frac{L}{3\pi^2 \delta^3}\, .
\end{equation}
This divergence is removed by adding a cosmological constant counterterm to the defect
\begin{equation}
    \delta S_\text{eff} = \left(\frac{1}{2}a_{\partial_i\phi}^2r^2\right)\frac{1}{3\pi^2 \delta^3} \int ds\,\mathds{1}\, ,
\end{equation}
which can be determined by looking at the OPE of $\partial_i \phi(s) \partial_j \phi(0)$ and is absorbed into $a_0$.
This will suffice to remove UV divergences in what will follow.

\paragraph*{Three-point functions}

\begin{figure}[t]
    \centering
    \input{Figures/DDphi2_MasslessScalar.tikz}
    \caption{Diagrams for computing the three-point function $\langle \phi^2(x)\mathcal D_{h_1}[r] \mathcal{D}_{h_2}[0]\rangle$. This is to be compared to the fusion prediction at $O(r)$ for extracting $a_{\phi^2}$.}
    \label{fig:MasslessScalar_DDphi2}
\end{figure}
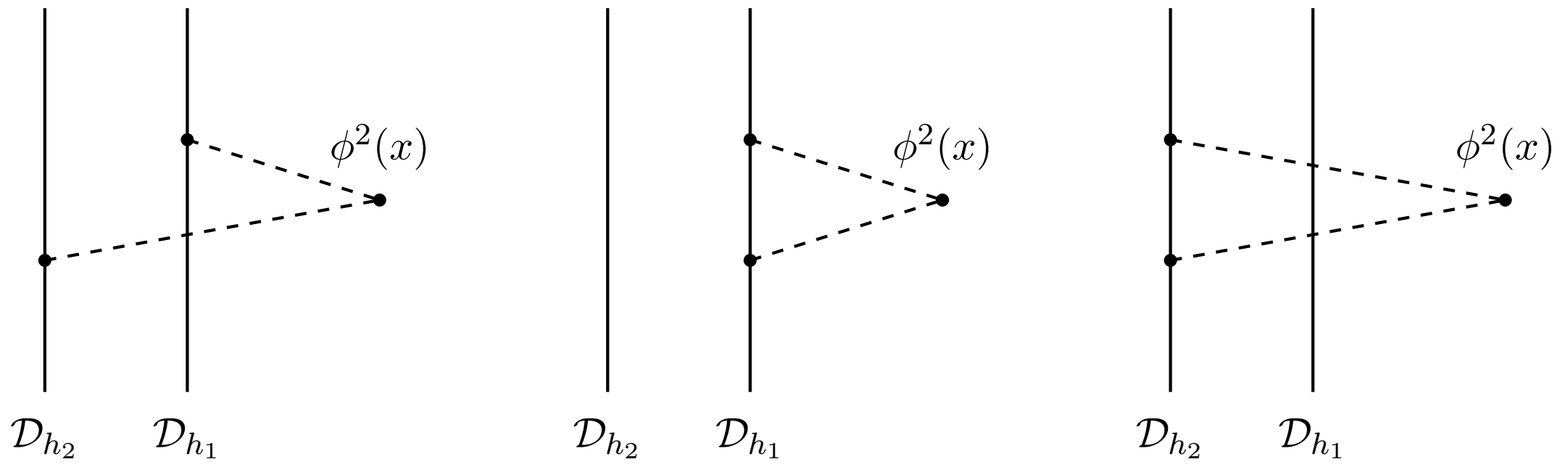
To fix the coefficients $a_i$, we should compute three-point functions of an operator in the bulk at a position $x \gg r$ away from our two defects, as shown in figure~\ref{fig:MasslessScalar_DDphi2} for $\phi^2$. 
We will choose $x$ to be aligned with the separation $r$ for simplicity, though this is of course arbitrary. 
To fix the coefficients up to $O(r^2)$, we need to compute three-point functions inserting $\mathds 1$, $\phi(x)$, $\phi^2(x)$, and $\phi^3(x)$. 
The coefficients of operators with derivatives will follow from these immediately.
The first of these, $\langle \mathds 1 \mathcal D_{h_1} \mathcal D_{h_2}\rangle$, is what we computed in~\eqref{eq:MasslessScalarDefect2Pt}.
The $\phi(x)$ and $\phi^2(x)$ three-point functions are given by\footnote{For all our three-point function calculations, we divide by the two-point function $\langle \mathcal{D}_{h_1}(r) \mathcal D_{h_2}(0)\rangle$ to remove the ladder diagram contributions that independently exponentiate. In an interacting theory, this would remove webs that are disconnected from the inserted operator $\mathcal{O}(x)$.}
\begin{subequations}
\begin{align}\label{eq:phiDD_3pt_massless}
    \frac{\langle \phi(x) \mathcal D_{h_1}[r] \mathcal D_{h_2}[0]\rangle}{\langle \mathcal{D}_{h_1}[r] \mathcal D_{h_2}[0]\rangle} &= \frac{h_1 + h_2}{4\pi x} + \frac{h_1}{4\pi}\left(\frac{r}{x^2}+\frac{r^2}{x^3}+O(r^3)\right)\,.\\\label{eq:r1DD_3pt_massless}
    \frac{\langle \phi^2(x) \mathcal D_{h_1}[r] \mathcal D_{h_2}[0]\rangle}{\langle \mathcal{D}_{h_1}[r] \mathcal D_{h_2}[0]\rangle} &= \frac{(h_1+h_2)^2}{16\pi^2 x^2}+\frac{h_1(h_1+h_2)r}{8\pi^2 x^3}+\frac{(3h_1^2+2h_1h_2)r^2}{16\pi^2 x^4} + O(r^3)\,.
\end{align}
\end{subequations}

To extract the coefficients $a_i$ we must compare each of these three-point functions to the correlation function after the fusion.
Following~\eqref{eq:prescription_defect_fusion}, the matching is given by
\begin{align}\label{eq:MasslessScalarFullDefectExpansion}
    \langle \mathcal O(x) \mathcal{D}_{h_1}[r] \mathcal D_{h_2}[0] \rangle \simeq & \,
    e^{\frac{a_0}{r}\int ds}\bigg\{\langle  \mathcal O(x)\mathcal  D_{h_{1}+h_{2}} \rangle \nonumber\\
    &- r\int ds\bigg[a_{\partial_i \phi} \langle \mathcal O(x) \mathcal D_{h_{1}+h_{2}} \partial_i\phi(s)\rangle+a_{\phi^2} \langle \mathcal O(x) \mathcal D_{h_{1}+h_{2}} \phi^2(s)\rangle\bigg] \nonumber\\
    &-r^2 \int ds \bigg[a_{\phi^3} \langle \mathcal O(x) \mathcal D_{h_{1}+h_{2}} \phi^3(s)\rangle+a_{\partial_i \phi^2} \langle \mathcal O(x) \mathcal D_{{h_1+h_2}}\partial_i \phi^2(s)\rangle\nonumber\\
    &\qquad +a_{\partial_i \partial_j \phi} \langle \mathcal O(x) \mathcal D_{{h_1+h_2}} \partial_i\partial_j \phi(s)\rangle\bigg]\\
    &+ \frac{r^2}{2}\int ds_1 \int ds_2 \bigg[a_{\partial_i \phi} a_{\partial_j \phi }\langle \mathcal O(x) \mathcal D_{{h_1+h_2}} \partial_i \phi(s_1)\partial_j \phi(s_2) \rangle\nonumber\\
    & \qquad + a_{\phi^2}^2 \langle \mathcal O(x) \mathcal D_{{h_1+h_2}} \phi^2(s_1)\phi^2(s_2) \rangle\nonumber\\
    &\qquad +2a_{\partial_i \phi} a_{\phi^2}\langle \mathcal O(x) \mathcal D_{{h_1+h_2}} \partial_i \phi(s_1)\phi^2(s_2) \rangle\bigg]  +O(r^3)\bigg\}\,.\nonumber
\end{align}
We need to compute the correlation functions appearing on the RHS of this equation for each inserted operator $\mathcal O(x)$.

Consider first the identity, which will fix the Casimir $a_0$.
With the definition of the defect from \eqref{eq:MasslessScalarDefectRenorm}, the first correlator is $\langle \mathcal{D}_{{h_1+h_2}}\rangle=1$. All other correlators on the fused defect vanish. Indeed, by dimensional analysis, they must be proportional to 
\begin{equation}
    \langle \mathcal D_{h_1+h_2} \int \mathcal O_1\cdots\int \mathcal O_n\rangle \sim L^{\sum_{i=1}^n(1-\Delta_i)}\to0\,,
\end{equation} which vanishes in the limit of infinitely long defects. By comparing~\eqref{eq:MasslessScalarFullDefectExpansion} with~\eqref{eq:MasslessScalarDefect2Pt} we determine
\begin{equation}
    a_0 = \frac{h_1 h_2}{4\pi} \, .
\end{equation}
This already reproduces the full correlator \eqref{eq:MasslessScalarDefect2Pt} of infinitely long conformal defects in isolation. Subleading terms in the effective action are necessary when fusing the defects in more general configurations. We proceed to fix those.

\begin{figure}[t]
    \centering
    \input{Figures/Fused_DDphi2_MasslessScalar.tikz}
    \caption{Diagrams contributing to the fusion prediction for $\langle \phi^2(x)\mathcal D_{h_1}[r] \mathcal D_{h_2}[0]\rangle$ up to $O(r)$. 
    The first diagram fixes the fused defect to be $\mathcal{D}_{h_1+h_2}$, while the latter two fix $a_{\partial_i\phi}$ and $a_{\phi^2}$ when combined with $\langle \phi(x)\mathcal D_{h_1}[r] \mathcal D_{h_2}[0]\rangle$.
    Other diagrams have been removed by our CFT definition of $\phi^2$ on the defect.}
    \label{fig:MasslessScalar_fused_DDphi2}
\end{figure}
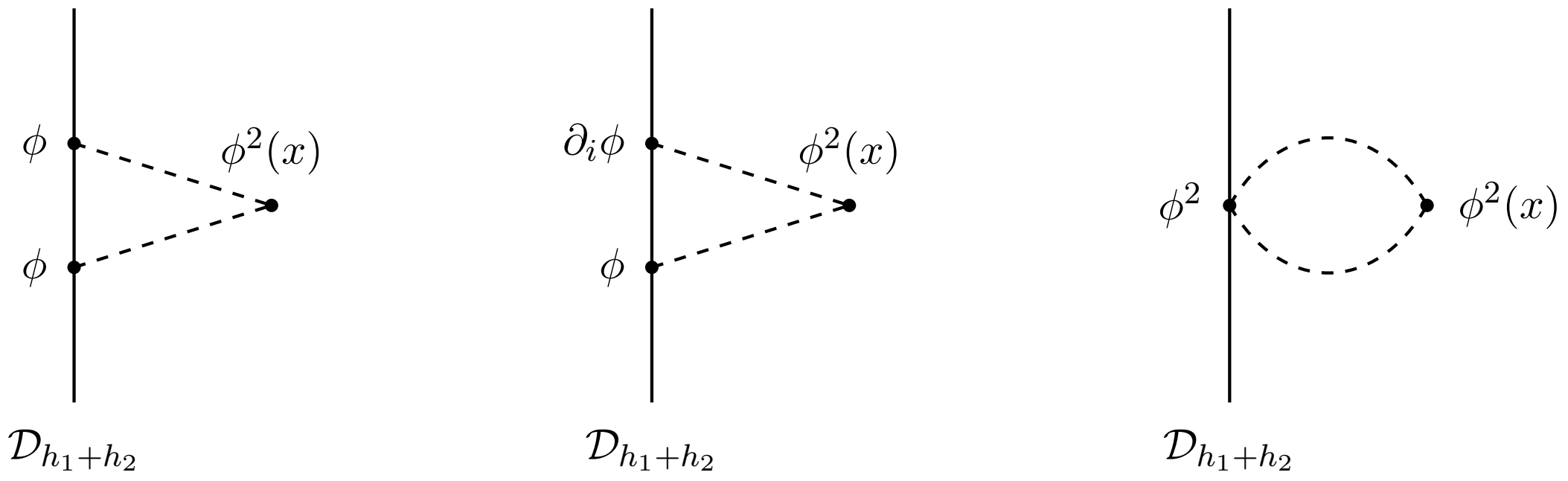

Consider now the insertions of $\phi(x), \phi^2(x)$ and equate (\ref{eq:phiDD_3pt_massless}, \ref{eq:r1DD_3pt_massless}) to \eqref{eq:MasslessScalarFullDefectExpansion}. With the correlators computed in appendix \ref{app:FreeCFTs}, we can write \eqref{eq:MasslessScalarFullDefectExpansion} for each insertion explicitly through order $O(r)$,\footnote{The $\phi^2$ contribution in the first equation vanishes after changing to the renormalized basis.}
\begin{subequations}
\begin{align}
    \frac{\langle \phi(x) \mathcal D_{h_{1}}[r]\mathcal D_{h_2}[0]\rangle}{\langle \mathcal D_{h_1}[r] \mathcal D_{h_2}[0]\rangle} &\simeq \frac{h_1+h_2}{4\pi x}-ra_{\partial_i \phi} \frac{x_i}{4\pi x^3}+O(r^2)\,, \\
    \frac{\langle \phi^2(x) \mathcal D_{h_{1}}[r]\mathcal D_{h_2}[0]\rangle}{\langle \mathcal D_{h_1}[r] \mathcal D_{h_2}[0]\rangle} &\simeq \frac{(h_1+h_2)^2}{16\pi^2 x^2}-ra_{\partial_i \phi}\frac{(h_1+h_2)x_i}{8\pi^2 x^4}-ra_{\phi^2}\frac{1}{16\pi^3 x^3}+O(r^2)\, .
\end{align}
\end{subequations}
For illustration, the diagrams producing the three terms in the second equation are depicted in figure~\ref{fig:MasslessScalar_fused_DDphi2}.
The denominator simply cancels the contribution from the Casimir. At order $O(r^0)$, we have the one-point functions $\langle  \mathcal O(x)\mathcal  D_{h_{1}+h_{2}} \rangle$, which confirm that the fused defect is indeed $\mathcal D_{h_1+h_2}$. 
At order $O(r)$, we have two equations for two unknowns: $a_{\partial_i \phi}$ and $a_{\phi^2}$. From the $\phi$ equation we immediately fix
\begin{equation}
    a_{\partial_i \phi} = -\delta_{i r} h_1\, .
\end{equation}
Plugging this into the $\phi^2$ equation and comparing to~\eqref{eq:r1DD_3pt_massless}, we see that we have
\begin{equation}
    a_{\phi^2} = 0\,.
\end{equation}
So at $O(r)$ for the free massless scalar, only the displacement operator $\partial_r\phi$ appears in the effective action.

Higher orders in $r$ follow the same logic. One just needs to consider equations from enough insertions to solve for all unknowns at any given order.
Since we will eventually want to go up to $O(r^3)$ for the massive scalar in section~\ref{sec:massive_scalar}, we also report here the coefficients $a_{\partial_i\partial_j \phi}$, $a_{\partial_i\phi^2}$, and $a_{\phi^3}$.
We can fix these three unknowns with the correlators $\langle \phi \mathcal D \mathcal D \rangle$, $\langle \phi^2 \mathcal D \mathcal D \rangle$, and $\langle \phi^3 \mathcal D \mathcal D \rangle$ through order $O(r^2)$. This requires computing all two-point functions appearing in the fusion~\eqref{eq:MasslessScalarFullDefectExpansion} for the three insertions at $x$, as well as the three-point function $\int \langle \mathcal O(x) \mathcal D_{{h_1+h_2}} \partial_i\phi \partial_i \phi\rangle$ entering at $O(r^2)$.
We leave the explicit calculations to appendix~\ref{app:FreeCFTs}.
Matching the full expression for each of them to the prediction from fusion gives
\begin{equation}
    a_{\phi^3}=0\,, \qquad a_{\partial_i \phi^2} = 0\,, \qquad a_{\partial_i \partial_j \phi}= -\frac{h_1}{2}\delta_{ir}\delta_{jr}\,.
\end{equation}
In fact, looking beyond $O(r^2)$, one readily finds that the only nonzero coefficients are\footnote{To see this, first note that our CFT definition for the operators $\mathcal{O}(s)$ in $S_\text{eff}$ sets to zero all diagrams with propagators connecting $\mathcal{O}(s)$ to the fused defect $\mathcal{D}_{{h_1+h_2}}$, for example $\langle \phi(x) \mathcal D_{{h_1+h_2}} \phi^n(s) \rangle $. 
Likewise, all integrated higher point functions such as $\langle \phi(x) \mathcal{D}_{{h_1+h_2}}\mathcal{O}_1(s_1)\mathcal{O}(s_2) \rangle $ set to zero after identification of the proper CFT operator basis in $S_\text{eff}$. 
The only remaining nonzero correlation functions are $\langle \phi(x)\mathcal{D}_{{h_1+h_2}} \partial_{i_1}\partial_{i_2} \ldots \partial_{i_n}\phi\rangle$, which yield the coefficients $a_{\partial_{i_1}\partial_{i_2} \ldots \partial_{i_n}\phi}$.
Plugging these into the correlation function for any other probe operator such as $\langle \phi^n(x) \mathcal{D}_{h_{1}}[r]\mathcal{D}_{h_2}[0]\rangle$ reproduces the full matching, giving zero for all other $a_i$'s.
We also could have anticipated this result from the multipole expansion, which we comment on in the next section.}
\begin{equation}
    a_{\partial_{i_1}\partial_{i_2} \ldots \partial_{i_n}\phi} = -\frac{h_1}{n!}\prod_{j=1}^n\delta_{i_j r}\, .
\end{equation}

\paragraph*{Curvature corrections}
So far, we have considered only parallel infinite lines where curvature corrections manifestly vanish.
However, for an arbitrary path $\gamma$, we expect the effective action to also include operators involving the extrinsic curvature $K^i$, as discussed in section \ref{sec:curvature}.
The easiest way to extract these is to consider circles of radius $L/2\pi$ centered at $x_2=x_3=0$ instead of infinite lines for our defects.
First, we consider the one-point function of the defect
\begin{align}
\begin{split}
    \log \langle\mathcal D_{h}\rangle_\text{circ} &= \frac{h^2}{8\pi^2} \int ds_1 ds_2  \frac{1}{|s_1-s_2|^2} - \frac{h^2 L}{4\pi^2 \delta} \\
    &= \frac{h^2}{8\pi} \int_{\delta_\theta}^{2\pi-\delta_\theta} d\theta \frac{1}{1-\cos\theta} - \frac{h^2 L}{4\pi^2 \delta}\\
    &= 0\,,
\end{split}
\end{align}
where we have regulated using the same scheme as before.
We see that our CFT normalization has rendered $\langle \mathcal{D}_{h}\rangle_\text{circ}=1$ as in the infinite line case.

\begin{figure}[t]
    \centering
    \input{Figures/DD_circ_MasslessScalar.tikz}
    \caption{The sole diagram contributing to the correlation function $\langle \mathcal D_{h_1}(r) \mathcal D_{h_2}(0) \rangle_\text{circ} $ for the free massless scalar with circular defect geometries.}
    \label{fig:MasslessScalar_DD_circ}
\end{figure}
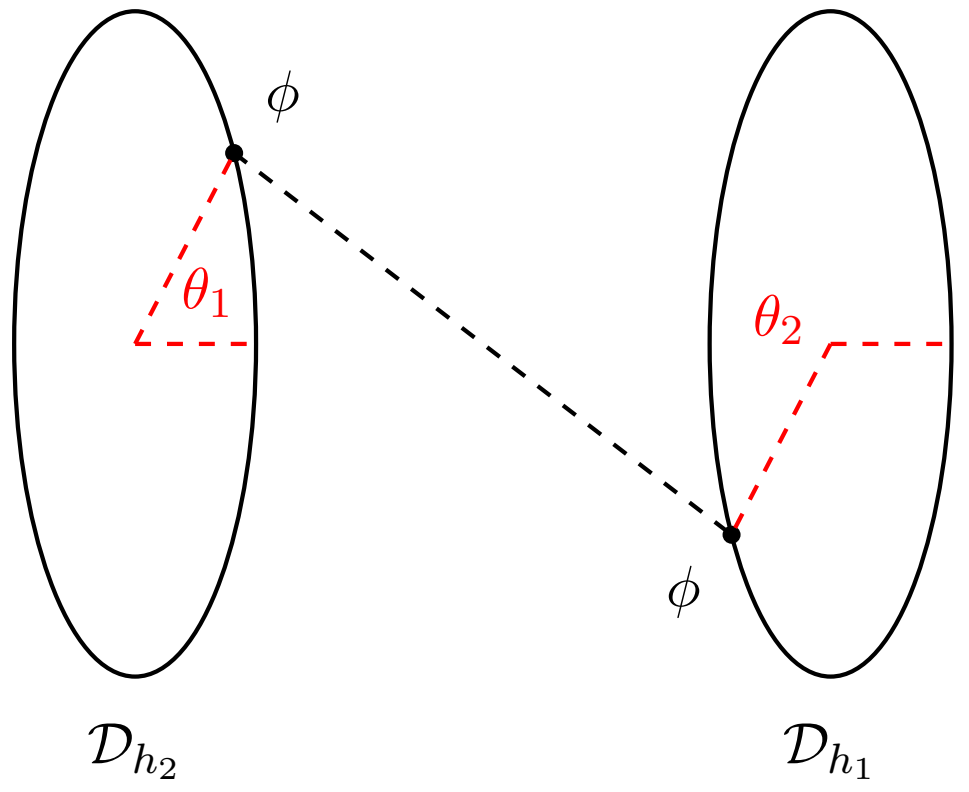
Since $\langle \mathcal{D}_{h}\rangle_\text{circ} =1$, the two-point function with this geometry is just the single diagram in figure~\ref{fig:MasslessScalar_DD_circ}, which gives
\begin{align}\label{eq:MasslessScalarDefect2PtLoops}
\begin{split}
    \log\langle \mathcal D_{h_1}[r] \mathcal D_{h_2}[0] \rangle_\text{circ} &= \frac{h_1h_2 L^2}{8\pi^2}\int_0^{2\pi}  \int_0^{2\pi} \frac{d\theta_1 d\theta_2}{2\pi^2 r^2+L^2[1-\cos(\theta_1-\theta_2)]} \\
    &= \frac{h_1h_2L^2}{4\pi^2 r\sqrt{r^2+L^2/\pi^2}}\\
    &=\frac{h_1h_2 L}{4\pi r}-\frac{h_1 h_2 \pi r}{8L}+\frac{3h_1h_2\pi^3r^3}{32L^3}+O(r^5)\, .
\end{split}
\end{align}
The new terms in the series in $r/L$ are to be matched by the curvature corrections. From the example in section \ref{sec:curvature}, we know that up to $O(r^3)$, two nonzero curvature terms may appear in our effective action for the circular geometry,
\begin{equation}
    S_\text{eff} \supset \int ds \left[ra_{K^2}K_i K^i+r^3a_{K^4}(K_i K^i)^2+O(r^5)\right]\, .
\end{equation}
Their coefficients can be fixed again by equating correlators before and after fusing the defects, as in \eqref{eq:MasslessScalarFullDefectExpansion}, but now including the new curvature terms.

Before we can read off their coefficients, however, we need to compute the correlation functions on the RHS of \eqref{eq:MasslessScalarFullDefectExpansion} in the new geometry, which generically develop new divergences.
For example, using the same scheme as in the infinite line case, $\delta_{\theta} = 2\sin^{-1}\left(\frac{\pi \delta}{L}\right)$, we have for the integrated two-point function of $\langle \partial \phi \partial \phi \rangle$
\begin{align}
\begin{split}
    \int ds_1 ds_2\langle \partial_r \phi(s_1) \partial_r \phi(s_2) \rangle &= \frac{\pi}{L^2} \int_{\delta_\theta}^{2\pi-\delta_\theta} d\theta\frac{1}{(1-\cos\theta)^2} -\frac{L}{3\pi^2 \delta^3}\\
    &= \frac{1}{2L\delta}\, .
\end{split}
\end{align}
This is removed by a counterterm proportional to the curvature $K^2$,
\begin{equation}
    \delta S_\text{eff} = \frac{h_1^2r^2}{16\pi^2 \delta} \int ds\, K_i K^i\, ,
\end{equation}
rendering the correlation function zero.
A similar counterterm is necessary for the one-point function $\langle \mathcal{D}\partial_r^2\phi \rangle $, which renders it zero as well.
We can anticipate that this pattern will hold to all orders: intermediate correlation functions in the fusion expansion of $\langle\mathcal{D}_{h_1}[r] \mathcal{D}_{h_2}[0]\rangle$ will develop new divergences, but after removing them with appropriate counterterms they will all be rendered zero.

Let us now extract the curvature coefficients.
For a circle, we recall from equation \eqref{eq:K2int} that $\int ds\, K^2 = 4\pi^2/L$, which lets us read off from the second term in~\eqref{eq:MasslessScalarDefect2PtLoops}
\begin{equation}
    a_{K^2} = \frac{h_1 h_2}{32\pi}\, .
\end{equation}
In fact, for this simple free theory, we can read off all coefficients $a_{K^{2n}}$ from \eqref{eq:MasslessScalarDefect2PtLoops}. The reason is that the correlation functions of dynamical operators $\partial_r^n\phi$ on the fused defect vanish even in the circular geometry, as mentioned above, and so the terms in \eqref{eq:MasslessScalarDefect2PtLoops} are generated only by the operators $(K_i K^i)^n$.
Explicitly, by comparing the series expansion~\eqref{eq:MasslessScalarDefect2PtLoops} with the integral \eqref{eq:K2int}, we readily obtain
\begin{equation}\label{eq:aK2n}
    a_{K^{2n}} = \frac{(-1)^{n+1}h_1 h_2}{2^{4n}4\pi}\binom{2n}{n}\, .
\end{equation}

To go to $O(r^3)$, which will be relevant for the massive scalar in section~\ref{sec:massive_scalar}, one also needs the coefficients $a_{\phi K^2}$, $a_{\partial_i \phi K^2}$ and $a_{\phi^2 K^2}$.
The steps are now familiar, requiring us to compute $\langle \phi \mathcal D \mathcal D \rangle$ as well as $\langle \phi^2 \mathcal D \mathcal D \rangle$ in our new circular configuration.
We relegate the necessary correlation functions to appendix~\ref{app:FreeCFTs} and simply report
\begin{equation}
     a_{\phi K^2} = 0 \, , \qquad a_{\partial_i \phi K^2} = 0\, , \qquad a_{\phi^2 K^2} = 0\, .
\end{equation}

\paragraph*{Summary}
In summary, for the free massless $4d$ scalar we have considered the fusion of two pinning field defects $\mathcal D_h = \exp[h \int_\gamma ds\, \phi(s)]\,$ with couplings $h_1$ and $h_2$ placed at $x=r$ and $x=0$, respectively, in the limit $r\to 0$.
Using the fusion effective action for conformal defects \cite{Diatlyk:2024zkk,Diatlyk:2024qpr,Kravchuk:2024qoh,Cuomo:2024psk}, we have demonstrated that the defects fuse to a pinning field defect with coupling $h_1+h_2$ dressed by an effective action $S_\text{eff}$ of irrelevant operators living on the defect.
After identifying the correct CFT basis of operators in $S_\text{eff}$ by removing UV divergences, we have computed the coefficients of all operators appearing through $O(r^3)$ for both infinite lines and circular defects by inserting probe operators at a position $x\gg r$, yielding
\begin{equation}\label{eq:MasslessScalar-Summary}
\begin{aligned}
    S_\text{eff} = \int ds \bigg[& -\frac{h_1h_2}{4\pi r} +r \frac{h_1 h_2}{32\pi} K_i K^i -rh_1 \partial_r \phi -r^2\frac{h_1}{2!}\partial_r^2 \phi\\
    &-r^3\frac{3h_1 h_2}{512\pi}\left(K_i K^i\right)^2 - r^3 \frac{h_1}{3!} \partial_r^3 \phi + O(r^4)\bigg] \, .
\end{aligned}
\end{equation}
For more general geometries beyond circles additional curvature terms may appear, but we will content ourselves with the above. 
This action is the starting point for the more complicated case where a relevant deformation is turned on, as we will see for the massive scalar in section~\ref{sec:massive_scalar}.

\subsubsection{Free Maxwell theory}\label{sec:Maxwell}

It is also instructive to work out the fusion of Wilson lines in free Maxwell theory, which will serve as a useful starting point for the Schwinger model and QCD.
We work in $\mathbb{R}^4$ with an action $S = \frac{1}{4} \int d^4x F^{\mu\nu}F_{\mu\nu}$, where we work with canonically normalized fields. 
We consider the fusion of two Wilson lines with charges $q_1$ and $q_2$ defined as
\begin{equation}
    \tilde W_{q_i} \equiv \exp \left( ie q_i\int_\gamma dx_\mu\, A^\mu(x) \right)\, ,
\end{equation}
where the tilde again denotes the ``bare'' defect before subtracting UV divergences.
We place the defects at positions $r$ and $0$ so that we have the fusion
\begin{equation}\label{eq:WW-fusion}
    W_{q_1}[\gamma,r] W_{q_2}[\gamma,0]  \simeq W_{q_1+q_2}[\gamma,0] e^{-S_{\mathrm{eff}}}\,,
\end{equation}
where the defects fuse to a Wilson line of charge $q_1+q_2$.
Up to $O(r^3)$, the effective action is given by
\begin{align}\label{eq:FreeMaxwellSeff_initial}
\begin{split}
    S_\text{eff} = \int ds \bigg[&-\frac{a_0}{r}+ra_{F_{\mu\nu}}F^{\mu\nu}+r^2a_{\partial_i F^{\mu\nu}}\partial_i F^{\mu\nu}\\
    &+ r^3 (a_{F_{\mu\nu}F_{\rho\sigma}} F^{\mu\nu}F^{\rho\sigma} +a_{\partial_i \partial_j F^{\mu\nu}}\partial_i\partial_j F^{\mu\nu}) + O(r^4)\bigg]\, .
\end{split}
\end{align}
Just as in the massless scalar, our goal is to fix the coefficients $a_i$ appearing in this effective action.

\paragraph*{Tensor structures and the $O(2)$ symmetry}
Before applying our prescription, we fix the tensor structure of the coefficients as much as possible. 
Let us start with $a_{F_{\mu\nu}}$. 
The general form of this coefficient may be written as
\begin{equation}
    a_{F_{\mu\nu}} = a_{F_1,i}v_{[\mu} n^i_{\nu]} + a_{F_2,ij}n^i_{[\mu } n^j_{\nu]}\,,
\end{equation}
where $n^i_{\mu}$ is the $i$th vector normal to $\gamma$, introduced in appendix~\ref{app:curvature}, and $v_\mu$ the vector tangent to it (denoted by $e_s^\mu$ there). Since the fusion coefficients $a_i$ can be extracted in any geometry that we like,\footnote{As long as the operators that we target are not manifestly zero in that geometry, like with curvature corrections in infinite straight lines.} it is convenient to choose a very symmetric one. We again consider infinite straight lines with a transverse separation $r$. This configuration enjoys an $O(2)$ symmetry along the two directions transverse to the plane defined by the Wilson lines. Since the left-hand side of \eqref{eq:WW-fusion} is $O(2)$ invariant, the effective action on the straight line should also be so. We can use this symmetry to argue that several terms in the action vanish, which will continue to hold on any other curve $\gamma$.

For $F^{\mu\nu}$, the $O(2)$ symmetry implies that only the component $F^{sr}$, with indices along $\gamma$ and the transverse separation $r$,   survives.
This means that we have 
\begin{equation}
    a_{F_{\mu\nu}} = {a_F} t_{\mu\nu} \equiv a_F v_{[\mu} n^r_{\nu]} \, ,
\end{equation}
with only one unknown coefficient.
We may use a similar argument for the other operators featuring in~\eqref{eq:FreeMaxwellSeff_initial}. 
For $\partial_i F^{\mu\nu}$, the $O(2)$ symmetry allows for $\partial_r F^{sr}$, $\partial_I F^{Is}$ and $\partial_I F^{I r}$, where $I$ is an $O(2)$ vector index. But, by the bulk equations of motion $\partial_\mu F^{\mu\nu}=0$, the second operator becomes redundant with the first one, and the third one a total derivative along $\gamma$ and can be dropped. We are thus only left with
\begin{align}
    a_{\partial_i F_{\mu\nu}} = a_{\partial F} t_{\mu\nu}n_i^r\,.
\end{align}

In fact, additionally using the equations of motion to remove operators such as $\partial_r (\partial_IF_{Is}) $ and the Bianchi identity to remove $\partial^2_\perp F_{\mu\nu}$, the full tower of operators of this kind is fixed to be of the form
\begin{align}
\begin{split}
    a_{\partial_i \ldots \partial_j F_{\mu\nu}} 
    &= a_{\partial_r^n F} t_{\mu\nu} n_{i}^r\ldots n_j^r
    \,.
\end{split}
\end{align}

The final operator to consider is $F_{\mu\nu}F_{\rho\sigma}$. It decomposes into the stress tensor $T_{\mu\nu}$, the singlet $F^2$ and the component $F_{sr}^2$. The remaining singlet $\epsilon^{\mu\nu\rho\sigma}F_{\mu\nu}F_{\rho\sigma}$ is removed by the $O(2)$ symmetry.
The general form of the coupling to the stress tensor compatible with the $O(2)$ and tracelessness is given by
\begin{align}
\begin{split}
    a_{T_{\mu\nu}} &= a_{T_{rr}}n^r_{\mu}n^r_{\nu}+a_{T_{vv}} v_{\mu}v_\nu + a_{T_{rv}}n^r_{(\mu}v^{}_{\nu)}\, ,
\end{split}
\end{align}
where a contribution $\sim n^I_{\mu}n^I_\nu$ was removed by the tracelessness condition.
Combining pieces, the effective action becomes
\begin{align}\label{eq:FreeMaxwellSeff}
\begin{split}
    S_\text{eff} = \int ds \bigg[&-\frac{a_0}{r} + r a_{F} t_{\mu\nu}F^{\mu\nu} + r^2 a_{\partial F} t_{\mu\nu} \partial_r F^{\mu\nu} +r^3 a_{F^2}F^2  \\
    &+r^3 a_{F^2_{sr}}(t_{\mu\nu} F^{\mu\nu})^2 + r^3 a_{\partial_r^2 F} t_{\mu\nu} \partial_r^2 F^{\mu\nu}   + r^3a_{T_{\mu\nu}}T^{\mu\nu}+ O(r^4)\bigg]\,.
\end{split}
\end{align}

\paragraph*{Extracting coefficients}
We are now in a position to begin computing and extracting the coefficients $a_i$.
We continue in the geometry of infinite straight lines. Using Feynman gauge for the gauge field propagators, we have the two-point function
\begin{align}\label{eq:WW-Maxwell}
\begin{split}
    \log \langle \tilde W_{q_1}[r] \tilde W_{q_2}[0] \rangle &= \frac{(i e)^2 }{4\pi^2} \int ds_1\int ds_2\, v_\mu v_\nu g^{\mu\nu}\left[ \frac{q_1 q_2}{(s_1-s_2)^2+r^2} +\frac{q_1^2+q_2^2}{2(s_1-s_2)^2}\right]\\
    &= -\frac{e^2q_1q_2L}{4\pi r} -\frac{e^2(q_1^2+q_2^2)L}{4\pi^2 \delta} \, .
\end{split}
\end{align}
The second term is to be absorbed into the proper definition of our conformal defects,
\begin{equation}
    W_{q_i} \equiv \tilde W_{q_i} \exp \left(\frac{e^2 q_i^2}{4\pi^2 \delta}\int ds \right)\, .
\end{equation}

Like in the previous example, we will fix the coefficients $a_i$ by comparing three-point functions $\langle \mathcal O(x) W_{q_1}[r] W_{q_2}[0]\rangle$ before and after fusing the lines for various insertions $\mathcal O$ at a position $x\gg r$ away from the defects. 
The fusion prediction for the identity insertion is
\begin{equation}
    \langle W_{q_1}[r] W_{q_2}[0] \rangle \simeq e^{\frac{a_0}{r}\int ds} \langle W_{{q_{1}+q_{2}}}[0]\rangle\,,
\end{equation}
as all other correlation functions on the infinitely-long conformal defect vanish. Using that $\langle W_{{q_{1}+q_{2}}}\rangle = 1$ for the regulated defect, we see by comparing to \eqref{eq:WW-Maxwell} that
\begin{equation}
    a_0 = -\frac{e^2 q_1 q_2}{4\pi }\, .
\end{equation}

To extract higher fusion coefficients $a_i$, we consider nontrivial insertions $\mathcal O(x)$.
We choose to align $x$ with the $r$ direction to simplify calculations.
For the first coefficient, $a_F$, it is enough to consider $F_{\mu\nu}(x)$. Before fusing the defects, we have
\begin{align}\label{eq:FWW_FreeMaxwell}
    \frac{\langle F_{\mu\nu}(x) W_{q_1}[r] W_{q_2}[0] \rangle }{\langle W_{q_1}[r] W_{q_2}[0] \rangle} &= iet_{\mu\nu}\left[\frac{q_1+q_2}{2\pi x^2} +\frac{q_1}{\pi}\frac{r}{x^3} + \frac{3q_1}{2\pi}\frac{r^2}{x^4}+O(r^3) \right]\,.
\end{align}
This is to be compared to the fusion prediction
\begin{align}\label{eq:FreeMaxwell_FWW_fusion}
    \frac{\langle F_{\mu\nu}(x) W_{q_1}[r]W_{q_2}[0] \rangle}{\langle  W_{q_1}[r]W_{q_2}[0]\rangle } \simeq & \  \langle F_{\mu\nu}(x) W_{{q_{1}+q_{2}}}\rangle - ra_{F}\int ds \  t^{\alpha\beta} \langle F_{\mu\nu}(x) W_{{q_{1}+q_{2}}} F_{\alpha \beta}(s)\rangle + O(r^2)
\end{align}
after we compute each of these correlators. At order $O(r^0)$, the matching of the one-point function
\begin{equation}
     \langle F_{\mu\nu}(x) W_{{q_{1}+q_{2}}}\rangle = \frac{ie(q_1+q_2)}{2\pi x^2} t_{\mu\nu}
\end{equation}
with the constant term in \eqref{eq:FWW_FreeMaxwell} confirms that the fused defect is $W_{q_1+q_2}$, as we anticipated. At order $O(r)$, using
\begin{equation}
    \int ds \ t^{\alpha \beta} \langle F_{\mu\nu}(x) W_{{q_{1}+q_{2}}} F_{\alpha\beta}(s)\rangle = -\frac{1}{\pi x^3} t_{\mu\nu}\, ,
\end{equation}
and comparing to \eqref{eq:FWW_FreeMaxwell}, we can read off
\begin{equation}
    a_{F} = ieq_1\, .
\end{equation}

By inspecting~\eqref{eq:FWW_FreeMaxwell} at higher orders (see appendix~\ref{app:FreeCFTs}) we can guess the form of all coefficients $a_{\partial_{i_1}\ldots \partial_{i_n}F_{\mu\nu}}$, since only $\partial_{i_1}\ldots \partial_{i_n} F_{\mu\nu}$ contribute to the three-point function $\langle F_{\mu\nu}(x) W_{q_{1}}[r] W_{q_2}[0]\rangle $ as $r\to 0$,
\begin{equation}
    a_{\partial_{i_1}\ldots \partial_{i_n}F_{\mu\nu}} = \frac{ie q_1}{(n+1)!}t_{\mu\nu}\prod_{j=1}^n n_{i_{j}}^r\, .
\end{equation}

It should hopefully be clear now how to continue the systematic determination of the remaining $a_i$ coefficients to higher orders. For completeness, we report the remaining coefficients in \eqref{eq:FreeMaxwellSeff} to the order shown, which happen to vanish,
\begin{equation}
    a_{F^2}=0\, , \qquad  a_{F^2_{sr}} = 0\, , \qquad a_{T_{\mu\nu}} = 0\, .
\end{equation}
These follow from inserting $F^2$, $F^2_{sr}$ and $T_{\mu\nu}$ at $x$. The reader can find the details in appendix~\ref{app:FreeCFTs}.

It is worth pointing out that since this is a free theory, we should expect to reproduce the classical multipole expansion, aside from the Casimir contribution. 
This follows just from expanding the delta functions in the Wilson line definition
\begin{align}
\begin{split}
        \int dx ds &v^\mu [q_1 A_{\mu} \delta(x-r)+q_2 A_\mu \delta(x)] = \int ds v^\mu \left[(q_1+q_2)A_\mu + q_1\sum_n \frac{r^n}{n!} \partial^n_r A_\mu\right]\, .
\end{split}
\end{align}
For our configuration, this is equivalent to expanding point sources $\rho(x) = q_1\delta(x-r) + q_2 \delta(x)$ as $r \to 0$.
Rewriting this tower of generated irrelevant operators $\partial_r^n A_\mu$ in terms of $F_{\mu\nu}$ and including the $-ie$ prefactor immediately gives us $a_{\partial_r^n F_{\mu\nu}} = \frac{ieq_1 }{(n+1)!}t_{\mu\nu}$, as we found above.
This observation has been known and widely used in the QCD literature since the late 1970's~\cite{Voloshin:1978hc,PhysRevLett.40.598,Peskin:1979va}, though there it only gives the leading contribution to this specific set of coefficients, as they receive higher order corrections in perturbation theory.

\paragraph*{Summary}
In summary, we have considered the fusion of Wilson lines $W_{q_1}[r]$ and $W_{q_2}[0]$ as $r\to 0$ in free Maxwell theory.
The fusion ansatz predicts that these fuse to a Wilson line $W_{q_1+q_2}$ dressed by an effective action $S_\text{eff}$ of operators living on the fused line.
We first fixed the tensor structure of the operators appearing in $S_\text{eff}$ by using the $O(2)$ transverse symmetry, and identified the correct CFT basis by removing UV divergences.
By inserting probe operators at a position $x \gg r$, we then extracted all coefficients $a_i$ in $S_\text{eff}$ that may contribute for circular or infinite straight defects up to $O(r^3)$, obtaining the effective action
\begin{equation}\label{eq:Seff_FreeMaxwell}
\begin{aligned}
    S_\text{eff} = \int ds \bigg[ & \frac{e^2 q_1 q_2}{4\pi r}+r\left(ieq_1 t_{\mu\nu} F^{\mu\nu} +\frac{3e^2q_1q_2}{32\pi}K_i K^i\right)+r^2\frac{ieq_1}{2} t_{\mu\nu} \partial_r F^{\mu\nu}\\
    & + r^3 \left( \frac{ieq_1}{6}t_{\mu\nu} \partial_r^2 F^{\mu\nu} - \frac{5e^2q_1q_2}{512\pi} (K_i K^i)^2\right) +O(r^4)\,\bigg] .
\end{aligned}
\end{equation}
Here $t_{\mu\nu} = v^{}_{[\mu}n^r_{\nu]}$, with $v_{\mu}$ the tangent vector to the defect and $n^r_{\mu}$ the normal vector in the $r$ direction.
In~\eqref{eq:Seff_FreeMaxwell} we have additionally included the coefficients of the operators appearing in the circular geometry, which are computed in appendix~\ref{app:FreeCFTs}.

\subsection{Massive scalar}\label{sec:massive_scalar}

Having worked out explicitly the fusion of defects in the conformal limit, we now turn to massive theories. As a first test of the framework introduced in section~\ref{sec:Lines}, we consider the same example from section \ref{sec:OPE_massive_scalar}: the free massive $4d$ scalar. We take the same pinning field (conformal) defects $\mathcal D_{h}$ from section \ref{sec:FreeCFT-scalar}, but we now turn on a relevant deformation in the bulk as in \eqref{eq:g-defSec3} with $g\varPhi \equiv \frac{m^2}{2}\phi^2$. We do not turn any relevant deformations localized on the defects. 

Consider the correlator $\langle \mathcal D_{h_1}[r] \mathcal D_{h_2}[0]\rangle_\text{QFT}$ of parallel defects in the shape of infinite straight lines with a transverse separation $r$. Since the theory is free, we can compute this correlator exactly. It is given by the same three diagrams as the massless scalar in figure~\ref{fig:MasslessScalar_DD}, but with the full massive propagator. The second and third diagrams evaluate to
\begin{align}\label{eq:MassiveScalar_Dhi}
\begin{split}
    \log \langle \mathcal{D}_{h_i}\rangle_\text{QFT} &= \frac{h_i^2 L}{8\pi^2}\int_{-\infty}^{\infty}ds\frac{mK_1(m |s|)}{|s|}-\frac{h_i^2 L}{4\pi^2\delta} = -\frac{h_i^2 L m}{8\pi}\, ,
\end{split}
\end{align}
where we have regulated the Bessel function integral with the same regulator $|s|>\delta$ as in the massless case, and used the definition of the conformal defect including the cosmological constant counterterm from~\eqref{eq:MasslessScalarDefectRenorm}. Combining this result with the first diagram, we obtain
\begin{align}
    \log \langle \mathcal{D}_{h_1}[r]\mathcal D_{h_2}[0] \rangle_\text{QFT} &= \frac{h_1 h_2m}{4\pi^2}\int  ds_1 \int ds_2 \frac{K_1(m\sqrt{r^2+(s_1-s_2)^2})}{\sqrt{r^2+(s_1-s_2)^2}}-\frac{(h_1^2+h_2^2) L m}{8\pi} \nonumber\\
    &= \frac{h_1 h_2 L}{4\pi r}e^{-mr} -\frac{(h_1^2+h_2^2) L m}{8\pi}\, ,
\end{align}
which is just the dimensionally reduced propagator with the additional $\langle \mathcal D_{h_i} \rangle $ contributions. 

We are interested in the short-distance limit, $r\to 0$, of this correlator,
\begin{align}
    \langle \mathcal{D}_{h_1}[r]\mathcal D_{h_2}[0] \rangle_\text{QFT} =\,& e^{\frac{h_1h_2L}{4\pi r}-\frac{(h_1+h_2)^2Lm}{8\pi}} \bigg[1+\frac{h_1 h_2 L}{8\pi}m^2r\nonumber\\
    &+\frac{1}{2}\left(\frac{h_1h_2L}{8\pi}\right)^2 m^4 r^2-\frac{h_1 h_2 L}{24\pi}m^3 r^2 \\
    &+\frac{1}{3!}\left(\frac{h_1h_2L}{8\pi}\right)^3m^6r^3+\frac{h_1h_2 L }{96\pi}m^4r^3-\frac{h_1^2 h_2^2 L^2}{192\pi^2}m^5 r^3+O(r^4)\bigg]\,.\nonumber
\end{align}
We aim to reproduce this behavior from the UV fixed point of the flow, i.e.\ the massless theory. Clearly, naive conformal perturbation theory cannot do the job, as the relevant coupling is $g\sim m^2$, and so odd powers of $m$ are manifestly non-perturbative. We will show that these terms come from IR condensates of the fused defect. For later convenience, 
it will be useful to regroup even and odd powers of $m$ in the short-distance expansion in the following form:
\begin{align}\label{eq:DD_massive_scalar_full}
    \langle \mathcal D_{h_1}[r] \mathcal D_{h_2}[0] \rangle_\text{QFT} = e^{\frac{h_1h_2L}{4\pi r}\cosh {m r}}e^{-\frac{(h_1+h_2)^2Lm}{8\pi}} \bigg[1-\frac{h_1 h_2 L}{24\pi}m^3 r^2+O(r^4)\bigg]\, .
\end{align}

According to the discussion from section~\ref{sec:Lines}, the above short-distance expansion should be captured by a defect OPE of the form
\begin{equation}\label{eq:dOPE-Mass-scalar}
    \mathcal D_{h_1}[\gamma,r] \mathcal D_{h_2}[\gamma,0] \simeq \mathcal{D}_{h_1+h_2}[\gamma,0]e^{-S_{\text{eff}}[\gamma]}\,,
\end{equation}
with an effective action nearly identical to the one from the massless case \eqref{eq:MasslessScalarSeff}, but with coefficients $a_i$ promoted to functions of $m^2r^2$ with an expansion in integer powers of $m^2$,
\begin{align}\label{eq:MassiveScalarSeff}\nonumber
    S_\mathrm{eff}
    = \int ds \bigg[& -\frac{a_0}{r}+a_{\phi} \phi + r\left(a_{\partial_i  \phi} \partial_i {\phi} + a_{\phi^2}  \phi^2\right)+r^2\left(a_{\partial_i\partial_j  \phi} \partial_i \partial_j  \phi + a_{ \phi^3}  \phi^3 + a_{\partial_i \phi^2}\partial_i ( \phi^2)\right) \\\nonumber
    &+r^3\Big(a_{\partial_i\partial_j\partial_k \phi}\partial_i\partial_j\partial_k\phi + a_{\phi \partial_i \partial_j\phi}\phi \partial_i\partial_j \phi + a_{\partial_i \phi^3}\partial_i(\phi^3)+ a_{\phi^4}\phi^4 +a_{T_{\mu\nu}}T^{\mu\nu} \\
    &+ a_{(\partial \phi)^2} (\partial_s \phi)^2\Big)+O(r^4)\bigg]\, .
\end{align}
Note the presence of the marginal operator $\phi$ here, as its coefficient $a_{\phi}$ may also in general experience corrections proportional to $m^2 r^2$. We will determine the fusion coefficients in this action using conformal perturbation theory, and show that \eqref{eq:dOPE-Mass-scalar} reproduces \eqref{eq:DD_massive_scalar_full} upon supplementing the condensates of the fused defect.

Expanding out each $a_i$ coefficient in $m^2$, and using our knowledge from the CFT limit \eqref{eq:MasslessScalar-Summary}, the action through order $O(r^3)$ becomes
\begin{align}\label{eq:MassiveScalarSeff_Expanded}
\begin{split}
    S_\text{eff}= \int ds \bigg[&-\frac{h_1h_2}{4\pi r}\mathds 1+r\left(-a_0^{(1)}\mathds 1-h_1 \partial_r \phi\right)+r^2\left(a^{(1)}_{\phi} \phi-\frac{h_1}{2!}\partial_r^2 \phi\right) \\
    &+ r^3\left(-a_0^{(2)}\mathds 1 - \frac{h_1}{3!} \partial_r^3 \phi +a_{\partial_i  \phi}^{(1)} \partial_i {\phi} + a^{(1)}_{\phi^2}\phi^2  \right)+O(r^4)\bigg]\, ,
\end{split}
\end{align}
where $a^{(n)}$ is proportional to $m^{2n}$ and will enter at order $n$ in CPT.
Note that the $a_i^{(n)}$'s may contain logarithms of $r$.
We see that we will need to compute $a_{\phi}$, $a_{\partial_i\phi}$, and $a_{\phi^2}$ to $O(m^2)$ as well as $a_{0}$ to $O(m^4)$ to determine $S_\text{eff}$ to $O(r^3)$. The computation will proceed order by order in CPT, matching correlator functions of the defects in the presence of bulk insertions before and after the fusion, as in \eqref{eq:prescription_defect_fusion}. We will confirm that fusion coefficients are local functions of $r$; free from IR divergences and independent of global details of the defects such as their length $L$.

\subsubsection{Fusion coefficients}
The $0$th order matching in CPT amounts to fusion in the massless limit, which we already addressed in section~\ref{sec:FreeCFT-scalar}. At first order in CPT, we have the unknowns $a_0^{(1)}$, $a_\phi^{(1)}$, $a_{\partial_i \phi}^{(1)}$ and $a_{\phi^2}^{(1)}$. We will extract them from the correlation functions $\langle \mathcal O_k(x) \mathcal D_{h_1}[r] \mathcal D_{h_2}[0]\rangle_\text{CPT}^{(1)}$ for the insertions $\mathcal O_k = \mathds 1$, $\phi$ and $\phi^2$.

\begin{figure}[t]
    \centering
    \input{Figures/DD_MassiveScalar_CPT_1.tikz}
    \caption{Diagrams contributing to the $\langle \mathcal{D}_{h_1}[r]\mathcal{D}_{h_2}[0] \rangle_\text{CPT}^{(1)} $ correlator at first order in CPT. Red crosses indicate insertions of the relevant deformation $\frac{-m^2}{2}\int d^4 y\, \phi^2(y)$.}
    \label{fig:MassiveScalar_DD_CPT_1}
\end{figure}
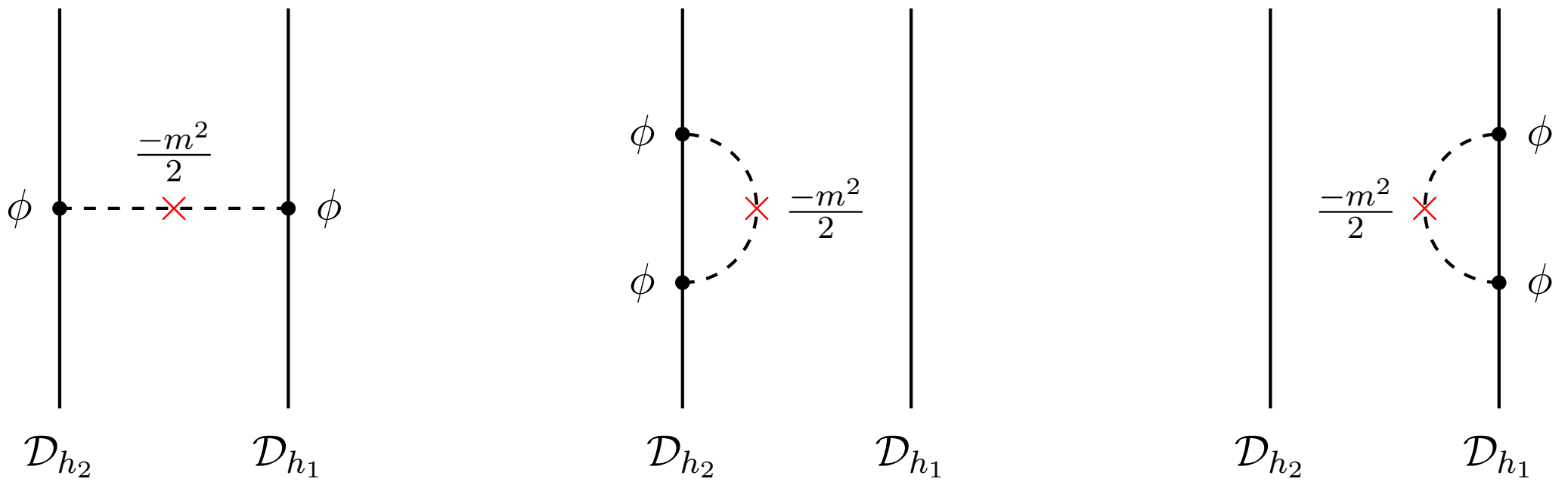

Let us start with the identity. The relevant diagrams are given in figure~\ref{fig:MassiveScalar_DD_CPT_1}, where the crosses indicate insertions of the relevant deformation $\frac{-m^2}{2}\int d^4y \,\phi^2(y)$. As discussed in section~\ref{sec:Lines}, we restrict the relevant deformation inside a ball of radius $R$ to regulate IR divergences. We regulate integrals $\int ds$ along the defects by a separate length $L$. While the matching can be performed at finite $R,L$, we work in the limit of $R\sim L \gg r$ to simplify expressions. Fusion coefficients will be IR-finite objects, independent of both $R$ and $L$.

The second and third diagrams evaluate to
\begin{align}
\begin{split}
    \log \langle \mathcal{D}_h \rangle_\text{CPT}^{(1)} &= \frac{h^2}{(4\pi^2)^2} \left(\frac{-m^2}{2}\right)\int ds_1 \int ds_2 \int d^4 y \frac{1}{|y-s_1 \hat{z}|^2}\frac{1}{|s_2 \hat{z}-y|^2} \\
    &= -\frac{h^2 m^2 L^2}{32\pi^2}\left(3+2\log\frac{2R}{L}\right)\,.
\end{split}
\end{align}
Combining them with the first diagram, we obtain
\begin{align}\label{eq:DD_CPT_1}
\begin{split}
    \log \langle \mathcal D_{h_1}[r] \mathcal D_{h_2}[0] \rangle_\text{CPT}^{(1)} &= \log\langle \mathcal{D}_{h_1} \rangle_\text{CPT}^{(1)} + \log\langle \mathcal D_{h_2} \rangle_\text{CPT}^{(1)} \\
    & \quad- \frac{h_1 h_2 m^2L}{16\pi^4}\int ds \int d^4 y \frac{1}{|y-r\hat{x}-s\hat{z}|^2 }\frac{1}{|y|^2}\\
    &= \frac{h_1h_2 Lm^2 r}{8\pi} -\frac{(h_1+h_2)^2 m^2 L^2}{32\pi^2}\left(3+2\log \frac{2R}{L}\right)\, ,
\end{split}
\end{align}
where we have dropped terms in the brackets that vanish as $L \to \infty$ after pulling out the extensive factor of $L$ from one of the $s$ integrals. This is to be reproduced by the expansion of the fusion effective action \eqref{eq:MassiveScalarSeff_Expanded}, computed at the same order in CPT,
\begin{align}\label{eq:MassiveScalarIdentityExpansion_1}
\begin{split}
    \log\langle \mathcal{D}_{h_1}[r] \mathcal D_{h_2}[0] \rangle_\text{CPT}^{(1)} \simeq & \ \log\langle  \mathcal  D_{{h_1+h_2}} \rangle^{(1)}_\text{CPT} +a_0^{(1)}rL + O(r^2)\,.
\end{split}
\end{align}
Here we dropped a contribution from $\langle \mathcal{D}_{h_1+h_2}\partial_r \phi\rangle_\text{CPT}^{(1)}$, which vanishes by rotation invariance around the fused defect. Matching the two expansions, we see that the IR divergent term is precisely canceled by $\log \langle \mathcal D_{h_1+h_2}\rangle $, letting us read off 
\begin{equation}
    a_0^{(1)} = \frac{h_1 h_2 m^2}{8\pi}\, .
\end{equation}

Note that the perturbative defect two-point function \eqref{eq:DD_CPT_1} is valid to all orders in $r$, while in \eqref{eq:MassiveScalarIdentityExpansion_1} we truncated the expansion of the fusion effective action to order $O(r)$. The absence of $r^2$ and higher terms in \eqref{eq:DD_CPT_1} is reproduced from the fused side by the vanishing of all remaining perturbative correlation functions on the fused defect. For instance, at order $O(r^2)$ both $\langle \mathcal D_{h_1+h_2} \partial_r^2 \phi\rangle_\text{CPT}^{(1)} $ and $\langle \mathcal D_{h_1+h_2} \partial_r \phi\partial_r \phi\rangle_\text{CPT}^{(1)} $ appear. 
The naive diagrams are UV divergent,
\begin{subequations}
\begin{align}
    \int ds \frac{\langle \mathcal{D}_{{h_1+h_2}} \partial_r^2\phi(s) \rangle^{(1)}_\text{CPT} }{\langle \mathcal D_{{h_1+h_2}} \rangle_\text{CPT} }
    & = \frac{(h_1+h_2)m^2L}{4\pi^2 \delta} \,,\\
    \int ds_1 \int ds_2\frac{\langle \mathcal D_{{h_1+h_2}} \partial_r \phi(s_1) \partial_r \phi(s_2) \rangle^{(1)}_\text{CPT}}{\langle \mathcal D_{{h_1+h_2}} \rangle_\text{CPT} }
    &= -\frac{m^2 L}{4\pi^2 \delta} \, ,
\end{align}
\end{subequations}
where we used the same UV cutoff $\delta$ as in previous sections. The first divergence is removed by defining the renormalized operator
\begin{equation}\label{eq:dr2phi-Renorm}
    [\partial_r^2 \phi(s)]_\text{r} \equiv \partial_r^2 \phi(s) - \frac{(h_1+h_2)m^2}{4\pi^2 \delta} \mathds 1 \, .
\end{equation}
The second is removed by adding to the action the counterterm
\begin{equation}\label{eq:drphidrphi-Renorm}
    \delta S_\text{eff}^{(1)} = -\frac{h_1^2m^2 r^2}{8\pi^2\delta} \int ds\,\mathds{1}\, ,
\end{equation}
where we have included the $\frac{1}{2}h_1^2 r^2$ prefactor that multiplies the correlator. After renormalization, both correlators are rendered zero. This pattern continues to higher orders,\footnote{This follows from dimensional analysis. The only scales in the fused defect are $R,L$ and $m$. CPT fixes the $m^2$ dependence, resulting in $\tfrac{1}{L}\langle \mathcal D_{h_1+h_2} \int \mathcal O_1\cdots\int \mathcal O_n\rangle_\text{CPT}^{(1)} \sim f(R/L)m^2 L^{1+\sum_{i=1}^n(1-\Delta_i)}\to0$, where $f(R/L)$ is an unknown dimensionless function and correlators vanish because the remaining operators have large enough dimension. Notice that these zeros are due to our simplifying limit $L\sim R\to \infty$. Both \eqref{eq:DD_CPT_1} and \eqref{eq:MassiveScalarIdentityExpansion_1} would have additional matching terms when working at finite $L,R$. 
} and so the matching of \eqref{eq:DD_CPT_1} and \eqref{eq:MassiveScalarIdentityExpansion_1} does not fix any further coefficients.

Next, we insert $\phi(x)$. The first-order three-point function is given by
\begin{align}\label{eq:Massless_Scalar_phiDD_CPT}
\begin{split}
    \frac{\langle \phi(x)\mathcal D_{h_1} [r]\mathcal D_{h_2}[0]\rangle^{(1)}_\text{CPT}}{\langle \mathcal D_{h_1} [r]\mathcal D_{h_2}[0]\rangle_\text{CPT}} =& \  \frac{(h_1+h_2)m^2}{16\pi^2}\left[2\pi x-L\left(3+2\log\frac{2R}{L}\right)\right]-\frac{h_1 m^2r}{8\pi}\,,
\end{split}
\end{align}
where we again are working in the simplifying limit $r \ll x\ll L\sim R$.\footnote{This limit places the bulk insertion inside the CPT ball from figure \ref{fig:CPTball}. This is not an issue per se, the insertion can be anywhere. One just needs to remember that bulk operators inside the ball might need renormalization. $\phi(x)$ does not.}
Upon fusing the defects, this should be reproduced by
\begin{align}\label{eq:MassiveScalarIdentityExpansion_2}
\begin{split}
    \frac{\langle \phi(x)\mathcal{D}_{h_1}[r] \mathcal D_{h_2}[0] \rangle_\text{CPT}^{(1)} }{\langle \mathcal D_{h_1} [r]\mathcal D_{h_2}[0]\rangle_\text{CPT}}\simeq & \ \frac{\langle  \phi(x)\mathcal  D_{{h_1+h_2}} \rangle^{(1)}_\text{CPT}}{\langle \mathcal D_{h_1+h_2} \rangle_\text{CPT}} +r h_1\int ds \frac{\langle  \phi(x)\mathcal  D_{{h_1+h_2}}\partial_r \phi(s) \rangle^{(1)}_\text{CPT}}{\langle \mathcal D_{h_1+h_2} \rangle_\text{CPT}}
    \\
    &\ -r^2 a_\phi^{(1)}\int ds\frac{\langle  \phi(x)\mathcal  D_{{h_1+h_2}} \phi(s) \rangle^{(0)}_\text{CPT}}{\langle \mathcal D_{h_1+h_2} \rangle_\text{CPT}}+O(r^3)\,,
\end{split}
\end{align}
where we dropped the vanishing correlators $\langle  \phi\mathcal  D_{{h_1+h_2}} \partial_r^2\phi \rangle^{(1)}_\text{CPT}$ and $\langle  \phi\mathcal  D_{{h_1+h_2}} \int\partial_r\phi\partial_r\phi\rangle^{(1)}_\text{CPT}$.
Feeding in the correlation functions on the fused defect,
\begin{subequations}
\begin{align}
    \frac{\langle \phi(x)\mathcal{D}_{{h_1+h_2}}\rangle^{(1)}_\text{CPT}}{ \langle \mathcal D_{{h_1+h_2}}\rangle_{\text{CPT}} }&= \  \frac{(h_1+h_2)m^2}{16\pi^2}\left[2\pi x-L\left(3+2\log\frac{2R}{L}\right)\right]\, ,\\
     \int ds\frac{\langle \phi(x)\mathcal{D}_{{h_1+h_2}}\partial_r\phi(s)\rangle^{(1)}_\text{CPT}}{\langle\mathcal D_{{h_1+h_2}}\rangle_{\text{CPT}} } &= -\frac{m^2}{8\pi}\, ,\\
     \int ds\frac{\langle \phi(x)\mathcal{D}_{{h_1+h_2}}\phi(s)\rangle^{(0)}_\text{CPT}}{\langle\mathcal D_{{h_1+h_2}}\rangle_{\text{CPT}} } &= \frac{1}{4\pi x}\, ,
\end{align}
\end{subequations}
we see that the first two terms fully account for \eqref{eq:Massless_Scalar_phiDD_CPT}, leaving us with
\begin{equation}
    a_\phi^{(1)} = 0\,.
\end{equation}

Continuing to $O(r^3)$, the steps are now familiar.
We first consider higher orders in\eqref{eq:MassiveScalarIdentityExpansion_2}, giving us access to the coefficient of $\partial_i\phi$.
Likewise, the coefficient of $\phi^2$ follows from the matching of the correlation function $\langle \phi^2(x) \mathcal{D}_{h_1}[r] \mathcal{D}_{h_2}[0]\rangle$ to the same order.
We relegate these computations to appendix~\ref{app:MassScalar}, finding
\begin{equation}
    a_{\partial_i\phi}^{(1)}=0 \, , \qquad a_{\phi^2}^{(1)} = 0\, .
\end{equation}

To fix the potential through $O(r^3)$, we also need the Casimir coefficient at second order in CPT.
The extraction follows the same steps as at first order, coming from the correlation function of $\langle\mathcal{D}_{h_1}[r]\mathcal{D}_{h_2}[0]\rangle^{(2)}$ (see appendix~\ref{app:MassScalar}), yielding
\begin{align}
    a_{0}^{(2)} = \frac{h_1 h_2 m^4}{96\pi}\, .
\end{align}

From this, we can notice a pattern starting to emerge.
At each order in CPT for $\log \langle \mathcal D_{h_1}[r] \mathcal D_{h_2}[0] \rangle^{(n)}_\text{CPT}$, there is a scheme-independent highest order in $r$ term that reproduces the expansion of the $\cosh mr$ from the full result \eqref{eq:DD_massive_scalar_full}.\footnote{Any potential terms at a higher order in $r$, but the same order in CPT, vanish in the $R,L\to \infty$ limit by dimensional analysis.} 
On the fused defect side, these terms are matched by $a_0^{(n)}$, while subleading terms contain IR divergences matched by correlation functions of $\partial_r^k\phi(s)$ on the defect. This lets us guess the full form
\begin{align}
    a_0(m^2r^2) = \frac{h_1 h_2}{4\pi}\left(1+\frac{1}{2}m^2 r^2 + \frac{1}{24}m^4 r^4 + \ldots\right)= \frac{h_1 h_2}{4\pi} \cosh(m r)\, .
\end{align}

\subsubsection{Condensates}
Having fixed all necessary coefficients up to $O(r^3)$, we are in a position to use the defect OPE \eqref{eq:dOPE-Mass-scalar} to predict the short-distance limit of the defect two-point function in the full QFT. We have
\begin{align}\label{eq:MassiveScalar-CPT-prediction}
    \langle \mathcal{D}_{h_1}[r] \mathcal D_{h_2}[0] \rangle_\text{QFT} \simeq & \,
    e^{\frac{h_1h_2L}{4\pi r}\cosh {m r}}\bigg[\langle  \mathcal  D_{{h_1+h_2}} \rangle_\text{QFT} + \frac{h_1}{2} r^2\int ds\,  \langle \mathcal D_{{h_1+h_2}} [\partial_r^2 \phi(s)]_\text{r}\rangle_\text{QFT}\nonumber\\
    &+ \frac{h_1^2}{2} r^2 \int ds_1 \int ds_2\,\langle \mathcal D_{{h_1+h_2}} [\partial_r \phi(s_1)\partial_r \phi(s_2)]_\text{r} \rangle_\text{QFT}+O(r^4)\bigg]\,.
\end{align}
This is as far as conformal perturbation theory can take us. We see that this expansion correctly captures all the $r$-dependence of the full result \eqref{eq:DD_massive_scalar_full}, but its coefficients are non-perturbative condensates on the fused defect.

Of course, calling condensates in a Gaussian theory ``non-perturbative'' is somewhat amusing. We mean that they lie beyond the reach of CPT around the massless theory. They become perturbative in terms of the massive propagator and can be evaluated explicitly.
We already evaluated the one-point function $\langle\mathcal D_{h_1+h_2}\rangle_\text{QFT}$ in \eqref{eq:MassiveScalar_Dhi}.
For the next condensate, we have
\begin{align}
    \frac{\langle \mathcal D_{{h_1+h_2}} [\partial_r^2 \phi(0)]_\text{r} \rangle_\text{QFT}}{\langle\mathcal D_{h_1+h_2}\rangle_\text{QFT}} =& \ (h_1+h_2) \left[-\frac{m^2}{4\pi^2}\int ds \frac{K_2(m|s|)}{|s|^2} +\frac{1}{3\pi^2 \delta^3} - \frac{m^2}{4\pi^2 \delta}\, \right]\nonumber\\
    &= -\frac{(h_1+h_2)m^3}{12\pi}\, ,
\end{align}
where the $1/\delta^3$ term came from our CFT definition for $\partial_r^2 \phi$ and the $1/\delta$ term came from the renormalization \eqref{eq:dr2phi-Renorm}.
The divergences in the Bessel function integral are precisely canceled by the counterterms we determined before, yielding a finite answer.

For the final condensate, we have a similar integral
\begin{align}
\begin{split}
    \int ds\, \frac{\langle \mathcal D_{{h_1+h_2}} [\partial_r \phi(s) \partial_r \phi(0)]_\text{r} \rangle_\text{QFT}}{\langle\mathcal D_{h_1+h_2}\rangle_\text{QFT}} &=  \left[\frac{m^2}{4\pi^2}\int ds\frac{K_2(m|s|)}{|s|^2} - \frac{1}{3\pi^2 \delta^3} + \frac{m^2}{4\pi^2 \delta}\right]\\
    &= \frac{m^3}{12\pi}\, ,
\end{split}
\end{align}
where the $1/\delta^3$ term again came from the CFT definition and the $1/\delta$ term from the counterterm \eqref{eq:drphidrphi-Renorm}. Notice that the condensates satisfy the relation
\begin{equation}\label{eq:dpdp=ddp-relation}
    (h_1+h_2) \int ds\, \langle \mathcal D_{{h_1+h_2}} [\partial_r \phi(s) \partial_r \phi(0)]_\text{r} \rangle_\text{QFT} = -\langle \mathcal D_{{h_1+h_2}} [\partial_r^2 \phi(0)]_\text{r} \rangle_\text{QFT} \,.
\end{equation}
This follows from translation invariance, as discussed in section~\ref{sec:translation_invariance}. The displacement operator on $\mathcal D_{h_1+h_2}$ is $D_i = (h_1+h_2) \partial_i \phi$, so this relation is a particular example of the identity~\eqref{eq:Di-Ward-id}.

We can finally plug the condensates into the CPT prediction \eqref{eq:MassiveScalar-CPT-prediction} to obtain
\begin{align}
\begin{split}
    \langle \mathcal{D}_{h_1}[r] \mathcal D_{h_2}[0] \rangle_\text{QFT}
    \simeq e^{\frac{h_1h_2L}{4\pi r}\cosh {m r}}e^{-\frac{(h_1+h_2)^2Lm}{8\pi}}\left[1-\frac{h_1 h_2 L}{24\pi}m^3 r^2 + O(r^4)\right]\, ,
\end{split}
\end{align}
beautifully reproducing the short-distance expansion of the full result \eqref{eq:DD_massive_scalar_full}. 
Note the cancellation between the two $\sim h_1^2$ terms at order $O(r^2)$.
If we had chosen a different configuration for our defects, such as $\mathcal D_{h_1}[r/2]\mathcal D_{h_2}[-r/2]$, the individual coefficients $a_{\partial_r \phi}$ and $a_{\partial_r^2 \phi}$ would change in such a way that this cancellation always gives a result independent of the overall location of the defects, as discussed in section~\ref{sec:translation_invariance}.
Generally, only when all terms contributing at any given order in $r$ are included does the result become independent of scheme and location choices.

\subsubsection{Curvature corrections}
Now let us fix the curvature corrections to the effective action \eqref{eq:MassiveScalarSeff_Expanded} using circular defects. We follow the same strategy as in section \ref{sec:FreeCFT-scalar}.
First, we have the expectation value of the defect itself,
\begin{align}
\begin{split}
    \log \langle \mathcal{D}_h \rangle_\text{circ}^\text{QFT} &=\frac{h^2 m L}{16\pi^2} \int_{\delta_\theta}^{2\pi-\delta_\theta} \hspace{-20pt }d\theta \hspace{5pt} \,\frac{K_1(\frac{mL}{\pi}|\sin\frac{\theta}{2}|)}{|\sin\frac{\theta}{2}|} -\frac{h^2 L}{4\pi^2 \delta}\\
    &= -\frac{h^2 m^2 L^2}{16\pi^2}\left[I_0\left(\frac{mL}{2\pi}\right)K_0\left(\frac{mL}{2\pi}\right)+I_1\left(\frac{mL}{2\pi}\right)K_1\left(\frac{mL}{2\pi}\right)\right]\,,
\end{split}
\end{align}
where we have subtracted the divergent contribution absorbed in the definition of the defect \eqref{eq:MasslessScalarDefectRenorm},\footnote{We must be careful to regulate the divergent part of the $\theta$ integral using the same choice of $\delta$ as in the infinite line case, i.e. $\delta_\theta = 2\sin^{-1}(\frac{\pi \delta}{L})$.}
and $I_\nu(z),K_\nu(z)$ are modified Bessel functions.
Next, the full two-point function of circular defects is given by
\begin{align}
\begin{split}
    \log\frac{\langle \mathcal D_{h_1}[r] \mathcal D_{h_2}[0] \rangle_\text{circ}^\text{QFT} }{\langle \mathcal D_{h_1} \rangle_\text{circ}^\text{QFT}\langle \mathcal D_{h_2} \rangle_\text{circ}^\text{QFT}}
    &= \frac{h_1 h_2 L^2m}{8\pi^3} \int_0^{2\pi} \hspace{-10pt}d\theta \, \frac{K_1\left(m\sqrt{r^2+\frac{L^2}{\pi^2}\sin^2\frac{\theta}{2}}\right)}{\sqrt{r^2+\frac{L^2}{\pi^2}\sin^2\frac{\theta}{2}}} 
    \equiv \frac{h_1 h_2 L^2 m}{8\pi^3}f(r)\,,
\end{split}
\end{align}
where we defined $f(r)$ to be the Bessel function integral (with $\rho\equiv \sqrt{r^2+L^2/\pi^2}$)
\begin{equation}
    f(r) = \frac{2\pi}{r \rho}\left[\frac{\rho+r}{2}I_0\left(m\frac{\rho-r}{2}\right)K_1\left(m\frac{\rho+r}{2}\right)+\frac{\rho-r}{2}I_1\left(m\frac{\rho-r}{2}\right)K_0\left(m\frac{\rho+r}{2}\right)\right]\,.
\end{equation}

Expanding the correlator for small values of $r$ gives
\begin{align}\label{eq:Massless_Scalar_DD_Circle_Full}
\begin{split}
    \log\langle \mathcal D_{h_1}[r] &\mathcal D_{h_2}[0] \rangle_\text{circ}^\text{QFT} \\
    \simeq& \ \frac{h_1 h_2L}{4\pi r}-\frac{(h_1 + h_2)^2 L^2 m^2 }{16\pi^2}\left[I_0K_0+I_1K_1\right]\\
    &+h_1 h_2 r\left[\frac{L m^2}{8\pi}-\frac{\pi}{8L}\right]\\
    &+h_1 h_2 r^2\bigg[\frac{m^2}{12} I_1 K_1+\frac{m^3 L}{48\pi}\left(I_1K_0-I_0K_1\right)-\frac{m^4 L^2}{48\pi^2}\left(I_0K_0+I_1K_1\right)\bigg]\\
    &+h_1 h_2 r^3 \left[\frac{Lm^4}{96\pi}-\frac{m^2 \pi}{48L}+\frac{3\pi^3 }{32L^3}\right]+O(r^4)\,,
\end{split}
\end{align}
where we have suppressed the common argument of the Bessel functions $I_{\nu} \to I_{\nu}(\frac{m L}{2\pi})$ and $K_{\nu}\to K_{\nu}(\frac{m L}{2\pi})$ for brevity.
This expression is already highly suggestive; the Bessel function contributions will be identified with operator insertions from $S_\text{eff}$, while the other terms are purely perturbative.
One can check that in the limit $L\to \infty$ this expression reduces to the infinite line result~\eqref{eq:DD_massive_scalar_full}.

Let us extract the coefficients of the curvature corrections and verify our intuition.
Up to $O(r^3)$, the defect OPE gives for this geometry
\begin{align}\label{eq:Massive_scalar_DD_matching_circ}
    \langle \mathcal{D}_{h_1}[r] \mathcal D_{h_2}[0] \rangle_{\text{circ}} \simeq& \,
    \exp\left[\int_\gamma ds\left(\frac{a_0}{r}- a_{K^2} r K_i K^i- a_{K^4} r^3 (K_i K^i)^2\right)+O(r^4)\right]\\
    &\times\bigg[\langle \mathcal  D_{{h_1+h_2}} \rangle_{\text{circ}} + \frac{h_1h_2}{2(h_1+h_2)}r^2\int_\gamma ds\,\langle \mathcal D_{{h_1+h_2}} [\partial_r^2 \phi(s)]_\text{r} \rangle_{\text{circ}} +O(r^4)\bigg]\,,\nonumber
\end{align}
where we used the previously determined coefficients $a_{\partial_i\phi}, a_{\partial_{i}\partial_j \phi}$  and implemented the relation \eqref{eq:dpdp=ddp-relation}, which holds for any defect geometry. We further set $a_{\phi K^2}=0$ invoking the CFT result because its first $m^2$ correction enters at $O(r^4)$.
From the CFT limit \eqref{eq:aK2n} we also know the leading order prediction $a_{K^{2n}}^{(0)} = \frac{(-1)^{n+1}h_1 h_2}{2^{4n}4\pi}\binom{2n}{n}$, so the only unknown coefficient at $O(r^3)$ is the first correction $a_{K^2}^{(1)}$.
To extract this coefficient, we simply need to compute at first order in CPT
\begin{align}
    \log \langle \mathcal D_{h_1}[r]\mathcal D_{h_2}[0]\rangle_\text{circ}^{(1)} &= \log \langle  \mathcal D_{h_1} \rangle_\text{circ}^{(1)}\langle \mathcal D_{h_2} \rangle_\text{circ}^{(1)}-\frac{h_1h_2m^2L^2}{32\pi^3}\int d\theta\left[1+\log \frac{R^2}{r^2+\frac{L^2}{\pi^2}\sin^2\frac{\theta}{2}}\right] \nonumber \\
    &=-\frac{(h_1+h_2)^2  m^2L^2}{32\pi^2}\left(1+2\log \frac{2\pi R}{L} \right) \\
    & \quad\, + \frac{h_1 h_2 m^2L}{8\pi}r-\frac{h_1h_2 m^2 \pi}{48 L}r^3 +O(r^5)\,. \nonumber
\end{align}
The first term is just $\log \langle \mathcal D_{{h_1+h_2}} \rangle_\text{circ}^{(1)}$ and the second term is already included in the $\cosh mr$ we computed before for $a_0(m^2r^2)$. 
Matching the $r^3$ term to~\eqref{eq:Massive_scalar_DD_matching_circ} and using $\int ds K_i K^i = \frac{4\pi^2}{L}$ immediately gives the first correction to $a_{K^2}$,
\begin{equation}
    a_{K^2}(m^2r^2) = \frac{h_1 h_2}{32\pi} + \frac{h_1 h_2m^2 r^2 }{192\pi} + O(m^4 r^4)\, .
\end{equation}

We can now evaluate condensates and compare the fusion prediction with the expansion of the full result~\eqref{eq:Massless_Scalar_DD_Circle_Full}.
We already computed $\langle \mathcal D_{{h_1+h_2}} \rangle_\text{circ}^\text{QFT} $, the remaining condensate is
\begin{align}
\begin{split}
    \int ds&\frac{\langle \mathcal D_{{h_1+h_2}} [\partial_r^2 \phi(s)]_\text{r}\rangle_{\text{circ}}^\text{QFT}}{\langle \mathcal D_{{h_1+h_2}}\rangle_\text{circ}^\text{QFT}} \\
    &\quad= (h_1+h_2) \left[\frac{-m^2}{8\pi}\int_{\delta_\theta}^{2\pi-\delta_\theta}d\theta \frac{K_2(\frac{mL}{\pi}|\sin\frac{\theta}{2}|)}{\sin^2\frac{\theta}{2}} +\frac{L}{3\pi^2 \delta^3} +\frac{1}{2L\delta}- \frac{m^2L}{4\pi^2 \delta}\, \right]\\
    & \quad= \frac{1}{6}(h_1+h_2)m^2 \left[I_1K_1+\frac{m L}{4\pi}(I_1K_0-I_0K_1)-\frac{ m^2 L^2}{4\pi^2}(I_0K_0+I_1K_1)\right]\, ,
\end{split}
\end{align}
where we have suppressed the argument $\frac{mL}{2\pi} $ of the Bessel functions and included the $K^2$ counterterm mentioned in section~\ref{sec:FreeCFT-scalar}.
Plugging these pieces into~\eqref{eq:Massive_scalar_DD_matching_circ}, the fusion prediction readily reproduces the full result~\eqref{eq:Massless_Scalar_DD_Circle_Full}.

\subsubsection{Summary}

Before finishing this example, we summarize what we have accomplished.
Starting from the CFT results derived in section~\ref{sec:free_CFTs}, we turned on a mass deformation for the scalar field.
Since the theory remains free, we have access to the full $\langle \mathcal D_{h_1}[r]\mathcal D_{h_2}[0] \rangle_\text{QFT}$ to compare with our prediction.
Promoting each coefficient in the effective action to a perturbatively calculable function of the deformation, $a_i \to a_i(m^2r^2)$, we saw that we would have to determine the CPT corrections of four coefficients to obtain the action to $O(r^3)$.
By computing the correlators $\langle \phi\mathcal D_{h_1} \mathcal{D}_{h_2} \rangle $ and $\langle \phi^2\mathcal D_{h_1} \mathcal{D}_{h_2}\rangle $ at first order in CPT, and $\langle \mathcal D_{h_1} \mathcal{D}_{h_2}\rangle $ to second order, we were able to extract the $m^2$ corrections of the coefficients of the identity, $\phi$, $\partial_i \phi$ and $\phi^2$.
Computing the condensates in the full theory and plugging them into the defect OPE, we then showed that our prediction reproduces the exact result up to $O(r^3)$.

We finally repeated the procedure for circular defects, finding a nonzero correction to the coefficient of $K_i K^i$, and demonstrating that the defect OPE reproduces the full result for circular defects through order $O(r^3)$. All in all, the final result for the fusion effective action up to this order is
\begin{equation}
\begin{aligned}
    S_\text{eff} = \int ds \bigg[& \left(-\frac{h_1h_2}{4\pi r}-r\frac{h_1h_2}{8\pi}m^2-r^3\frac{h_1h_2}{96\pi }m^4\right)\mathds 1 -rh_1 \partial_r \phi \\
    &+\left(r \frac{h_1 h_2}{32\pi} +r^3\frac{h_1h_2}{192\pi}m^2\right) K_i K^i-r^2\frac{h_1}{2!}\partial_r^2 \phi\\
    &-r^3\frac{3h_1 h_2}{512\pi}\left(K_i K^i\right)^2 - r^3 \frac{h_1}{3!} \partial_r^3 \phi + O(r^4)\bigg] \, ,
\end{aligned}
\end{equation}
where additional curvature terms that are turned on in more complicated geometries have been left unfixed. The independence of the fusion coefficients $a_i(m^2r^2)$ from global data such as the length of the defects or the CPT cutoff $R$, together with the agreement with the exact results, serve as nontrivial checks of the formalism of section \ref{sec:Lines}.

\subsection{Schwinger model}\label{sec:Schwinger}

For our final solvable example, we consider the massless Schwinger model \cite{Schwinger:1962tp}.\footnote{The authors thank Diego Delmastro for suggesting this example to test our methods.}
The logic remains the same as in previous sections, though we will take this chance to demonstrate dimensional regularization as our choice of IR regularization scheme, discussing some new subtleties and showing that everything works out in the end.
For completeness, the local operator OPE for this model is worked out in appendix~\ref{app:Schwinger}, also in dimensional regularization.

The Schwinger model consists of a $U(1)$ gauge field coupled to $N_f$ massless Dirac fermions in $2d$,
\begin{equation}\label{eq:Schwinger-action}
    S = \int d^2x\, \bigg(\frac{1}{4} F_{\mu\nu}F^{\mu\nu} +  \sum_{i=1}^{N_f} \bar \psi_i \gamma^\mu(\partial_\mu - ieA_\mu)\psi_i\bigg)\,,
\end{equation}
where we are using canonical normalization since it is the natural choice when doing CPT.
This model is exactly solvable using a chiral transformation~\cite{Kim:1999wn} (see also \cite[App.\ A.5]{Delmastro:2026dpw}), which we review in appendix~\ref{app:Schwinger}.
Since we are in $d=2$, the gauge coupling $e$ has mass dimension~$1$, serving as our relevant deformation.

We will consider the fusion of two infinite Wilson lines of charge $q$ and $-q$, respectively, placed at positions $x=r$ and $x=0$ and stretching along the $y$-axis. 
Using the exact solution, a Wilson line may be written as
\begin{align}
    W_q[\gamma] =&\, \exp\left({iq e\int_\gamma A_\mu dx^\mu}\right) = \exp\left({iq \int_\gamma \epsilon_{\mu\nu}\partial^\nu\phi}dx^\mu\right)\nonumber\,,
\end{align}
where $\phi$ is a free scalar field with kinetic term $\frac{1}{2e^2} (\partial^2 \phi)^2 + \frac{N_f}{2\pi}(\partial_\mu \phi)^2$. Notice that, unlike Wilson lines in $d=4$, where the deformation is marginal, in $2d$ these defects are defined as relevant deformations on top of the trivial line. We are thus in a scenario where both bulk and defect relevant deformations are turned on, as in \eqref{eq:g-def} and \eqref{eq:h-def} with $g\sim h\sim e$.

Using the scalar field propagator
\begin{equation}
    \langle \phi(x)\phi(0) \rangle_\text{QFT} =-\frac{1}{2N_f}\log|x| + C - \frac{1}{2N_f} K_0\left(\sqrt{\tfrac{N_f}{\pi}}e|x|\right)\, ,
\end{equation}
where $C$ is an unphysical constant,
we have the expectation value of a single Wilson line,
\begin{align}
    \log\langle W_q\rangle_\text{QFT} &= -\frac{(iq)^2}{2}L\lim_{x\to 0}\int ds \,\partial^2_x\left[\frac{-1}{2N_f}\log\rho + C - \frac{1}{2N_f} K_0\left(\sqrt{\tfrac{N_f}{\pi}}e\rho\right) \right] \nonumber\\ 
    &=-\frac{Leq^2\sqrt{\pi}}{4\sqrt{N_f}}\, ,
\end{align}
where we defined $\rho \equiv \sqrt{x^2 + s^2}$.
We can then compute the correlator of parallel Wilson lines, and study its short-distance~limit,
\begin{align}\label{eq:Schwinger_WW_full}
    \log \langle W_{q}[r]W_{-q}[0]\rangle_\text{QFT} &= \log \langle W_{q}\rangle_\text{QFT} \langle W_{-q}\rangle_\text{QFT} + q^2 \int ds_1 \int ds_2 \,\langle\partial_x \phi(r, s_1) \partial_x \phi(0, s_2)\rangle\nonumber\\
    &= -\frac{Leq^2\sqrt{\pi}}{2\sqrt{N_f}}\left[1-e^{-\sqrt{\frac{N_f}{\pi}}er}\right]\\
    &= -\frac{e^2q^2L r}{2}+\frac{e^3q^2\sqrt{N_f}Lr^2}{4\sqrt{\pi}}+O(r^3)\, .\nonumber
\end{align}
This is the exact result. We will show how it is reproduced order by order by the OPE of Wilson lines. We will see that the first term is fully perturbative while the second one involves a condensate contribution.

\subsubsection{Fusion prediction from CPT}

Since the Wilson lines are relevant deformations on top of the trivial line, the conformal (well, topological) defect in the OPE \eqref{eq:dOPE2} is again the trivial line,
\begin{equation}
    W_{q}[r] W_{-q}[0] \simeq \mathds{1} e^{-S_\text{eff}}\, .
\end{equation}
The fusion effective action is given by
\begin{equation}\label{eq:Seff1-Schwinger}
    S_\text{eff} =\int ds \left[-\frac{a_0}{r}+\frac{a_{A}}{r}v^\mu A_\mu+a_F \epsilon^{\mu\nu}F_{\mu\nu} +r a_{F^2} F^2+ra_{\partial F}\epsilon^{\mu\nu}\partial_r F_{\mu\nu}+r a_{T_{\mu\nu}}T^{\mu\nu}+O(r^2)\right] ,
\end{equation}
where $v^\mu$ is a vector tangent to the line.
Following the discussion from section \ref{sec:Lines}, we have included in $S_\text{eff}$ the relevant operator $A_\mu$ and the marginal operator $F_{\mu\nu}$. Their coefficients $a_{A},a_F$ die in the CFT limit, but they can have perturbative contributions starting at order $O(e^1)$.
The leading order result $a_A\sim er$ defines the Wilson line that the defects fuse to. Since we are considering lines of opposite charge, we expect them to fuse to the trivial Wilson line, and so we expect this coefficient to vanish. This is consistent with the absence of an $r$-independent constant term in the full result \eqref{eq:Schwinger_WW_full}, but we will show it explicitly.

The operators shown in \eqref{eq:Seff1-Schwinger} will suffice for computations up to $O(r^2)$ because each of the coefficients $a_i$ enters at least with one factor of $e r$. The reason for this is that in the CFT limit $e\to 0$ the defects trivialize, as the coupling of the photon to the Wilson line contains a factor of $e$, and the effective action vanishes.
In fact, we can be slightly smarter: operators with an even number of fields must enter with even powers of $e$ in CPT, while operators with an odd-number of fields must enter with an odd power of $e$.
This follows from the $\mathbb{Z}_2$ symmetry properties of each operator as $A_\mu \to - A_\mu$, where $e$ can be thought of as a spurion of this $\mathbb{Z}_2$ symmetry.

Using this fact and our notation that $a^{(n)}_i$ refers to the $O(e^n)$ CPT correction, we can rewrite the action as
\begin{equation}\label{eq:Seff2-Schwinger}
    S_\text{eff} =\int ds \left[a_{A}^{(1)}v^\mu A_\mu+r\left(-a_0^{(2)}+a_F^{(1)} \epsilon^{\mu\nu}F_{\mu\nu} \right)+r^2\left(a_{A}^{(3)}v^\mu A_\mu+a_{\partial F}^{(1)}\epsilon^{\mu\nu}\partial_r F_{\mu\nu}\right)+O(r^3)\right] ,
\end{equation}
where $F^2$ and $T_{\mu\nu}$ cannot contribute at $O(r^2)$ since their coefficients must enter first at $O(e^2r^2)$. 
Our goal, as always, is to extract these coefficients and demonstrate that they reproduce the full result~\eqref{eq:Schwinger_WW_full}.
We will content ourselves with the matching through order $O(r^2)$ for this comparison. We see that we need to go to third order in $a_A$, second order in $a_0$ and first order in $a_{\partial F}, a_{F}$.

\paragraph*{Dimensional regularization}

Before we set up the system of equations and solve for the coefficients $a_i$, let us make some brief comments on using dimensional regularization to handle IR divergences.
We will always use a physical length $L$ for the IR regulator of the final integral over the Wilson line, since that scale is physical away from the infinite line case and any cancellations must occur independent of how we treat it.
We will use dimensional regularization for bulk insertions in CPT as well as for UV divergences coming from the integral over the line.

The most important difference between the previously used cutoff regularization and dimensional regularization is that only logarithmic divergences appear in the latter. For example, 
\begin{equation}
    \int_{\delta<|y|<R} \hspace{-23pt}d^4 y \hspace{6pt}  \frac{1}{|y|^4} \sim \log \frac{R}{\delta} \, , \qquad \qquad \int d^{4-2\epsilon} y \frac{1}{|y|^4} \sim  \frac{1}{\epsilon_\text{UV}}-\frac{1}{\epsilon_\text{IR}} \, ,
\end{equation}
where one has to distinguish between an IR and UV $\epsilon$. 
Power divergences, on the other hand, are set to zero, for example
\begin{equation}
    \int_{|y|<R} \hspace{-16pt}d^4 y\, \frac{1}{|y|^2} \sim R^2 \, , \qquad \qquad \int d^{4-2\epsilon} y \frac{1}{|y|^2} = 0 \, .
\end{equation}

The same is true for UV divergences, where the power divergences $\sim 1/\delta^n$ seen in previous sections are simply set to zero by dimensional regularization. 
In this way, dimensional regularization serves both as a regulator and a renormalization scheme choice for power UV divergences, where no counterterm needs to be explicitly specified.
For example, in the CFT predictions of section~\ref{sec:free_CFTs}, none of the UV divergences would have shown up, making the step of identifying the CFT basis unnecessary.
Likewise, CPT contributions proportional to $R^n$ are rendered zero.
In general, one must be careful to ensure that dimensional regularization has not hidden uncanceled IR divergences, but in practice it simplifies computations for the $a_i$'s, which we know are free from IR divergences.

\paragraph*{Fixing the coefficients}
Let us start computing the various correlation functions that we need to fix the fusion coefficients. Rather than the exact solution from appendix~\ref{app:Schwinger}, we now use standard perturbation theory of the gauge theory \eqref{eq:Schwinger-action}, which we regard as conformal perturbation theory in the (relevant) gauge coupling $e$.
We will take Feynman gauge throughout, where the photon propagator is $\langle A_{\mu}(x)A_{\nu}(0)\rangle^{(0)}_\text{CPT} = -\frac{\delta_{\mu\nu}}{2\pi}\log \mu |x| $ with $\mu$ an arbitrary scale, and work to $O(e^2)$.
We first need the single Wilson line expectation value, given by
\begin{align}
\begin{split}
    \log \langle W_q \rangle^{(2)}_\text{CPT} &= -\frac{(ieq)^2}{4\pi}L\int ds \log \mu |s| = \frac{e^2q^2}{4\pi} L^2\left(\log \frac{L\mu}{2}-1\right)\, .
\end{split}
\end{align}
The two-point function follows immediately,
\begin{align}\label{eq:Schwinger_WW_CPT_1}
\begin{split}
    \log \langle W_q[r]W_{-q}[0] \rangle^{(2)}_\text{CPT} &= \log \langle W_{q}\rangle^{(2)}_\text{CPT} \langle W_{-q}\rangle^{(2)}_\text{CPT} -\frac{e^2q^2}{2\pi}L\int ds \log \mu \sqrt{r^2+s^2}\\
    &= -\frac{e^2q^2Lr}{2}\,.
\end{split}
\end{align}

Let us first extract the coefficient $a_{A}^{(1)}$, which we expect to vanish, as mentioned above.
We can do this by looking at the $r$-independent term in the fusion expansion,
\begin{equation}
    \log \langle W_{q}[r] W_{-q}[0] \rangle^{(2)}_\text{CPT} \simeq \log \langle e^{-\int a_A^{(1)} v_\mu A^\mu ds} \rangle^{(2)}_\text{CPT} + O(r) \, .
\end{equation}
It is tempting to immediately set $a_{A}^{(1)} = 0$ since~\eqref{eq:Schwinger_WW_CPT_1} has no $r$-independent contribution, but we first need to check that the correlation function on the RHS is nonzero.
We have
\begin{align}
\begin{split}
    \log\langle e^{-\int a_A^{(1)} v_\mu A^\mu ds} \rangle^{(2)}_\text{CPT} &= \frac{(a_A^{(1)})^2}{2} L\int ds \, v_\mu  v_\nu \langle A^\mu(s) A^\nu (0) \rangle^{(0)}_\text{CPT} = \frac{(a_A^{(1)})^2}{4\pi}L^2\left(1-\log \frac{L\mu}{2}\right),
\end{split}
\end{align}
where we know $a_{A}^{(1)} \sim e$ by the previously discussed $\mathbb{Z}_2$ symmetry, so this is at the same order in $e$ as~\eqref{eq:Schwinger_WW_CPT_1}.
We can now safely conclude that
\begin{equation}
    a_A^{(1)} = 0 \, .
\end{equation}

We continue with $\langle W_{q}W_{-q} \rangle^{(2)}_\text{CPT}$ at subleading orders in $r$.
Dropping contributions that vanish by symmetry, the prediction from the effective action up to $O(r^2)$ is
\begin{align}\label{eq:Schwinger_WW_fusion}
\begin{split}
    \log \langle W_{q}[r] W_{-q}[0] \rangle^{(2)}_\text{CPT} \simeq &  \ ra_0^{(2)}L +\frac{r^2}{2} (a_F^{(1)})^2L\int ds\, \epsilon^{\mu\nu} \epsilon^{\alpha\beta}\langle F_{\mu\nu}(s)F_{\alpha\beta}(0)\rangle^{(0)}_\text{CPT} + O(r^3) \,,
\end{split}
\end{align}
where we have evaluated one $s$-integral in each term using translation invariance.
Matching the $O(r)$ term to \eqref{eq:Schwinger_WW_CPT_1} allows us to extract
\begin{equation}
    a_0^{(2)} = -\frac{e^2 q^2}{2} \, .
\end{equation}
The absence of $O(r^2)$ terms in~\eqref{eq:Schwinger_WW_CPT_1} does not fix the $a_F^{(1)}$ coefficient.
Just like in the massive scalar above, it turns out that at this order in CPT the $\langle FF\rangle$ two-point function is a pure power divergence and vanishes after renormalization.
To see this, we can write
\begin{equation}
    \int ds\,\langle F^{\mu\nu}(s)F^{\alpha\beta}(0) \rangle^{(0)}_\text{CPT} \sim \int ds\int d^d p\, e^{i p \cdot x} = \int ds \int dp_y\, e^{ip_y s} \int d^{d-1}p_i = 0\,,
\end{equation}
where dimensional regularization directly sets the transverse momentum $p_i$ integral to zero.

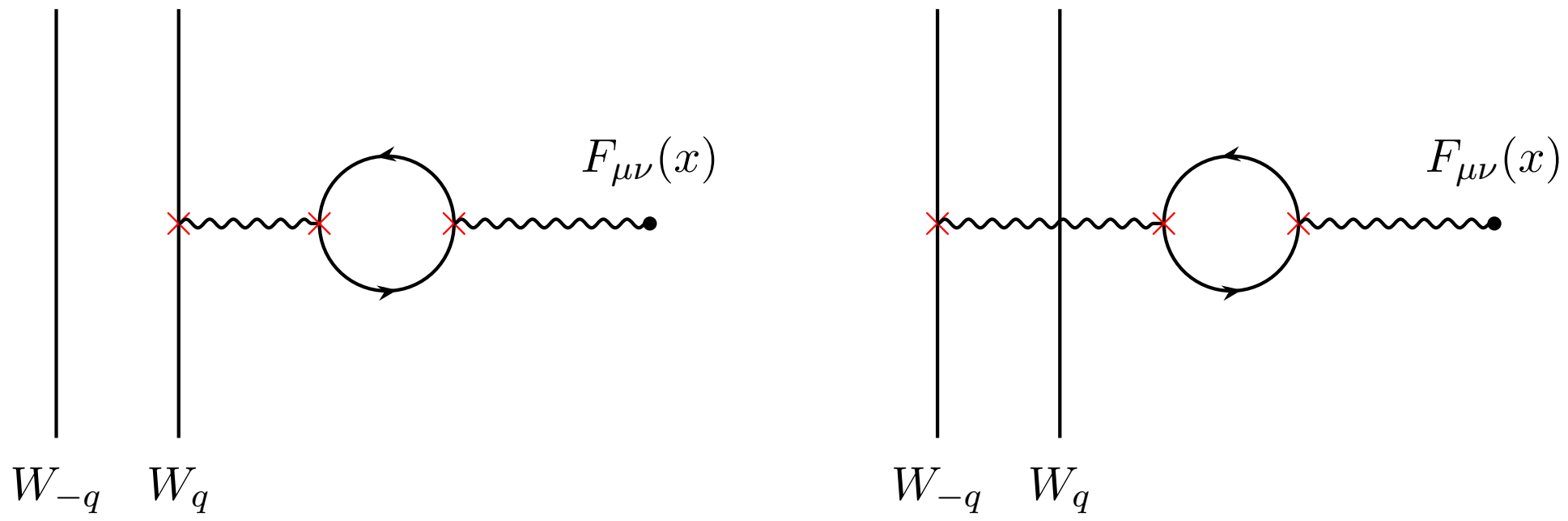
\begin{figure}[t]
    \centering
    \input{Figures/Schwinger_FWW.tikz}
    \caption{The two diagrams contributing to $\langle F_{\mu\nu}(x) W_{q}[r]W_{-q}[0] \rangle $ at third order in CPT. Red crosses indicate insertions of the relevant coupling $e$.}
    \label{fig:Schwinger_FWW}
\end{figure}
Of course, the $\langle F F\rangle$ two-point function might contribute non-perturbatively, and so we still need to determine the coefficient $a_F^{(1)}$ to match~\eqref{eq:Schwinger_WW_full}.
We can do this by considering the three-point function $\langle F W_{q}W_{-q}\rangle$ with $F_{\mu\nu}(x)$ inserted at $x \gg r$.
In $d=2$, the result is zero at tree-level, so to do the matching we must compute all quantities at third order in CPT through the two diagrams shown in figure~\ref{fig:Schwinger_FWW}.
These are just insertions of the vacuum polarization
\begin{align}
\begin{split}
    \Pi_{\mu\nu}(p) &=  e^2 N_f\mu^{2-d}\int \frac{d^d k}{(2\pi)^d} \frac{\mathrm{tr}(\gamma_\mu \slashed{k}\gamma_\nu (\slashed{p}+\slashed{k}))}{k^2(p+k)^2} = \frac{e^2 N_f}{\pi}\left(g_{\mu\nu}-\frac{p_\mu p_\nu}{p^2}\right)\, .
\end{split}
\end{align}
Inserting this into the full correlator and regulating the final $p$ integral with dimensional regularization, we have\footnote{In obtaining this expression, we have performed the $s$ integral first before the momentum integral. 
This is equivalent to performing the computation in $\mathbb{R}^1 \times S^1$ after compactifying the $y$-direction in the limit $L \to \infty$. This choice avoids potential boundary terms and simplifies the computation.}
\begin{align}
\begin{split}
\frac{\langle F_{\mu\nu}(x) W_{q}[r] W_{-q}[0]\rangle^{(3)}_\text{CPT}}{\langle W_{q}[r]W_{-q}[0]\rangle_\text{CPT}} =& \  -ieq \int ds\int \frac{d^2 p}{(2\pi)^2}[e^{ip_x(x-r)+ip_y s}-e^{ip_xx+ip_ys}] \\
    & \ \qquad \times \frac{i(p_\mu g_{\nu \alpha}-p_\nu g_{\mu \alpha})}{p^2}\Pi^{\alpha \beta}(p) \frac{g_{\beta \rho}v^\rho}{p^2}\\
    = & \  i\frac{e^3 N_f q}{4\pi} \epsilon_{\mu\nu}(2rx-r^2)\,.
\end{split}
\end{align}

This is to be compared to the fusion prediction
\begin{align}\label{eq:FWW-3CPT}
\begin{split}
    \frac{\langle F_{\mu\nu}(x) W_{q}[r] W_{-q}[0]\rangle^{(3)}_\text{CPT}}{\langle W_{q}[r]W_{-q}[0]\rangle_\text{CPT}} \simeq & \ -ra_F^{(1)}\int ds \,\epsilon_{\alpha\beta}\langle F_{\mu\nu}(x) F^{\alpha\beta}(s)\rangle^{(2)}_\text{CPT} \\
    & \ -r^2a_{\partial F}^{(1)}\int ds \,\epsilon_{\alpha\beta}\langle F_{\mu\nu}(x) \partial_r F^{\alpha\beta}(s)\rangle^{(2)}_\text{CPT} \\
    & \ -r^2a_{A}^{(3)}\int ds\, v_\alpha\langle F_{\mu\nu}(x)A^\alpha(s)\rangle_\text{CPT}^{(0)} + O(r^3)\, ,
\end{split}
\end{align}
so we only need to compute the $\langle FF\rangle$ two-point function at second order in CPT to extract $a_F^{(1)}$. 
Using the same regularization scheme as before, this integrated correlation function is given by
\begin{align}
\begin{split}
    \int ds \,\epsilon_{\alpha \beta} \langle F_{\mu\nu}(x) F^{\alpha \beta}(s)\rangle^{(2)}_\text{CPT} 
    = & \ \frac{e^2 N_f}{\pi} \epsilon_{\mu\nu}x\,.
\end{split}
\end{align}
Plugging this into \eqref{eq:FWW-3CPT}, we see that we have
\begin{equation}
    a_{F}^{(1)} = -\frac{i e q}{2}\, .
\end{equation}

From the $O(r^2)$ term in the above three-point function we would like to also extract $a_{\partial F}$. 
Of course, $\langle F_{\mu\nu}(x) \partial_r F_{\alpha\beta}(s) \rangle^{(2)}_\text{CPT} $ follows just from taking a derivative of the above two-point function in the $-r$ direction. 
However, the $A_\mu$ contribution $\langle F_{\mu\nu}(x) A_{\alpha}(s)\rangle^{(0)}_\text{CPT}$ turns out to give an identical prediction.
In fact, the operator $v^\mu A_\mu$ is redundant with $\epsilon^{\mu\nu}\partial_r F_{\mu\nu}\sim e^2 v^\mu A_\mu$ by the full equations of motion of the massless Schwinger model following the exact solution in appendix~\ref{app:Schwinger}. We can therefore always set $a_A^{(n)}=0$ (for $n\geq 3$) to remove this operator beyond leading order, giving us
\begin{equation}
    a_{\partial F}^{(1)} = -\frac{i e q}{4}\, ,
\end{equation}
as in the free Maxwell case from section~\ref{sec:free_CFTs}. 
We should emphasize that this is unique to the massless Schwinger model. In general, one would have to find another operator insertion to obtain a new equation to disentangle $a^{(1)}_{\partial F}$ and $a_{A}^{(3)}$.

\paragraph*{Fusion prediction}
Having determined the coefficients in the effective action \eqref{eq:Seff2-Schwinger} through $O(r^2)$, we can give the short-distance limit of the Wilson line two-point function as predicted by the defect OPE:
\begin{align}
\begin{split}
    \log \langle W_{q}[r] W_{-q}[0] \rangle_\text{QFT} \simeq &  \ -\frac{e^2 q^2}{2}rL -\frac{e^2 q^2}{8}r^2L\int ds\, \epsilon^{\mu\nu} \epsilon^{\alpha\beta}\langle F_{\mu\nu}(s)F_{\alpha\beta}(0)\rangle_\text{QFT} + O(r^3) \,.
\end{split}
\end{align}
It correctly reproduces the first term in \eqref{eq:Schwinger_WW_full}, and it captures the correct $r$-dependence of the second term, but its coefficient depends on a non-perturbative condensate.

In this simple model, the condensate can be computed exactly using the solution reviewed in appendix~\ref{app:Schwinger},
\begin{equation}
    \epsilon^{\mu\nu}\epsilon^{\alpha\beta} \int ds\,\langle F_{\mu\nu}(s)F_{\alpha\beta}(0)\rangle_\text{QFT} = -2\sqrt{\frac{N_f}{\pi}}e\,.
\end{equation}
Plugged into the expansion above, we see that it indeed reproduces the second term in \eqref{eq:Schwinger_WW_full}, confirming the validity of the defect OPE through order $O(r^2)$.
This example is quite similar to the free massive scalar from section~\ref{sec:massive_scalar}, where this is the analogue of the contribution from the operator $\partial_i \phi$. 
We further expect the odd powers of $r$ in~\eqref{eq:Schwinger_WW_full} to come from the perturbative expansion of $a_0(er)$ while the even powers will come from IR condensates of integrated operators from the effective action.
Higher order coefficients in the effective action can be systematically computed following similar steps to that example.

\subsubsection{Summary}

For our final solvable example, we have demonstrated the fusion prediction in the massless Schwinger model where the gauge coupling is itself a relevant deformation. 
We considered the fusion of infinite Wilson lines in the $y$-direction separated by a distance $r$ in the $x$-direction as $r\to 0$, both perturbatively and using the exact solution.
We found the effective action 
\begin{equation}
    S_\text{eff} = \int ds \bigg[r\frac{e^2q^2}{2} \mathds 1 -r\frac{i e q}{2} \epsilon_{\mu\nu}F^{\mu\nu} -r^2\frac{ieq}{4}\epsilon_{\mu\nu}\partial_rF^{\mu\nu}+O(r^3)\bigg]\,,
\end{equation}
where we have not determined curvature terms, but it should be straightforward to do so.
We demonstrated that this action reproduces the short-distance expansion of the full $\langle W_{q}[r]W_{-q}[0]\rangle_\text{QFT}$ correlator through order $O(r^2)$.

\section{Ising model}\label{sec:IsingModel}

Having demonstrated the success of our proposal in solvable examples in the previous section, we now turn our focus to a non-perturbative check in the $2d$ Ising model.
We first compute our prediction for the asymptotic short-distance expansion of the potential between two magnetic impurities in sections~\ref{sec:Ising_DefectFusion} and~\ref{sec:IsingCPT}, arriving at a prediction in terms of non-perturbative IR condensates. 
In section~\ref{sec:IsingCondensates} (and appendix~\ref{app:TBA}) we show these condensates can be computed exactly with integrability techniques.
In section~\ref{sec:IsingDMRG} we use \texttt{DMRG} to independently compute the potential and the condensates.
The section culminates in figure~\ref{fig:fitpotential}, demonstrating the agreement between the asymptotic prediction developed in this paper and the numerically extracted potential.

\subsection{Effective action}\label{sec:Ising_DefectFusion}

We consider the fusion of parallel infinite pinning field $\sigma$-defects on top of the critical theory
\begin{equation}\label{eq:Ising_Action}
    \mathcal{D}_{h} = \exp\left(h \int \sigma(s) ds\right) \, ,
\end{equation}
where the defects are infinite in the $y$-direction and placed at $x=r$ and $x=0$.
We think of these new interactions as relevant deformations living on top of the trivial defect, with couplings of mass dimension $[h] = 7/8$.
The two defects fuse to another trivial defect dressed by an effective action
\begin{equation}
    \mathcal{D}_{h}[r] \mathcal{D}_{h}[0] \simeq \mathds{1}\, e^{-S_\text{eff}} \qquad (\text{as } r\to 0)\,,
\end{equation}
where $S_\text{eff}$ consists of all operators present in the bulk theory, including relevant and marginal operators,
\begin{align}
S_\text{eff} = \int ds \bigg[& -\frac{a_0}{r} + \frac{a_\sigma}{r^{7/8}}\sigma + a_\epsilon \epsilon +a_{\partial \sigma} r^{1/8}\partial_x \sigma + a_{\partial \epsilon} r\partial_x \epsilon \nonumber\\ &+ a_T rT + a_{\bar T} r \overline T + a_{\partial^2\sigma} r^{9/8} \partial_x^2 \sigma+\ldots\bigg]\, .
\end{align}
Here each coupling $a_i$ is a function of $hr^{7/8}$.
Our goal will be to determine our prediction for the potential between the defects up to $O(r^{5/2})$.
Before we start computing in CPT, let us apply dimensional analysis to determine which couplings we need to compute and at which order.

In the absence of the relevant deformation $h$, we expect all couplings to vanish, since then we are simply fusing trivial defects. 
We know then that each coupling must start at least at $O(h r^{7/8})$. 
We can do a bit better if we think about the symmetry properties of the operators. 
Our defects break the $\mathbb{Z}_2$ spin-flip symmetry under which $\sigma \to -\sigma$ and $\epsilon \to \epsilon$.
Treating the coupling $h$ as a spurion of this symmetry, coefficients of operators proportional to $\sigma$ must be proportional to odd powers of $h r^{7/8}$, while operators that do not transform under the $\mathbb{Z}_2$ must be proportional to even powers.
Writing each coefficient as an expansion $a_i = \sum_{n}(r^{7/8})^na_i^{(n)}$ where $a_i^{(n)}$ may still contain logs of $r$, we can re-expand the effective action in a more useful form
\begin{align}\label{eq:Ising_Seff}
\begin{split}
S_\text{eff} =\int ds \bigg[& a_\sigma^{(1)} \sigma  -r^{3/4}a_0^{(2)}  \mathds 1 + r a_{\partial \sigma}^{(1)}\partial_x \sigma +  r^{7/4}\left(a_{\sigma}^{(3)} \sigma + a_{\epsilon}^{(2)} \epsilon\right) \\ &+ r^2 a_{\partial^2 \sigma}^{(1)} \partial_x^2 \sigma - r^{5/2}a_0^{(4)} \mathds 1 + O(r^{11/4})\bigg]\, .
\end{split}
\end{align}

\subsection{Fusion coefficients}

Our goal will be to determine all of the coefficients in~\eqref{eq:Ising_Seff} following our prescription from section~\ref{sec:prescription_lines}, i.e. by computing three-point functions $\langle \mathcal{O}_k(x) \mathcal{D}_{h}[r] \mathcal{D}_{h}[0] \rangle_\text{CPT}$ before and after fusing the defects to obtain a system of equations for the $a_i$'s.
The first object we will need is the expectation value of one defect, 
\begin{align}
\begin{split}
    \langle \mathcal{D}_{h} \rangle_\text{CPT}^{(2)} &= \frac{h^2}{2}L\int ds \langle \sigma(s) \sigma(0) \rangle
    = \frac{2^{5/4}}{3}h^2 L^{7/4}\,,
\end{split}
\end{align}
where the superscript $(2)$ on the correlation function indicates the order in the deformation $h$.
The correlation function $\langle \sigma(s) \sigma(0) \rangle$ on the RHS is evaluated in the critical theory.\footnote{
Note our regularization prescription: whenever possible, we perform one integral by pulling out a factor of $L$ using translation invariance, and then we perform the remaining integrals over $[-L/2,L/2]$, usually dropping all terms that vanish as $L \to \infty$ for simplicity.
We will do this for all integrals that follow.}

Using this, we can consider the correlator $\langle \mathcal{D}_h[r]\mathcal{D}_h[0] \rangle$ and start extracting the $a_i$'s.
At second order in CPT, we have a correlation function
\begin{align}
\begin{split}
    \langle \mathcal{D}_{h}[r] \mathcal{D}_{h}[0] \rangle_\text{CPT}^{(2)} &= 2\langle \mathcal{D}_{h} \rangle^{(2)}_\text{CPT} +   h^2 L\int ds \frac{1}{(s^2+r^2)^{1/8}} \\
    &= 4\frac{2^{5/4}}{3}h^2L^{7/4}+\frac{\sqrt{\pi } h^2 L r^{3/4} \Gamma \left(-\frac{3}{8}\right)}{\Gamma \left(\frac{1}{8}\right)} + O({\tfrac{r^2}{L^{1/4}}})\,.
\end{split}
\end{align}
The OPE prediction for this same correlation function is given by
\begin{align}
\begin{split}
\langle \mathds 1 e^{-S_\text{eff}} \rangle_\text{CPT}^{(2)} &= \frac{(a_\sigma^{(1)})^2}{2} L \int ds \langle \sigma(s) \sigma(0)\rangle + r^{3/4}a_0^{(2)} L + \ldots\\
&= (a_{\sigma}^{(1)})^2 \frac{2^{5/4}}{3}L^{7/4}+r^{3/4}a_0^{(2)}L + \ldots\,,
\end{split}
\end{align}
where the neglected terms proportional to operators of higher scaling dimension vanish as $L\to \infty$ by dimensional analysis at this order in CPT.
Matching powers of $r$, we see that
\begin{align}
    (a^{(1)}_\sigma)^2 &= (2h)^2\, , \qquad \qquad
    a^{(2)}_{0} = h^2\frac{\sqrt{\pi }  \Gamma \left(-\frac{3}{8}\right)}{\Gamma \left(\frac{1}{8}\right)} \equiv h^2 C_0 \, ,
\end{align}
where the sign of $a^{(1)}_\sigma$ remains unfixed at this stage.

To fix the sign and extract the coefficients $a_{\partial^n \sigma}^{(1)}$, we must insert a $\mathbb{Z}_2$-odd operator at the position $x$, which we will choose to be $\sigma(x)$. 
Working to first order in CPT, the three-point function prediction is
\begin{align}\label{eq:Ising_sigmaDD}
\begin{split}
    \langle \sigma(x) \mathcal{D}_{h}[r]\mathcal{D}_{h}[0] \rangle_\text{CPT}^{(1)} &= h \int ds \left[\langle \sigma(x,0) \sigma(r,s) \rangle +\langle \sigma(x,0) \sigma(0,s) \rangle\right]\\
    &=  h\int ds \left[2\langle \sigma(x,0) \sigma(0,s)\rangle+\sum_{n\geq1} \frac{r^n}{n!} \langle\sigma(x,0) \partial_x^n\sigma(0,s) \rangle \right]\,,
\end{split}
\end{align}
where we can expand in $r$ before integrating since the full series expansion remains regular as $r\to 0$ for all $s$.
This is to be matched by the OPE prediction given by
\begin{align}
    \langle \sigma(x) \mathds 1 e^{-S_\text{eff}} \rangle_\text{CPT}^{(1)}  &=\hspace{-1pt} -\hspace{-3pt}\int \hspace{-3pt} ds \bigg[a_\sigma^{(1)}\langle \sigma(x)\sigma(s) \rangle+ ra_{\partial \sigma}^{(1)} \langle \sigma(x) \partial_x\sigma(s) \rangle + r^2 a_{\partial^2\sigma}^{(1)} \langle \sigma(x) \partial^2_x \sigma(s)\rangle + \ldots \bigg].
\end{align}
Matching order by order in $r$, we see that we can fix all three of these coefficients,
\begin{align}
    a_\sigma^{(1)} = -2h \, , \qquad \qquad 
    a_{\partial \sigma}^{(1)} = -h\, , \qquad \qquad 
    a_{\partial^2 \sigma}^{(1)} = -\frac{h}{2}\, ,
\end{align}
where the sign of $a_\sigma^{(1)}$ is now fixed. 
The fact that $a_\sigma^{(1)}= -2h$ is simply the statement that at leading order in $r$ the two defects fuse to a new $\sigma$-defect with a coupling that is the sum of the two, as one would expect.

The next coefficient we will target is $a_{\epsilon}^{(2)}$. 
Since it enters at an even order in CPT, we may extract it by inserting $\epsilon(x)$ in the bulk. 
The extraction is easiest looking at the leading term in the limit $x\to \infty$ before integrating.
We have\footnote{Using the shorthand $\langle \mathcal{O}(\infty) \ldots \rangle \equiv \lim_{|x|\to \infty} |x|^{2\Delta_\mathcal{O}}\langle \mathcal{O}(x)\ldots\rangle$ as in section~\ref{sec:OPE_massive_scalar}.}
\begin{align}
    \langle \epsilon(\infty) \mathcal{D}_{h}[r] \mathcal{D}_{h}[0]\rangle_\text{CPT}^{(2)} &= h^2\int ds_1 ds_2  \bigg[\langle \epsilon(\infty) \sigma(s_1)\sigma(s_2) \rangle +\langle \epsilon(\infty) \sigma(r,s_1)\sigma(0,s_2) \rangle  \bigg] \nonumber \\
    &= h^2 L^{11/4}\frac{2^{5/4}}{7}+h^2 r^{7/4} L\frac{\sqrt{\pi} \Gamma(-\tfrac{7}{8})}{2\Gamma(-\tfrac{3}{8})}+O(r^2)\,.
\end{align}
After fusing, our prediction is given by (dropping terms that vanish as $x\to \infty$)
\begin{align}
    \langle \epsilon(\infty) \mathds 1 e^{-S_\text{eff}} \rangle_\text{CPT}^{(2)} = & \ \int ds_1 ds_2 \frac{(a_{\sigma}^{(1)})^2}{2} \langle \epsilon(\infty) \sigma(s_1)\sigma(s_2) \rangle - r^{7/4} a_{\epsilon}^{(2)}\int ds \langle \epsilon(\infty)\epsilon(s)\rangle +O(r^2)\, .
\end{align}
The first term is immediately reproduced by plugging in our previously determined $a_\sigma^{(1)}$.
Using $\int \langle \epsilon(\infty) \epsilon(s)\rangle ds = L$, we find
\begin{equation}
    a_{\epsilon}^{(2)} = - h^2\frac{\sqrt{\pi } \Gamma \left(-\frac{7}{8}\right)}{2\Gamma \left(-\frac{3}{8}\right)} \equiv -h^2 C_{\epsilon}\, .
\end{equation}

Let us move on to the extraction of $a_{\sigma}^{(3)}$, which requires going to third order in CPT. 
We will consider once again the insertion $\sigma(x)$, this time with $x\to \infty$ for simplification.
The three-point function is given by
\begin{align}\label{eq:Ising_sigmaDD_order3}
    \frac{\langle \sigma(\infty) \mathcal{D}_{h}[r]\mathcal{D}_{h}[0]\rangle_\text{CPT}^{(3)}}{\langle \mathcal{D}_h[r]\mathcal{D}_h[0] \rangle_\text{CPT} }&=\  h^3\int ds_1 ds_2 ds_3 \bigg[ \frac{1}{3}\langle \sigma(\infty) \sigma(0,s_1)\sigma(0,s_2)\sigma(0,s_3) \rangle_c \nonumber \\ \nonumber
    & \qquad \qquad \qquad \qquad \quad+ \langle \sigma(\infty) \sigma(r,s_1)\sigma(0,s_2)\sigma(0,s_3) \rangle_c+O(r^2)\bigg]\\ \nonumber
    &= h^3 \bigg[\frac{4}{3} \int ds_1 ds_2 ds_3\langle \sigma(\infty) \sigma(0,s_1)\sigma(0,s_2)\sigma(0,s_3) \rangle_c \\
    & \qquad \qquad + 2r^{7/4}L\left(C_\epsilon \log \frac{L}{r} - C_\sigma\right) +O(r^2)\bigg]\,,
\end{align}
where taking $x\to \infty$ has turned four terms into two and dividing by $\langle \mathcal{D}_h[r]\mathcal{D}_h[0]\rangle$ has removed disconnected pieces, denoted by the subscript $c$. 
Here we have defined the constant
\begin{equation}
    C_\sigma \approx 5.60020\,,
\end{equation}
which is derived in appendix~\ref{app:Ising}.

This is to be matched by the fusion prediction from expanding the effective action
\begin{align}\label{eq:Ising_sigmaDD_order3_fused}
\begin{split}
    \frac{\langle \sigma(\infty) \mathds 1 e^{-S_\text{eff}} \rangle_\text{CPT}^{(3)}}{\langle \mathds 1 e^{-S_\text{eff}} \rangle_\text{CPT} } = & \ -\frac{(a_\sigma^{(1)})^3}{3!}\int ds_1 ds_2 ds_3 \langle \sigma(\infty) \sigma(s_1)\sigma(s_2)\sigma(s_3)\rangle_c \\
    & \ +  a_\sigma^{(1)} a_\epsilon^{(2)}r^{7/4}\int ds_1 ds_2 \langle \sigma(\infty) \sigma(s_1)\epsilon(s_2)\rangle_c \\
    & \ - a_{\sigma}^{(3)} r^{7/4}\int ds \langle \sigma(\infty) \sigma(s)\rangle_c  + O(r^2)\, ,
\end{split}
\end{align}
where all other terms have been set to zero after removing disconnected pieces and taking $x\to \infty$.
The first term is immediately reproduced using $a_\sigma^{(1)}=-2h$ from earlier. 
Before we can match the $O(r^{7/4})$ terms, we must renormalize $\epsilon(s)$ since the second line of~\eqref{eq:Ising_sigmaDD_order3_fused} is UV divergent:
\begin{equation}
    \int ds_1 ds_2 \langle \sigma(\infty) \sigma(s_1)\epsilon(s_2)\rangle_c =L \int_{\delta/2}^{L/2} ds \frac{1}{|s|} = L\log \frac{L}{\delta} \, .
\end{equation}
This divergence is removed by defining the renormalized operator on the defect
\begin{equation}
    [\epsilon(s)]_\text{r} = \epsilon(s) - a_\sigma^{(1)}  \log (\mu \delta) \sigma(s) \,,\label{eq:Ising_eps_ren}
\end{equation}
where $\mu$ is a scheme-dependent scale. Using the renormalized $[\epsilon(s)]_\text{r}$ now in~\eqref{eq:Ising_sigmaDD_order3_fused}, we can immediately extract
\begin{equation}
    a_{\sigma}^{(3)} = 2h^3 (C_\epsilon  \log r \mu+C_\sigma )\, .
\end{equation}

Finally, we need to extract $a_0^{(4)}$ to obtain our prediction to $O(r^{5/2})$.
The details are provided in appendix~\ref{app:Ising}, and we simply report here the numerical result
\begin{equation}
    a_0^{(4)} = -h^4 C_0^{(4)}\, , \qquad \qquad C_0^{(4)} \approx 5.2678 \, .
\end{equation}

Expanding out the effective action in the correlator $\langle \mathcal{D}_{h}[r]\mathcal{D}_{h}[0]\rangle$, we arrive at our prediction for the short-distance behavior of the potential. Defining
\begin{equation}
    \langle\mathcal O\rangle_{2h}
    \equiv
    \frac{\langle\mathcal D_{2h}\mathcal O\rangle}
         {\langle\mathcal D_{2h}\rangle}\, , \label{eq:fused_condensate}
\end{equation}
we find
\begin{equation}\label{eq:Ising_potential_OPE}
    \boxed{
\begin{aligned}
    V_{h,h}(r) =& \ V_0 - h^2 C_0r^{3/4}+2h^3\left(C_\epsilon \log r \mu + C_{\sigma} \right)\langle \sigma\rangle_{2h}\,r^{7/4}\\
    & \ -h^2  C_\epsilon\langle [\epsilon ]_\text{r}\rangle_{2h}\, r^{7/4} - \frac{h}{4}  \langle \partial^2_x\sigma \rangle_{2h}\,r^2 + h^4 C_0^{(4)} r^{5/2}+O(r^{11/4})\,,
\end{aligned}}
\end{equation}
where $V_0=V_{h,h}(0)$ is a scheme dependent constant not determined by CPT, and we have used translation invariance to write the $O(r^2)$ term as a single condensate, see section~\ref{sec:translation_invariance}.\footnote{\label{fn:dx2sigma_ren}Here and below, $\partial_x^2\sigma$ on the defect is understood with its power divergences subtracted, the leading one being $\partial_x^2\sigma \to \partial_x^2\sigma +\frac{4h}{5\delta^{5/4}}$, where we regulated $|s|>\delta$.}
The constants are given by
\begin{equation}\label{eq:Ising_constants}
     C_0 = \frac{\sqrt{\pi }\, \Gamma \left(-\frac{3}{8}\right)}{\Gamma \left(\frac{1}{8}\right)} \, , \qquad C_\epsilon = \frac{\sqrt{\pi }\, \Gamma \left(-\frac{7}{8}\right)}{2\,\Gamma \left(-\frac{3}{8}\right)}\, , \qquad C_\sigma \approx 5.60020\, , \qquad C_0^{(4)} \approx 5.2678\,.
\end{equation}
\label{sec:IsingCPT}
\subsection{Exact condensates}\label{sec:IsingCondensates}
{The prediction~\eqref{eq:Ising_potential_OPE} gives the asymptotic expansion of the potential of magnetic pinning fields to order $r^{5/2}$ in terms of three condensates: $\langle\sigma\rangle_{2h}$, $\langle[\epsilon]_\text{r}\rangle_{2h}$ and $\langle\partial_x^2\sigma\rangle_{2h}$. These require non-perturbative methods to compute. For example, they could be measured in the Ising lattice model, extrapolated to the continuum. We will do exactly that in section~\ref{sec:IsingDMRG}. Here, instead, we point out that these condensates (and many others) can be computed exactly, in the continuum, thanks to integrability. See e.g. \cite{Fateev:1997nn,Dorey:2000eh,Bajnok:2013eaa} for related work.}

{The simplest condensate to compute is $\langle \sigma \rangle_h$.  For that purpose, let us place the theory in $\mathbb{R} \times S_1^\beta$, with the defect wrapping the thermal circle at $x=0$. The defect partition function}
\begin{equation}
    G(\lambda) = \left \langle \exp\left({ h \int_0^\beta \sigma(s) ds}\right) \right \rangle
\end{equation}
is then a function of $\lambda = \beta h^{8/7}$.\footnote{This is up to a cosmological constant counterterm, which we fix by removing any linear term in $\beta$ at small $\beta$.} At large $\lambda$ we have
\begin{equation}
   \log G(\lambda) = -\beta e(h) + O(\lambda^0), \label{eq:Gasymp}
\end{equation}
and 
\begin{equation}
    \langle \sigma \rangle_h = -\partial_h e(h). \label{eq:sigmafrome}
\end{equation}
Here, $e(h) = - C h^{8/7}$ is the shift in the energy of the ground-state due to the presence of the defect in the impurity channel. Our task is thus to compute $G(\lambda)$ and extract the asymptotics (\ref{eq:Gasymp}).

Following \cite{Lamacraft:2008yr}, consider the high-temperature expansion:
\begin{equation}
\begin{split}
    {G(\lambda)} &{= 1+\sum_{n\ge1} h^{2n}\int_{0<s_1<\dots<s_{2n}<\beta} ds_1\cdots ds_{2n}\, \langle \sigma(s_{2n})\cdots\sigma(s_1)\rangle_\beta\, ,}\\
    {\langle \sigma(s_{2n})\cdots\sigma(s_1)\rangle_\beta} &{= \prod_{i<j} d_{ij}^{\,2\Delta_\sigma q_iq_j}\, ,\qquad q_i=(-1)^i\, ,\qquad d_{ij}=\frac{\beta}{\pi}\left|\sin\frac{\pi(s_i-s_j)}{\beta}\right| ,} \label{eq:isingnpoint}
\end{split}
\end{equation}
where we use the result for the $2n$-point $\sigma$ correlator \cite{Kadanoff:1970kz}. Note that the correlator is almost that of vertex operators of a free boson!
 Indeed, recall that, with $V_{q_j} =  {:}e^{i q_j \sqrt{2\Delta_\sigma}\,\varphi}{:}$,
\begin{equation}
    \left\langle \prod_{j} V_{q_j}(s_j)\right\rangle_{\beta} = \prod_{i<j} d_{ij}^{\,2\Delta_\sigma q_iq_j} \label{eq:vertex}\qquad (\text{free-boson})
\end{equation}
if $\sum_j q_j=0$, and zero otherwise. 

One might be tempted to identify   $\sigma  \stackrel{?}{=} V_+ + V_-$ but the $2n$-point correlator of the right-hand side would produce a sum over \textit{all} orderings of $V_+$ and $V_-$. For example, at $n=2$ we would have the contributions
\begin{equation}
    (++--) + (--++)+(+-+-) +(-+-+) +(+--+) + (-++-),
\end{equation}
while in (\ref{eq:isingnpoint}) only one alternating term is present. To reproduce the alternating product we introduce an auxiliary two-state Hilbert space, so that
\begin{equation}
    \sigma \rightarrow S^+ V_+ + S^- V_-,
\end{equation}
where $S^+$ and $S^-$ are the $SU(2)$ raising and lowering operators, $(S^\pm)^2=0$. Up to a factor of two corresponding to $\pm \leftrightarrow \mp$, we find
\begin{equation}
     \label{eq:Kondo} G(\lambda) = \frac{1}{2} \text{tr}\left\langle \mathcal{P} \exp\left( h \int_0^\beta ds\, \left(S^+ V_+ + S^- V_-\right) \right)\right\rangle\,,
\end{equation}
where the correlator is now evaluated in the free-boson theory.  This is the partition function of the anisotropic Kondo model with impurity spin $s=\tfrac{1}{2}$ \cite{Bazhanov:1994ft,Fendley:1995kj,Bazhanov:1996dr,LeClair:1997gz,Lamacraft:2008yr}.

In a pair of beautiful papers \cite{Bazhanov:1994ft,Bazhanov:1996dr}, Bazhanov, Lukyanov and Zamolodchikov (BLZ) studied (for greater reasons) exactly this 
class of operators. They showed that, for any value of $\Delta$, the operators (now viewed as acting on the circle Hilbert space)
\begin{equation}
  \mathbf{T}_j(\lambda) \equiv \text{tr}_j \mathcal{P} \exp\left( h \int_0^\beta S_j^+ V^\Delta_+ + S_j^- V^\Delta_- \right), \label{eq:transfer}
\end{equation}
with $V_\pm^\Delta \equiv  {:}e^{\pm i \sqrt{2\Delta}\,\varphi}{:}$, define one-parameter families of commuting transfer matrices:\footnote{Here $S_j^\pm$ are the raising and lowering generators of $U_q(\text{sl}_2)$ in the spin-$j$ representation, with $q=e^{i\pi\Delta}$. For $j=\tfrac12$ these are the usual $S^\pm$.}
\begin{equation}
[\mathbf{T}_j(\lambda),\mathbf{T}_{j'}(\lambda')]=0,
\end{equation}
see also \cite{Runkel:2007wd,Runkel:2010ym,Gaiotto:2020fdr} for related work.
As in any integrable theory, the transfer matrix serves as a generating function of local conserved charges. Indeed, this is achieved by expanding $\log \mathbf{T}_j$ at large $\lambda$. For example, 
\begin{equation}
\log \mathbf{T}_{1/2}(\lambda) \simeq m\,\frac{\lambda}{2\pi} \;-\; \sum_{n\ge1} C_n \left(\frac{\lambda}{2\pi}\right)^{1-2n} \mathbf{I}_{2n-1}\, ,
\qquad \lambda = \beta\, h^{\frac{1}{1-\Delta}}\, , \label{eq:Tasymp}
\end{equation}
where $\mathbf{I}_{2n-1}$ are higher-spin local charges.\footnote{One might point out that 
$\mathbf{T}_j$ itself, at different values of $\lambda$, can be used as families of conserved
 charges. The key point is that the $\mathbf{I}_{2n-1}$ are local.} The usual Hamiltonian of the (auxiliary free-boson) theory is nothing but
 $\mathbf{H} = \frac{2\pi} {\beta}\mathbf{I}_1 $.\footnote{ We note that, since all of the $\mathbf{I}_{2n-1}$ commute, we could have picked any (linear combination) of them as the Hamiltonian, which would define a different integrable theory diagonalized by the same eigenstates! We will use precisely this trick below to compute  $\langle{\partial_x^2\sigma}\rangle$.}
 The authors then proceeded to solve for the coefficients $(m, C_n)$ using integrability techniques. They obtained
\begin{equation} \label{eq:BLZresult}
\begin{split}
    m &= \frac{2\sqrt\pi\,\Gamma\!\left(\frac12-\frac\xi2\right)}{\Gamma\!\left(1-\frac\xi2\right)}\,\Gamma(1-\Delta)^{1+\xi}\, ,\\
    C_n &= \frac{\sqrt\pi\,(1+\xi)}{n!}\,\Delta^{n}\,
    \frac{\Gamma\!\big((n-\frac12)(1+\xi)\big)}{\Gamma\!\big(1+(n-\frac12)\xi\big)}\,\Gamma(1-\Delta)^{-(2n-1)(1+\xi)}\, ,
\end{split}
\end{equation}
where $\xi = \Delta/(1-\Delta)$.

This is sufficient for our purposes. Using (\ref{eq:Kondo}), we find
\begin{equation}
    G(\lambda) = \tfrac{1}{2}\langle \mathbf{T}_{1/2}(\lambda)\rangle\big\vert_{\Delta = \Delta_\sigma}\,,
\end{equation}
where the right-hand side is computed in the zero-momentum vacuum of the free boson on a circle. We can then finally read off, by matching (\ref{eq:Tasymp}) and (\ref{eq:Gasymp}),
\begin{equation}
e(h) = - \frac{m}{2\pi} h^{8/7} \implies \langle \sigma \rangle_h = \frac{8}{7}\frac{\Gamma\!\left(\frac37\right)\Gamma\!\left(\frac78\right)^{8/7}}{\sqrt\pi\,\Gamma\!\left(\frac{13}{14}\right)} h^{1/7}.
\end{equation}

Next, let us consider the condensate $\langle [\epsilon]_\text{r}\rangle_h$. The trick here 
is to note that $\epsilon$ is an exactly marginal deformation on the trivial defect. Its effect is to move the trivial line along a conformal manifold in which $\sigma$ correlators are again of the form (\ref{eq:isingnpoint}), but with nontrivial anomalous dimensions. Let us understand this statement to first
order in the $\epsilon$ deformation, which is all that we need to compute $\langle [\epsilon]_\text{r}\rangle_h$.

We consider again the high-temperature expansion (\ref{eq:isingnpoint}), but now with an $\epsilon$ insertion integrated:
\begin{equation}
\begin{split}
    &\int_0^\beta du\,\Big\langle \epsilon(u)\, e^{h\int_0^\beta \sigma(s)\,ds}\Big\rangle_\beta = \sum_{n\ge1} h^{2n}\int_{0<s_1<\dots<s_{2n}<\beta}\hspace{-50pt} ds_1\cdots ds_{2n}\hspace{2pt}
    \int\limits_{|u-s_j|>\tfrac{\delta}{2}} du\,\langle \epsilon(u)\,\sigma(s_{2n})\cdots\sigma(s_1)\rangle_\beta\, . \label{eq:epsiloninsertion}
\end{split}
\end{equation}
Here, as in the CPT computations of section \ref{sec:IsingCPT}, we introduce a UV cutoff window $\delta/2$ around each $\sigma$ insertion. The correlation function in (\ref{eq:epsiloninsertion}) can be computed in terms of those in (\ref{eq:isingnpoint}) via the OPE
\begin{equation}
    \sigma(u + \delta)\sigma(u) = \delta^{-1/4} \left( 1 + \frac{1}{2}\delta \epsilon(u) + \dots\right).
\end{equation}
After a straightforward computation, we find
\begin{equation}
    \langle\epsilon(u)\,\sigma(s_{2n})\cdots\sigma(s_{1})\rangle_\beta
    =\frac{s(u)}{2}\partial_u\left(\sum_{j=1}^{2n} {q_j}\log d_{u,j}\;
   \right) \langle\sigma(s_{2n})\cdots\sigma(s_{1})\rangle_\beta\, ,
\end{equation}
where $s(u) = (-1)^k$, with $k$ the number of insertions with $s_j>u$. We see that, in each segment of the alternating chains of (\ref{eq:epsiloninsertion}), the $\epsilon$ operator acts as a total derivative. Integrating in $u$ produces
\begin{align}
    \int\limits_{|u-s_j|>\frac{\delta}{2}} du\,
    \langle\epsilon(u)\,\sigma(s_{2n})\cdots\sigma(s_{1})\rangle_\beta
    &=-\left(2n\log\frac{\delta}{2}+2\sum_{i<j}q_iq_j\log d_{ij}\right)
    \langle\sigma(s_{2n})\cdots\sigma(s_{1})\rangle_\beta\, \nonumber\\
    &=-\left(\partial_\Delta+2n\log\frac{\delta}{2}\right)
    \prod_{i<j} d_{ij}^{\,2\Delta q_iq_j}\,\bigg|_{\Delta=\Delta_\sigma}\, .
\end{align}

Thus, in each term of the high-temperature expansion, the $\epsilon$ insertion amounts to a shift in the dimension of $\sigma$, up to counterterms! Writing $2n h^{2n} = h \partial_h h^{2n}$, we can now resum (\ref{eq:epsiloninsertion}):
\begin{equation}
     \int_0^\beta du\,\Big\langle \epsilon(u)\, e^{h\int_0^\beta \sigma(s)\,ds}\Big\rangle_\beta = - \left( \partial_\Delta + \log\frac{\delta}{2} h\partial_h\right) G_\Delta(\lambda)|_{\Delta=\Delta_\sigma}\,, \label{eq:epsresult}
\end{equation}
where, by the arguments above,
\begin{equation}
    G_\Delta(\lambda) = \frac{1}{2} \langle \mathbf{T}_{1/2}(\lambda) \rangle\,,
\end{equation}
with $\Delta$ generic as in (\ref{eq:transfer}). Normalizing (\ref{eq:epsresult}) by the magnetic pinning partition function, taking large $\lambda$ and dividing by $\beta$, we conclude
\begin{equation}
\langle \epsilon \rangle_h = \partial_\Delta e_\Delta(h)|_{\Delta = \Delta_\sigma}  - \log \frac{\delta}{2} h \langle \sigma \rangle_h.
\end{equation}
Here, $e_\Delta(h) = -m h^{\frac{1}{1-\Delta}}/2\pi$ with $m$ given in (\ref{eq:BLZresult}). The final step is to renormalize $\epsilon$ as in equation (\ref{eq:Ising_eps_ren})
\begin{equation}
    [\epsilon]_\text{r} = \epsilon + h \log(\mu \delta) \sigma\,,
\end{equation}
which gives
\begin{equation}
    \langle[\epsilon]_\text{r}\rangle_h
    =\partial_\Delta e_\Delta(h)\Big|_{\Delta=\Delta_\sigma}
    +\log(2\mu)\,h\,\langle\sigma\rangle_h\, .
\end{equation}
Plugging in the results, we conclude
\begin{equation}
    \langle[\epsilon]_\text{r}\rangle_h
    =h^{8/7}\left[-C'+\frac{8}{7}\,C\,\log\left(\frac{2\mu}{h^{8/7}}\right)\right] ,
\end{equation}
where
\begin{equation}
    C =\frac{\Gamma\!\left(\frac37\right)\Gamma\!\left(\frac78\right)^{8/7}}{\sqrt\pi\,\Gamma\!\left(\frac{13}{14}\right)}\, ,
    \qquad C' =C \left[\frac{32}{49}\left(\psi\!\left(\tfrac{13}{14}\right)-\psi\!\left(\tfrac37\right)+2\log\Gamma\!\left(\tfrac78\right)\right)-\frac87\,\psi\!\left(\tfrac78\right)\right] ,
\end{equation}
with $\psi$ the digamma function.  

The computation of $\langle \partial^2_x \sigma \rangle_h$ is more involved, but quite interesting, especially for the integrability aficionado. In appendix \ref{app:TBA} we compute it via a ($g$-function) thermodynamic Bethe ansatz (TBA) on a deformed theory, where the Hamiltonian $\mathbf{H}$ is replaced by $\mathbf{H} + \kappa\, \mathbf{I}_3$, to first order in $\kappa$ \cite{Fioretto:2009yq, Mossel:2012vp,Downing:2021mfw,Downing:2023lnp}. The condensate $\langle \partial^2_x \sigma \rangle_h$ is then obtained from the spin-3 charge stored on the defect at low temperature, in the same way that $\langle\sigma\rangle_h$ was obtained from the energy $e(h)$.
Here we simply quote the result
\begin{equation}
    \langle \partial^2_x \sigma \rangle_h
    =\frac{64\,\Gamma\!\left(\frac27\right)\Gamma\!\left(\frac78\right)^{24/7}}{5\sqrt\pi\,\Gamma\!\left(\frac{11}{14}\right)}\;h^{17/7}\, . \label{eq:hardcondensate}
\end{equation}
Along the way, we rederive $\langle \sigma \rangle_h$ in TBA language.

\subsection{Defect \texorpdfstring{\texttt{DMRG}}{DMRG}}
\label{sec:IsingDMRG}

The goal of this section is to provide a non-perturbative check
of the framework outlined in section~\ref{sec:Lines}.
To this end, we use \texttt{DMRG} \cite{White:1992zz,White:1993zza} to measure the potential between two pinning defects of equal strength $h$,
\begin{equation}
V_{h,h}(r)
\equiv -\lim_{L\to\infty}\frac{1}{L}
\log\left\langle
\mathcal D_h[r]\,\mathcal D_h[0]
\right\rangle \, ,
\end{equation}
and compare it with the short-distance prediction (\ref{eq:Ising_potential_OPE}). Taken together, the CPT coefficients of section~\ref{sec:IsingCPT} and the exact condensates of section~\ref{sec:IsingCondensates} fix this prediction completely, up to an overall additive constant. We find a striking agreement between the two computations, see figure \ref{fig:fitpotential}.

In a second, independent \texttt{DMRG} simulation, we consider a single pinning defect and measure the condensates $\langle \sigma_D\rangle_{2h}$ and $\langle \epsilon_D\rangle_{2h}$ directly.\footnote{In this subsection we write $\sigma_D$ and $\epsilon_D$ for the defect operators $\sigma$ and $[\epsilon]_\text{r}$ of section~\ref{sec:IsingCPT}, to distinguish them from bulk insertions.}
The condensate measurements were in fact obtained first, before we realized that they could be computed exactly. For this reason, and because it is what one would have to do in more complicated applications (e.g.\ QCD) where nothing is known analytically, we keep our discussion below deliberately agnostic about the exact formulas (other than for comparing results). The condensates can thus be extracted in three ways: (i) analytically, via integrability; (ii) simulating a single defect with insertions; (iii) fitting the asymptotics of the potential. The three methods agree to several digits.

Let us now discuss how the computations were performed.  We consider the critical Ising chain with Hamiltonian
\begin{equation}
    \widehat H
    = -\sum_{j=1}^{N-1} Z_j Z_{j+1}
      -\sum_{j=1}^{N} X_j
      -\sum_{\alpha} g Z_{j_\alpha}, \label{eq:latticeH}
\end{equation}
where $X_j$ and $Z_j$ are Pauli operators, $N$ is the number of
sites, and $g$ is the lattice pinning strength at site
$j_\alpha$.  We study three configurations: the chain without
defects, a single pinning defect, and two pinning defects.
The zero-defect simulations fix the bulk operator normalizations,
the one-defect simulations determine the condensates, and the
two-defect simulations determine the potential. All computations were performed in the finite lattice, and then extrapolated to the continuum. Let us consider them in order.

    \subsubsection{Zero-defect}

    The first step is to relate the lattice operators to their continuum counterparts.
    This is an EFT matching problem. We treat the lattice as the UV theory, and expand the operators in terms of the operators in the IR theory, the Ising CFT deformed by irrelevant operators.

    For the $\mathbb{Z}_2$-odd lattice operator $Z_j$, we have\footnote{Odd-level descendants are excluded by parity.}
    \begin{equation}\label{eq:Z-matching}
        Z_j=a^{1/8}\left[
        Z_\sigma\,\sigma(x_j)
        +a^2\sum_i c_{2,i}\,
            \mathcal O_{\sigma,2,i}(x_j)
        +a^4\sum_i c_{4,i}\,
            \mathcal O_{\sigma,4,i}(x_j)
        +\cdots
        \right],
    \end{equation}
    where $a$ is the lattice spacing and $\mathcal O_{\sigma,\ell,i}$ are (Virasoro) descendants in the sigma trajectory at level $\ell$.\footnote{ Note that the Wilson coefficients $c_{m,i}$ are constants here, by dimensional analysis. In the presence of bulk deformations, these are promoted to functions of $m a$ where $m$ is the mass scale of the flow. The computations in this section can be extended to that case with minimal modifications by taking this into account, with matching being performed at $a \ll n a \ll m^{-1}$.}

    Consider the two-point function
    $\langle Z_jZ_{j+n}\rangle$. We substitute the operator
    matching above and evaluate the resulting continuum correlators
    in the IR EFT. To leading order in $1/n$, we obtain the power-law decay of the $\langle\sigma \sigma\rangle$ correlator in the CFT. Corrections arise both from
    subleading terms in the operator matching \eqref{eq:Z-matching} and from irrelevant
    terms in the IR effective action. Combining these contributions,
    we obtain the large-separation expansion
    \begin{equation}
        \langle Z_jZ_{j+n}\rangle
        =\frac{1}{n^{1/4}}\left[
        Z_\sigma^2+\frac{b_2}{n^2}+\frac{b_3}{n^3}+\frac{b_4}{n^4}+\cdots \label{eq:ZEFT}
        \right].
    \end{equation}

    To extract the Wilson coefficient $Z_\sigma$, we compute
    $\langle Z_jZ_{j+n}\rangle$ via \texttt{DMRG}\footnote{We use the Julia
    packages \texttt{ITensors.jl} and \texttt{ITensorMPS.jl}~\cite{Fishman_2022}. Throughout, errors from truncated bond dimensions and sweep convergence are negligible.}
    at fixed $n$ for eight chain sizes $N\in[960,3200]$,
    and extrapolate to infinite volume via the fit
    \begin{equation}
        \langle Z_jZ_{j+n}\rangle^\text{fit}_N
        =\langle Z_jZ_{j+n}\rangle
        +\sum_{k=1}^{3}\frac{a_{k,n}}{N^k}.
    \end{equation}
    With the infinite-volume estimate in hand for all $n \in [6,32]$, we fit (\ref{eq:ZEFT})
    truncated to the terms displayed, obtaining\footnote{Here and below, the uncertainty estimates come from infinite volume extrapolation errors, varying the fit ranges and omitting data points. When applicable, we also include errors from the continuum extrapolation. In general, they should be understood as a rough estimate of numerical stability.}
    \begin{equation}
        Z_\sigma \simeq 0.80312(4). \label{eq:Zsigma}
    \end{equation}

    For the $\mathbb Z_2$-even operator $X_j$, the matching includes
    an identity contribution,
    \begin{equation}
        X_j=Z_\mathbf{1} \mathbf 1-a Z_\epsilon\epsilon(x_j)
        +\sum_{\Delta=2}^{5}a^\Delta
            \sum_i c^X_{\Delta,i}\, \label{eq:Xmatching}
            \mathcal O_{\Delta,i}(x_j)
        +\cdots,
    \end{equation}
    where the minus sign ensures $C_{\sigma\sigma\epsilon}$ is positive with
    $Z_\epsilon>0$. The sum includes (Virasoro) descendants of $\mathbf{1}$ and $\epsilon$.
    $Z_\mathbf{1}$ can be determined by extrapolating $\langle X_j\rangle$ to infinite volume. To extract $Z_\epsilon$, we instead
    use the connected two-point function and fit $\langle X_jX_{j+n}\rangle-\langle X_j\rangle\langle X_{j+n}\rangle$ in a similar manner to the computation outlined above. The final result is
    \begin{equation}
        Z_\mathbf{1} \simeq 0.63661978(4), \label{eq:Zmatch}
        \qquad Z_\epsilon \simeq0.3183(4).
    \end{equation}

    \subsubsection{One-defect}

    We can now proceed to the determination of the defect condensates
    $\langle \sigma_D \rangle$ and $\langle \epsilon_D \rangle$.
    To this end, we use \texttt{DMRG} to determine the ground state of the spin
    chain with a single pinning defect at site $j=N/2$, for various
    values of $g$ and $N$.

    To obtain a defect with finite coupling strength $h$ in the continuum we must scale $g = \gamma a^{7/8}$ with $\gamma$ kept fixed as $a \rightarrow 0$ in physical units determined by $h$.
    Along this trajectory, the defect action approaches the magnetic pinning
    defect of the continuum theory, with corrections of the form
    \begin{equation}
        S_D= -h\int d\tau\,\sigma_D + \int d\tau\left[
            \gamma^{8/7}c_{\mathbf1}(g)\,\mathbf1
            +\gamma c_\sigma(g)\,\sigma_D \label{eq:defectactoinmatch}
            +c_\epsilon(g)\,\epsilon_D
            +\cdots
        \right],
    \end{equation}
    where, for  $\epsilon_D$ defined with a fixed subtraction scale in units of $h$ as in (\ref{eq:Ising_eps_ren}), we have
    \begin{equation}
        \begin{aligned}
            c_{\mathbf1}(g)
            &=c_{\mathbf1}^{(0)}g^{6/7}+\cdots,\\
            c_\sigma(g)
            &= g^2\left[c_\sigma^{(2)}
            +c_\sigma^{(\log)}\log g\right]+\cdots,\\
            c_\epsilon(g)
            &=c_\epsilon^{(2)}g^2+\cdots,
        \end{aligned}
    \end{equation}
    as follows from symmetry considerations and dimensional analysis.  Note that the logarithmic term in $c_\sigma$ is allowed due to the additive mixing between $\epsilon_D$ and $\sigma_D$. Thus, as $g\to0$, only the magnetic perturbation survives.

    Next, we match $Z_j$ onto local operators on the defect.
    \begin{equation}
        Z_j
        =a^{1/8}d_\sigma(g)\,\sigma_D
        +d_{\mathbf1}(g)\,\mathbf1
        +a\,d_\epsilon(g)\,\epsilon_D
        +\cdots, \label{eq:matchingwithscale}
    \end{equation}
    where, as above,
    \begin{equation}
        \begin{aligned}
            d_{\mathbf1}(g)
            &=d_{\mathbf1}^{(1)}g+\cdots,\\
            d_\sigma(g)
            &=Z_\sigma
            +g^2\left[d_\sigma^{(2)}
            +d_\sigma^{(\log)}\log g\right]+\cdots,\\
            d_\epsilon(g)
            &=d_\epsilon^{(1)}g+\cdots.
        \end{aligned}
    \end{equation}

    To extract $\langle \sigma_D \rangle$,
    we evaluate, using \texttt{DMRG}, the expectation value $\langle Z_j \rangle$ on the ground state of the Hamiltonian with a single pinning defect.
    Taking into account the matching (\ref{eq:matchingwithscale}), the defect action (\ref{eq:defectactoinmatch}) and the irrelevant bulk deformations, we fit to the truncated ansatz
    \begin{equation}
        \langle Z_j\rangle_g^{\mathrm{fit}}
        =g^{1/7}\left[
            Z_\sigma\gamma^{-1/7}\langle\sigma_D\rangle_h
            +m_{6/7}g^{6/7}
            +g^2\left(m_2+m_{\log}\log g\right)
        \right],
    \end{equation}
    where, as above, we first extrapolate to infinite volume at fixed $g$, and then fit $g \rightarrow0$. We use six lattice spacings with $10^{-3}<g<10^{-1}$. Note that $Z_\sigma$ was determined in (\ref{eq:Zsigma}). We obtain
    \begin{equation}
        \gamma^{-1/7} \langle \sigma_D \rangle_h = 1.23331(13). \label{eq:spin-condensate}
    \end{equation}

    It remains to fix $\gamma$ in terms of $h$ and determine $\langle \epsilon_D \rangle$. For that purpose, we consider an insertion of $\epsilon$ separated from the defect by a distance $r$. CPT predicts the following bulk-to-defect OPE:
    \begin{equation}
    \begin{aligned}
        \label{eq:energy-profile} \langle\epsilon(r)\rangle_h
        ={}&\langle\epsilon_D\rangle_h-h\log r\,\langle\sigma_D\rangle_h+c_{\mathbf1}h^2r^{3/4}\\
        &+h^2r^{7/4}\left[c_\epsilon\langle\epsilon_D\rangle_h+h(c_\sigma-c_\epsilon\log r)\langle\sigma_D\rangle_h\right]\\
        &+r^2\left[\tfrac12\langle\epsilon_{D,2}\rangle_h+h(c_{\sigma,2}-4\log r)\langle\sigma_{D,2}\rangle_h\right]+\cdots,
    \end{aligned}
    \end{equation}
    where $\mathcal O_{D,2}\equiv[\partial_\perp^2\mathcal O]_D$, as worked out in appendix~\ref{app:bulk-to-defect}. Above we fix
    \begin{equation}
        c_{\mathbf1}
        =\frac{\Gamma\!\left(\frac78\right)\Gamma\!\left(-\frac38\right)^3}
        {4\sqrt{\pi}\,\Gamma\!\left(-\frac34\right)}
        =1.7797467013...\,,
    \end{equation}
    see equation (\ref{eq:Ising_BulkBound_c1}). The remaining powers have non-perturbative coefficients. As usual, the logarithmic terms signal the additive mixing of $\epsilon_D$ and $\epsilon_{D,2}$ with $\sigma_D$ and $\sigma_{D,2}$ respectively, see e.g.\ (\ref{eq:Ising_eps_ren}).

We fit (\ref{eq:energy-profile}) to the \texttt{DMRG} data\footnote{Thanks to locality, the \texttt{DMRG} data for $\langle\epsilon(r)\rangle_h$ at fixed separation $r$ can be obtained using the bulk matching \eqref{eq:Xmatching}, with coefficients \eqref{eq:Zmatch}. Corrections in the matching due to the presence of the defect are negligible in the continuum limit.} to extract
$\langle\epsilon_D\rangle_h$ from the constant term and
$h/\gamma$ from the  $\log r$ coefficient, using our result (\ref{eq:spin-condensate}) for  $\langle\sigma_D\rangle_h$.
We initially fix $\gamma$ to set reference units, assigning
each lattice pair $(g,n)$ the separation $r=n(g/\gamma)^{8/7}$.
The fitted ratio $h/\gamma$ then converts distances to units
set by $h$:
\begin{equation}
    h^{8/7}r
    =\left(\frac{h}{\gamma}\right)^{8/7}ng^{8/7}.
\end{equation}

As before, we first extrapolate to infinite volume and then take the continuum
limit using the same six lattice spacings. The short-distance fits cover $0.01<rh^{8/7}<0.15$. For the results shown in figure~\ref{fig:fitpotential} and equations~(\ref{eq:condensatesresultDMRG}), (\ref{eq:fit1result}), (\ref{eq:fit2result}) and (\ref{eq:fit3result}),
we use units in which $h=1$, with the fused defect having coupling $2h$, and set $\mu=1$. In sum, we obtain
\begin{equation}
h = 0.401(11) \gamma,\qquad \langle \sigma_D\rangle_{2h}^{\texttt{DMRG}} = 1.552(7) h^{1/7},\qquad \langle \epsilon_D\rangle_{2h}^{\texttt{DMRG}} = -6.04(26)h^{8/7}. \label{eq:condensatesresultDMRG}
\end{equation}
These are to be compared with the exact values
\begin{equation}
\langle \sigma\rangle_{2h}^{\text{exact}} = 1.5512...\, h^{1/7},\qquad \langle [\epsilon]_\text{r}\rangle_{2h}^{\text{exact}} = -6.057...\,h^{8/7}, \label{eq:condensatesresultExact}
\end{equation}
which follow from the results of section~\ref{sec:IsingCondensates} evaluated at coupling $2h$ and $\mu=1$.

\subsubsection{Two-defects}

Finally, we compute the ground-state energy $V_{h,h}(r)$. The first step is to convert lattice energies.
This is achieved by matching the pinning term $g Z_{j_\alpha}$ in (\ref{eq:latticeH})
 to the continuum Hamiltonian term $h \sigma_D$. Using (\ref{eq:Zsigma}) and
 (\ref{eq:condensatesresultDMRG}) we obtain\footnote{A number of parameters in this section are known exactly from the exact solution of the lattice model, but we leave the discussion generic to illustrate the method. Here, the exact conversion factor is $1/(2a)$.}
\begin{equation}
    V_{h,h}(r)-V_{h,h}(r_0)
    ={0.499(14)} a^{-1}
    \left(V^\text{lat}_{h,h}(r)-V^\text{lat}_{h,h}(r_0)\right).
\end{equation}
Here, we subtract lattice ground-state energies at fixed
$g$, $N$. This cancels the extensive vacuum energy, corresponding
to a cosmological constant counterterm.

\begin{figure}[t]
    \centering
    \includegraphics[width=0.95\linewidth]{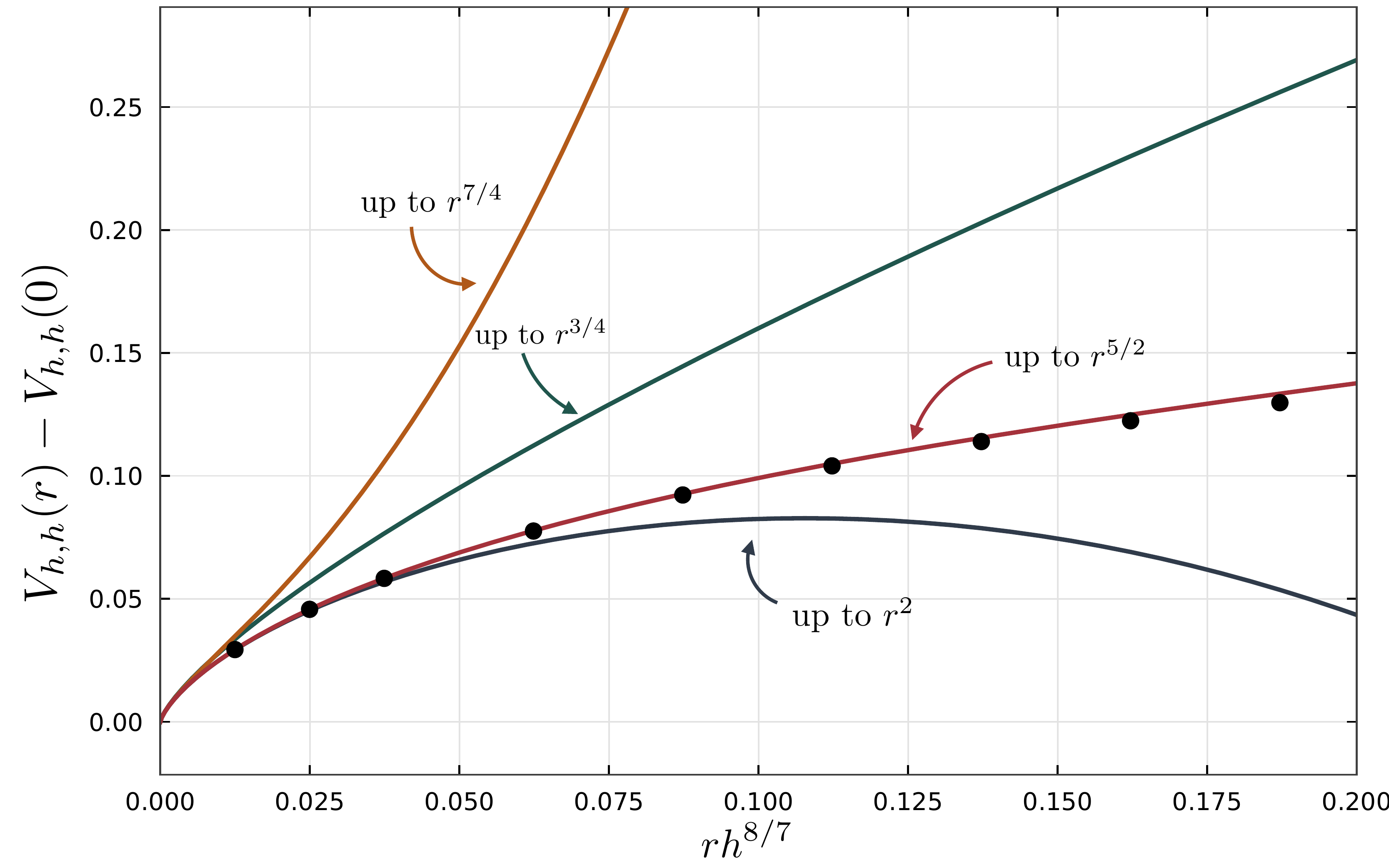}
\caption{
The potential $V_{h,h}(r)-V_{h,h}(0)$ in units $h=1$
(black points),
compared with truncations of
\eqref{eq:Ising_potential_OPE} through orders $r^{3/4}$ (green),
$r^{7/4}$ (orange), $r^2$ (blue), and $r^{5/2}$ (red).
All coefficients are fixed by (\ref{eq:Ising_constants}) and the exact condensates of section~\ref{sec:IsingCondensates}. The constant $V_0=V_{h,h}(0)$ is estimated by fitting to the two smallest separations.
}
    \label{fig:fitpotential}
\end{figure}

In the lattice, we insert two defects with coupling $g$ separated by $n$ sites, and measure the energies for various values of $N$.
After extrapolating to infinite volume and the continuum, we obtain the potential $V_{h,h}(r)$ for nine separations $r \in (0.01,0.2)h^{-8/7}$.
The result is illustrated in figure \ref{fig:fitpotential}. The prediction (\ref{eq:Ising_potential_OPE}) holds up to an additive constant, which we estimate by fitting to the two smallest separations.

Overall, we find excellent agreement between the short-distance asymptotics
in the potential and the predictions advocated by the framework of section
\ref{sec:Lines}. Figure \ref{fig:fitpotential} further compares the results for the potential with  the prediction (\ref{eq:Ising_potential_OPE}) truncated at successive orders. As is clear from the picture, it is essential to include the higher terms of order $r^2$  and $r^{5/2}$ in order to properly reproduce the potential in the $r$ range considered.

As mentioned, we can also extract the condensates from the numerical potential itself. For this we write the prediction (\ref{eq:Ising_potential_OPE}) as
\begin{equation}
    V_{h,h}(r)
    = V_0 - C_0 h^2 r^{3/4}
    + \bigl[A \log(r) + B\bigr]r^{7/4}
    + D r^2 + E r^{5/2} + O(r^{11/4})\,. \label{eq:potentialfit}
\end{equation}
The coefficients $A$ and $B$ are related to the condensates through
\begin{equation}
    A = 2h^3 C_\epsilon\langle \sigma\rangle_{2h} \,  ,\qquad B = 2h^3 C_\sigma  \langle \sigma\rangle_{2h}-h^2C_\epsilon \langle\epsilon\rangle_{2h}\, . \label{eq:ABcondensates}
\end{equation}
We fit the data according to
(\ref{eq:potentialfit}).  Keeping all parameters $A,B,D,E$
free makes the problem quite ill-conditioned due to the strong correlation among the fit parameters
over the available range in $r$. Instead, we perform three exercises. First, fixing $A$ with the $\langle \sigma_D \rangle_{2h}$ condensate determination (\ref{eq:condensatesresultDMRG}) and leaving $D$ and $E$ free,
we obtain the fit prediction for $B$ and, in turn
\begin{equation}
    \langle \epsilon_D\rangle_{2h}^\text{fit} =  -6.07(11) h^{8/7}, \label{eq:fit1result}
\end{equation}
which should be compared with the direct condensate determination of $\langle \epsilon_D \rangle$ (\ref{eq:condensatesresultDMRG}).

Similarly, fixing $B$ according to the condensate predictions (\ref{eq:condensatesresultDMRG}), which only uses the linear combination (\ref{eq:ABcondensates}), allows us to fit for $A$ and, in turn, obtain
\begin{equation}
 \langle \sigma_D\rangle_{2h}^\text{fit} = 1.550(8) h^{1/7}, \label{eq:fit2result}
\end{equation}
to be compared with the determination of $ \langle \sigma_D\rangle_{2h}$ in (\ref{eq:condensatesresultDMRG}).

Finally, fixing both $A$ and $B$ to the central values of the condensate determinations (\ref{eq:condensatesresultDMRG}), and $E=h^4C_0^{(4)}$ from (\ref{eq:Ising_potential_OPE}), we fit for $D = -\tfrac{h}{4}\langle \partial_x^2\sigma\rangle_{2h}$ and, in turn, obtain
\begin{equation}
 \langle \partial_x^2\sigma_D\rangle_{2h}^\text{fit} = 139.3(2)\, h^{17/7}, \label{eq:fit3result}
\end{equation}
to be compared with the exact value $\langle \partial_x^2\sigma_D\rangle_{2h}^\text{exact} = 139.19\, h^{17/7}$ that follows from (\ref{eq:hardcondensate}).

\begin{figure}[t]
    \centering
    \includegraphics[width=0.95\linewidth]{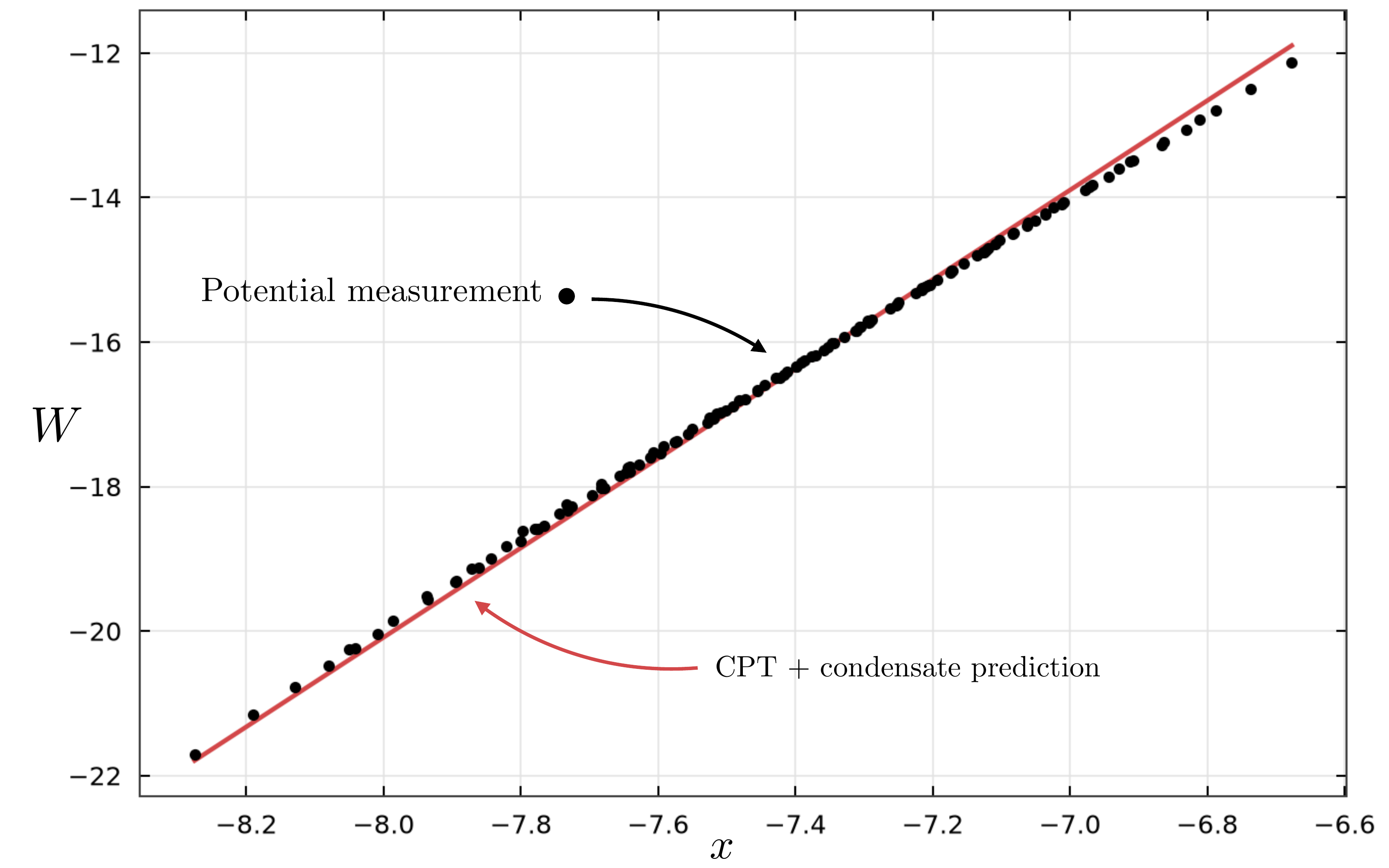}
    \caption{
    Black dots: the linear combination $W$ as a function of $x$ for all
    $\binom{9}{4}$ quadruplets built out of the \texttt{DMRG} data.
    Red line: CPT prediction for $W(x)$ up to $o(r^{5/2})$ corrections in the potential, with coefficients determined from the independent
    condensate determinations (\ref{eq:condensatesresultDMRG}).
    The maximum relative deviation is $1.98\%$.
    }
    \label{fig:fitline}
\end{figure}

Before concluding, let us briefly point out a separate strategy to tackle the ill-conditioning mentioned above.  Choose any four distinct points $r_i$ in the data and define
\begin{equation}
    W = \sum_{i=1}^4 w_i (V(r_i) + C_0 h^2 r_i^{3/4} ). \label{eq:Wdef}
\end{equation}
Here, $w_i$ are the solution to the linear system
\begin{equation}
    \sum_{i=1}^{4} w_i = 0,\qquad
    \sum_{i=1}^{4} w_i r_i^2 = 0,\qquad
    \sum_{i=1}^{4} w_i r_i^{5/2} = 0,\qquad
    \sum_{i=1}^{4} w_i r_i^{7/4} = 1.
\end{equation}
Crucially, the weights are chosen so that, when plugging the asymptotic (\ref{eq:potentialfit}) into $W$, the result is independent of $V_0$, $D$ and $E$. Up to higher CPT corrections, we find
\begin{equation}
W = A \sum_{i=1}^4 w_i r_i^{7/4 }\log r_i     + B.
\end{equation}

The independent condensate determination (\ref{eq:condensatesresultDMRG})
predicts that $W$ is a completely fixed linear function of $x = \sum_{i=1}^4 w_i r_i^{7/4 }\log r_i $ if $r_i$ are taken asymptotically small.
Figure~\ref{fig:fitline} shows this prediction, compared with the data obtained by computing $W$
with the  $\binom{9}{4}$ different quadruplets of data points.\footnote{These are, of course, highly correlated.} Again, we observe great agreement between the potential at short separations and the predictions from the framework of section \ref{sec:Lines}, with the maximum relative deviation in figure \ref{fig:fitline} of less than $2\%$.

\section{\boldmath Conclusions and \texorpdfstring{$q\bar q$-potential}{quark-antiquark potential}}\label{sec:conclusions}

In this paper we have introduced the OPE of line operators in massive QFTs. It is described by an effective action comprising an infinite tower of corrections on the fused defect, with Wilson coefficients computable order by order in CPT around the UV fixed point. We have shown how to compute these coefficients and tested the validity of our proposal in a variety of theories.
In the solvable examples of section \ref{sec:examples} we found perfect agreement between exact results and predictions from the OPE. In the non-perturbative example of section~\ref{sec:IsingModel} we also found remarkable agreement with numerical results.
We take this as strong evidence in favor of our proposal.

This new tool enables predictions for the short-distance limit of potentials between static probes. More generally, for any correlation function with parallel defects, it yields an order-by-order asymptotic expansion in the separation $r$ of the defects. This expansion achieves a sharp separation between UV and IR degrees of freedom. On the UV side are the fusion coefficients, computable in CPT; on the IR side, the correlation functions of the fused defect. Since the $r$-dependence enters only through fusion coefficients, the OPE completely fixes the functional form of the asymptotic expansion, to any order in $r$. Only the non-perturbative condensates of the fused defect remain undetermined.

This UV-IR connection makes the uses of the OPE twofold. It can be used as a fitting function to extract IR condensates from short-distance measurements. Or, if the condensates have been measured directly, they can be plugged into the OPE to determine the UV limit of correlation functions. Another important feature of these expansions is universality.
If a fused defect features in the OPEs of different (pairs of) defects, its IR condensates will be shared by the corresponding correlators, which will only differ by the perturbative fusion coefficients.
Likewise, since the OPE is an operator statement, the same fusion coefficients apply in correlation functions of two given defects irrespective of distant insertions. Moreover, since fusion coefficients are local, they are also independent of the geometry of the defects. This leads to universal predictions e.g.\ for the temperature dependence of the UV limit of defect correlation functions.

When it comes to applications of the OPE, in this paper we have but considered the simplest examples. Many exciting new avenues await exploration. A promising one, which we are currently investigating, is the static quark-antiquark potential in pure Yang-Mills (YM) theory. It is defined from the correlator of infinitely-long straight Wilson lines,
\begin{equation}
    V_{q\bar q}(r) \equiv -\lim_{L\to\infty} \frac{1}{L}\log\langle W_{\Box}[r]W_{\overline{\Box}}[0]\rangle\,.
\end{equation}
Wilson lines in YM theory naturally fall in the class of defects considered in this paper. They are deformations of trivial lines in the free theory as in \eqref{eq:g-def} and \eqref{eq:h-def} with $g,h$ both proportional to the gauge coupling $g_\text{YM}$.

In $d=3$, the gauge coupling $g_\text{YM}$ is relevant, with dimensions of $\sqrt{\text{mass}}$, and our framework directly applies. In contrast, in $d=4$ the gauge coupling is only marginally relevant, and the vanishing classical dimension complicates things. We still expect the defect OPE as outlined in this paper to apply, but some aspects related to dimensional analysis will need care. Let us discuss in broad strokes the structure of the short-distance $q\bar q$-potential that follows from our framework, both for $d=3,4$. Further details will be provided in upcoming work.

Since $\Box \otimes \overline{\Box} = 1 \oplus \text{adj}$, the two Wilson lines fuse to a non-simple defect, $1 \oplus W_\text{adj}$. The effective action therefore has the block-matrix structure discussed in section~\ref{sec:non-simple}. Listing the lowest-lying operators, we can write
\begin{equation}
    e^{-S_\text{eff}} = \mathcal P \exp \left[-\int ds \begin{pmatrix}
        \frac{-a_0^{(1)}}{r}  & a_{F^{\mu\nu}}r F_{\mu\nu}^c \\
         a_{F_{\mu\nu}}r F^{\mu\nu}_b & \hspace{5pt}\frac{-a^{(\text{adj})}_0\delta_{b}^c}{r} - i v^\mu A_\mu^a {f_{ab}}^c + a_{F^{\mu\nu}}^{(\text{adj})} r F_{\mu\nu}^a {t_{ab}}^c 
    \end{pmatrix}
    +O(r^2)\right]\,,
\end{equation}
where $a,b,c$ are $SU(N)$ adjoint indices and ${t_{ab}}^c$ is an invariant tensor in the two-dimensional space of $\text{adj}\otimes \text{adj}\to \text{adj}$ intertwiners spanned by ${f_{ab}}^c$ and ${d_{ab}}^c$. We are using conventions where $A_\mu^a$ has mass dimension 1. The first entry of the matrix contains singlet operators, starting with $\text{tr} F^2$. The last entry of the matrix contains operators living on the adjoint Wilson line, hence the bi-adjoint tensor structures. Off-diagonal entries contain endpoint operators for the adjoint Wilson line, the first being $F_{\mu\nu}^a$.

Taking the expectation value, we obtain the short-distance expansion
\begin{align}\label{eq:WW-prediction}
    \langle W_{\Box}[r]W_{\overline{\Box}}[0]\rangle \simeq&\, e^{a_0^{(1)}L/r} + e^{a_0^{(\text{adj})}L/r} \bigg(\langle W_\text{adj}\rangle 
    -  a^{(\text{adj})}_{F_{\mu\nu}}r L \langle W_\text{adj} F^{\mu\nu}(0)\rangle  
    +\cdots
    \bigg)\\
    & \hspace{-3pt}+ e^{a_0^{(1)}L/r} a_{F_{\mu\nu}}a_{F^{\rho\sigma}} r^2 L\int ds \, e^{(a_0^{(\text{adj})}-a_0^{(1)})s/r}\langle F^{\mu\nu}_b(0) W_\text{adj}[0,s]_c^b F^c_{\rho\sigma}(s)\rangle +\cdots. \nonumber
\end{align}
According to the framework that we have put forward in this paper, the fusion coefficients $a_i$ should be dimensionless functions of $r$ computable in CPT. Note that CPT here is none other than standard perturbation theory in the gauge coupling $g_\text{YM}$. On the other hand, the expectation values of the adjoint Wilson line are non-perturbative IR condensates. It is easy to see that the Casimir term of the trivial line dominates the one of the adjoint line, so the second term in \eqref{eq:WW-prediction} is comparatively exponentially suppressed and can be dropped.

Taking the log, we reach the following prediction for the short-distance limit of the static $q\bar q$-potential:
\begin{equation}\label{eq:Vqq-prediction}
    V_{q\bar q}(r) \simeq -\frac{a_0^{(1)}}{r} - a_{F_{\mu\nu}}a_{F^{\rho\sigma}} r^2 \int ds \, e^{(a_0^{(\text{adj})}-a_0^{(1)})s/r}\langle F^{\mu\nu}_b(0) W_\text{adj}[0,s]_c^b F^c_{\rho\sigma}(s)\rangle + \cdots\,.
\end{equation}
After the discovery of asymptotic freedom \cite{Gross:1973id,Politzer:1973fx}, the quark-antiquark potential was one of the early targets of perturbation theory \cite{Susskind:LesHouches1976,Fischler:1977yf,Appelquist:1977tw}. At the first few orders, the perturbative computation of $V_{q\bar q}(r)$ amounts to the computation of $a_0^{(1)}(r)$. For instance, the leading contribution is the Coulomb potential: $a_0^{(1)}(r) \sim g_\text{YM}^2 r \log r$ in $d=3$, and $a_0^{(1)}(r) \sim 1/\log r$ in $d=4$. Soon enough, though, the computation of $V_{q\bar q}(r)$ attempts to reproduce the condensate $\langle FW_\text{adj}F\rangle$, which is non-perturbative and thus full of IR divergences in perturbation theory. This results in what was deemed a ``breakdown of perturbation theory'' in~\cite{Appelquist:1977es}.

With our framework, the resolution to this classic puzzle is quite transparent. One should only compute the $a_i(r)$ in perturbation theory, and leave the condensates as undetermined non-perturbative quantities to be fixed by other means. We are working towards verifying this numerically in conjunction with lattice QCD experts.
A similar resolution was achieved using potential non-relativistic QCD (pNRQCD) \cite{Brambilla:1999qa,Brambilla:1999xf}, an effective field theory for non-relativistic heavy quark-antiquark systems. We expect the defect OPE to provide a reformulation of this EFT directly in terms of UV QCD operators, enabling a systematic computation of subleading corrections to (\ref{eq:Vqq-prediction}), but we leave a detailed comparison of the two methods for future work.
    
As with pNRQCD, we anticipate that the results for the $q\bar q$-potential will find applications in heavy quarkonium, i.e.\ mesons made of heavy quarks such as the charmonium $J/\psi$. They will also provide subleading corrections for the UV behavior of confining strings. In~\cite{Albert:2026fqj}, some of us pointed out that the Polyakov loop correlator can be regarded as a cylinder partition function for confining strings. By channel duality, the limit where the loops collide controls the high-energy states in the closed-string channel. Using the leading (Coulomb) behavior of the $q\bar q$-potential, we extracted, among other results, the leading closed-string spectral density. The prediction \eqref{eq:Vqq-prediction} will enable the systematic order-by-order determination of subleading corrections.

Beyond Yang-Mills and QCD, the OPE of lines unlocks the door to the UV of defect correlators in many more quantum field theories. Some immediate extensions of our framework that would be interesting to pursue are generalizations to higher-dimensional defects, as well as applications to the fusion of interfaces and boundaries. We look forward to these and many other applications of the OPE of lines beyond CFT.

\section*{Acknowledgments}
We are grateful to Simon Caron-Huot, Yichul Choi, Gabriel Cuomo, Diego Delmastro, Anna-Maria Gl\"uck, Zohar Komargodski, Ian Moult, Chintan Patel, Benjamin S\o gaard and Joan Soto for useful discussions. JA and MF acknowledge partial support from Simons Foundation grant 917464 (Simons Collaboration
on Confinement and QCD Strings). This material is based upon work supported by the U.S. Department of Energy, Office of Science, Office of High Energy Physics, under Award Number DE-SC0011632. This research was supported in part by grant NSF PHY-2309135 to the Kavli Institute for Theoretical Physics (KITP). AH thanks the KITP for generous hospitality while this work was carried out.

\section*{AI/Human statement}

All ideas behind the OPE of lines in QFT and its applications are our own. So is the text. AI models (GPT-5.6 Sol, GPT-6 Astra, Claude Opus 5.5, Fable 5.2) were used to assist with CPT computations throughout the paper, all of which were reproduced by hand. They were also used to write and run code for the \texttt{DMRG} computations, as well as for proofreading, figure-making, and minor editing of the manuscript.

After this work was complete, with the Ising potential matched against (\ref{eq:Ising_potential_OPE}) with the \texttt{DMRG}-extracted condensates, Claude Opus 5.5 and GPT-6 Astra pointed out that the magnetic pinning defect condensates could also be computed with integrability (initially via \cite{Lamacraft:2008yr}). Section \ref{sec:IsingCondensates} and appendix \ref{app:TBA} are the outcome of following that observation through, with most computations carried out by agents under interactive direction of the authors, who set the strategy, presentation and verified the results.

\appendix
\addtocontents{toc}{\protect\setcounter{tocdepth}{1}}

\section{Curvature corrections}\label{app:curvature}

This appendix reviews some background material on the construction of curvature invariants for the discussion in section~\ref{sec:curvature}. Although we are ultimately interested in lines, the discussion is phrased for defects of general dimension. We follow \cite{Capovilla:1994bs} and \cite[Ch.\ 8]{Lee}. Consider a $p$-dimensional manifold $\Sigma_p$ embedded in a $d$-dimensional spacetime manifold $M_d$, with embedding $X^\mu(\sigma^a)$ (where $\mu = 1,\dots, d$, $a=1,\dots,p$). Let $g_{\mu\nu}$ be a metric on $M_d$. On a given point $\sigma^a$ of $\Sigma_p$, a basis for its tangent space is given by the $p$ vectors
\begin{equation}
	e_a^\mu(\sigma) \equiv \frac{\partial X^\mu(\sigma)}{\partial \sigma^a}\,.
\end{equation}
The product of these vectors defines the induced metric on $\Sigma_p$,
\begin{equation}
	h_{ab} \equiv e_a \cdot e_b = e_a^\mu e_b^\nu g_{\mu\nu}\,.
\end{equation}
Out of the induced metric $h$, one can construct a Riemann tensor $\mathcal R_{abcd}$ in the usual way. This captures the \textit{intrinsic curvature} of $\Sigma_p$.

The manifold $\Sigma_p$ has further geometric information describing the embedding into $M_d$. It can be accessed by studying the space of vectors normal to $\Sigma_p$. At a point $\sigma\in \Sigma_p$, a basis for this space is given by a set of $d-p$ vectors $n_i^\mu(\sigma)$ satisfying
\begin{equation}
	e_a \cdot n_i = 0\,, \qquad n_i \cdot n_j = \delta_{ij}\,.
\end{equation}
We will raise/lower tangent indices $a,b,...$ with $h_{ab}$ and normal indices $i,j,...$ with $\delta_{ij}$. One can ask how the basis vectors $\{e_a, n_i\}$ change as we move slightly along $\Sigma_p$ by taking bulk covariant derivatives ($D_\mu v^\nu = \partial_\mu v^\nu +\Gamma^\nu_{\mu\rho} v^\rho$) along the tangent directions, i.e.\ $e_a^\mu D_\mu$. The normal and tangent components of these derivatives are fixed by the Gauss-Weingarten equations,
\begin{equation}\label{eq:GWeqs}
	e_a^\mu D_\mu e_b^\nu = \gamma_{ab}^ce_c^\nu - K_{ab}^in_i^\nu\,, \qquad
	e_a^\mu D_\mu n^{i\nu} = K_{ab}^ie^{b\nu} + \omega_a^{ij}n_j^\nu\,.
\end{equation}
These define the tensor $K_{ab}^i$, known as the \textit{second fundamental form}, which captures the \textit{extrinsic curvature} of $\Sigma_p$ in $M_d$. It is symmetric in the tangent indices $ab$, and it transforms as a tensor under diffeomorphisms of $\Sigma_p$ and local rotations of the normal vectors $n^i(\sigma)$. In turn, $\gamma_{ab}^c$ and $\omega_a^{ij}$ are, respectively, coefficients of the connections on the tangent and normal bundles (the former being the Christoffel symbols relative to the induced metric $h$). They can be used to define a covariant derivative $\nabla_a$ on $\Sigma_p$ acting on tangent indices as $\nabla_a v^b = \partial_a v^b + \gamma^b_{ac}v^c$, and on normal indices as $\nabla_a v^i = \partial_a v^i - \omega_a^{ij}v_j$. They also define another curvature tensor,
\begin{equation}
    \Omega_{ab}^{ij} = \partial_b \omega_a^{ij} - \partial_a \omega_b^{ij} - \omega_b^{ik}{{\omega_a}_k}^j + \omega_a^{ik}{{\omega_b}_k}^j\,.
\end{equation}

Curvature corrections for an action on $\Sigma_p$ are given by local integrals, $\int_{\Sigma_p} d^p \sigma \sqrt{\text{det} h}$, of invariants under $\Sigma_p$-diffeomorphisms and local rotations of the normal bundle, built out of $h_{ab}, K_{ab}^i,\mathcal R_{abcd}, \Omega_{ab}^{ij},\nabla_a$ as well as the spacetime curvature $R_{\mu\nu\rho\sigma}$ of $M_d$. Not all of these ingredients are independent. They are related by a set of equations carrying combinations of the names of Gauss, Codazzi, Mainardi, and Ricci (see e.g.\ \cite{Lee,Capovilla:1994bs}). We do not reproduce them here, as we will not have need for them. Integration by parts can also relate curvature corrections which appear distinct on first sight.

Let us restrict to the case of line defects ($p=1$) in $M_d = \mathbb R^d$, with $\Sigma_1 \equiv \gamma$ and $\sigma^1 \equiv s$. We note that $\mathcal R_{ssss}, \Omega_{ss}^{ij}, R_{\mu\nu\rho \sigma} = 0$.\footnote{By a slight abuse of notation, we denote the tangent index by the same name as the coordinate, $s$.} The induced metric is a single coefficient $h_{ss}$ and, by defining $K^i \equiv K_{ss}^i/h_{ss}$, we are left only with singlet contractions of this vector and the covariant derivative $\nabla\equiv (1/\sqrt{h_{ss}})\nabla_s$. The only curvature corrections to consider are thus products of $\nabla^a K_i \nabla^b K^i$, as we pointed out in section~\ref{sec:curvature}.

As a simple example, we evaluate curvature corrections in the case where $\gamma$ is a circle of length $L$ embedded in $\mathbb R^d$. This is the only nontrivial geometry that we will consider in this paper. Choosing suitable Cartesian coordinates on $\mathbb R^d$, we can write the embedding function as
\begin{equation}
	X^\mu(s) = \frac{L}{2\pi}\left(\cos \frac{2\pi s}{L}, \sin \frac{2\pi s}{L},0,\ldots,0 \right)\,, \qquad s\in [0,L)\,.
\end{equation}
It follows that the tangent vector and the induced metric are
\begin{equation}
	e_s^\mu(s) =\partial_s X^\mu(s) = \left(-\sin \frac{2\pi s}{L}, \cos \frac{2\pi s}{L},0,\ldots,0 \right)\,, \qquad h_{ss} = e_s\cdot e_s = 1\,.
\end{equation}
A set of normal vectors is given by
\begin{equation}
	n^{1\mu} = \left(\cos \frac{2\pi s}{L}, \sin \frac{2\pi s}{L},0,\ldots,0 \right)\,, \quad  n^{j\mu} = \left(0,0,0,\ldots,1\ldots,0 \right)\,, \quad (1<j\leq d-1)\,.
\end{equation}

In Cartesian coordinates, the flat-space Christoffel symbols vanish, so the bulk covariant derivative in the direction tangent to $\gamma$, $e_s^\mu D_\mu$, is simply the partial derivative $\partial_s$. The variation of the basis vectors along the $s$ direction is then given by
\begin{subequations}
\begin{align}
	e_s^\mu D_\mu e_s^\nu =&\, \frac{2\pi}{L}\left(-\cos \frac{2\pi s}{L}, -\sin \frac{2\pi s}{L},0,\ldots,0 \right)\,,\\
	e_s^\mu D_\mu n^{1\nu} =&\, \frac{2\pi}{L}\left(-\sin \frac{2\pi s}{L}, \cos \frac{2\pi s}{L},0,\ldots,0 \right)\,,\\
	e_s^\mu D_\mu n^{j\nu} =&\, 0\,.
\end{align}
\end{subequations}

Comparing these expressions with the Gauss-Weingarten equations \eqref{eq:GWeqs}, we determine the second fundamental form: $K_{ss}^1 = \frac{2\pi}{L}$, $K_{ss}^j = 0$ ($1<j\leq d-1$). Upon dividing by the induced metric, we have the normal vector
\begin{equation}
	K^i = \left(2\pi/L,0,\ldots,0\right)\,.
\end{equation}
The induced Christoffel symbols $\gamma^s_{ss}$ vanish, in agreement with the computation from the induced metric $\gamma^s_{ss} = \tfrac{1}{2}h_{ss}^{-1}\partial_s h_{ss}$, and so does the connection on the normal bundle, $\omega_s^{ij} = 0$. The covariant derivative on $\gamma$ is then also just a partial derivative, $\nabla_s = \partial_s$, and we find that $\nabla K^i=0$. This simplification is special to the circular geometry that we are considering. We conclude that the nontrivial curvature invariants evaluate to
\begin{equation}\label{eq:K2int}
	\int_\gamma ds \sqrt{h_{ss}} (K_iK^i)^{n} = \int_0^{L} \hspace{-5pt} ds\, \left(\frac{4\pi^2}{L^2}\right)^{n} = L\left(\frac{2\pi}{L}\right)^{2n}\,.
\end{equation}

\section{Free CFTs}\label{app:FreeCFTs}

Here we provide supplemental details for the calculations in section~\ref{sec:free_CFTs} for the massless scalar and free Maxwell theory.

\subsection{Free massless scalar}

We start by writing out the relevant integrals necessary for the case of parallel infinite lines.
First, we have the one-point functions of operators at $x$ in the presence of the defect
\begin{subequations}
\begin{align}
    \langle \phi^n(x) \mathcal D_{{h_1+h_2}}\rangle &= \left(\frac{h_1+h_2}{4\pi x}\right)^n\,,
\end{align}
\end{subequations}
which are the first terms in the three-point function matching in the presence of the defect.
Now also including operators brought down from the effective action, we have the correlation functions\footnote{Recall that the operators appearing on the defect have been rotated to remove divergences, i.e. $\phi^2 \to \tilde{\phi}^2-\frac{A}{\delta} \tilde{\phi}$.}
{\allowdisplaybreaks
\begin{subequations}
\begin{align}
    \int ds \langle \phi^n(x) \mathcal{D}_{{h_1+h_2}} \partial_i \phi(s)\rangle &=n \frac{x_i}{4\pi x^3}\left(\frac{h_1+h_2}{4\pi x}\right)^{n-1}\,,\\\nonumber
    \int ds \langle \phi^{n+m}(x) \mathcal D_{{h_1+h_2}} \phi^{n}(s) \rangle & = \frac{(n+m)!}{m!}\left(\frac{h_1+h_2}{4\pi x}\right)^m \int ds\left[\frac{1}{4\pi^2(x^2+s^2)}\right]^n\\
    &= \frac{(n+m)!}{m!}\frac{\sqrt{\pi} \Gamma(n-\frac{1}{2})}{(4\pi^2)^n \Gamma(n)x^{2n-1}}\left(\frac{h_1+h_2}{4\pi x}\right)^m\,,\\\nonumber
    \int ds \langle \phi^n(x) \mathcal{D}_{{h_1+h_2}} \partial_{i_1}\partial_{i_2}\ldots\partial_{i_m} \phi(s)\rangle &=n \left(\frac{h_1+h_2}{4\pi x}\right)^{n-1}(-1)^m \partial_{i_1}\partial_{i_{2}}\ldots \partial_{i_m} \frac{1}{4\pi x}\\
    &= \frac{n(2m-1)!!x_{\langle{i_1}}x_{i_2}\ldots x_{i_m \rangle}}{4\pi x^{2m+1}}\left(\frac{h_1+h_2}{4\pi x}\right)^{n-1}\,,
\end{align}
\end{subequations}
}
where all other combinations were rendered zero after rotating bases to eliminate the divergences.
In the second correlation function we have $m\ge 0$, for $m<0$ the result is zero.
For the third we have in the numerator the symmetric traceless combination $x_{\langle {i_1}}x_{i_2}\ldots x_{i_m \rangle}$. For the first couple cases used in the main text this integral becomes
\begin{subequations}
\begin{align}
        \int ds \langle \phi^n(x) \mathcal{D}_{{h_1+h_2}} \partial_i \phi(s)\rangle &=n \frac{x_i}{4\pi x^3}\left(\frac{h_1+h_2}{4\pi x}\right)^{n-1}\,,\\
        \int ds \langle \phi^n(x) \mathcal{D}_{{h_1+h_2}} \partial_i \partial_j\phi(s)\rangle &=n \frac{1}{4\pi}\left(\frac{3x_ix_j}{ x^5}-\frac{\delta_{ij}}{ x^3}\right)\left(\frac{h_1+h_2}{4\pi x}\right)^{n-1}\,.
\end{align}
\end{subequations}
Finally, in the main text we also needed the correlation functions
\begin{subequations}
\begin{align}
    \int ds_1 ds_2 \langle \phi^n(x) \mathcal D_{h_1+h_2} \partial_i \phi (s_1) \partial_j \phi(s_2)\rangle &= \frac{n(n-1)}{16\pi^2}\left(\frac{h_1+h_2}{4\pi x}\right)^{n-2} \frac{x_ix_j}{ x^6} \,,\\
    \int ds_1 ds_2 \langle \phi^n(x) \mathcal D_{h_1+h_2} \partial_i \phi (s_1) \partial_j \partial_k \phi(s_2)\rangle &= \frac{n(n-1)}{16\pi^2}\left(\frac{h_1+h_2}{4\pi x}\right)^{n-2} \frac{x_i}{x^3} \left[\frac{3x_{j}x_{k}}{x^5}-\frac{\delta_{jk}}{x^3} \right]\,,
\end{align}
\end{subequations}
which follow from the above. These suffice to extract all coefficients $a_i$ up to $O(r^3)$ for the case of infinite parallel defects.

\paragraph*{Curvature corrections}

To have the full result at $O(r^2)$ in the circular geometry used in section~\ref{sec:FreeCFT-scalar}, we also need to consider the operators $\phi K_i K^i$ and $\partial^i\phi K_i $.\footnote{Here this $\phi$ is the redefined $\phi = \tilde \phi - \frac{h_1+h_2}{2\pi^2 \delta}\mathds 1$, whose shift was absorbed into the cosmological constant counterterm in \eqref{eq:MasslessScalarDefectRenorm}.}
We may extract them from a three-point function as before.
Using an external $\phi(x)$ insertion, we have
\begin{align}\label{eq:MasslessScalarPhiDD_loops}
    &\frac{\langle \phi(x) \mathcal{D}_{h_1}[r]\mathcal{D}_{h_2}[0]\rangle_\text{circ}}{\langle \mathcal{D}_{h_1}[r]\mathcal{D}_{h_2}[0]\rangle_\text{circ}} = \frac{h_2 L}{L^2+4\pi^2 x^2}+\frac{h_1 L}{L^2+4\pi^2(r-x)^2}\\
    & \qquad \qquad \qquad\quad= \frac{(h_1  +h_2)L}{L^2+4\pi^2 x^2}+\frac{8 \pi^2 h_1  xL r}{(L^2+4\pi^2 x^2)^2}+\frac{4\pi^2h_1 (-L^2+12\pi^2 x^2)L r^2}{(L^2+4\pi^2 x^2)^3}+O(r^3)\,. \nonumber
\end{align}
We want to match this with the prediction from the effective action. 
Since $K^2$ is $s$-independent for our geometry and it evaluates to $4\pi^2/L^2$, we have (dropping $a_i$'s that we found were 0 already)
\begin{align}\label{eq:MasslessScalarDefectExpansionCurvature}
    \langle \phi(x) \mathcal{D}_{h_1}[r] \mathcal D_{h_2}[0] &\rangle_{\text{circ}} \simeq \,
    \exp\left[\int ds\left(\frac{a_0}{r}- a_{K^2} r K_\mu K^\mu+O(r^3)\right)\right]\bigg\{\langle  \phi(x)\mathcal  D_{{h_1+h_2}} \rangle\nonumber\\
    &- r\int ds\bigg[a_{\partial_i \phi} \langle \phi(x) \mathcal D_{{h_1+h_2}} \partial_i\phi(s)\rangle\bigg] \nonumber\\
    &-r^2 \int ds \bigg[a_{\partial_i \partial_j \phi} \langle \phi(x) \mathcal D_{{h_1+h_2}} \partial_i\partial_j \phi(s)\rangle\\
    &+\frac{4\pi^2}{L^2}a_{\phi K^2} \langle \phi(x) \mathcal D_{{h_1+h_2}} \phi(s) \rangle+a_{\partial_i\phi K^i} \langle \phi(x) \mathcal D_{{h_1+h_2}} K^i \partial_i\phi(s) \rangle \bigg]\nonumber\\
    &+ \frac{r^2}{2}\int ds_1 \int ds_2 a_{\partial_i \phi} a_{\partial_j \phi }\langle \phi(x) \mathcal D_{{h_1+h_2}} \partial_i \phi(s_1)\partial_j \phi(s_2) \rangle +O(r^3)\bigg\}\nonumber\,.
\end{align}

Dividing by the identity insertion removes the exponential prefactor, leaving only $a_{\phi K^2}$ and $a_{\partial_i \phi K^i}$ as unknowns.
The correlation functions on the RHS of~\eqref{eq:MasslessScalarDefectExpansionCurvature} must be evaluated in our new geometry.
The three-point function $\langle \phi(x) \mathcal D_{{h_1+h_2}} \partial_i \phi \partial_j \phi \rangle $ vanishes since we removed the divergent $\langle \partial_i \phi \partial_j \phi \rangle $ contribution already, while the remaining two-point functions are given by
\begin{subequations}
\begin{align}
    \int ds \langle \phi(x) \mathcal D_{{h_1+h_2}} \phi(s) \rangle_\text{circ} &= \frac{L}{L^2+4\pi^2x^2}\,,\\
    \int ds \langle \phi(x) \mathcal D_{{h_1+h_2}} \partial_r \phi(s) \rangle_\text{circ} &= \frac{8L\pi^2 x}{(L^2+4\pi^2x^2)^2}\,,\\
    \int ds \langle \phi(x) \mathcal D_{{h_1+h_2}} K^i\partial_i \phi(s) \rangle_\text{circ} &= -\frac{8L\pi^2}{(L^2+4\pi^2x^2)^2}\,,\\
    \int ds \langle \phi(x) \mathcal D_{{h_1+h_2}} \partial_r^2 \phi(s) \rangle_\text{circ} &= -\frac{8L\pi^2(L^2-12 \pi^2 x^2)}{(L^2+4\pi^2 x^2)^3}\,.
\end{align}
\end{subequations}
Inserting these into the defect fusion along with the $a_i$'s found earlier, we immediately get
\begin{equation}
    a_{\phi K^2} = 0\, , \qquad a_{\partial_i \phi K^i} = 0 \, .
\end{equation}

In the circular geometry, for the extraction of $a_{\partial_i \phi K^2}$ and $a_{\phi^2 K^2}$ we first need to compute the contributions to $\langle \phi \mathcal D \mathcal D\rangle$ and $\langle \phi^2 \mathcal D \mathcal D\rangle$ from the correlation functions of operators in the effective action after fusion.
Here we specialize to $x$ on the $x$-axis.
For the first, the only result missing from the main text that we need is given by
\begin{equation}
    \int ds\langle \phi(x) \mathcal{D}_{h_1+h_2} \partial_x^3\phi(s) \rangle_\text{circ} = -\frac{384 \pi ^4 L x \left(L^2-4 \pi ^2 x^2\right)}{\left(L^2+4 \pi ^2 x^2\right)^4}\,,
\end{equation}
which reproduces the $O(r^3)$ term in the expansion when combined with its coefficient, yielding $a_{\partial_i \phi K^2} =0$.

For the $\phi^2(x)$ insertion, we have several contributions to consider.
Our fusion must reproduce the full expression given by
\begin{align}
\begin{split}
\frac{\langle \phi^2(x) \mathcal D_{h_1}(r) \mathcal D_{h_2}(0)\rangle_\text{circ}}{\langle \mathcal D_{h_1}(r) \mathcal D_{h_2}(0)\rangle_\text{circ}} &= \frac{L^2 \left(h_1 \left(L^2+4 \pi ^2 x^2\right)+h_2 \left(L^2+4 \pi ^2 (r-x)^2\right)\right)^2}{ \left(L^2+4 \pi ^2 x^2\right)^2 \left(L^2+4 \pi ^2 (r-x)^2\right)^2}\,,
\end{split}
\end{align}
order by order in $r$. 
The condensate contributions $\int ds \langle \phi^n(x) \mathcal{D}_{h_1+h_2} \mathcal{O}_i(s) \rangle$ are all just the $\int ds \langle \phi(x) \mathcal{D}_{h_1+h_2} \mathcal{O}_i(s) \rangle$ contributions we computed in the main text and above multiplied by the factor $n\left(\frac{L(h_1+h_2)}{L^2+4\pi^2x^2}\right)^{n-1}$, so we will not repeat them.
We do need two contributions that did not contribute when inserting $\phi$ which are given by
\begin{subequations}
\begin{align}
\int ds_1 ds_2 \langle \phi^2(x) \mathcal D_{h_1+h_2} \partial_x \phi (s_1) \partial_x \phi (s_2)\rangle_\text{circ} &= 2\left[\frac{8 \pi ^2 L x}{\left(L^2+4 \pi ^2 x^2\right)^2}\right]^2 \,, \\
\int ds_1 ds_2 \langle \phi^2(x) \mathcal D_{h_1+h_2} \partial_x \phi (s_1) \partial_x^2 \phi (s_2)\rangle_\text{circ} &= -\frac{128 \pi ^4 L^2 x \left(L^2-12 \pi ^2 x^2\right)}{\left(L^2+4 \pi ^2 x^2\right)^5}\,,
\end{align}
\end{subequations}
which trivially follow from integrals we already computed.
One can check that plugging these in with the $a_i$'s that we computed in section~\ref{sec:FreeCFT-scalar} reproduces the full correlator, yielding ${a_{\phi^2 K^2} = 0}$.

\subsection{Free Maxwell theory}
Here we provide details omitted from the main text for $4d$ free Maxwell theory, starting with the infinite line case.
For the extraction of the coefficients, we first need the one-point functions in the presence of the defect, contributing at $O(r^0)$ on the fused defect side,
\begin{subequations}
\begin{align}
    \langle F_{\mu\nu}(x) W_{q}\rangle &= -\frac{ieq}{2\pi} \left[\frac{x_{[\mu} v_{\nu]}}{x^3}\right]\,,\\
\langle F^2(x) W_{q}\rangle &= -\frac{e^2q^2}{8\pi^2x^4}\,,\\
    \langle T_{\mu\nu}(x) W_{q}\rangle &= -\frac{e^2q^2}{32\pi^2 x^4}\left[2\delta_{\mu x} \delta_{\nu x} +2\delta_{\mu z} \delta_{\nu z}-\delta_{\mu\nu}\right]\,.
\end{align}
\end{subequations}
To extract the coefficient $a_{F}$, we need the integrated two-point function enterring at $O(r)$
\begin{subequations}
\begin{align}
    \int ds \ t^{\alpha\beta}\langle F_{\mu\nu}(x) W_qF_{\alpha\beta}(s) \rangle &= \int ds t^{\alpha \beta} \left[\frac{I_{\mu \alpha}I_{\nu \beta}-I_{\nu \alpha}I_{\mu \beta}}{\pi^2(x^2+s^2)^2}\right]
    =-\int ds \frac{2v^{}_{[\mu}n^r_{\nu]}}{\pi^2(x^2+s^2)^2} \nonumber\\
    &= -\frac{1}{\pi x^3}t_{\mu\nu}\,,
\end{align}
\end{subequations}
where $t_{\mu\nu} = v_{[\mu} n^r_{\nu]}$ as discussed in section~\ref{sec:free_CFTs}, and $I_{\mu\nu} = \delta_{\mu\nu} -2 \frac{(x \hat{x}-s \hat{z})_\mu (x \hat{x}-s \hat{z})_\nu}{x^2+s^2}$. Here we have specialized to $x$ on the $x$-axis.
For the two-point functions $\langle F_{\mu\nu}(x) W_{q_{1}+q_2}\partial_r^n F_{\alpha\beta}\rangle$, we can just take derivatives of the above.
These are sufficient to fix $a_{F}$ in the main text.

With this in hand, we can move on to $O(r^2)$ where we need the two remaining correlation functions in~\eqref{eq:FreeMaxwell_FWW_fusion}.
Before we compute these, we must remove UV divergences. 
In particular, we have
\begin{align}
\begin{split}
    \int ds \langle W_q \partial_r \tilde{F}_{\mu\nu}(s)\rangle &= i e q Lv_\alpha \int ds \langle A_\alpha \partial_r \tilde{F}_{\mu\nu}\rangle = \frac{i e q L}{2\pi^2}v_\alpha \int ds \frac{2\delta_{[\mu\alpha} \delta_{\nu r]}}{|s|^4}\\ 
    &= \frac{2ieq L}{3\pi^2 \delta^3}t_{\mu\nu}\,,
\end{split}
\end{align}
which is removed by changing basis for $\partial_r F_{\mu\nu}$:
\begin{equation}
    \partial_r F_{\mu\nu} = \partial_r \tilde{F}_{\mu\nu} -\frac{2 i e (q_1+q_2)}{3\pi^2 \delta^3}t_{\mu\nu}\mathds{1}\, ,
\end{equation}
removing contractions between $W_{{q_{1}+q_{2}}}$ and $\partial_r F_{\mu\nu}$. 
We will mean the redefined operator in all of the following.
The two-point function $\langle W_{{q_{1}+q_{2}}} F_{\mu\nu}(s_1)F_{\alpha\beta}(s_2)\rangle $ is also divergent:
\begin{equation}
    \frac{1}{2}r^2a_{F}^2\int d{s} \ t^{\mu\nu}t^{\alpha\beta}\langle W_{{q_{1}+q_{2}}} F_{\mu\nu}(s)F_{\alpha\beta}(0)\rangle = \frac{e^2q_1^2r^2}{3\pi^2\delta^3}\,,
\end{equation}
which is removed by adding to the action the counterterm
\begin{equation}
    \delta S_\text{eff} = \frac{e^2q_1^2r^2}{3\pi^2\delta^3} \int ds\, \mathds{1} \, .
\end{equation}

After these subtleties are handled, the first correlation function $\langle F_{\mu\nu}(x)W_{{q_{1}+q_{2}}}F_{\alpha\beta}F_{\rho \lambda} \rangle$ is set to zero by our counterterm.
We then only need the correlation function
\begin{equation}
    \int ds \ t^{\alpha\beta}\langle F_{\mu\nu}(x)W_{{q_{1}+q_{2}}}\partial_rF_{\alpha\beta}(s)\rangle \, = -\frac{3}{\pi x^4}t_{\mu\nu}\,.
\end{equation}
Plugging this into~\eqref{eq:FreeMaxwell_FWW_fusion} and comparing to the full result lets us extract the coefficient
\begin{equation}
    a_{\partial F} = \frac{i eq_1}{2}\, ,
\end{equation}
unsurprisingly following the same pattern as in the free scalar case. 

Finally, at $O(r^3)$ we have several coefficients to extract. 
Extraction of $a_{\partial_r^2 F}$ is immediate following the main text and the equations above, and so we need to fix $a_{F^2}$, $a_{F^2_{sr}}$ and $a_{T_{\mu\nu}}$.
For $a_{F^2}$, we need to insert $F^2(x)$.
The result for the relevant three-point function is given by
\begin{align}\label{eq:F2WW}
\begin{split}
     \frac{\langle F^2(x)W_{q_1}[r]W_{q_2}[0] \rangle}{\langle W_{q_1}[r]W_{q_2}[0]\rangle} =& \  -\frac{e^2(q_1+q_2)^2}{8\pi^2x^4}-\frac{e^2 q_1(q_1+q_2) r}{2\pi^2x^5}-\frac{e^2q_1(5q_1+3q_2)r^2}{4\pi^2x^6}\\
     & \ -\frac{e^2q_1(5q_1+2q_2)r^3}{2\pi^2x^7}+O(r^4)\,,
\end{split}
\end{align}
which is to be reproduced by our fusion prediction
{
\begin{align}\label{eq:FreeMaxwell_F2DD_fusion}
    \frac{\langle F^2(x)W_{q_1}[r]W_{q_2}[0] \rangle}{\langle W_{q_1}[r]W_{q_2}[0]\rangle} \simeq & \ \langle F^2(x) W_{{q_{1}+q_{2}}}\rangle  \\
    & \ - ra_{F}\int ds \ t^{\mu\nu}\langle F^2(x) W_{{q_{1}+q_{2}}}F_{\mu\nu}(s)\rangle \nonumber\\
    & \ - r^2 a_{\partial F}\int ds  \ t^{\mu\nu}\langle F^2(x) W_{{q_{1}+q_{2}}}\partial_r F_{\mu\nu}(s)\rangle \nonumber\\
    & \ +\frac{1}{2}r^2 a_{F}^2\int ds_1 \int ds_2 t^{\mu\nu} t^{\alpha\beta} \langle F^2(x)W_{{q_{1}+q_{2}}}F_{\mu\nu}(s_1)F_{\alpha\beta}(s_2) \rangle \nonumber\\
    & \ +r^3 a_{\partial F} a_{F}\int ds_1 \int ds_2 t^{\mu\nu} t^{\alpha\beta} \langle F^2(x)W_{{q_{1}+q_{2}}}\partial_rF_{\mu\nu}(s_1) F_{\alpha \beta}(s_2)\rangle \nonumber\\
    & \ -r^3 a_{\partial_r^2 F} \int ds \ t^{\mu\nu} \langle F^2(x)W_{{q_{1}+q_{2}}}\partial_r^2F_{\mu\nu}(s) \rangle \nonumber\\
    & \ -r^3 a_{F^2} \int ds \langle F^2(x)W_{{q_{1}+q_{2}}}F^2(s) \rangle \nonumber\\
    & \ -r^3 a_{F^2_{sr}} \int ds \langle F^2(x)W_{{q_{1}+q_{2}}}(t_{\mu\nu} F^{\mu\nu})^2 \rangle 
    +O(r^4)\, , \nonumber
\end{align}
}
where we have dropped the $\sim r^3\langle F^2(x) W_{{q_{1}+q_{2}}}F_{\mu\nu}(s_1)F_{\alpha\beta}(s_2)F_{\rho\lambda}(s_3) \rangle $ contribution which vanishes by Wick contractions after removing the divergences as $s_i\to s_j$.
We have also removed $\langle F^2(x) W_{q_1+q_2} T^{\mu\nu}(s) \rangle $ which vanishes in the CFT by orthogonality of $F^2$ and $T^{\mu\nu}$.

The relevant correlation functions needed to insert and extract $a_{F^2}$ are
\begin{subequations}
\begin{align}
    \int ds \ t^{\mu\nu}\langle F^2(x) W_{{q_{1}+q_{2}}}F_{\mu\nu}(s)\rangle &= -\frac{ie(q_1+q_2)}{2\pi^2 x^5} \,,\\
    \int ds  \ t^{\mu\nu}\langle F^2(x) W_{{q_{1}+q_{2}}}\partial^n_r F_{\mu\nu}(s)\rangle &= -\frac{i e (q_1+q_2)(n+2)!}{4\pi^2 x^{5+n}}\,,\\
    \int ds_1 \int ds_2 t^{\mu\nu} t^{\alpha\beta} \langle F^2(x)W_{{q_{1}+q_{2}}}F_{\mu\nu}(s_1)F_{\alpha\beta}(s_2) \rangle &= \frac{1}{\pi^2 x^6}\,,\\
    \int ds_1 \int ds_2 t^{\mu\nu} t^{\alpha\beta} \langle F^2(x)W_{{q_{1}+q_{2}}}\partial^n_rF_{\mu\nu}(s_1) F_{\alpha \beta}(s_2)\rangle &= \frac{(n+2)!}{2\pi^2 x^{6+n}}\,,\\
    \int ds \langle F^2(x)W_{{q_{1}+q_{2}}}F^2(s) \rangle&= \frac{15}{\pi^3 x^7} \,,\\
    \int ds \langle F^2(x)W_{{q_{1}+q_{2}}}(t_{\mu\nu} F^{\mu\nu})^2 \rangle&= \frac{5}{4\pi^3 x^7} \,.
\end{align}
\end{subequations}
The first four correlation functions, together with the computed coefficients, already reproduce the full result \eqref{eq:F2WW}. 
To disentangle the final two operators, one can insert $(t_{\mu\nu} F^{\mu\nu})^2$ at $x$ and diagonalize the system. Doing so yields
\begin{equation}
    a_{F^2} = 0 \, , \qquad a_{F^2_{sr}} = 0\,.
\end{equation}

To fix $a_{T_{\mu\nu}}$, we insert $T_{\mu\nu}(x)$.
The full result is given by
\begin{align}\label{eq:FreeMaxwell_TDD}
\begin{split}
     \frac{\langle T^{\mu\nu}(x)W_{q_1}[r]W_{q_2}[0] \rangle}{\langle W_{q_1}[r]W_{q_2}[0]\rangle} = & -\frac{1}{4} \bigg[\frac{e^2(q_1+q_2)^2}{8\pi^2x^4}+\frac{e^2 q_1(q_1+q_2) r}{2\pi^2x^5}+\frac{e^2q_1(5q_1+3q_2)r^2}{4\pi^2x^6}\\
     & \ +\frac{e^2q_1(5q_1+2q_2)r^3}{2\pi^2x^7}+O(r^4)\bigg] \times (2\delta_{\mu x}\delta_{\nu x}+2\delta_{\mu z} \delta_{\nu z} - \delta_{\mu\nu})\,.
\end{split}
\end{align}
The fusion prediction comes from inserting $T_{\mu\nu}$ instead of $F^2$ in~\eqref{eq:FreeMaxwell_F2DD_fusion}.
The result for the relevant correlation functions are the same as above multiplied by $\frac{1}{4} p_{\mu\nu}$
where $p_{\mu\nu} = 2\delta_{\mu x}\delta_{\nu x} +2\delta_{\mu z}\delta_{\nu z} - \delta_{\mu\nu}$. The only exception is the stress-tensor two-point function, which is given by 
\begin{subequations}
\begin{align}
    \int ds n_{\alpha}^rn_{\beta}^r \langle T^{\mu\nu}(x)W_{{q_{1}+q_{2}}}T_{\alpha\beta}(s) \rangle&= \frac{1}{16\pi^3x^7}\left[13\delta_{\mu r}\delta_{\nu r}+7\delta_{\mu z}\delta_{\nu z}-5\delta_{\mu \nu}\right]\,,\\
    \int ds v_\alpha v_\beta \langle T^{\mu\nu}(x)W_{{q_{1}+q_{2}}}T_{\alpha\beta}(s) \rangle&= \frac{1}{16\pi^3x^7}\left[7\delta_{\mu r}\delta_{\nu r}+13\delta_{\mu z}\delta_{\nu z}-5\delta_{\mu \nu}\right] \,,\\
    \int ds n_{(\alpha}^rv^{}_{\beta)} \langle T^{\mu\nu}(x)W_{{q_{1}+q_{2}}}T_{\alpha\beta}(s) \rangle&= -\frac{3}{16\pi^3 x^7}(\delta_{\mu x}\delta_{\nu z}+\delta_{\mu z}\delta_{\nu x})\,,
\end{align}
\end{subequations}
where we have written out the three different possible contractions relevant for $a_{T_{\mu\nu}}$ explicitly.
Since these are nonzero and different, matching eq.~\eqref{eq:FreeMaxwell_TDD} gives us 
\begin{equation}
    a_{T_{\mu\nu}} = 0
\end{equation}
for all three possible contractions.
This has now fixed all coefficients appearing for infinite lines up to $O(r^3)$.

\paragraph*{Curvature corrections}
Let us conclude our free Maxwell example by also computing the curvature corrections $a_{K^2}$ and $a_{K^4}$, changing our geometry to two parallel circles of radius $L/2\pi$.
Including the subtracted divergence, the one-point function for each Wilson line is given by
\begin{equation}
    \log \langle W_{q} \rangle_\text{circ} = \frac{(i eq)^2}{4\pi^2}\int ds_1 \int ds_2 \frac{v_1 \cdot v_2}{2|s_1-s_2|^2} + \frac{e^2q^2L}{4\pi^2 \delta} = \frac{e^2q^2}{4} \, ,
\end{equation}
which is nonzero, in contrast to the scalar case.
Our two-point function with this setup is then given by
\begin{align}
\begin{split}
    \log\langle W_{q_1}[r] W_{q_2}[0]\rangle_\text{circ} &=\log\langle W_{q_1} \rangle_\text{circ}\langle W_{q_2} \rangle_\text{circ} - \frac{e^2q_1q_2}{4\pi^2}\int d\theta_1 d\theta_2 \frac{\cos(\theta_1-\theta_2)}{\frac{4\pi^2r^2}{L^2}+2[1-\cos(\theta_1-\theta_2)]}\\
    &= -\frac{e^2q_1q_2L}{4\pi r}+\frac{e^2(q_1+q_2)^2}{4}-\frac{3e^2q_1q_2\pi r}{8L}+\frac{5e^2q_1q_2 \pi^3 r^3}{32L^3}+O(r^5)\,,
\end{split}
\end{align}
where the UV divergences in $\langle W_{q_i}\rangle $ were already removed.
The new terms are to be matched by the curvature corrections $a_{K^{2n}}\int ds (K_i K^i)^n$ as usual. 
Evaluating these integrals with the results from section~\ref{sec:curvature}, we find
\begin{align}
    a_{K^2} = \frac{3 e^2q_1q_2}{32\pi}\, , \qquad a_{K^4} = -\frac{5e^2q_1q_2}{512\pi} \, .
\end{align}

Finally, for all operators that may contribute up to $O(r^3)$, we also need to compute the coefficient of $v_\mu K^\nu F_{\mu\nu}$, $F_{\mu\nu} K_i K^i$, and $K^i  \partial_i F_{\mu\nu}$.
To extract the associated coefficients, we must consider $\langle F_{\mu\nu}  W_{q_1}[r] W_{q_2}[0]\rangle_\text{circ} $ both before and after the fusion.
We have
\begin{align}
\begin{split}
    \frac{\langle F_{\mu\nu}(x) W_{q_1}[r]W_{q_2}[0] \rangle_\text{circ}}{\langle W_{q_1}[r] W_{q_2}[0]\rangle_\text{circ}} = & \ \delta_{[\mu y}\delta_{\nu z]} \frac{i e L^2}{2\pi^3}\Bigg[ \frac{q_1 +q_2}{(\frac{L^2}{4\pi^2}+x^2)^2}+\frac{4q_1 x r}{(\frac{L^2}{4\pi^2}+x^2)^3}\\
    & \ -\frac{2q_1(\frac{L^2}{4\pi^2}-5x^2)r^2}{(\frac{L^2}{4\pi^2}+x^2)^4}
     -\frac{4 q_1 xr^3 (3 \frac{L^2}{4\pi^2}-5 x^2)}{(\frac{L^2}{4\pi^2}+x^2)^5}\Bigg] \, .
\end{split}
\end{align}
For the matching we additionally need the correlation functions:
\begin{subequations}
\begin{align}
    \tfrac{1}{\langle W_{q_1+q_2} \rangle}\int ds \ t^{\alpha \beta}\langle F_{\mu\nu}(x) W_{q_1+q_2} F_{\alpha\beta}(s) \rangle_\text{circ} &=-\frac{2 x L^2}{\pi^3 \left(x^2+\frac{L^2}{4\pi^2}\right)^3} \delta_{[\mu y}\delta_{\nu z]}\,,\\
    \tfrac{1}{\langle W_{q_1+q_2} \rangle}\int ds \ t^{\alpha \beta}\langle F_{\mu\nu}(x) W_{q_1+q_2} \partial_rF_{\alpha\beta}(s) \rangle_\text{circ} &= \frac{2L^2(\frac{L^2}{4\pi^2}-5x^2)}{\pi^3(\frac{L^2}{4\pi^2}+x^2)^4}\delta_{[\mu y}\delta_{\nu z]}\,,\\
    \tfrac{1}{\langle W_{q_1+q_2} \rangle}\int ds \ v^\alpha \langle F_{\mu\nu}(x) W_{q_1+q_2} K^\beta F_{\alpha\beta}(s) \rangle_\text{circ} &= \frac{L^2-4\pi^2 x^2}{\pi^3(\tfrac{L^2}{4\pi^2}+x^2)^3}\delta_{[\mu y}\delta_{\nu z]}\,,\\
    \tfrac{1}{\langle W_{q_1+q_2} \rangle}\int ds \ t^{\alpha \beta}\langle F_{\mu\nu}(x) W_{q_1+q_2} K^i \partial_i F_{\alpha\beta}(s) \rangle_\text{circ} &= \frac{2(5L^2\pi^5 x - 4\pi^7 x^3)}{\pi^8(\tfrac{L^2}{4\pi^2}+x^2)^4}\delta_{[\mu y}\delta_{\nu z]}\,,\\
    \tfrac{1}{\langle W_{q_1+q_2} \rangle}\int ds \ t^{\alpha \beta}\langle F_{\mu\nu}(x) W_{q_1+q_2} \partial^2_r F_{\alpha\beta}(s) \rangle_\text{circ} &= \frac{12 L^2 x (3\frac{L^2}{4\pi^2}-5x^2)}{\pi^3(\frac{L^2}{4\pi^2}+x^2)^5}\delta_{[\mu y}\delta_{\nu z]}\,.
\end{align}
\end{subequations}
Inserting the above into the defect fusion formula with the coefficients we already determined reproduces the full result up to $O(r^3)$, yielding 
\begin{equation}
    a_{FK^2}=0\, , \qquad a_{FK}=0\, , \qquad a_{K\partial F}=0\, .
\end{equation}

\section{Massive scalar}\label{app:MassScalar}

Here we provide some more details for the integrals appearing in section~\ref{sec:massive_scalar}.
The most common integral that appears is the first CPT correction to the integrated two-point function $\int ds \langle \phi(s)\phi(0)\rangle$ in CPT
\begin{align}
\begin{split}
    \int ds \int d^4 y \frac{1}{|y-s\hat{z}|^2}\frac{1}{|y|^2}
    &=2\pi^2 \int_{-L/2}^{L/2} ds \left(\frac{1}{2}+\log \frac{R}{|s|}\right)=L \pi^2 \left(3+2\log \frac{2R}{L}\right)\, .
\end{split}
\end{align}
We also need the related correlation function for $\partial_i \phi$,
\begin{align}
\begin{split}
    \int ds\langle \phi(x)\partial_x \phi(s)\rangle_\text{CPT}^{(1)}  &= m^2\left(\frac{1}{4\pi^2}\right)^2\int ds\int d^4 y  \left[\partial_x\frac{1}{|y-x|^2}\frac{1}{|y-s\hat{z}|^2}\right]\\
    &= -\frac{m^2}{8\pi^2}\int_{-\infty}^{\infty} ds \left(\frac{x}{x^2+s^2}\right) = -\frac{m^2}{8\pi} \, .
\end{split}
\end{align}

For the extraction of $a_{\phi^2}$, we consider the insertion of $\phi^2(x)$. 
The three-point function at $O(m^2)$ to all orders in $r$ is given by\footnote{The diagrams contributing to this correlator are those from figure~\ref{fig:MasslessScalar_DDphi2} but with an $m^2\phi^2$ insertion on one propagator. Note that we only introduced the IR regulator $L$ in divergent $s$-integrals along the defects.}
\begin{align}\label{eq:phi2Dh1Dh2}
    \frac{\langle [\phi^2(x)]_\text{r}\mathcal D_{h_1} [r]\mathcal D_{h_2}[0]\rangle^{(1)}_\text{CPT}}{\langle \mathcal D_{h_1} [r]\mathcal D_{h_2}[0]\rangle_\text{CPT}} =& \  
    \frac{-m^2 \log R\mu}{8\pi^2}+
    \frac{(h_{1}+h_2)^2 m^2}{16\pi^2}\left[1-\frac{1}{2\pi x}L\left(3+2\log\frac{2R}{L}\right)\right]\nonumber\\
    &\ -\frac{h_1 (h_1+h_2)m^2 r}{32\pi^3 x^2} L\left(3+2\log \frac{2R}{L}  \right)\\
    &\ +\left[\frac{h_1 h_2m^2}{16\pi^2}-\frac{h_1(h_1+h_2)m^2}{32\pi^3x}L\left(3+2\log \frac{2R}{L}  \right)\right]\left(\sum_{n=2}^\infty \frac{r^n}{x^n}\right)\,,\nonumber
\end{align}
where we need to use the renormalized operator from \eqref{eq:phi2Ren} inside the ball.
From the fused defect side, the first term is reproduced by the disconnected $\langle[\phi^2]_\text{r}\rangle_\text{CPT}^{(1)}\langle \mathcal D_{h_1+h_2}\rangle_\text{CPT}^{(0)}$ contribution, and the second and third terms by the connected correlation functions
\begin{subequations}
\begin{align}
    \frac{\langle [\phi^2(x)]_\text{r}\mathcal{D}_{{h_1+h_2}}\rangle^{(1)}_\text{CPT}}{\mathcal \langle \mathcal D_{{h_1+h_2}}\rangle_{\text{CPT}} }&= \frac{(h_{1}+h_2)^2 m^2}{16\pi^2}\left[1-\frac{1}{2\pi x}L\left(3+2\log\frac{2R}{L}\right)\right] \, ,\\
     \int ds\frac{\langle [\phi^2(x)]_\text{r}\mathcal{D}_{{h_1+h_2}}\partial_r \phi(s)\rangle^{(1)}_\text{CPT}}{\mathcal \langle\mathcal D_{{h_1+h_2}}\rangle_{\text{CPT}} } &= -\frac{(h_1+h_2)m^2}{32\pi^3 x^2}L\left(3+2\log \frac{2R}{L}\right)\, .
\end{align}
\end{subequations}

\begin{figure}[t]
    \centering
    \input{Figures/Fused_DDphi2_CPT_1.tikz}
    \caption{The only nonzero connected diagrams in the fused $\langle  [\phi^2(x)]_\text{r} \mathcal{D}_{h_1+h_2} e^{-S_\text{eff}}\rangle_\text{CPT}^{(1)} $ calculation at first order in CPT. Diagrams where the mass insertion is on the leg attached to $\partial_i^n\phi$ (with $n\ge2$ derivatives) vanish as $L\to \infty$. 
    In the first diagram, the $\phi$ insertion represents the defect $\mathcal D_{{h_1+h_2}}\sim e^{(h_1+h_2)\int\phi}$ itself.
    }
    \label{fig:Fused_DDphi2_CPT_1}
\end{figure}
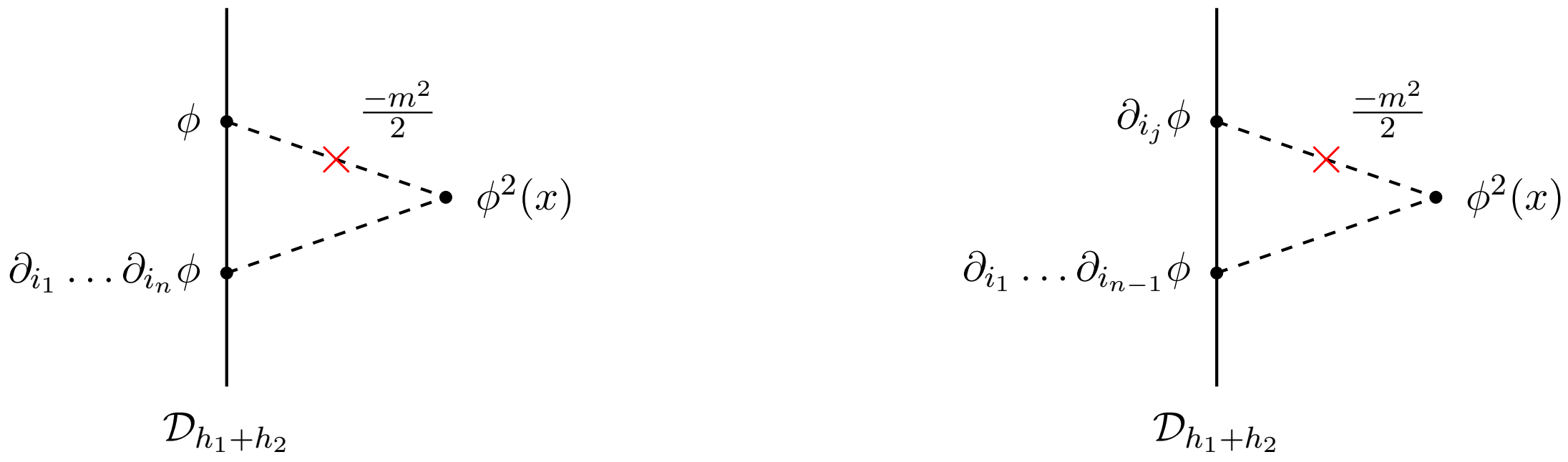
For the terms starting at $O(r^2)$, we have contributions from $ \partial_{i_1}\ldots \partial_{i_n}\phi$.
The only nonzero contributions in this limit are given by the diagrams in figure~\ref{fig:Fused_DDphi2_CPT_1}.
At order $O(m^2)$, they can be computed in closed form. Specializing to $i_n = r$ for simplicity, they are given~by
\begin{subequations}
\begin{align}
    &\frac{\int ds\langle \phi^2(x) \mathcal D_{{h_1+h_2}} \partial^n_{r}\phi(s) \rangle^{(1)}_\text{CPT}}{\mathcal \langle \mathcal D_{{h_1+h_2}}\rangle_{\text{CPT}} }=  (h_1+h_2)\left[\frac{m^2 }{16\pi^2}-\frac{m^2}{32\pi^3 x}L\left(3+2\log \frac{2R}{L}\right)\right]\left(\frac{n!}{x^n}\right)\, ,\\
     &\frac{\int ds_1 ds_2\langle \phi^2(x) \mathcal D_{{h_1+h_2}} \partial_r\phi(s_1)\partial_{r}^{n-1}\phi(s_2)\rangle^{(1)}_\text{CPT}}{\mathcal \langle\mathcal D_{{h_1+h_2}}\rangle_{\text{CPT}} } = -\frac{m^2(n-1)!}{16\pi^2 x^n}\left(1+\delta_{n,2}\right)\, ,
\end{align}
\end{subequations}
where we left the necessary renormalizations implicit. The Kronecker delta accounts for the extra symmetry factor when $n=2$, and both equations are for $n\geq 2$.
Plugging these into the expansion of the effective action with $a_{\partial_r^n \phi}=-\frac{h_1}{n!}$, we immediately see that they reproduce the full \eqref{eq:phi2Dh1Dh2} to all orders in $r$. Since $\langle \phi^2 \mathcal D_{h_1+h_2} \phi^2\rangle_\text{CPT}^{(0)}$ is nonzero, but its contribution is not needed, we conclude that
\begin{equation}
    a_{\phi^2}^{(1)} = 0\, .
\end{equation}

\begin{figure}[t]
    \centering
    \input{Figures/DD_MassiveScalar_CPT_2.tikz}
    \caption{Diagrams contributing to the $\langle \mathcal{D}_{h_1}[r]\mathcal{D}_{h_2}[0] \rangle_\text{CPT}^{(2)} $ correlator at second order in CPT. Red crosses indicate insertions of the relevant deformation $\frac{-m^2}{2}\int d^4 y\, \phi^2(y)$. These are all to be multiplied by the CFT Casimir contribution $\exp(a_0^{(0)} L/r)$. Only diagrams with two insertions on the same propagator contribute to $\log \langle \mathcal{D}_{h_1}[r]\mathcal{D}_{h_2}[0] \rangle_\text{CPT}^{(2)} $ in~\eqref{eq:DD_massive_scalar_CPT}.}
    \label{fig:MassiveScalar_DD_CPT_2}
\end{figure}
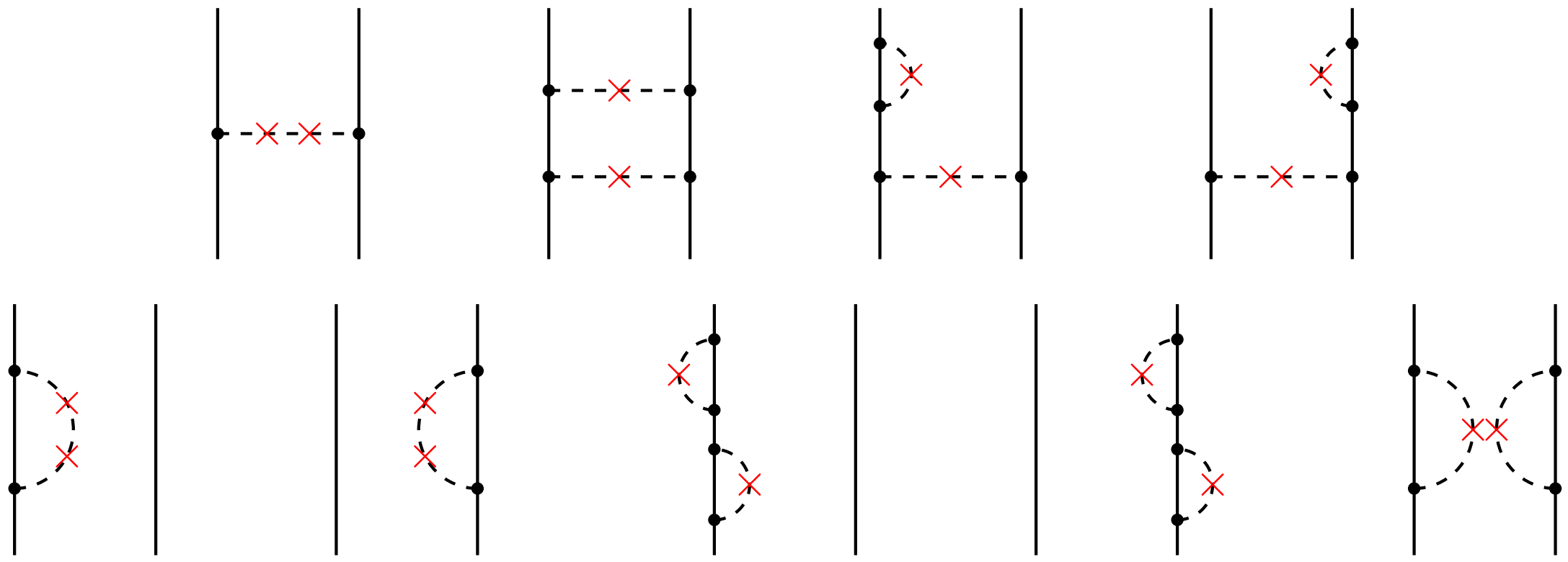

We next target $a_0^{(2)}$ by graduating to second order in CPT. The identity insertion suffices to fix this coefficient. The defect two-point function is given by the same diagrams as in figure~\ref{fig:MassiveScalar_DD_CPT_1} but with two insertions on each propagator, requiring the calculation of $\langle \phi(x) \phi(s) \rangle^{(2)} $.
Stripping off the prefactor, we first need the integral
\begin{align}
    \int d^4 y_1 d^4y_2 \frac{1}{|y_1|^2}\frac{1}{|y_1-y_2|^2}\frac{1}{|y_2-x|^2}
   & = \pi^2\int d^4y_2 \frac{2\log \frac{R}{|y_2|} + 1}{|y_2-x|^2} \nonumber\\ 
    &=\frac{\pi^4}{4}\bigg[8R^2-|x|^2\left(5+4\log \frac{R}{|x|}\right)\bigg]\,.
\end{align}
Now substituting $|x|=\sqrt{s^2+r^2}$ and integrating over $s$ from $-L/2$ to $L/2$, we arrive at
\begin{align}
\begin{split}
    \int_{-L/2}^{L/2} ds \frac{\pi^4}{4}\bigg[8R^2-|x|^2\left(5+4\log \frac{R}{|x|}\right)\bigg] =& \   \frac{\pi ^4 L}{144}  \left(288 R^2-19 L^2+12 L^2 \log \frac{L}{2 R}\right) \\
    & \ -\frac{\pi ^4 L}{4} r^2 \left(7+4\log \frac{2 R}{L}\right)+ \frac{2 \pi ^5 r^3}{3}\,.
\end{split}
\end{align}
With this in hand, we find the result\footnote{Note the log, which has removed disconnected diagrams from the product $\langle \mathcal D_{h_1}\mathcal{D}_{h_2}\rangle^{(1)}\langle \mathcal{D}_{h_1}\mathcal{D}_{h_2}\rangle^{(1)}$. The full set of diagrams contributing to $\langle \mathcal{D}_{h_1}\mathcal{D}_{h_2}\rangle^{(2)} $ is shown in figure~\ref{fig:MassiveScalar_DD_CPT_2}. The ones that survive the log are those which cannot be split by cutting only defect lines.}
\begin{align}\label{eq:DD_massive_scalar_CPT}
\begin{split}
    \log \langle \mathcal D_{h_1}[r] \mathcal D_{h_2}[0] \rangle^{(2)}_\text{CPT}
    =& \ \log\langle \mathcal{D}_{{h_1+h_2}} \rangle^{(2)}_\text{CPT} -\frac{h_1h_2 L^2 m^4 r^2 \left(4 \log \frac{2 R}{L}+7\right)}{256 \pi ^2} +\frac{h_1 h_2m^4r^3L}{96\pi}.
\end{split}
\end{align}
On the fused defect side, at second order in CPT we have
\begin{align}\label{eq:MassiveScalarIdentityExpansion_RHS}
\begin{split}
    \log \langle \mathcal{D}_{h_1}[r] \mathcal D_{h_2}[0] \rangle_\text{CPT}^{(2)} \simeq & \ \log \langle  \mathcal  D_{{h_1+h_2}} \rangle^{(2)}_\text{CPT} \\
    &\ +\frac{h_1^2}{2}r^2\int ds_1 ds_2\, \frac{\langle \mathcal D_{h_1+h_2} \partial_r \phi(s_1) \partial_r \phi(s_2)\rangle_\text{CPT}^{(2)}}{\langle  \mathcal  D_{{h_1+h_2}} \rangle_\text{CPT}}\bigg|_\text{con.}\\
    & \ +\frac{h_1}{2}r^2\int ds\, \frac{\langle \mathcal D_{h_1+h_2} \partial_r^2 \phi(s)\rangle_\text{CPT}^{(2)}}{\langle  \mathcal  D_{{h_1+h_2}} \rangle_\text{CPT}}\bigg|_\text{con.}\\
    & \ +a_0^{(2)}r^3L +O(r^4)\,,
\end{split}
\end{align}
where we used our knowledge that $a_\phi^{(1)}=a_{\phi^2}^{(1)}=0$, as well as symmetries of the fused defect, to drop vanishing contributions. The subscript ``con.'' instructs us to keep only connected diagrams (with respect to the defect).

The $O(r^2)$ contribution in~\eqref{eq:DD_massive_scalar_CPT} may be rewritten in a highly suggestive integral form,
\begin{equation}
    \frac{h_1h_2m^4}{128 \pi ^6}r^2 L\int ds \int d^4{y_1} \int d^4y_2\bigg[\frac{1}{|y_1|^2}\frac{1}{|y_1-y_2|^2}\partial_r^2\frac{1}{|y_2-r \hat{x} - s\hat{z}|^2}\, \bigg]_{r=0}.
\end{equation}
This corresponds to the integrated correlator
\begin{equation}
    \frac{h_1 h_2}{2}r^2\int ds_1 ds_2 \langle \phi(s_1)\partial_r^2 \phi(s_2) \rangle^{(2)}_\text{CPT}\,,
\end{equation}
which is precisely what the two $O(r^2)$ terms in \eqref{eq:MassiveScalarIdentityExpansion_RHS} reduce to.
Comparing the $O(r^3)$ terms, we determine
\begin{align}
    a_{0}^{(2)} = \frac{h_1 h_2 m^4}{96\pi}\, .
\end{align}

\section{Schwinger model}\label{app:Schwinger}

For our final solvable example in section~\ref{sec:Schwinger}, we consider the Schwinger model with $N_f$ massless fermions. 
The purpose of this appendix is to show the exact example and demonstrate the OPE of local operators in analogy with section~\ref{sec:OPE_massive_scalar}, this time using dimensional regularization.

\subsection{Solution}

The massless Schwinger model is exactly solvable \cite{Schwinger:1962tp,Lowenstein:1971fc}. One convenient way to do so is by performing a chiral rotation~\cite{Kim:1999wn}, which allows us to compute the full non-perturbative predictions for comparison with our OPE predictions.
We start with a $U(1)$ gauge field and $N_f$ complex Dirac fermions in $2d$
\begin{equation}
    S = \int d^2x\, \left(\frac{1}{4} F_{\mu\nu}F^{\mu\nu} +  \sum_{i=1}^{N_f} \bar \psi_i \gamma^\mu(\partial_\mu - ieA_\mu)\psi_i\right)\,.
\end{equation}
We think of $\psi_i$ and $\bar \psi_i$ as independent fields, and under a gauge transformation, we have
\begin{equation}
    \psi_i(x) \to e^{i\theta(x)} \psi_i(x)\,, \qquad \bar\psi_i(x) \to e^{-i\theta(x)} \bar\psi_i(x)\,, \qquad A_\mu(x) \to A_\mu(x) + \frac{1}{e}\partial_\mu \theta(x)\,.
\end{equation}
Now we use Hodge decomposition to write the gauge field as $a=\star d \phi + d\chi$.\footnote{Here we drop an additional harmonic piece which takes care of global issues of the gauge field configuration that do not enter in $\mathbb{R}^2$.} In components,
\begin{equation}
    eA_\mu = \epsilon_{\mu\nu}\partial^\nu \phi + \partial_\mu \chi\,.
\end{equation}
Note that upper and lower indices are raised with the Euclidean metric $g_{\mu\nu} = \delta_{\mu\nu}$.
The second part is pure gauge, so it will drop from any physical observables.
Now we do a field redefinition to
\begin{equation}
    \xi_i \equiv e^{-\gamma_c \phi -i \chi}\psi_i\,,\qquad \bar\xi_i \equiv \bar \psi_ie^{-\gamma_c \phi +i \chi}\,,
\end{equation}
where $\gamma_c$ is the chirality matrix. Equivalently, $\psi_i = e^{\gamma_c \phi + i \chi}\xi_i$ and $\bar\psi_i = \bar \xi e^{\gamma_c \phi - i \chi}$.

This brings the action to
\begin{equation}
    S = \int d^2x\, \left(\frac{1}{2e^2} (\partial^2 \phi)^2 +  \sum_{i=1}^{N_f} \bar \xi_i \gamma^\mu(\partial_\mu + \gamma_c \partial_\mu \phi -i\epsilon_{\mu\nu}\partial^\nu \phi)\xi_i\right)\,.
\end{equation}
We can pick a basis for the gamma matrices $\gamma_1 = \sigma_z$, $\gamma_2 = -\sigma_y$ so that they are hermitian, satisfy the Clifford algebra $\{\gamma_\mu,\gamma_\nu\} = 2\delta_{\mu\nu}$, and they anticommute with $\gamma_c \equiv -i\gamma_1\gamma_2 =\sigma_x$ (which is also hermitian). Then,
\begin{equation}
    \gamma^\mu \gamma_c \partial_\mu = \begin{pmatrix}
        i\partial_2 &  \partial_1 \\
         -\partial_1  & -i\partial_2
    \end{pmatrix}\,, \qquad 
    \gamma^\mu \epsilon_{\mu\nu}\partial^\nu = \begin{pmatrix}
        \partial_2 & -i\partial_1 \\
        i\partial_1 & -\partial_2
    \end{pmatrix}\,,
\end{equation}
or $i\gamma^\mu \epsilon_{\mu\nu} = \gamma_\nu \gamma_c$. The gauge field is thus completely canceled in the Dirac piece of the action,
\begin{equation}
    S = \int d^2x\, \left(\frac{1}{2e^2} (\partial^2 \phi)^2 +  \sum_{i=1}^{N_f} \bar \xi_i \gamma^\mu\partial_\mu\xi_i\right)\,.
\end{equation}
Note that $\chi$ dropped directly so that the action is gauge invariant, as it should be.

This is not yet the whole story. The chiral rotation that we did is anomalous, so there is an extra piece coming from the Jacobian of the field transformation in the path integral measure. We just note from \cite[Appendix B]{Kim:1999wn} that the measure changes by
\begin{equation}
    D\psi D\bar \psi = D\xi D\bar \xi e^J\,, \qquad \frac{\delta J}{\delta\phi} =  -\frac{N_fe}{2\pi} \epsilon_{\mu\nu} F^{\mu\nu} = \frac{N_f}{\pi} \partial^2 \phi\,.
\end{equation}
So $J = -\int d^2 x\frac{N_f}{2\pi}(\partial_\mu \phi)^2$. The number of flavors appears because each fermion contributes one factor to the anomaly. 
Putting this change in the measure into the action, we obtain
\begin{equation}
    S[\xi,\phi] = \int d^2x\, \left(\frac{1}{2e^2} (\partial^2 \phi)^2 + \frac{N_f}{2\pi}(\partial_\mu \phi)^2 +  \sum_{i=1}^{N_f} \bar \xi_i \gamma^\mu\partial_\mu\xi_i\right)\,.
\end{equation}
We have rewritten the theory as a free theory of $N_f$ Dirac fermions $\xi_i,\bar \xi_i$ and a boson $\phi$ with a higher-derivative kinetic term.

Any correlation function can then be computed in terms of these new fields through Wick contractions. We just need their propagators. For the fermion, we have the standard two-point function
\begin{align}
    \langle \xi_i(x) \bar \xi_j(0)\rangle = -\gamma^\mu \partial_\mu \int \frac{d^2k}{(2\pi)^2}\frac{e^{-ik\cdot x}}{k^2} \delta_{ij} 
    =\frac{x^\mu \gamma_\mu}{2\pi x^2} \delta_{ij}\,.
\end{align}
For the boson, the two-point function is given by
\begin{align}\label{eq:PhiPhi}
    \langle \phi(x)\phi(0)\rangle =&\,  \int \frac{d^2k}{(2\pi)^2}\frac{e^{-ik\cdot x}}{\frac{1}{e^2}k^4 + \frac{N_f}{\pi} k^2} 
    =   -\frac{1}{2N_f}\log|x| + C - \frac{1}{2N_f}K_0\left(\sqrt{\tfrac{N_f}{\pi}} e|x|\right) \nonumber \,,
\end{align}
where $\sqrt{\tfrac{N_f}{\pi}}e$ plays the role of a mass.
Here $C$ is a scheme-dependent constant coming from regulating the diverging Fourier transform. We will see that $C$ drops out of the physical correlation functions we will consider, as it should.

\subsection{OPE of local operators}

Here we will work out the OPE of the field strength $F_{12}F_{12}$, following essentially the same steps as for the massive scalar in section~\ref{sec:OPE_massive_scalar}.
Since this is an interacting theory in terms of the original $U(1)$ gauge field and fermions, this is a somewhat more nontrivial example. 
We will also use this as an opportunity to demonstrate using dimensional regularization to handle divergences rather than the hard cutoffs mostly used in the main text.

We will be interested in considering the two-point function
\begin{align}
    \langle F_{12}(x) F_{12}(0)\rangle =& \  \frac{1}{e^2}\langle \partial^2 \phi(x) \partial^2\phi(0)\rangle\,\\
    =&\,\frac{1}{e^2}\partial_x^2\partial_y^2\left(-\frac{1}{2N_f}\log|x-y| + C - \frac{1}{2N_f}K_0\left(\sqrt{\tfrac{N_f}{\pi}} e|x-y|\right)\right)_{y=0}\nonumber\\
    =&\,- \frac{1}{2N_f e^2}\left(\sqrt{\tfrac{N_f}{\pi}} e\right)^4 K_0\left(\sqrt{\tfrac{N_f}{\pi}} e|x|\right) \nonumber\\
    =& \  -\frac{N_f e^2}{2\pi^2}K_0\left(\sqrt{\tfrac{N_f}{\pi}} e|x|\right)\,,\nonumber
\end{align}
which is valid for $x \neq 0$. 
At $x=0$ additional contact terms $\sim \delta^{(2)}(0)$ appear that will not matter for our purposes.
As $x \to 0$, the expansion of this two-point function is given by
\begin{align}\label{eq:<FF>asym}
    \langle F_{12}(x) F_{12}(0)\rangle\simeq &\,(\gamma +\log (\tfrac{m |x|}{2}))\frac{N_fe^2}{2\pi^2} + \frac{1}{4} (-1+\gamma +\log (\tfrac{m |x|}{2})) \frac{N_fe^2}{2\pi^2} m^2|x|^2 \\
    &  \, + \frac{1}{128}(-3+2\gamma +2 \log (\tfrac{m |x|}{2}))\frac{N_fe^2}{2\pi^2} m^4|x|^4 + O(|x|^6)\,,\nonumber
\end{align}
where we have defined $m\equiv \sqrt{e^2N_f/\pi}$.
Our goal is to reproduce this expansion order by order as an OPE.
This already looks quite similar to the massive scalar result~\eqref{eq:MassScalarUVasym} in structure, with logarithms of $e$ that will be reproduced by non-perturbative condensates and logarithms of $|x|$ that will be computed as coefficients in CPT.
The OPE will be of the form
\begin{equation}
    F_{12}(x)F_{12}(0)\simeq \frac{c_{\mathds 1}(e|x|)}{|x|^2}\mathds 1 + c_{F^2}(e|x|)F_{12}^2(0) 
    + \ldots \,.
\end{equation}
We will compute and check just the leading $O(|x|^0)$ contribution for demonstration, as higher orders follow systematically.
We will also assume $N_f \geq 2$ such that the operator $\det \bar\psi\psi$ does not enter until higher order in $|x|$.

\subsubsection{Two-point functions in CPT}
Our procedure will be similar to section~\ref{sec:OPE_massive_scalar}. 
We will first compute the necessary two- and three-point functions in CPT, this time using dimensional regularization, renormalizing the operator $F^2_{12}$ as it shows up.
Solving the resulting system of two equations lets us extract the leading contributions to the OPE coefficients $c_\mathds{1}$ and $c_{F^2}$.
We will then compute the condensates in the full theory and plug them in, demonstrating that the OPE is correct.

We will consider inserting at a finite position $y$ the operators $\mathcal{O}_k = \mathds 1, F^2_{12}$. We will start with the RHS, computing the two-point functions $\langle \mathcal{O}_k(y) \mathcal{O}_i(0)\rangle$.
Starting with the identity, we need the correlation function $\langle F^2_{12}(0)\rangle$.
The leading contribution in CPT is both IR and UV divergent, giving us
\begin{align}
    \langle F^2_{12}\rangle_\text{CPT} =& 
    \ \frac{e^2N_f}{4\pi^2}\left[-\frac{1}{\epsilon_\text{UV}}-\frac{1}{\epsilon_\text{IR}}\right]\, +O(e^4)\,.
\end{align}
The IR divergence emerges from regulating the integral in $d=2+2\epsilon_\text{IR}$ dimensions, while the UV divergence comes from regulating in $d=2-2\epsilon_\text{UV}$ dimensions.
We must track them separately.

The UV divergence is removed by defining a renormalized operator $[F^2_{12}]_r \equiv F^2_{12} + \frac{e^2 N_f}{4\pi^2 \epsilon_\text{UV}}$ so that the renormalized correlation function becomes
\begin{equation}
    \langle [F^2_{12}]_r \rangle_\text{CPT} = 
    \ \frac{e^2N_f}{4\pi^2}\left[-\frac{1}{\epsilon_\text{IR}}\right]\, +O(e^4)\,.
\end{equation}

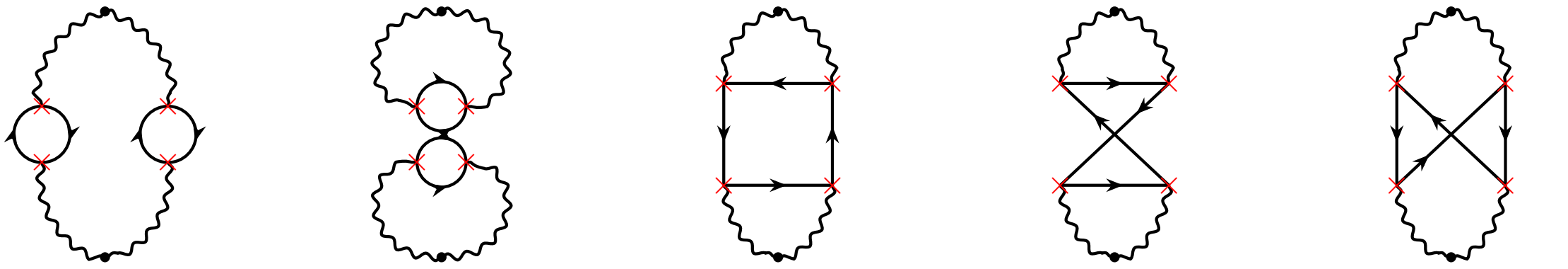
\begin{figure}[t]
    \centering
    \input{Figures/Schwinger_F2F2.tikz}
    \caption{Diagrams contributing to the two-point function $\langle F^2_{12}(y)F^2_{12}(0)\rangle $ for the extraction of the OPE coefficient $c_{F^2}$. Diagrams with a tree-level $\langle F_{12}(x) F_{12}(y) \rangle $ correlator vanish, making these the leading contributions in CPT. The three box diagrams vanish for massless fermions in $2d$ in dimensional regularization, making the first two diagrams the only nonzero contributions.}
    \label{fig:Schwinger_F2F2}
\end{figure}
The second two-point function we need is from inserting $\mathcal{O}_k(y) = F^2_{12}(y)$. 
This is immediately zero at tree-level and first order in CPT since the tree-level $\langle FF \rangle $ two-point function vanishes at separated points.
At second order, we have the five diagrams in figure~\ref{fig:Schwinger_F2F2}.
In fact, it turns out that the box integrals all vanish in $2d$ for massless fermions, as the momenta integrals are scaleless and free of logarithmic divergences.
This is unique to the massless Schwinger model, and is why the theory is solvable in the first place.
The first two diagrams give a nonzero result\footnote{Here we've neglected some evanescent terms which do not contribute to the final result.}
\begin{equation}
    \langle [F^2_{12}]_r(y) [F^2_{12}]_r(0) \rangle_\text{CPT} = \frac{e^4N_f^2}{8\pi^4}\left[-\frac{1}{\epsilon_\text{IR}}+\gamma+\log |y|^2 \mu^2 \pi\right]^2 + \frac{e^4N_f^2 }{16\pi^4 \epsilon_\text{IR}^2} + O(e^6)\,.
\end{equation}

\subsubsection{Three-point functions in CPT}

We will now consider the two three-point functions $\langle \mathcal{O}_k(y) F_{12}(x)F_{12}(0)\rangle$ where $\mathcal{O}_k = \mathds 1, F^2_{12}$. 
Starting with the identity, the tree-level contribution vanishes at separated points, forcing us to go to first order in CPT.
We have\footnote{Note that we have enforced $p_1^2+p_2^2 = p^2$ in $2d$ in the integrand here, which we do everywhere. 
Enforcing this instead in $d$-dimensions gives an evanescent difference.}
\begin{align}
\begin{split}
    \langle F_{12}(x) F_{12}(0) \rangle_\text{CPT} =& \  - \mu^{2-d}\int \frac{d^d p}{(2\pi)^d}e^{ip \cdot x} \\
    & \ \qquad \times \frac{i(p_1 g_{2 \alpha}-p_2 g_{1 \alpha})}{p^2}\Pi_{\alpha \beta}(p) \frac{-i(p_1 g_{2 \beta}-p_2 g_{1 \beta})}{p^2}\\ 
    =& \ -\frac{e^2 N_f}{\pi}\mu^{2-d}\int \frac{d^d p}{(2\pi)^d}e^{ip \cdot x} \frac{1}{p^2}\\
    =& \ \frac{e^2N_f}{4\pi^2}\left[-\frac{1}{\epsilon_\text{IR}}+\gamma+\log |x|^2 \mu^2 \pi\right]\, .
\end{split}
\end{align}

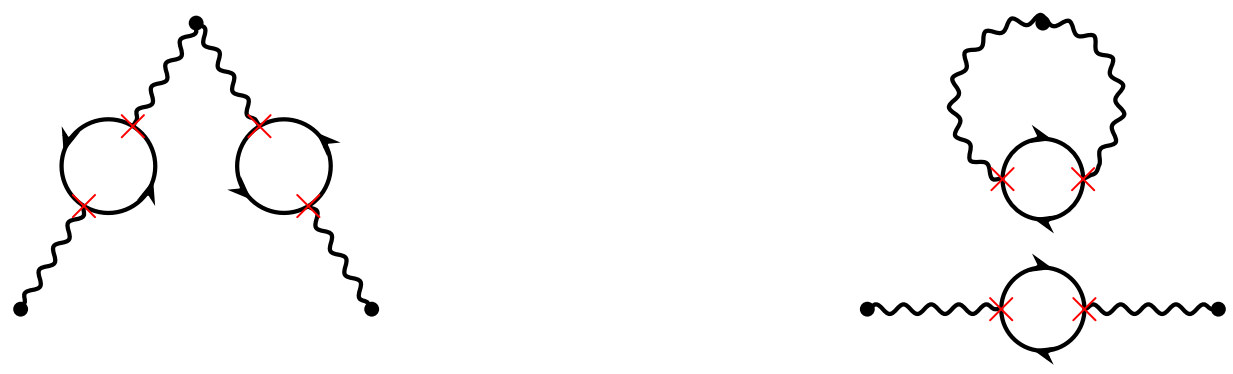
\begin{figure}[t]
    \centering
    \input{Figures/Schwinger_F2FF.tikz}
    \caption{Diagrams contributing to the three-point function $\langle F^2_{12}(y) F_{12}(x)F_{12}(0)\rangle $. All other $O(e^4)$ contributions vanish.}
    \label{fig:Schwinger_F2FF}
\end{figure}

The second three-point function we need is inserting the operator $F^2_{12}(y)$. 
Since the tree-level $\langle F_{12}(x)F_{12}(y)\rangle$ correlator is zero at separated points, it turns out that the three-point function $\langle F^2_{12}(y) F_{12}(x)F_{12}(0)\rangle $ vanishes at both tree-level and first order in CPT immediately after Wick contractions.
The first nonzero contributions come from the diagrams in figure~\ref{fig:Schwinger_F2FF}, where all other contributions vanish at this order in CPT.
This gives us a result
\begin{align}
    \langle [F_{12}^2]_r(y) F_{12}(x)F_{12}(0)\rangle_\text{CPT} 
    = & \ \frac{e^4N_f^2}{8\pi^4}\left[-\frac{1}{\epsilon_\text{IR}}+\gamma+\log |y|^2 \mu^2 \pi\right]^2 \nonumber \\
    & \ - \frac{e^4N^2_f}{16\pi^4 \epsilon_\text{IR}}\left[-\frac{1}{\epsilon_\text{IR}}+\gamma+\log |x|^2 \mu^2 \pi\right] +O(|x|)\, ,
\end{align}
where we have series-expanded in $|x|$. 

\subsubsection{OPE predictions}

Combining ingredients, we have a system of two equations of the form given in~\eqref{eq:3=2CPT}, where we plug in all of the above condensates.
Solving for the coefficients, we find a result independent of IR divergences
\begin{subequations}
\begin{align}
    c_{\mathds 1} &= \frac{e^2 |x|^2N_f}{4\pi^2}(\gamma + \log |x|^2 \mu^2 \pi)+ O(e^4|x|^4)\, ,\\
    c_{F^2} &= 1 + O(e^2|x|^2)\, .
\end{align}
\end{subequations}
Having established the regulator independence of these coefficients, we can plug them back into the OPE to get the QFT prediction for the two-point function
\begin{equation}
\boxed{
    \langle F_{12}(x)F_{12}(0)\rangle_\text{QFT} \simeq \frac{e^2N_f}{4\pi^2}\left(\gamma+\log |x|^2\mu^2 \pi\right) +\langle[F_{12}^2]_r\rangle_\text{QFT}+ O(|x|^2) \, .}
\end{equation}

Let us conclude by computing the condensate $\langle [F^2_{12}]_r\rangle$ in the full theory and inserting it into the OPE to demonstrate that it reproduces~\eqref{eq:<FF>asym}.
We have
\begin{align}
\begin{split}
    \langle [F^2_{12}]_r \rangle_\text{QFT} &= \frac{1}{e^2}\langle (\partial^2 \phi)^2 \rangle +\frac{e^2 N_f}{4\pi^2\epsilon_\text{UV}}\\
    &= \mu^{2-d}\int \frac{d^d k}{(2\pi)^d}\frac{k^4}{k^4+\frac{e^2N_f}{\pi}k^2}+\frac{e^2 N_f}{4\pi^2\epsilon_\text{UV}}\\
    &=\frac{e^2N_f}{4\pi^2}\left(\gamma-\log \frac{4\pi \mu^2}{m^2}\right)\,.
\end{split}
\end{align}
Inserting this into the OPE along with the computed coefficients, we find through $O(|x|^0)$
\begin{equation}
    \langle F_{12}(x)F_{12}(0) \rangle_\text{QFT} \simeq \frac{e^2N_f}{2\pi^2}\left(\gamma+\log \frac{m|x|}{2}\right)+O(|x|^2)\,,
\end{equation}
reproducing the expected result, with no scheme-dependent $\mu$ remaining.
Higher orders in $|x|$ can be systematically computed in a straightforward manner by applying the same techniques as we have shown here.

\section{Ising model}\label{app:Ising}

Here we provide some details for the calculation of the correlation functions in section~\ref{sec:IsingModel} for the fusion of pinning field $\sigma$-defects that were omitted from the main text.

\subsection{Lines}

First, we discuss the CPT computations for the extraction of the fusion coefficients $a_\sigma^{(3)}$ and $a_0^{(4)}$, contribution at $O(r^{7/4})$ and $O(r^{5/2})$, respectively.

\paragraph*{Extraction of $a_\sigma^{(3)}$}

For the extraction of $a_{\sigma}^{(3)}$, we need to consider the $O(r^{7/4})$ contribution to the correlation function $\int ds_1 ds_2 ds_3\langle \sigma(\infty)\sigma(r,s_1)\sigma(0,s_2)\sigma(0,s_3)\rangle_c $.
After performing an integral over $s_3$ using translation invariance, we can isolate the relevant region by defining $s_2 = r(u_1+u_2)$ and $s_1 = r u_2$ and expanding as $r\to 0$, giving the contribution
\begin{align}
    \int ds_1 ds_2 ds_3\langle \sigma(\infty)\sigma(r,s_1)\sigma(0,s_2)\sigma(0,s_3) \rangle_c^{O(r^{7/4})} =  r^{7/4}L \int du_1 du_2 [f(u_1,u_2) -f_h(u_1,u_2)]\,,
\end{align}
where the function $f(u_1,u_2)$ is given by
\begin{align}\label{eq:asig3_f}
   f(u_1,u_2)=\bigg[\frac{1}{2}\frac{\rho_1+\rho_2+|u_1+u_2|}{(\rho_1\rho_2|u_1+u_2|)^{1/2}}\bigg]^{1/2}-\frac{1}{\rho_1^{1/4}}-\frac{1}{\rho_2^{1/4}}-\frac{1}{|u_1+u_2|^{1/4}}\,,
\end{align}
where $\rho_i = (1+u_i^2)^{1/2}$.
The function $f_h(u_1,u_2)$ subtracts the hard region overlap and is obtained by replacing $\rho_i \to |u_i|$ in~\eqref{eq:asig3_f}.

As written, the integral is logarithmically divergent as $u_i\to \infty$. 
These divergences can be cured by subtracting an appropriate kernel, then adding it back using a regulator.
An appropriate kernel is
\begin{equation}
    g(u_i,u_j) = \frac{\rho_i^{3/4}-|u_i|^{3/4}}{4 \rho_j}\, .
\end{equation}
These let us arrive at the finite integral given by
\begin{equation}
   C_\sigma = -\frac{1}{2}\int_{-\infty}^\infty du_1 du_2 [f(u_1,u_2)-f_h(u_1,u_2)-g(u_1,u_2)-g(u_2,u_1)] \approx 5.60020\,,
\end{equation}
where we have included a factor of $-\frac{1}{2}$ for later convenience. 
To get the full result, we need to add back the subtracted kernels after integrating them with an appropriate regulator $L/(2r)$, which gives
\begin{equation}
    2\int_{-\infty}^{\infty} du_i \int^{L/(2r)}_{-L/(2r)} du_j g(u_i,u_j) = 2C_\epsilon \log \frac{L}{r}\, .
\end{equation}
Combining pieces and including the relevant prefactors, we have the result at $O(r^{7/4})$
\begin{equation}
    \int ds_1 ds_2 ds_3\langle \sigma(\infty) \sigma(r,s_1)\sigma(0,s_2)\sigma(0,s_3) \rangle_c^{O(r^{7/4})} = 2r^{7/4}L\left[C_\epsilon \log \frac{L}{r} - C_\sigma\right]\, .
\end{equation}

\paragraph*{Extraction of $a_0^{(4)}$}

Let us now discuss the extraction of the coefficient $a_0^{(4)}$ for the $r^{5/2}$ contribution to the potential.
We need to compute up to this order the correlator 
\begin{align}\label{eq:Ising_DD_cpt4}
\begin{split}
    \log\langle \mathcal{D}_{h}[r] \mathcal{D}_{h}[0]\rangle^{(4)} =& \  \frac{h^4}{12}L\int ds_1 ds_2 ds_3 \langle\sigma(0,s_1)\sigma(0,s_2)\sigma(0,s_3)\sigma(0,0) \rangle_c \\
    & \ + \frac{h^4}{3}L\int ds_1 ds_2 ds_3 \langle \sigma(0,s_1)\sigma(0,s_2)\sigma(0,s_3)\sigma(r,0) \rangle_c \\
    & \ + \frac{h^4}{4} L \int ds_1 ds_2 ds_3 \langle \sigma(r,s_1)\sigma(r,s_2)\sigma(0,s_3)\sigma(0,0)\rangle_c\,,
\end{split}
\end{align}
where these are connected correlators evaluated in the critical theory.
The first line is $r$-independent, and so the problem boils down to computing the latter two lines. 
The matching is to the condensates
\begin{align}
\begin{split}
    \frac{1}{L}\log \langle \mathds 1 e^{-S_\text{eff}} \rangle^{(4)} =& \ \frac{1}{L}\log\langle \mathcal{D}_{2h}  \rangle^{(4)} +r^{7/4} h^2 C_\epsilon \langle \mathcal{D}_{2h}[\epsilon]_\text{r}\rangle^{(2)}_c\\
    & \ - r^{7/4} 2h^3 [C_\epsilon \log r\mu +C_\sigma]\langle \mathcal{D}_{2h} \sigma\rangle_c^{(1)}\\
    & \ + r^2 \frac{h}{4} \langle \mathcal{D}_{2h} \partial^2_x \sigma \rangle_c^{(3)}+ r^{5/2}a_0^{(4)}  + O(r^{11/4})\,,
\end{split}
\end{align}
where we have plugged in the known coefficients already.
Through $O(r^2)$ it is already reproduced by known condensates, and our goal is to extract the $O(r^{5/2})$ alone.

The explicit expressions are long and not illuminating, so we will just sketch the procedure to extract the constant $C_0^{(4)}$.
We have to compute the final two lines of~\eqref{eq:Ising_DD_cpt4}.
The relevant region comes from when all $s_i \sim r$ simultaneously, after subtracting overlapping regions from lower orders in $r$. 
Let us start with the second line of~\eqref{eq:Ising_DD_cpt4}. 
The result can be written as a finite integral
\begin{align}
\begin{split}
    \frac{1}{r^{5/2}}\int &ds_1 ds_2 ds_3\langle \sigma(0,s_1)\sigma(0,s_2)\sigma(0,s_3) \sigma(r,0)\rangle_c^{O(r^{5/2})} \\
    &   =\int_{-\infty}^{\infty} du_1 du_2 du_3\bigg[F_{1,\text{bare}}(u_i)-F_{1,\text{hard}}(u_i)-F_{1,\text{trip}}(u_i)-F_{1,\text{pair}}(u_i)\bigg]\\
    & \equiv C_{0,1}^{(4)} \approx -10.019547\,,
\end{split}
\end{align}
where these are all functions of all three $u_i$'s, where $u_i = s_i/r$. 
$F_{1,\text{bare}}(u_i)$ is the bare result of setting $s_i \to r u_i$ and expanding to leading order in $r$.

The hard subtraction $F_{1,\text{hard}}(u_i)$ comes from expanding the correlation function in $r$ before making the $s_i \to r u_i$ substitution. 
The leading term removes the contribution from $\langle \mathcal{D}_{2h}\rangle^{(4)}$, while the subleading term of $O(r^2)$ removes the contribution from $\langle \mathcal{D}_{2h} \partial_x^2\sigma\rangle^{(3)}$ on the RHS of the fusion.

The final two terms $F_{1,\text{trip}}$ and $F_{1,\text{pair}}$ subtract the overlap with the $r^{7/4}$ region.
They correspond to the collision of three and two $\sigma$'s as $r\to 0$, respectively.
For the triple-collision regions, there are three kernels, corresponding to $u_3 \gg u_1,u_2$ and permutations. 
They can be explicitly written using
\begin{subequations}
\begin{equation}
    F_{1,\text{trip}}= f_{1,\text{trip}}(u_1,u_2,u_3) + f_{1,\text{trip}}(u_1,u_3,u_2) + f_{1,\text{trip}}(u_3,u_2,u_1)\, ,
\end{equation}
\begin{equation}\label{eq:f1trip}
    f_{1,\text{trip}}(u_1,u_2,u_3)= \lim_{\lambda\to \infty} \lambda^{1/4} [F_{1,\text{bare}}(u_1,u_2,\lambda  u_3) - F_{1,\text{hard}}(u_1,u_2,\lambda u_3) ]\, ,
\end{equation}
\end{subequations}
where the $\lambda^{1/4}$ just removes the leading divergence.
Subleading terms in $\lambda$ do not contribute at $O(r^{5/2})$, but would at higher orders.

We finally need to remove the pair collision region, where one $u_i \to 0$ with the other two fixed.
This can be determined just like the triple region from
\begin{subequations}
\begin{equation}
    F_{1,\text{pair}}= f_{1,\text{pair}}(u_1,u_2,u_3) + f_{1,\text{pair}}(u_2,u_1,u_3) + f_{1,\text{pair}}(u_3,u_2,u_1)\, ,
\end{equation}
\begin{align}\label{eq:f1pair}
\begin{split}
    f_{1,\text{pair}}(u_1,u_2,u_3)
    = \lim_{\lambda\to \infty} \lambda^{5/4} [&F_{1,\text{bare}}(u_1,\lambda u_2,\lambda  u_3)  \\
    &- F_{1,\text{hard}}(u_1,\lambda u_2,\lambda u_3)-F_{1,\text{trip}}(u_1,\lambda u_2,\lambda u_3) ]\, .
\end{split}
\end{align}
\end{subequations}

The procedure for the third line of~\eqref{eq:Ising_DD_cpt4} is the same; expand in $r$ with all $s_i \to r u_i$ and subtract overlapping regions.
We have
\begin{align}
\begin{split}
    \frac{1}{r^{5/2}}\int &ds_1 ds_2 ds_3\langle \sigma(r,s_1)\sigma(r,s_2)\sigma(0,s_3) \sigma(0,0)\rangle_c^{O(r^{5/2})} \\
    &   =\int_{-\infty}^{\infty} du_1 du_3 du_3\bigg[F_{2,\text{bare}}(u_i)-F_{2,\text{hard}}(u_i)-F_{2,\text{trip}}(u_i)-F_{2,\text{pair}}(u_i)\bigg]\\
    &  \equiv C_{0,2}^{(4)} \approx -7.711723\,.
\end{split}
\end{align}
The bare and hard kernels $F_{2,\text{bare}}$, $F_{2,\text{hard}}$ follow from the same steps as in the previous correlation function.
The triple-collision and pair-collision regions require slightly more care, since this correlation function is not fully symmetric under $u_i \leftrightarrow u_j $.

For the triple collision contribution, there are four regions corresponding to the combinations of three $\sigma$'s colliding with at least one at $x=r$ and one at $x=0$. 
These can be obtained with appropriate mappings of the integration variables as
\begin{align}\label{eq:F2trip}
\begin{split}
    F_\text{2,trip}(u_i) =& \  f_{2,\text{trip}}(u_1,u_2,u_3) +f_{2,\text{trip}}(u_2,u_1,u_3)\\
    & \ + f_{2,\text{trip}}(u_1-u_3,u_2-u_3,-u_3)+f_{2,\text{trip}}(u_2-u_3,u_1-u_3,-u_3)\,,
\end{split}
\end{align}
where we have derived $f_{2,\text{trip}}(u_1,u_2,u_3)$ from the region where $u_3 \gg u_1,u_2$, the same as~\eqref{eq:f1trip} with $F_{1} \to F_{2}$.
Likewise, the pair subtraction kernel $F_{2,\text{pair}}$ follows from replacing $f_{2,\text{trip}} \to f_{2,\text{pair}} $ in~\eqref{eq:F2trip} and defining $f_{2,\text{pair}}$ in analogy to~\eqref{eq:f1pair} after changing the subscript $1\to 2$.

Including the relevant prefactors in~\eqref{eq:Ising_DD_cpt4} and matching to $r^{5/2}a_0^{(4)}$, we arrive at the prediction for the coefficient $a_0^{(4)}$,
\begin{subequations}
\begin{align}
    a_0^{(4)} = h^4\left[ \frac{1}{3}C_{0,1}^{(4)} +\frac{1}{4}C_{0,2}^{(4)}\right] = -h^4 C_0^{(4)} \, , \qquad C_0^{(4)} \approx 5.2678 \, .
\end{align}
\end{subequations}

\subsection{Bulk to defect}\label{app:bulk-to-defect}

In order to fix the scheme of the condensate $\langle \mathcal{D}_{h}[\epsilon]_r \rangle $ in our numerical simulations in section~\ref{sec:IsingDMRG}, we require the bulk-to-defect OPE of $\epsilon(r) \mathcal{D}_h[0]$ as $r\to 0$, as described in section~\ref{sec:bulk-to-defect}.
We only need the leading few terms.
The prediction for this OPE is given by
\begin{equation}
    \epsilon(r) \mathcal{D}_h[0] \simeq \frac{b_{\mathds{1}}}{r} \mathds{1}_D + \frac{b_\sigma}{r^{7/8}}\sigma_D +b_{\epsilon} \epsilon_D + r^{1/8}b_{\partial \sigma} \partial_x \sigma_D + r b_{\partial \epsilon} \partial_x \epsilon_D + \ldots\,,
\end{equation}
where we have used the subscript $D$ to indicate the operators on the defect.
The coefficients are computable using CPT as a series in $hr^{7/8}$, just like for lines as before.
Using the $\mathbb{Z}_2$ symmetry properties of the operators, we can write this in a more helpful form as
\begin{align}\label{eq:Ising_EpsD_BulkBoundCPT}
\begin{split}
    \epsilon (r) \mathcal{D}_h[0] \simeq & \ \epsilon_{D}+b^{(1)}_\sigma \sigma_D + r^{3/4}b^{(2)}_{\mathds 1} \mathds{1}_D +  r \left[b_{\partial \sigma}^{(1)} \partial_x \sigma_D +\partial_x \epsilon_D\right]\\
    & \ + r^{7/4}\left[b_{\epsilon}^{(2)}\epsilon_D + b_\sigma^{(3)} \sigma_D\right] +  r^2 \left[b_{\partial^2 \sigma}^{(1)} \partial^2_x \sigma_D + \tfrac{1}{2} \partial^2_x \epsilon_D\right] + \ldots\,,
\end{split}
\end{align}
where we have used $b^{(0)}_{\partial^n\epsilon} = 1/n!$.
We will only be using the $O(r^0)$ and $O(r^{3/4})$ terms in this expression, so we need to fix $b_\sigma^{(1)}$ and $b_{\mathds{1}}^{(2)}$.
We can isolate $b_\sigma^{(1)}$ by inserting $\sigma(\infty)$ and computing at first order in CPT, giving us the matching for $\langle \sigma(\infty) \epsilon(r) \mathcal{D}_h[0] \rangle^{(1)} $
\begin{align}
\begin{split}
    h\int ds\,\langle \sigma(\infty) \epsilon(r) \sigma(s)\rangle \simeq h\int ds\,\langle\sigma(\infty)\epsilon(0)\sigma(s) \rangle +b_\sigma^{(1)} \langle \sigma(\infty) \sigma(0) \rangle + O(r)\,,
\end{split}
\end{align}
where we have used the shorthand $\langle \mathcal{O}(\infty) \ldots \rangle \equiv \lim_{|x|\to \infty} |x|^{2\Delta_\mathcal{O}}\langle \mathcal{O}(x)\ldots\rangle$, and all these correlators are evaluated in the critical theory.
Computing both sides gives us
\begin{align}
    h \log \frac{L}{r} \simeq h \log \frac{L}{\delta} + b_\sigma^{(1)}\,,
\end{align}
where we have a UV divergence from the $\langle\sigma  \epsilon \sigma \rangle $ correlator on the RHS, to be removed by renormalizing $\epsilon$. Using the same scheme as in the main text, we define the renormalized operator
\begin{equation}
    [\epsilon_D]_\text{r} = \epsilon_D + h \log (\mu \delta)\sigma_D \, ,
\end{equation}
which lets us extract the coefficient
\begin{equation}
    b^{(1)}_\sigma = -h \log \mu r \, .
\end{equation}

To extract the coefficient $b_{\mathds{1}}^{(2)}$, we need to go to second order in CPT. 
The simplest correlator to consider is
\begin{equation}
    \langle \epsilon(r)\mathcal{D}_h[0]  \rangle^{(2)} = \frac{h^2}{2}\int ds_1 ds_2\langle \epsilon(r) \sigma(s_1)\sigma(s_2) \rangle\,.
\end{equation}
We only need the $O(r^{3/4})$ contribution, coming from the region where $s_1 \sim s_2 \sim r$.
We can write an explicit integral expression for it by taking this limit and subtracting the overlap with the other regions
\begin{equation}
    b_{\mathds{1}}^{(2)} = \frac{h^2}{4} \int_{-\infty}^\infty du_1 du_2 \bigg[\frac{|u_1-u_2|^{3/4}}{\rho_1 \rho_2} -\frac{|u_1-u_2|^{3/4}}{|u_1||u_2|} - \bigg(\frac{\rho_2^{-1}-|u_2|^{-1}}{|u_1|^{1/4}} + (u_1 \leftrightarrow u_2)\bigg)\bigg],
\end{equation}
where $\rho_i = (1+u_i^2)^{1/2}$, the second term is the hard region where $s_1,s_2 \gg r$, and the third is from where $s_i \gg s_j\sim r$.
The integral can be done analytically, giving us
\begin{equation}\label{eq:Ising_BulkBound_c1}
    b_{\mathds{1}}^{(2)} = h^2 c_\mathds{1}\, , \qquad \qquad c_\mathds{1} \equiv \frac{\Gamma(\tfrac{7}{8})\Gamma(-\tfrac{3}{8})^3}{4\sqrt{\pi}\Gamma(-\tfrac{3}{4})} \approx 1.7797467013\,.
\end{equation}

\section{TBA and exact defect condensates}
\label{app:TBA}
In this appendix we describe the computation of $\langle \partial_x^2 \sigma \rangle_h$. It will be useful to fold the theory along the defect, so that we can treat it as a boundary for the double copy of the Ising CFT \cite{LeClair:1997gz}. We will then be able to import various pieces of technology from the literature on integrable boundary conditions \cite{Ghoshal:1993tm,LeClair:1995uf,Fendley:1995kj,LeClair:1997gz,Dorey:2004xk}.

In the main text, the main trick to compute $\langle \sigma \rangle_h$ was to note that, in the presence of the magnetic impurity, the Hamiltonian gets modified as $\mathbf{H} = \mathbf{H}_0 - h \sigma(0)$. Here $x=0$ is the position of the impurity. It was then sufficient to compute the energy $e(h)$ stored on the defect, from which we extracted $\langle \sigma \rangle_h$ via (\ref{eq:sigmafrome}). 

As discussed in the main text, besides the Hamiltonian, 
the integrable bulk theory has a number of higher-spin local integrable 
charges $\mathbf{I}_{2n+1}^0$. Here we show that, in the presence of the defect, the spin-3 integrable charge gets modified by a local insertion on the defect as 
\begin{equation}
\mathbf{I}_3 = \mathbf{I}_3^0 + 6h \partial_x^2 \sigma(0) + \partial_s \mathcal O(0)\, . \label{eq:deformedq3}
\end{equation}
 The last term is a total (time) derivative, and will drop in our computations. In analogy with the Hamiltonian case, we will then be able to extract $\langle \partial_x^2 \sigma \rangle_h$ from the charge $e_3(h)$ stored on the defect.

The first step is to establish (\ref{eq:deformedq3}). Our goal is to construct $\mathbf{I}_3 = \mathbf{I}_3^0 - h \Theta$ such that
\begin{equation}
    [\mathbf{H},\mathbf{I}_3] = 0. \label{eq:wantHI}
\end{equation}
Given that $[\mathbf{H}_0,\mathbf{I}_3^0] = 0$, this is achieved if 
\begin{equation}
    h[\mathbf{I}_3^0,\sigma(0)] - h[\mathbf{H}_0,\Theta] + h^2 [\sigma(0),\Theta(0)]=0\,.
\end{equation}
By dimensional analysis, the last term (computed with a point-split regulator) is a pure local power divergence, since there are no operators with the correct dimensions to make up its finite part. It can thus be removed by adding a counterterm in the definition of $\Theta(0)$ on the defect, and so the $h^2$ term can be dropped. This is the same kind of subtraction of power divergences that is implicit in $\partial^2_x \sigma$ on the defect, see e.g.\ footnote~\ref{fn:dx2sigma_ren}. We thus see that (\ref{eq:wantHI}) is satisfied if
\begin{equation}
    [\mathbf{I}_3^0,\sigma(0)] = \partial_s \Theta(0)\,, \label{eq:want2}
\end{equation}
for some $\Theta$. 

The undeformed folded theory actually possesses three spin-4 chiral currents. These are $:T_1^2:$, $:T_2^2:$ and $T_1T_2$, where $T_i$ is the chiral stress tensor in the $i$-th Ising copy. We thus write a family of spin-3 charges following the standard boundary recipe \cite{Ghoshal:1993tm,LeClair:1997gz}:
\begin{equation}
    \mathbf{I}_3^0 = \frac{\alpha}{2\pi}\int_0^\infty dx :T_1^2:(x) + :T_2^2:(x) + k T_1 T_2(x) + (T_i \rightarrow \bar{T}_i)\,,
\end{equation}
for any $k$, and with $\alpha$ a normalization which we fix later for convenience. It is sufficient to find a pair ($k$,$\Theta$) so that (\ref{eq:want2}) holds. We will fix both $k$ and $\Theta$ simultaneously. Indeed, computing $ [\mathbf{I}_3^0,\sigma(0)] $  with the OPE, we find
\begin{equation}
   \tfrac{1}{\alpha}[  \mathbf{I}_3^0  ,\, \sigma(0)]
    =\frac{2+k}{16}\big(L_{-3}+\bar L_{-3}\big)\sigma
    +2\big(L_{-2}L_{-1}+\bar L_{-2}\bar L_{-1}\big)\sigma
    +k\big(L_{-1}\bar L_{-2}+L_{-2}\bar L_{-1}\big)\sigma\,,
\end{equation}
where the barred generators correspond to the generators of the second copy in the folded picture. Using the level-2 null vector of the Ising minimal model, 
\begin{equation}
    L_{-2}\,\sigma=\frac43\,L_{-1}^2\,\sigma\, ,\qquad
    \bar L_{-2}\,\sigma=\frac43\,\bar L_{-1}^2\,\sigma\, ,
\end{equation}
we find
\begin{equation}
   \tfrac{1}{\alpha} \big[\,\mathbf{I}_3^0\, ,\, \sigma(0)\,\big]
    =\Big(\frac{2+k}{16}-2\Big)\big(L_{-3}+\bar L_{-3}\big)\sigma
    +\frac83\big(L_{-1}^3+\bar L_{-1}^3\big)\sigma
    +\frac{4k}{3}\,L_{-1}\bar L_{-1}\big(L_{-1}+\bar L_{-1}\big)\sigma\, .
\end{equation}
The two last terms are total time derivatives, as desired by (\ref{eq:want2}). The first one cancels if $k=30$. We conclude that (\ref{eq:wantHI}) holds if
\begin{equation}
    (k,\Theta) = \left(30, 8 \alpha \partial_x^2 \sigma + \frac{32\alpha}{3}\partial_s^2 \sigma\right).
\end{equation}
Finally, we fix $\alpha$ in order to match the normalization of \cite{Bazhanov:1994ft,Bazhanov:1996dr} so that we can later import the results (\ref{eq:BLZresult}) which we will need. This can be achieved by acting with the charge on the cylinder vacuum (i.e. in the opposite quantization),
\begin{equation}
    \mathbf{I}_3^0 |\Omega\rangle_\beta =   \left(\frac{2 \pi}{\beta}\right)^3  \frac{31 \alpha}{1440}|\Omega\rangle_\beta\,, \label{eq:i3vev}
\end{equation}
while in \cite{Bazhanov:1994ft} one has $ \mathbf{I}_3^0 |\Omega\rangle_{2\pi} = -\frac{31}{1920} |\Omega\rangle_{2\pi}$, see (11) there.\footnote{Note that what we refer to here as the $\mathbf{I}_3^0$ of \cite{Bazhanov:1994ft} corresponds to $\mathbf{I}_3$ in the main text (\ref{eq:Tasymp}). There, $\mathbf{I}_3$ is fixed by requiring it commutes with the pinning defect $\mathbf{T}_{1/2}(\lambda)$. This is equivalent to our condition (\ref{eq:want2}) upon integrating the $\sigma$ insertion around the circle and exponentiating. The commutator then vanishes since the right-hand side of (\ref{eq:want2}) is a total derivative. Hence both results must match.}  We conclude $\alpha  = -3/4$, which implies (\ref{eq:deformedq3}).

Before proceeding with the determination of $\langle  \partial_x^2 \sigma \rangle$, let us slightly reframe the computation of $\langle \sigma \rangle$. In the main text, it was determined from the partition function of the pinning defect, computed as the expectation value of the transfer matrix operator (\ref{eq:transfer}).
Alternatively, we could have computed the same partition function as a thermal trace:
\begin{equation}
    G(\lambda) = \frac{\text{Tr}_h e^{-\beta \mathbf{H}}}{\text{Tr} e^{-\beta \mathbf{H}_0}}\,, \label{eq:thermalratios}
\end{equation}
where the numerator is traced over the pierced Hilbert space (as opposed to the denominator). The computation of $e_3(h)$ can be performed, in analogy, by defining\footnote{Note that at finite $\kappa$ the trace might not be well defined since $\mathbf{I}_3$ is not necessarily a positive operator. For our purposes, it is sufficient to think of  (\ref{eq:chargeratio}) as a generating function for $\mathbf{I}_3$ insertions. We work to linear order.}
\begin{equation}
    G_\kappa(\lambda) = \frac{\text{Tr}_h e^{-\beta (\mathbf{H} + \kappa \mathbf{I}_3)}}{\text{Tr} e^{-\beta (\mathbf{H}_0 + \kappa \mathbf{I}_3^0)}}. \label{eq:chargeratio}
\end{equation}
We then have
\begin{equation}
    \log G_\kappa(\lambda) = - \beta \left( e(h) + \kappa e_3(h)\right) + O(\beta^0,\kappa^2)\,,
\end{equation}
and, finally,
\begin{equation}
\partial_h e_3=6\langle\partial_x^2\sigma\rangle_h  \label{eq:e3result}.
\end{equation}
In this last step, we used that $\langle\partial_s \mathcal{O}\rangle_h =  \langle [\mathbf{H},\mathcal{O}]\rangle_h=0$.  
We can also use that  $e_3(h) = c_3 h^{24/7}$ by dimensional analysis to rewrite (\ref{eq:e3result}) as
\begin{equation}
\langle\partial_x^2\sigma\rangle_h = \frac{4}{7}\frac{e_3(h)}{h}.
\end{equation}

The thermal traces above can be computed via thermodynamic Bethe ansatz \cite{Zamolodchikov:1989cf,LeClair:1995uf}. The first step is to specify the scattering description of the folded theory. As explained in \cite{LeClair:1997gz}, the folded theory (for which $c=1$) with the magnetic pinning can be regarded as (an orbifold of) massless boundary sine-Gordon theory at the particular coupling $\frac{\beta_\text{SG}^2}{8\pi} = \Delta_\sigma$. The magnetic pin boundary then defines a completely diagonal reflection amplitude, and the boundary TBA is simple.

The bulk particle spectrum at this value of the sine-Gordon coupling $\beta_\text{SG}$ is composed of the two (self-conjugate) kinks ($\pm$) and six breathers ($b=1,\dots, 6$) where
\begin{equation}
    {E_\pm(\theta)=e^{\theta}\,,\qquad E_b(\theta)=\mu_b\,e^{\theta}\,,\qquad \mu_b=2\sin\frac{b\pi}{14}\,.}
\end{equation}
The scattering of these particles is purely diagonal. In terms of the function
\begin{equation}
    F_\alpha(\theta)=\frac{\tanh\tfrac12\!\left(\theta+i\pi\alpha\right)}{\tanh\tfrac12\!\left(\theta-i\pi\alpha\right)}\, ,
\end{equation}
the two-particle $S$-matrices are \cite{LeClair:1997gz}
\begin{equation}
\begin{split}
    S_{++}(\theta)&=S_{--}(\theta)=S_{+-}(\theta)=F_{\frac17}(\theta)\,F_{\frac27}(\theta)\,F_{\frac37}(\theta)\, ,\\
    S_{ab}(\theta)&=F_{\frac{|a-b|}{14}}(\theta)\left[\prod_{k=1}^{\min(a,b)-1}F_{\frac{|a-b|+2k}{14}}(\theta)\right]^2 F_{\frac{a+b}{14}}(\theta)\, ,\\
    S_{\pm a}(\theta)&=(-1)^a\prod_{k=0}^{a-1}F_{\frac{7-a+2k}{14}}(\theta)\, ,
\end{split}
\end{equation}
with $a,b=1,\dots,6$ labelling the breathers and $\theta$ the rapidity difference. As mentioned, the magnetic pinning boundary reflects each particle diagonally, with amplitudes \cite{LeClair:1997gz}
\begin{equation}
    R_\pm(\theta)=\tanh\!\left(\frac{\theta-\theta_B}{2}-\frac{i\pi}{4}\right) ,\qquad
    R_a(\theta)=F_{-\frac{a}{14}}(\theta-\theta_B)\, ,
\end{equation}
where $T_B\equiv e^{\theta_B} = \tau h^{8/7}$ is a scale set by the defect. $\tau$ will be fixed soon. Only the phase-shifts enter the TBA. We define
\begin{equation}
    \varphi_{bc}(\theta)=-i\,\partial_\theta\log S_{bc}(\theta)\, ,\qquad
    \mathcal{K}_b(\theta-\theta_B)=-i\,\partial_\theta\log R_b(\theta)\, ,
\end{equation}
where $b,c$ now run over all eight species. Using
\begin{equation}
    -i\,\partial_\theta\log F_\alpha(\theta)=-\frac{2\sin\pi\alpha\,\cosh\theta}{\sinh^2\theta+\sin^2\pi\alpha}\, ,
\end{equation}
one finds in particular
\begin{equation}
    \mathcal{K}_\pm(\theta)=\frac{1}{\cosh\theta}\, ,\qquad
    \mathcal{K}_a(\theta)=\frac{2\sin\frac{a\pi}{14}\,\cosh\theta}{\sinh^2\theta+\sin^2\frac{a\pi}{14}}\, .
\end{equation}

To compute $e(h)$, we start from the standard boundary TBA in the ``g-function'' channel of \cite{LeClair:1995uf}. For the numerator in (\ref{eq:thermalratios}), the free energy is given by
\begin{equation}
    \log \text{Tr}_h\, e^{-\beta \mathbf{H}}
    = -\beta E_0(h)
    + R\sum_b\int\frac{d\theta}{2\pi}\,\mu_b\,e^{\theta}\,L_b(\theta)
    + \sum_b\int\frac{d\theta}{2\pi}\,\mathcal{K}_b(\theta-\theta_B)\,L_b(\theta)
    + \dots\, , \label{eq:boundaryTBA}
\end{equation}
with $L_b(\theta)\equiv\log\!\left(1+e^{-\varepsilon_b(\theta)}\right)$, where the pseudo-energies governing the particle densities in the thermal state are fixed by the usual thermodynamic saddle analysis:
\begin{equation}
    \varepsilon_b(\theta)=\beta\,\mu_b\,e^{\theta}-\sum_c\int\frac{d\theta'}{2\pi}\,\varphi_{bc}(\theta-\theta')\,L_c(\theta')\, . \label{eq:TBAappendix}
\end{equation}
Let us break down (\ref{eq:boundaryTBA}) and (\ref{eq:TBAappendix}). In (\ref{eq:boundaryTBA}), $E_0(h)$ is the ground state energy of the pierced Hilbert space at large volume $R$.
Note that the excited energies computed by Bethe ansatz are measured relative to this ground state energy. The second term corresponds to the extensive contribution to the free energy.
 It depends entirely on the bulk quantities. 
 
 The remaining terms contribute to order $R^0$. The third term corresponds to the boundary contribution to the free energy while the dots correspond to additional contributions due to the one-loop TBA saddle analysis \cite{Dorey:2004xk,Pozsgay:2010tv} (these terms were left out in \cite{LeClair:1995uf}). Crucially, these additional terms depend entirely on the bulk data.
 When constructing the ratio (\ref{eq:thermalratios}), these terms cancel against the denominator, which is obtained by setting $h \to 0$ in (\ref{eq:boundaryTBA}). We find
 \begin{equation}
    \log G(\lambda) = - \beta e(h)  + \sum_b\int\frac{d\theta}{2\pi}\,\mathcal{K}_b(\theta-\theta_B)\,L_b(\theta) - K\,, \label{eq:boundaryresidue}
 \end{equation}
where $e(h) = E_0(h) - E_0(0)$. Here, $K$ is a constant corresponding to the $h\to 0$ limit of the boundary term in (\ref{eq:boundaryTBA}), i.e. it is the boundary term associated to the transparent defect.
Finally, note that the pseudo-energies are also determined via (\ref{eq:TBAappendix}) by bulk data (since the $O(R^0)$ boundary effect does not affect the $O(R)$ saddle analysis).

To fix $e(h)$, we consider the $\lambda \to 0$ limit, where $G(\lambda)$ can be computed in the high-temperature expansion (\ref{eq:isingnpoint}). The high-temperature expansion is a series in powers of $h^2$. On the other hand, we will see that the integral in (\ref{eq:boundaryresidue}) produces a non-analytic term in $h$. Those must cancel against the other terms to recover the analytic expansion \cite{Fendley:1995kj}. This determines $e(h)$.

To perform this computation, it is useful to first rewrite (\ref{eq:TBAappendix}) slightly.
Define
\begin{equation}
    (f*g)(\theta)\equiv\int_{-\infty}^{\infty}\frac{d\theta'}{2\pi}\,f(\theta-\theta')\,g(\theta')\, . \label{eq:convolution}
\end{equation}
The key is the identity
\begin{equation}
    \mathcal{K}_b=\delta_{b1}\,s-s*\varphi_{1b}\, ,\qquad s(\theta)=\frac{7}{\cosh 7\theta}\, , \label{eq:kernelidentity}
\end{equation}
which can be easily verified in Fourier space. In words, the boundary amplitude can be replaced by a scattering process with the lightest breather. It originates from the property
\begin{equation}
    \left(R_b\!\left(\theta+\frac{i\pi}{14}\right)R_b\!\left(\theta-\frac{i\pi}{14}\right)\right)^2=\frac{1}{S_{1b}(\theta)^2}\, , \label{eq:RRS}
\end{equation}
valid for any particle $b$.

Convoluting $s$ against the lightest breather TBA equation (\ref{eq:TBAappendix}) for $b=1$ and using (\ref{eq:kernelidentity}), we find
\begin{equation}
    s*\varepsilon_1
    =s*\beta\mu_1e^{\theta}-\sum_b s*\varphi_{1b}*L_b
    =s*\beta\mu_1e^{\theta}-s*L_1+\sum_b\mathcal{K}_b*L_b\, .
\end{equation}
Solving for the boundary term,  we obtain
\begin{equation}
    \sum_b\int\frac{d\theta}{2\pi}\,\mathcal{K}_b(\theta-\theta_B)\,L_b(\theta)
    =\int\frac{d\theta}{2\pi}\,s(\theta-\theta_B)\Big[\log\!\big(1+e^{\varepsilon_1(\theta)}\big)-\beta\,\mu_1\,e^{\theta}\Big]\, . \label{eq:collapsed}
\end{equation}
The last term is elementary,
\begin{equation}
    \int\frac{d\theta}{2\pi}\,s(\theta-\theta_B)\,\beta\,\mu_1\,e^{\theta}=\beta\,T_B\tan\frac{\pi}{14}\, ,
\end{equation}
and so (\ref{eq:boundaryresidue}) becomes
\begin{equation}
    \log G(\lambda)=-\beta\left[e(h)+T_B\tan\frac{\pi}{14}\right]
    +\int\frac{d\theta}{2\pi}\,s(\theta-\theta_B)\,\log\!\big(1+e^{\varepsilon_1(\theta)}\big)-K\, . \label{eq:Gcollapsed}
\end{equation}
We will show that the bracket is the only non-analytic piece, and must vanish, determining $e(h)$.

It can be shown\footnote{Indeed, the TBA (\ref{eq:TBAappendix}) is equivalent to the one associated to the $D_8$ Dynkin diagram in the classification of \cite{Zamolodchikov:1991et}. It can be written as the $D_8$ $Y$-system
\begin{equation}
    Y_b\!\left(\theta+\frac{i\pi}{14}\right)Y_b\!\left(\theta-\frac{i\pi}{14}\right)=\prod_{c}\big(1+Y_c(\theta)\big)^{N_{bc}}\, ,\qquad Y_b\equiv e^{\varepsilon_b}\, ,
\end{equation}
where $N_{bc}$ is the incidence matrix of the diagram and $14$ is the Coxeter number of $D_8$. The solutions of such systems are periodic \cite{Zamolodchikov:1991et}, here with period $\frac{8\pi i}{7}$. } that the solutions $\varepsilon_j(\theta)$ of (\ref{eq:TBAappendix}) are periodic,
\begin{equation}
    \varepsilon_j(\theta) = \varepsilon_j\left(\theta + \frac{8 \pi i}{7}\right).
\end{equation}
Together with the fact that they depend on $\theta$ and $\beta$ only through $\beta e^{\theta}$, as is clear from (\ref{eq:TBAappendix}), this means that they are functions of $x \equiv (\beta e^{\theta})^{7/4}$ alone.

To fix $e(h)$ it is sufficient to consider the leading behavior of (\ref{eq:Gcollapsed}) at small $h$. In this limit, $\theta_B \to -\infty$, and the kernel has support at $\theta \to -\infty$, where $Y_1 = e^{\varepsilon_1}$ admits a regular expansion in powers of $x$ \cite{Zamolodchikov:1991et},
\begin{equation}
    \log\!\big(1+Y_1(\theta)\big)=f_0+O(x)\, . \label{eq:Yexpand}
\end{equation}
Since the kernel is normalized as $\int\frac{d\theta}{2\pi}\,s(\theta)=\frac12$, we find
\begin{equation}
    \int\frac{d\theta}{2\pi}\,s(\theta-\theta_B)\,\log\!\big(1+e^{\varepsilon_1(\theta)}\big)
    =\frac{f_0}{2}+O\big((\beta T_B)^{7/4}\big)\, .
\end{equation}
On the other hand, the high-temperature expansion (\ref{eq:isingnpoint}) gives $\log G(\lambda)=O(h^2\beta^{7/4})$. Both are of higher order than $\beta$. Comparing the terms of order $\beta^0$ and $\beta^1$ in (\ref{eq:Gcollapsed}), we conclude
\begin{equation}
    K=\frac{f_0}{2}\, ,\qquad e(h)=-T_B\tan\frac{\pi}{14}\, . \label{eq:eTB}
\end{equation}

The last step is to fix the map $T_B=\tau\,h^{8/7}$. Within the TBA approach, this can be done by matching one further order in the small $\lambda$ expansion: the term of order $(\beta T_B)^{7/4}$ in (\ref{eq:Gcollapsed}) must reproduce the first term, of order $h^2\beta^{7/4}$, of the high-temperature expansion (\ref{eq:isingnpoint}). The former is only known numerically (to us). To keep the discussion analytic, we instead import the main text result, $e(h)=-\frac{m}{2\pi}h^{8/7}$, and read off
\begin{equation}
    T_B=\frac{m}{2\pi\tan\frac{\pi}{14}}\,h^{8/7}
    =\frac{\Gamma\!\left(\frac{1}{14}\right)\Gamma\!\left(\frac78\right)^{8/7}}{\sqrt\pi\,\Gamma\!\left(\frac47\right)}\,h^{8/7}\,. \label{eq:TBh}
\end{equation}

Of course, the step (\ref{eq:TBh}) was completely circular. The point is that the TBA argument used to illustrate the computation of $e(h)$ generalizes straightforwardly to $e_3(h)$, where we reuse the same value of $\tau$.

Indeed, consider now the deformed ratio (\ref{eq:chargeratio}), to first order in $\kappa$. In the TBA derivation, the only modification is in the energy factor carried by the particles, which is now measured by the modified Hamiltonian \cite{Fioretto:2009yq, Mossel:2012vp,Downing:2021mfw,Downing:2023lnp}. Equation (\ref{eq:boundaryresidue}) is mostly unmodified and becomes
\begin{equation}
    \log G_\kappa(\lambda) = - \beta \big( e(h) + \kappa\, e_3(h) \big)
    + \sum_b\int\frac{d\theta}{2\pi}\,\mathcal{K}_b(\theta-\theta_B)\,L^{\kappa}_b(\theta) - K + O(\kappa^2)\, , \label{eq:boundaryresiduekappa}
\end{equation}
where the pseudo-energy in $L_b^\kappa\equiv\log(1+e^{-\varepsilon_b^\kappa})$ must now satisfy the deformed TBA equation 
\begin{equation}
    \varepsilon^\kappa_b(\theta)=\beta\left(\mu_b\,e^{\theta}+\kappa\,\lambda_3\,q_b\,e^{3\theta}\right)-\sum_c\left(\varphi_{bc}*L^\kappa_c\right)(\theta)\, . \label{eq:TBAdeformed}
\end{equation}
Here, we replaced the energy $\mu_b e^\theta$ by the eigenvalue of $\mathbf{H}_0+\kappa\,\mathbf{I}_3^0$ on a particle of type $b$. The exact formula is 
\begin{equation}
    \mathbf{I}_3^0\,|\theta\rangle_b= \lambda_3\,q_b\,e^{3\theta}\,|\theta\rangle_b\, ,\qquad q_a=-2\sin\frac{3a\pi}{14}\, ,\qquad q_\pm=1\, . \label{eq:chargeparticle}
\end{equation}
The dependence on the rapidity is determined from the spin of the charge. The relative charges $q_a$ can be fixed via integrable bootstrap \cite{Zamolodchikov:1978xm}: one starts with charge $q_\pm=1$, and computes the breather charges via fusion of the kinks. The overall normalization $\lambda_3$ will be fixed below.

The steps leading to (\ref{eq:collapsed}) follow unchanged, with only the Hamiltonian eigenvalue being modified inside the bracket. Its integral is also elementary,
\begin{equation}
    \int\frac{d\theta}{2\pi}\,s(\theta-\theta_B)\,\beta\left(\mu_1\,e^{\theta}+\kappa\,\lambda_3\,q_1\,e^{3\theta}\right)
    =\beta\,T_B\tan\frac{\pi}{14}-\kappa\,\beta\,\lambda_3\,T_B^{3}\tan\frac{3\pi}{14}\, ,
\end{equation}
so that (\ref{eq:Gcollapsed}) is replaced by
\begin{equation}
    \log G_\kappa(\lambda)=
    -\kappa\,\beta\left[e_3(h)-\lambda_3\,T_B^{3}\tan\frac{3\pi}{14}\right]
    +\int\frac{d\theta}{2\pi}\,s(\theta-\theta_B)\,\log\!\big(1+e^{\varepsilon^\kappa_1(\theta)}\big)-K+O(\kappa^2)\, . \label{eq:Gkappacollapsed}
\end{equation}
The bracket term is proportional to $h^{24/7}$, which is again not an integer power of $h^2$. One can again argue that the integral term comes as a series in $h^2$ at small $h$ (one more order in (\ref{eq:Yexpand}) is required in this case). Consistency with the high-temperature expansion then fixes
\begin{equation}
    e_3(h)=\lambda_3\,T_B^{3}\tan\frac{3\pi}{14}\, . \label{eq:e3TB}
\end{equation}

It remains to fix the normalization $\lambda_3$. Numerically, the determination is quite simple. We simply use the thermodynamic solution to compute the total spin-3 charge density carried by the gas of particles in the thermal state. This is
\begin{equation}
    j_3\equiv\frac{\langle\mathbf{I}_3^0\rangle_\beta}{R}
    =\sum_b\int d\theta\,\rho_b(\theta)\,\lambda_3\,q_b\,e^{3\theta}
    =\frac{3\lambda_3}{\beta}\sum_b q_b\int\frac{d\theta}{2\pi}\,e^{3\theta}\,L_b(\theta)\, , \label{eq:j3}
\end{equation}
where $\rho_b$ is the density of particles of type $b$. In the last step we used the TBA relation between $\rho_b$ and $\varepsilon_b$, which implies $\rho_b=-\frac{1}{2\pi\beta}\,\partial_\theta L_b$, and integrated by parts.

The left-hand side of (\ref{eq:j3}) was computed in the opposite channel\footnote{The two are related by Wick rotation, which for a spin-4 current produces a factor of $i^4=1$.}  in (\ref{eq:i3vev}), 
\begin{equation}
j_3=\frac{-1}{\beta}\left(\frac{2\pi}{\beta}\right)^3\frac{31}{1920}\, .
\end{equation}
The integral on the right-hand side of (\ref{eq:j3}) can be computed from the TBA solution, and we can solve for $\lambda_3$.

To proceed analytically, we note that the same integral controls $G(\lambda)$ at large $\lambda$. In this regime $\theta_B$ is far above the rapidities populated in the gas, and we can expand the kernels in (\ref{eq:boundaryresidue}) in powers of $e^{\theta-\theta_B}$,
\begin{equation}
    \mathcal{K}_b(\theta-\theta_B)=2\,\mu_b\,\frac{e^{\theta}}{T_B}-2\,q_b\,\frac{e^{3\theta}}{T_B^3}+\dots\, ,
\end{equation}
valid for any particle $b$. Using (\ref{eq:j3}), the cubic term contributes to $\log G(\lambda)$ as
\begin{equation}
    -\frac{2}{T_B^3}\sum_b q_b\int\frac{d\theta}{2\pi}\,e^{3\theta}\,L_b(\theta)=-\frac{2}{3\lambda_3 T_B^3}\,\beta\,j_3\, .
\end{equation}
On the other hand, the corresponding term in (\ref{eq:Tasymp}) is 
\begin{equation}
-C_2\left(\frac{\lambda}{2\pi}\right)^{-3}{}_\beta\langle\Omega|\mathbf{I}_3|\Omega\rangle_\beta = -C_2\,h^{-24/7}\,\beta\,j_3.
\end{equation}
Comparing the two expressions above we have\footnote{The same argument applied to the linear term of the kernels reproduces (\ref{eq:TBh}).}
\begin{equation}
    \lambda_3\,T_B^{3}=\frac{2}{3\,C_2}\,h^{24/7}\, ,
\end{equation}
with $C_2$ given in (\ref{eq:BLZresult}) at $\Delta=\Delta_\sigma$. 

We conclude
\begin{equation}
    e_3(h)=\frac{2}{3\,C_2}\tan\frac{3\pi}{14}\;h^{24/7}\, ,\qquad
    \langle\partial_x^2\sigma\rangle_h=\frac{4}{7}\frac{e_3(h)}{h}
    =\frac{64\,\Gamma\!\left(\frac27\right)\Gamma\!\left(\frac78\right)^{24/7}}{5\sqrt\pi\,\Gamma\!\left(\frac{11}{14}\right)}\;h^{17/7}\, ,
\end{equation}
which is the result quoted in the main text (\ref{eq:hardcondensate}). Uff!

\newpage
\bibliographystyle{ytphys}
\bibliography{refs}

\end{document}

%% file: Figures/Hab.tikz
\begin{tikzpicture}[line cap=round, line join=round]

\def\R{1.3}
\def\W{3}
\def\H{1}
\def\S{1}

\draw[thick, dashed] (-\R,0) -- (-\R,-\H);
\draw[thick, dashed] ( \R,0) -- ( \R,-\H);

\filldraw[
    fill=gray,
    fill opacity=0.2
]
    (-\W-\S,-\H) --
    (\W-\S,-\H) --
    (\W+\S,\H) --
    (-\W+\S,\H) -- cycle;

\draw[thick] (-\R,0) -- (-\R,2);
\draw[thick] ( \R,0) -- ( \R,2);
\draw[thick] (-\R,-2) -- (-\R,-\H-0.05);
\draw[thick] ( \R,-2) -- ( \R,-\H-0.05);

\fill (-\R,0) circle (1.5pt);
\fill ( \R,0) circle (1.5pt);

\draw[<->]
    (-\R+0.25,0) -- (\R-0.25,0)
    node[midway,above=3pt] {$r$};

\node[left, yshift=5pt, xshift = -1.5pt] at (-\R,-2) {$\mathcal D_b$};
\node[left, yshift=5pt, xshift = -1.5pt] at ( \R,-2) {$\mathcal D_a$};

\node at (\W+0.3,-0.65) {$\mathcal H_{ab}(r)$};

\end{tikzpicture}

%% file: Figures/RG.tikz
\begin{tikzpicture}[
    scale=0.9,
    line cap=round,
    line join=round,
    >=Stealth,
    every node/.style={font=\large},
    flow/.style={
        postaction={decorate},
        decoration={
            markings,
            mark=at position #1 with {\arrow{Stealth}}
        }
    }
]
\definecolor{flowblue}{RGB}{65,180,205}

\coordinate (IR)  at (-5,-2);

\draw[name path=curveA, thick, flow=.6, flow=.8,bend right=15]
    (3,0)
    to (IR);

\draw[name path=curveB, thick, flow=.35, flow=.60]
    (1,5)
    to[bend right=15] coordinate[pos=0.15] (D1D2) (0,-2);

  \path[name intersections={
    of=curveA and curveB,
    by={DCFT}
  }];

\fill (DCFT) circle[radius=3pt];
\node[below right] at (DCFT) {$\mathcal D_c^{\mathrm{CFT}}$};

\node[] at (D1D2) {\Large $\times$};
\node[right=4pt] at (D1D2) {$\mathcal D_a\mathcal D_b$};

\draw[gray, dashed]
    (DCFT) circle[radius=1.5];

\coordinate (massStart) at ($(D1D2)+(0.1,1)$);
\coordinate (turn)      at ($(DCFT)+(-.45,1)$);
\coordinate (merge)     at ($(DCFT)+(-1.8,-.25)$);

\draw[cyan, thick, flow=.2, flow=.45, flow=.75, flow=.92]
    (massStart)
    .. controls ($(massStart)+(-0.5,-1)$)
            and ($(turn)+(.25,1)$) .. (turn)
    coordinate[pos=0.27] (D1D2m)
    .. controls ($(turn)+(-.25,-1)$)
            and ($(merge)+(.65,.28)$) .. (merge)
    .. controls ($(merge)+(-.65,-.28)$)
            and ($(IR)+(1.2,.8)$) .. (IR);

\fill (IR) circle[radius=2.5pt];
\node[below left] at (IR) {$\mathrm{IR}$};

\node[] at (D1D2m) {\Large {\color{cyan} $\times$}};

\draw[ ->]
    (0.4,2.5) to[bend right=6] (0.25,2.5-0.75);
\node[below right] at (0.4,2.5) {\small $Er\to0$};

\draw[ ->]
    (-2.2,-.95) to[bend right=5] (-2.9,-1.25);
\node[below=6pt, xshift = 5pt] at (-2.2,-.95) {\small $E/m\to0$};

\end{tikzpicture}

%% file: Figures/CPTball.tikz
\begin{tikzpicture}[line cap=round, line join=round]

\begin{scope}
\def\d{0.5}
\def\R{1.2}

\draw[thick, dashed] (-\d,{sqrt(\R^2-\d^2)-0.09}) -- (-\d,-{sqrt(\R^2-\d^2)+0.05});
\draw[thick, dashed] ( \d,{sqrt(\R^2-\d^2)-0.09}) -- ( \d,-{sqrt(\R^2-\d^2)+0.05});

\draw[thick] (-\d,2.2) -- (-\d,{sqrt(\R^2-\d^2)});
\draw[thick] ( \d,2.2) -- ( \d,{sqrt(\R^2-\d^2)});
\draw[thick] (-\d,-{sqrt(\R^2-\d^2)}) -- (-\d,-2.2);
\draw[thick] ( \d,-{sqrt(\R^2-\d^2)}) -- ( \d,-2.2);

\draw[<->] (-\d+0.1,1.8) -- (\d-0.1,1.8)
    node[midway,above=3pt] {$r$};

\shade[ball color = gray!20, opacity = 0.3] (0,0) circle (\R);
\draw (0,0) circle (\R);
\draw (-\R,0)  arc[start angle=180,
      end angle=360,
      x radius=\R,
      y radius=0.35];
\draw[dashed] (\R,0) arc [start angle=0,
      end angle=180,
      x radius=\R,
      y radius=0.35];

\draw[->] (0,0) -- (0.66*\R,0.66*\R);
\node[above right] at (0.66*\R,0.66*\R) {$R$};
\end{scope}

\node at (2.3,0) {$\simeq$};

\begin{scope}[xshift=4.6cm]
\def\d{0.5}
\def\R{1.2}

\draw[thick, dashed] (0,{\R-0.09}) -- (0,-{\R+0.05});

\draw[thick] (0,2.2) -- (0,{\R});
\draw[thick] (0,-{\R}) -- (0,-2.2);

\shade[ball color = gray!20, opacity = 0.3] (0,0) circle (\R);
\draw (0,0) circle (\R);
\draw (-\R,0)  arc[start angle=180,
      end angle=360,
      x radius=\R,
      y radius=0.35];
\draw[dashed] (\R,0) arc [start angle=0,
      end angle=180,
      x radius=\R,
      y radius=0.35];

\draw[->] (0,0) -- (0.66*\R,0.66*\R);
\node[above right] at (0.66*\R,0.66*\R) {$R$};
\end{scope}

\end{tikzpicture}

%% file: Figures/CPTballLocal.tikz
\begin{tikzpicture}[line cap=round, line join=round]

\def\d{0.5}
\def\R{1.5}

\shade[ball color = gray!20, opacity = 0.3] (0,0) circle (\R);
\draw (0,0) circle (\R);
\draw (-\R,0)  arc[start angle=180,
      end angle=360,
      x radius=\R,
      y radius=0.5];
\draw[dashed] (\R,0) arc [start angle=0,
      end angle=180,
      x radius=\R,
      y radius=0.5];

\draw[->] (0,0) -- (0.66*\R,0.66*\R);
\node[above right] at (0.66*\R,0.66*\R) {$R$};

\fill (0,0) circle (1.5pt);
\node[below right=-3pt] at (0,0) {$\mathcal O_2$};
\fill (-0.55*\R,0.4*\R) circle (1.5pt);
\node[above right=-2pt] at (-0.55*\R,0.4*\R) {$\mathcal O_1$};
\fill (-4,0) circle (1.5pt);
\node[below] at (-4,0) {$\mathcal O_k(z)$};

\draw[->] (0,0) -- (-0.55*\R+0.08,0.4*\R-0.08)
    node[midway,below left=-2pt] {$x$};

\end{tikzpicture}

%% file: Figures/phiphiCPT.tikz
\tikzset{
  defect/.style={line width=0.8pt},
  scalar/.style={dashed, line width=0.8pt},
  insertion/.style={circle, fill=black, inner sep=1.3pt},
  mass insertion/.style={font=\large, inner sep=0pt}
}

\begin{tikzpicture}[baseline={(current bounding box.center)}]
  
  \node[insertion] at (-4,0) {};
  \node[left=3pt] at (-4,0) {$\phi$};
  \node[insertion] at (-2,0) {};
  \node[right=3pt] at (-2,0) {$\phi$};
  \draw[scalar] (-4,0) -- (-2,0);

  \node[] at (-1,0) {$+$};
  
  \node[insertion] at (0,0) {};
  \node[left=3pt] at (0,0) {$\phi$};
  \node[insertion] at (2,0) {};
  \node[right=3pt] at (2,0) {$\phi$};
  \draw[scalar] (0,0) -- (2,0);
  \node[mass insertion] at (1,0) {\textcolor{red}{$\times$}};
  \node[above=3pt] at (1,0) {$-\frac{m^2}{2}$};

  \node[] at (3,0) {$+$};

  \node[insertion] at (4,0) {};
  \node[left=3pt] at (4,0) {$\phi$};
  \node[insertion] at (6,0) {};
  \node[right=3pt] at (6,0) {$\phi$};
  \draw[scalar] (4,0) -- (6,0);
  \node[mass insertion] at (4.66,0) {\textcolor{red}{$\times$}};
  \node[above=3pt] at (4.66,0) {$-\frac{m^2}{2}\;\;$};
  \node[mass insertion] at (5.33,0) {\textcolor{red}{$\times$}};
  \node[above=3pt] at (5.33,0) {$\;\;-\frac{m^2}{2}$};

  \node[] at (7,0) {$+$};
  \node[right=13pt] at (7,0) {$\cdots$};
\end{tikzpicture}

%% file: Figures/phi2phiphiCPT.tikz
\tikzset{
  defect/.style={line width=0.8pt},
  scalar/.style={dashed, line width=0.8pt},
  insertion/.style={circle, fill=black, inner sep=1.3pt},
  mass insertion/.style={font=\large, inner sep=0pt}
}

\def\sep{0.37}
\pgfmathsetmacro{\step}{1.75+\sep}

\begin{tikzpicture}[
  baseline={(current bounding box.center)}
]

\begin{scope}[xshift={-0.2 cm}]

\node[insertion] (a) at (0,0) {};
\node[insertion] (b) at (0.7,1) {};
\node[insertion] (c) at (1.4,0) {};

\node[left=3pt]        at (a) {$\phi$};
\node[above right=2pt] at (b) {$\phi^2$};
\node[right=3pt]       at (c) {$\phi$};

\draw[scalar] (a) -- (b) -- (c);

\end{scope}

\begin{scope}[xshift={1*\step cm}]

\node[insertion] (a) at (0,0) {};
\node[insertion] (b) at (0.7,1) {};
\node[insertion] (c) at (1.4,0) {};

\draw[scalar]
  (a) -- (b)
  -- node[mass insertion, midway, rotate=45]
       {\textcolor{red}{$\times$}}
  (c);

\end{scope}

\begin{scope}[xshift={2*\step cm}]

\node[insertion] (a) at (0,0) {};
\node[insertion] (b) at (0.7,1) {};
\node[insertion] (c) at (1.4,0) {};

\draw[scalar] (a) -- (c);
\draw[scalar] (0.7,1.4) circle (0.4);

\node[mass insertion] at (0.7,1.8)
  {\textcolor{red}{$\times$}};

\end{scope}

\begin{scope}[xshift={3*\step cm}]

\node[insertion] (a) at (0,0) {};
\node[insertion] (b) at (0.7,1) {};
\node[insertion] (c) at (1.4,0) {};

\draw[scalar]
  (a) -- (b)
  -- node[mass insertion, pos=1/3, rotate=45]
       {\textcolor{red}{$\times$}}
     node[mass insertion, pos=2/3, rotate=45]
       {\textcolor{red}{$\times$}}
  (c);

\end{scope}

\begin{scope}[xshift={4*\step cm}]

\node[insertion] (a) at (0,0) {};
\node[insertion] (b) at (0.7,1) {};
\node[insertion] (c) at (1.4,0) {};

\draw[scalar]
  (a)
  -- node[mass insertion, midway, rotate=-45]
       {\textcolor{red}{$\times$}}
  (b)
  -- node[mass insertion, midway, rotate=45]
       {\textcolor{red}{$\times$}}
  (c);

\end{scope}

\begin{scope}[xshift={5*\step cm}]

\node[insertion] (a) at (0,0) {};
\node[insertion] (b) at (0.7,1) {};
\node[insertion] (c) at (1.4,0) {};

\draw[scalar] (a) -- (c);
\draw[scalar] (0.7,1.4) circle (0.4);

\node[mass insertion] at (1.1,1.4)
  {\textcolor{red}{$\times$}};
\node[mass insertion] at (0.3,1.4)
  {\textcolor{red}{$\times$}};

\end{scope}

\begin{scope}[xshift={6*\step cm}]

\node[insertion] (a) at (0,0) {};
\node[insertion] (b) at (0.7,1) {};
\node[insertion] (c) at (1.4,0) {};

\draw[scalar]
  (a)
  -- node[mass insertion, midway]
       {\textcolor{red}{$\times$}}
  (c);

\draw[scalar] (0.7,1.4) circle (0.4);

\node[mass insertion] at (0.7,1.8)
  {\textcolor{red}{$\times$}};

\end{scope}

\end{tikzpicture}

%% file: Figures/phi4phiphiCPT.tikz
\tikzset{
  defect/.style={line width=0.8pt},
  scalar/.style={dashed, line width=0.8pt},
  insertion/.style={circle, fill=black, inner sep=1.3pt},
  mass insertion/.style={font=\large, inner sep=0pt}
}

\def\sep{0.45}
\def\diagramwidth{1.9}
\pgfmathsetmacro{\step}{\diagramwidth+\sep}

\begin{tikzpicture}[
  baseline={(current bounding box.center)}
]

\begin{scope}[xshift={-0.1 cm}]

\node[insertion] (a) at (0,0) {};
\node[insertion] (b) at (0.7,1) {};
\node[insertion] (c) at (1.4,0) {};

\node[left=3pt] at (a) {$\phi$};
\node[right=4pt, yshift=-1pt] at (b) {$\phi^4$};
\node[right=3pt] at (c) {$\phi$};

\draw[scalar] (a) -- (b) -- (c);

\draw[scalar] (0.7,1.4) circle (0.4);

\node[mass insertion] at (0.7,1.8)
  {\textcolor{red}{$\times$}};

\end{scope}

\begin{scope}[xshift={1*\step cm}]

\node[insertion] (a) at (0,0) {};
\node[insertion] (b) at (0.7,1) {};
\node[insertion] (c) at (1.4,0) {};

\draw[scalar] (a) -- (b) -- (c);

\draw[scalar] (0.7,1.4) circle (0.4);

\node[mass insertion] at (1.1,1.4)
  {\textcolor{red}{$\times$}};
  
\node[mass insertion] at (0.3,1.4)
  {\textcolor{red}{$\times$}};

\end{scope}

\begin{scope}[xshift={2*\step cm}]

\node[insertion] (a) at (0,0) {};
\node[insertion] (b) at (0.7,1) {};
\node[insertion] (c) at (1.4,0) {};

\draw[scalar]
  (a) -- (b)
  -- node[mass insertion, midway, rotate=45]
       {\textcolor{red}{$\times$}}
  (c);

\draw[scalar] (0.7,1.4) circle (0.4);

\node[mass insertion] at (0.7,1.8)
  {\textcolor{red}{$\times$}};

\end{scope}

\begin{scope}[xshift={3*\step cm}]

\node[insertion] (a) at (0,0) {};
\node[insertion] (b) at (0.7,1) {};
\node[insertion] (c) at (1.4,0) {};

\draw[scalar] (a) -- (c);

\draw[scalar] (0.3,1) circle (0.4);

\draw[scalar] (1.1,1) circle (0.4);

\node[mass insertion] at (-0.1,1)
  {\textcolor{red}{$\times$}};
  
\node[mass insertion] at (1.5,1)
  {\textcolor{red}{$\times$}};

\end{scope}

\end{tikzpicture}

%% file: Figures/Defect_Prescription.tikz
\def\bracketVerticalScale{0.5}

\def\bracketDepth{0.18}

\def\defectHeight{2.25}

\def\lhsOperatorDistance{1.5}
\def\rhsOperatorDistance{2.5}

\def\lhsRightDefectX{1.2}
\def\rhsDefectX{7.4}
\def\lhsLeftBracketTipX{-0.75}
\def\rhsLeftBracketTipX{6.65}

\pgfmathsetmacro{\diagramMid}{0.5*\defectHeight}
\pgfmathsetmacro{\arrowHeight}{\diagramMid}

\pgfmathsetmacro{\lhsOperatorX}{\lhsRightDefectX+\lhsOperatorDistance}
\pgfmathsetmacro{\lhsRightBracketTipX}{\lhsOperatorX+0.75}
\pgfmathsetmacro{\lhsRightBracketEndX}{\lhsRightBracketTipX-\bracketDepth}
\pgfmathsetmacro{\lhsLeftBracketEndX}{\lhsLeftBracketTipX+\bracketDepth}

\pgfmathsetmacro{\rhsOperatorX}{\rhsDefectX+\rhsOperatorDistance}
\pgfmathsetmacro{\rhsRightBracketTipX}{\rhsOperatorX+0.75}
\pgfmathsetmacro{\rhsRightBracketEndX}{\rhsRightBracketTipX-\bracketDepth}
\pgfmathsetmacro{\rhsLeftBracketEndX}{\rhsLeftBracketTipX+\bracketDepth}

\pgfmathsetmacro{\simeqX}{0.5*(\lhsRightBracketTipX+\rhsLeftBracketTipX)}

\tikzset{
  defect/.style={line width=0.8pt},
  operator insertion/.style={circle, fill=black, inner sep=1.2pt},
  separation arrow/.style={line width=0.6pt}
}

\begin{tikzpicture}[baseline={(current bounding box.center)}]

  \begin{scope}[
    shift={(0,\diagramMid)},
    yscale=\bracketVerticalScale
  ]
    \draw[defect]
      (\lhsLeftBracketEndX,\diagramMid)
      -- (\lhsLeftBracketTipX,0)
      -- (\lhsLeftBracketEndX,-\diagramMid);
    \draw[defect]
      (\lhsRightBracketEndX,\diagramMid)
      -- (\lhsRightBracketTipX,0)
      -- (\lhsRightBracketEndX,-\diagramMid);
  \end{scope}

  \draw[defect] (0,0) -- (0,\defectHeight);
  \draw[defect] (\lhsRightDefectX,0)
    -- (\lhsRightDefectX,\defectHeight);

  \draw[separation arrow, <-]
    (0.05,\arrowHeight) -- (0.44,\arrowHeight);
  \node at (0.60,\arrowHeight) {$r$};
  \draw[separation arrow, ->]
    (0.76,\arrowHeight) -- (1.15,\arrowHeight);

  \node[below=3pt] at (0,0) {$\mathcal D_a$};
  \node[below=3pt] at (\lhsRightDefectX,0) {$\mathcal D_b$};

  \node[operator insertion] at (\lhsOperatorX,\diagramMid) {};
  \node[above=5pt] at (\lhsOperatorX,\diagramMid) {$\mathcal O(x)$};

  \node at (\simeqX,\diagramMid) {$\simeq$};

  \begin{scope}[
    shift={(0,\diagramMid)},
    yscale=\bracketVerticalScale
  ]
    \draw[defect]
      (\rhsLeftBracketEndX,\diagramMid)
      -- (\rhsLeftBracketTipX,0)
      -- (\rhsLeftBracketEndX,-\diagramMid);
    \draw[defect]
      (\rhsRightBracketEndX,\diagramMid)
      -- (\rhsRightBracketTipX,0)
      -- (\rhsRightBracketEndX,-\diagramMid);
  \end{scope}

  \draw[defect] (\rhsDefectX,0) -- (\rhsDefectX,\defectHeight);
  \node[below=3pt] at (\rhsDefectX,0) {$\mathcal D_c^\text{CFT}$};

  \node[right=7pt] at (\rhsDefectX,\diagramMid)
    {$e^{-S_{\mathrm{eff}}}$};

  \node[operator insertion] at (\rhsOperatorX,\diagramMid) {};
  \node[above=5pt] at (\rhsOperatorX,\diagramMid) {$\mathcal O(x)$};
\end{tikzpicture}

%% file: Figures/segment.tikz
\begin{tikzpicture}[very thick, line cap=round]

    \draw[]
        (-50:3 and 1) arc[start angle=-50, end angle=230,
                         x radius=3, y radius=1];
    \draw[cyan, dashed]
        (-50:3 and 1) arc[start angle=-50, end angle=-130,
                         x radius=3, y radius=1];

    \fill (-130:3 and 1) circle (2pt)
        node[below] {$\mathcal{O}_j^{21}(0)$};
    \fill (-50:3 and 1) circle (2pt)
        node[below] {$\mathcal{O}_i^{12}(s)$};

    \node at (0,1.35) {$\mathcal D_2$};
    \node at (0,-0.7) {$\mathcal D_1$};
\end{tikzpicture}

%% file: Figures/translation.tikz
\begin{tikzpicture}[
    line cap=round,
    line join=round
]

\def\xlength{4.0}
\def\ylength{4.0}
\def\r{2.0}
\def\R{1.8}

\def\depthX{0.4}
\def\depthY{0.25}

\def\platecolor{gray!20}

\begin{scope}[
    x={(\depthX cm,\depthY cm)},
    y={(0cm,1cm)},
    z={(1cm,0cm)}
]

\def\leftplate{
    ({-\xlength/2},{-\ylength/2},{-\r/2}) --
    ({ \xlength/2},{-\ylength/2},{-\r/2}) --
    ({ \xlength/2},{ \ylength/2},{-\r/2}) --
    ({-\xlength/2},{ \ylength/2},{-\r/2}) -- cycle
}

\def\rightplate{
    ({-\xlength/2},{-\ylength/2},{\r/2}) --
    ({ \xlength/2},{-\ylength/2},{\r/2}) --
    ({ \xlength/2},{ \ylength/2},{\r/2}) --
    ({-\xlength/2},{ \ylength/2},{\r/2}) -- cycle
}

\def\circlepath{
    plot[domain=0:360, samples=181, smooth]
      ({\R*cos(\x)},{\R*sin(\x)},0) -- cycle
}

\fill[\platecolor] \leftplate;

\draw[thick] \circlepath;

\fill[\platecolor] \rightplate;

\begin{scope}
    \clip \rightplate;
    \draw[thick, dashed] \circlepath;
\end{scope}

\draw[] \leftplate;
\draw[] \rightplate;

\foreach \angle/\label in {130/1, 165/2, 200/3}{
    \fill ({\R*cos(\angle)},{\R*sin(\angle)},0)
        circle (2pt)
        node[left=2pt] {$\mathcal{O}_{\label}$};
}

\end{scope}

\node at (-2.2,-2.2) {$\Sigma_L$};
\node at ( 1.4,-2.2) {$\Sigma_R$};

\end{tikzpicture}

%% file: Figures/bulk-to-defect.tikz
\begin{tikzpicture}[baseline=(current bounding box.center)]

\draw[thick] (0,-2) -- (0,2);
\node[right=2pt] at (0,-1.85) {$\mathcal D$};

\fill (-1.6,0) circle (2pt);
\node[below left] at (-1.5,0) {$\mathcal O_i$};

\draw[<->] (-1.5,0.3) -- (-0.1,0.3)
    node[midway,above] {$|x_\perp|$};

\node at (2,0) {$\displaystyle
    \simeq \hspace{5pt} \sum_{j}
    \frac{b_{ij}(m|x_\perp|)}
    {|x_\perp|^{\Delta_i-\Delta_j}}
$};

\draw[thick] (4,-2) -- (4,2);
\fill (4,0) circle (2pt);
\node[below right] at (4,0) {$\mathcal O_j$};
\node[right=2pt] at (4,-1.85) {$\mathcal D$};

\end{tikzpicture}

%% file: Figures/DD_MasslessScalar.tikz
\tikzset{
  defect/.style={line width=0.8pt},
  scalar/.style={dashed, line width=0.8pt},
  insertion/.style={circle, fill=black, inner sep=1.3pt}
}

\begin{tikzpicture}[baseline={(current bounding box.center)}]
  \draw[defect] (0,0) -- (0,3.5);
  \draw[defect] (2,0) -- (2,3.5);

  \node[insertion] at (0,1.75) {};
  \node[insertion] at (2,1.75) {};
  \draw[scalar] (0,1.75) -- (2,1.75);

  \node[left=3pt] at (0,1.75) {$\phi$};
  \node[right=3pt] at (2,1.75) {$\phi$};
  \node[below=3pt] at (0,0) {$\mathcal D_{h_2}$};
  \node[below=3pt] at (2,0) {$\mathcal D_{h_1}$};
\end{tikzpicture}
\hspace{2cm}
\begin{tikzpicture}[baseline={(current bounding box.center)}]
  \draw[defect] (0,0) -- (0,3.5);
  \draw[defect] (2,0) -- (2,3.5);

  \node[insertion] at (0,1.1) {};
  \node[insertion] at (0,2.4) {};
  \draw[scalar]
    (0,1.1) arc[start angle=-90,end angle=90,radius=0.65];

  \node[left=3pt] at (0,1.1) {$\phi$};
  \node[left=3pt] at (0,2.4) {$\phi$};
  \node[below=3pt] at (0,0) {$\mathcal D_{h_2}$};
  \node[below=3pt] at (2,0) {$\mathcal D_{h_1}$};
\end{tikzpicture}
\hspace{2cm}
\begin{tikzpicture}[baseline={(current bounding box.center)}]
  \draw[defect] (0,0) -- (0,3.5);
  \draw[defect] (2,0) -- (2,3.5);

  \node[insertion] at (2,1.1) {};
  \node[insertion] at (2,2.4) {};
  \draw[scalar]
    (2,1.1) arc[start angle=-90,end angle=-270,radius=0.65];

  \node[right=3pt] at (2,1.1) {$\phi$};
  \node[right=3pt] at (2,2.4) {$\phi$};
  \node[below=3pt] at (0,0) {$\mathcal D_{h_2}$};
  \node[below=3pt] at (2,0) {$\mathcal D_{h_1}$};
\end{tikzpicture}

%% file: Figures/DDphi2_MasslessScalar.tikz
\def\horizontalScale{1.3}

\tikzset{
  defect/.style={line width=0.8pt},
  propagator/.style={dashed, line width=0.8pt},
  insertion/.style={circle, fill=black, inner sep=1.2pt}
}

\begin{tikzpicture}[
  xscale=\horizontalScale,
  baseline={(current bounding box.center)}
]
  \begin{scope}[xshift=0cm]
    \draw[propagator] (0,1.20) -- (2.35,1.75);
    \draw[propagator] (1.0,2.30) -- (2.35,1.75);

    \draw[defect] (0,0) -- (0,3.5);
    \draw[defect] (1.0,0) -- (1.0,3.5);

    \node[insertion] at (0,1.20) {};
    \node[insertion] at (1.0,2.30) {};

    \node[insertion] at (2.35,1.75) {};
    \node[above=4pt] at (2.35,1.75) {$\phi^2(x)$};

    \node[below=3pt] at (0,0) {$\mathcal D_{h_2}$};
    \node[below=3pt] at (1.0,0) {$\mathcal D_{h_1}$};
  \end{scope}

  \begin{scope}[xshift=3.95cm]
    \draw[propagator] (1.0,1.20) -- (2.35,1.75);
    \draw[propagator] (1.0,2.30) -- (2.35,1.75);

    \draw[defect] (0,0) -- (0,3.5);
    \draw[defect] (1.0,0) -- (1.0,3.5);

    \node[insertion] at (1.0,1.20) {};
    \node[insertion] at (1.0,2.30) {};

    \node[insertion] at (2.35,1.75) {};
    \node[above=4pt] at (2.35,1.75) {$\phi^2(x)$};

    \node[below=3pt] at (0,0) {$\mathcal D_{h_2}$};
    \node[below=3pt] at (1.0,0) {$\mathcal D_{h_1}$};
  \end{scope}

  \begin{scope}[xshift=7.90cm]
    \draw[propagator] (0,1.20) -- (2.35,1.75);
    \draw[propagator] (0,2.30) -- (2.35,1.75);

    \draw[defect] (0,0) -- (0,3.5);
    \draw[defect] (1.0,0) -- (1.0,3.5);

    \node[insertion] at (0,1.20) {};
    \node[insertion] at (0,2.30) {};

    \node[insertion] at (2.35,1.75) {};
    \node[above=4pt] at (2.35,1.75) {$\phi^2(x)$};

    \node[below=3pt] at (0,0) {$\mathcal D_{h_2}$};
    \node[below=3pt] at (1.0,0) {$\mathcal D_{h_1}$};
  \end{scope}
\end{tikzpicture}

%% file: Figures/Fused_DDphi2_MasslessScalar.tikz
\def\horizontalScale{1.3}

\tikzset{
  defect/.style={line width=0.8pt},
  propagator/.style={dashed, line width=0.8pt},
  insertion/.style={circle, fill=black, inner sep=1.2pt}
}

\begin{tikzpicture}[
  xscale=\horizontalScale,
  baseline={(current bounding box.center)}
]
  \begin{scope}[xshift=0cm]
    \draw[propagator] (0,1.20) -- (1.35,1.75);
    \draw[propagator] (0,2.30) -- (1.35,1.75);

    \draw[defect] (0,0) -- (0,3.5);

    \node[insertion] at (0,1.20) {};
    \node[insertion] at (0,2.30) {};
    \node[left=3pt] at (0,1.20) {$\phi$};
    \node[left=3pt] at (0,2.30) {$\phi$};

    \node[insertion] at (1.35,1.75) {};
    \node[above=4pt] at (1.35,1.75) {$\phi^2(x)$};

    \node[below=3pt] at (0,0) {$\mathcal D_{h_1+h_2}$};
  \end{scope}

  \begin{scope}[xshift=3.95cm]
    \draw[propagator] (0,1.20) -- (1.35,1.75);
    \draw[propagator] (0,2.30) -- (1.35,1.75);

    \draw[defect] (0,0) -- (0,3.5);

    \node[insertion] at (0,1.20) {};
    \node[insertion] at (0,2.30) {};
    \node[left=3pt] at (0,1.20) {$\phi$};
    \node[left=3pt] at (0,2.30) {$\partial_i\phi$};

    \node[insertion] at (1.35,1.75) {};
    \node[above=4pt] at (1.35,1.75) {$\phi^2(x)$};

    \node[below=3pt] at (0,0) {$\mathcal D_{h_1+h_2}$};
  \end{scope}

  \begin{scope}[xshift=7.90cm]
    \draw[propagator]
      (0,1.75) .. controls (0.35,0.95) and (1.00,0.95) .. (1.35,1.75);
    \draw[propagator]
      (0,1.75) .. controls (0.35,2.55) and (1.00,2.55) .. (1.35,1.75);

    \draw[defect] (0,0) -- (0,3.5);

    \node[insertion] at (0,1.75) {};
    \node[left=3pt] at (0,1.75) {$\phi^2$};

    \node[insertion] at (1.35,1.75) {};
    \node[right=4pt] at (1.35,1.75) {$\phi^2(x)$};

    \node[below=3pt] at (0,0) {$\mathcal D_{h_1+h_2}$};
  \end{scope}
\end{tikzpicture}

%% file: Figures/DD_circ_MasslessScalar.tikz
\tikzset{
  defect loop/.style={draw=black, line width=0.8pt},
  propagator/.style={draw=black, dashed, line width=0.8pt},
  angular ray/.style={draw=red, dashed, line width=0.8pt},
  insertion/.style={circle, fill=black, inner sep=1.2pt}
}

\begin{tikzpicture}[baseline={(current bounding box.center)}]
  \coordinate (C1) at (0,0);
  \coordinate (C2) at (4.60,0);

  \draw[defect loop] (C1)
    ellipse[x radius=0.80cm,y radius=2.20cm];
  \draw[defect loop] (C2)
    ellipse[x radius=0.80cm,y radius=2.20cm];

  \coordinate (P1) at (0.655,1.262);
  \coordinate (P2) at (3.945,-1.262);

  \draw[propagator] (P1) -- (P2);
  \node[insertion] at (P1) {};
  \node[insertion] at (P2) {};
  \node[above right=2pt] at (P1) {$\phi$};
  \node[below left=2pt] at (P2) {$\phi$};

  \coordinate (X1) at (0.80,0);
  \draw[angular ray] (C1) -- (X1);
  \draw[angular ray] (C1) -- (P1);
  \node[red] at (0.48,0.34) {$\theta_1$};

  \coordinate (X2) at (5.40,0);
  \draw[angular ray] (C2) -- (X2);
  \draw[angular ray] (C2) -- (P2);
  \node[red] at (4.26,0.16) {$\theta_2$};

  \node[below=5pt] at (0,-2.20) {$\mathcal D_{h_2}$};
  \node[below=5pt] at (4.60,-2.20) {$\mathcal D_{h_1}$};
\end{tikzpicture}

%% file: Figures/DD_MassiveScalar_CPT_1.tikz
\tikzset{
  defect/.style={line width=0.8pt},
  scalar/.style={dashed, line width=0.8pt},
  insertion/.style={circle, fill=black, inner sep=1.3pt},
  mass insertion/.style={font=\large, inner sep=0pt}
}

\begin{tikzpicture}[baseline={(current bounding box.center)}]
  \draw[defect] (0,0) -- (0,3.5);
  \draw[defect] (2,0) -- (2,3.5);

  \node[insertion] at (0,1.75) {};
  \node[insertion] at (2,1.75) {};
  \draw[scalar] (0,1.75) -- (2,1.75);

  \node[mass insertion] at (1,1.75) {\textcolor{red}{$\times$}};
  \node[above=3pt] at (1,1.75) {$\frac{-m^2}{2}$};

  \node[left=3pt] at (0,1.75) {$\phi$};
  \node[right=3pt] at (2,1.75) {$\phi$};
  \node[below=3pt] at (0,0) {$\mathcal D_{h_2}$};
  \node[below=3pt] at (2,0) {$\mathcal D_{h_1}$};
\end{tikzpicture}
\hspace{2cm}
\begin{tikzpicture}[baseline={(current bounding box.center)}]
  \draw[defect] (0,0) -- (0,3.5);
  \draw[defect] (2,0) -- (2,3.5);

  \node[insertion] at (0,1.1) {};
  \node[insertion] at (0,2.4) {};
  \draw[scalar]
    (0,1.1) arc[start angle=-90,end angle=90,radius=0.65];

  \node[mass insertion] at (0.65,1.75) {\textcolor{red}{$\times$}};
  \node[right=3pt] at (0.65,1.75) {$\frac{-m^2}{2}$};

  \node[left=3pt] at (0,1.1) {$\phi$};
  \node[left=3pt] at (0,2.4) {$\phi$};
  \node[below=3pt] at (0,0) {$\mathcal D_{h_2}$};
  \node[below=3pt] at (2,0) {$\mathcal D_{h_1}$};
\end{tikzpicture}
\hspace{2cm}
\begin{tikzpicture}[baseline={(current bounding box.center)}]
  \draw[defect] (0,0) -- (0,3.5);
  \draw[defect] (2,0) -- (2,3.5);

  \node[insertion] at (2,1.1) {};
  \node[insertion] at (2,2.4) {};
  \draw[scalar]
    (2,1.1) arc[start angle=-90,end angle=-270,radius=0.65];

  \node[mass insertion] at (1.35,1.75) {\textcolor{red}{$\times$}};
  \node[left=3pt] at (1.35,1.75) {$\frac{-m^2}{2}$};

  \node[right=3pt] at (2,1.1) {$\phi$};
  \node[right=3pt] at (2,2.4) {$\phi$};
  \node[below=3pt] at (0,0) {$\mathcal D_{h_2}$};
  \node[below=3pt] at (2,0) {$\mathcal D_{h_1}$};
\end{tikzpicture}

%% file: Figures/Schwinger_FWW.tikz
\tikzset{
  defect/.style={line width=0.8pt},
  gauge/.style={
    line width=0.8pt,
    decorate,
    decoration={snake,amplitude=1.0pt,segment length=5.5pt}
  },
  red cross/.style={text=red,inner sep=0pt,font=\normalsize},
  operator insertion/.style={circle,fill=black,inner sep=1.15pt},
  fermion/.style={line width=0.8pt},
  centered arrow/.style={
    postaction={
      decorate,
      decoration={
        markings,
        mark=at position 0.55 with {\arrow{Stealth[length=4.5pt,width=3.5pt]}}
      }
    }
  }
}

\begin{tikzpicture}[baseline={(current bounding box.center)}]

  \begin{scope}[xshift=0cm]
    \draw[defect] (0,0) -- (0,3.5);
    \draw[defect] (1.0,0) -- (1.0,3.5);
    \node[below=3pt] at (0,0) {$W_{-q}$};
    \node[below=3pt] at (1.0,0) {$W_q$};

    \node[red cross] at (1.0,1.75) {$\times$};
    \draw[gauge] (1.0,1.75) -- (2.15,1.75);

    \draw[fermion,centered arrow]
      (3.25,1.75) arc[start angle=0,end angle=180,radius=0.55];
    \draw[fermion,centered arrow]
      (2.15,1.75) arc[start angle=180,end angle=360,radius=0.55];

    \node[red cross] at (2.15,1.75) {$\times$};
    \node[red cross] at (3.25,1.75) {$\times$};

    \draw[gauge] (3.25,1.75) -- (4.85,1.75);
    \node[operator insertion] at (4.85,1.75) {};
    \node[above=5pt] at (4.85,1.75) {$F_{\mu\nu}(x)$};
  \end{scope}

  \begin{scope}[xshift=7.20cm]
    \draw[defect] (0,0) -- (0,3.5);
    \draw[defect] (1.0,0) -- (1.0,3.5);
    \node[below=3pt] at (0,0) {$W_{-q}$};
    \node[below=3pt] at (1.0,0) {$W_q$};

    \node[red cross] at (0,1.75) {$\times$};

    \draw[gauge] (0,1.75) -- (1.85,1.75);

    \draw[fermion,centered arrow]
      (2.95,1.75) arc[start angle=0,end angle=180,radius=0.55];
    \draw[fermion,centered arrow]
      (1.85,1.75) arc[start angle=180,end angle=360,radius=0.55];

    \node[red cross] at (1.85,1.75) {$\times$};
    \node[red cross] at (2.95,1.75) {$\times$};

    \draw[gauge] (2.95,1.75) -- (4.55,1.75);
    \node[operator insertion] at (4.55,1.75) {};
    \node[above=5pt] at (4.55,1.75) {$F_{\mu\nu}(x)$};
  \end{scope}

\end{tikzpicture}

%% file: Figures/Fused_DDphi2_CPT_1.tikz
\def\horizontalScale{1.5}

\def\attachmentOffset{0.70}

\def\panelShift{5.10}

\pgfmathsetmacro{\lowerAttachment}{1.75-\attachmentOffset}
\pgfmathsetmacro{\upperAttachment}{1.75+\attachmentOffset}

\pgfmathsetmacro{\massInsertionX}{0.675}
\pgfmathsetmacro{\massInsertionY}{0.5*(\upperAttachment+1.75)}

\tikzset{
  defect/.style={line width=0.8pt},
  propagator/.style={dashed, line width=0.8pt},
  insertion/.style={circle, fill=black, inner sep=1.2pt},
  mass insertion/.style={font=\Large, inner sep=0pt}
}

\begin{tikzpicture}[
  xscale=\horizontalScale,
  baseline={(current bounding box.center)}
]
  \begin{scope}[xshift=-1cm]
    \draw[propagator]
      (0,\lowerAttachment) -- (1.35,1.75);
    \draw[propagator]
      (0,\upperAttachment) -- (1.35,1.75);

    \draw[defect] (0,0) -- (0,3.5);

    \node[insertion] at (0,\lowerAttachment) {};
    \node[insertion] at (0,\upperAttachment) {};
    \node[left=3pt] at (0,\lowerAttachment) {$\partial_{i_1}\ldots \partial_{i_{n}}\phi$};
    \node[left=3pt] at (0,\upperAttachment) {$\phi$};

    \node[insertion] at (1.35,1.75) {};
    \node[right=4pt] at (1.35,1.75) {$\phi^2(x)$};

    \node[mass insertion]
      at (\massInsertionX,\massInsertionY) {\textcolor{red}{$\times$}};
    \node[above right=2pt]
      at (\massInsertionX,\massInsertionY) {$\frac{-m^2}{2}$};

    \node[below=3pt] at (0,0) {$\mathcal D_{h_1+h_2}$};
  \end{scope}

  \begin{scope}[xshift=\panelShift cm]
    \draw[propagator]
      (0,\lowerAttachment) -- (1.35,1.75);
    \draw[propagator]
      (0,\upperAttachment) -- (1.35,1.75);

    \draw[defect] (0,0) -- (0,3.5);

    \node[insertion] at (0,\lowerAttachment) {};
    \node[insertion] at (0,\upperAttachment) {};
    \node[left=3pt] at (0,\lowerAttachment) {$\partial_{i_1}\ldots \partial_{i_{n-1}}\phi$};
    \node[left=3pt] at (0,\upperAttachment) {$\partial_{i_j}\phi$};

    \node[insertion] at (1.35,1.75) {};
    \node[right=4pt] at (1.35,1.75) {$\phi^2(x)$};

    \node[mass insertion]
      at (\massInsertionX,\massInsertionY) {\textcolor{red}{$\times$}};
    \node[above right=2pt]
      at (\massInsertionX,\massInsertionY) {$\frac{-m^2}{2}$};

    \node[below=3pt] at (0,0) {$\mathcal D_{h_1+h_2}$};
  \end{scope}
\end{tikzpicture}

%% file: Figures/DD_MassiveScalar_CPT_2.tikz
\newlength{\diagramsep}
\setlength{\diagramsep}{1.5cm}

\tikzset{
  defect/.style={line width=0.8pt},
  scalar/.style={dashed, line width=0.8pt},
  insertion/.style={circle, fill=black, inner sep=1.15pt},
  mass insertion/.style={text=red, font=\large, inner sep=0pt}
}

\newcommand{\masscrossat}[1]{%
  \node[mass insertion] at (#1) {$\times$};%
}

\makebox[\textwidth][c]{%
  \begin{tikzpicture}[baseline={(current bounding box.center)},scale=0.72]
  \draw[defect] (0,0) -- (0,3.2);
  \draw[defect] (1.8,0) -- (1.8,3.2);
  \node[insertion] at (0,1.6) {};
  \node[insertion] at (1.8,1.6) {};
  \draw[scalar] (0,1.6) -- coordinate[pos=0.35] (d1a) coordinate[pos=0.65] (d1b) (1.8,1.6);
  \masscrossat{d1a}
  \masscrossat{d1b}
\end{tikzpicture}
  \hspace{\diagramsep}%
  \begin{tikzpicture}[baseline={(current bounding box.center)},scale=0.72]
  \draw[defect] (0,0) -- (0,3.2);
  \draw[defect] (1.8,0) -- (1.8,3.2);
  \node[insertion] at (0,1.05) {};
  \node[insertion] at (1.8,1.05) {};
  \node[insertion] at (0,2.15) {};
  \node[insertion] at (1.8,2.15) {};
  \draw[scalar] (0,1.05) -- coordinate[pos=0.5] (d2a) (1.8,1.05);
  \draw[scalar] (0,2.15) -- coordinate[pos=0.5] (d2b) (1.8,2.15);
  \masscrossat{d2a}
  \masscrossat{d2b}
\end{tikzpicture}
  \hspace{\diagramsep}%
  \begin{tikzpicture}[baseline={(current bounding box.center)},scale=0.72]
  \draw[defect] (0,0) -- (0,3.2);
  \draw[defect] (1.8,0) -- (1.8,3.2);
  \node[insertion] at (0,1.05) {};
  \node[insertion] at (1.8,1.05) {};
  \node[insertion] at (0,1.95) {};
  \node[insertion] at (0,2.75) {};
  \draw[scalar] (0,1.05) -- coordinate[pos=0.5] (d3a) (1.8,1.05);
  \draw[scalar] (0,1.95) arc[start angle=-90,end angle=90,radius=0.40] coordinate[pos=0.5] (d3b);
  \masscrossat{d3a}
  \masscrossat{d3b}
\end{tikzpicture}
  \hspace{\diagramsep}%
  \begin{tikzpicture}[baseline={(current bounding box.center)},scale=0.72]
  \draw[defect] (0,0) -- (0,3.2);
  \draw[defect] (1.8,0) -- (1.8,3.2);
  \node[insertion] at (0,1.05) {};
  \node[insertion] at (1.8,1.05) {};
  \node[insertion] at (1.8,1.95) {};
  \node[insertion] at (1.8,2.75) {};
  \draw[scalar] (0,1.05) -- coordinate[pos=0.5] (d4a) (1.8,1.05);
  \draw[scalar] (1.8,1.95) arc[start angle=-90,end angle=-270,radius=0.40] coordinate[pos=0.5] (d4b);
  \masscrossat{d4a}
  \masscrossat{d4b}
\end{tikzpicture}
}

\vspace{0.35cm}

\makebox[\textwidth][c]{%
  \begin{tikzpicture}[baseline={(current bounding box.center)},scale=0.72]
  \draw[defect] (0,0) -- (0,3.2);
  \draw[defect] (1.8,0) -- (1.8,3.2);
  \node[insertion] at (0,0.85) {};
  \node[insertion] at (0,2.35) {};
  \draw[scalar] (0,0.85) arc[start angle=-90,end angle=90,radius=0.75]
    coordinate[pos=0.35] (d5a) coordinate[pos=0.65] (d5b);
  \masscrossat{d5a}
  \masscrossat{d5b}
\end{tikzpicture}
  \hspace{\diagramsep}%
  \begin{tikzpicture}[baseline={(current bounding box.center)},scale=0.72]
  \draw[defect] (0,0) -- (0,3.2);
  \draw[defect] (1.8,0) -- (1.8,3.2);
  \node[insertion] at (1.8,0.85) {};
  \node[insertion] at (1.8,2.35) {};
  \draw[scalar] (1.8,0.85) arc[start angle=-90,end angle=-270,radius=0.75]
    coordinate[pos=0.35] (d6a) coordinate[pos=0.65] (d6b);
  \masscrossat{d6a}
  \masscrossat{d6b}
\end{tikzpicture}
  \hspace{\diagramsep}%
  \begin{tikzpicture}[baseline={(current bounding box.center)},scale=0.72]
  \draw[defect] (0,0) -- (0,3.2);
  \draw[defect] (1.8,0) -- (1.8,3.2);
  \node[insertion] at (0,0.45) {};
  \node[insertion] at (0,1.35) {};
  \node[insertion] at (0,1.85) {};
  \node[insertion] at (0,2.75) {};
  \draw[scalar] (0,0.45) arc[start angle=-90,end angle=90,radius=0.45] coordinate[pos=0.5] (d7a);
  \draw[scalar] (0,1.85) arc[start angle=-90,end angle=-270,radius=0.45] coordinate[pos=0.5] (d7b);
  \masscrossat{d7a}
  \masscrossat{d7b}
\end{tikzpicture}
  \hspace{\diagramsep}%
  \begin{tikzpicture}[baseline={(current bounding box.center)},scale=0.72]
  \draw[defect] (0,0) -- (0,3.2);
  \draw[defect] (1.8,0) -- (1.8,3.2);
  \node[insertion] at (1.8,0.45) {};
  \node[insertion] at (1.8,1.35) {};
  \node[insertion] at (1.8,1.85) {};
  \node[insertion] at (1.8,2.75) {};
  \draw[scalar] (1.8,0.45) arc[start angle=-90,end angle=90,radius=0.45] coordinate[pos=0.5] (d8a);
  \draw[scalar] (1.8,1.85) arc[start angle=-90,end angle=-270,radius=0.45] coordinate[pos=0.5] (d8b);
  \masscrossat{d8a}
  \masscrossat{d8b}
\end{tikzpicture}
  \hspace{\diagramsep}%
  \begin{tikzpicture}[baseline={(current bounding box.center)},scale=0.72]
  \draw[defect] (0,0) -- (0,3.2);
  \draw[defect] (1.8,0) -- (1.8,3.2);
  \node[insertion] at (0,0.85) {};
  \node[insertion] at (0,2.35) {};
  \node[insertion] at (1.8,0.85) {};
  \node[insertion] at (1.8,2.35) {};
  \draw[scalar] (0,0.85) arc[start angle=-90,end angle=90,radius=0.75] coordinate[pos=0.5] (d9a);
  \draw[scalar] (1.8,0.85) arc[start angle=-90,end angle=-270,radius=0.75] coordinate[pos=0.5] (d9b);
  \masscrossat{d9a}
  \masscrossat{d9b}
\end{tikzpicture}
}

%% file: Figures/Schwinger_F2F2.tikz
\tikzset{
  photon/.style={line width=0.8pt,decorate,decoration={snake,amplitude=0.9pt,segment length=5.2pt}},
  operator insertion/.style={circle,fill=black,inner sep=1.15pt},
  interaction vertex/.style={text=red,inner sep=0pt,font=\normalsize},
  fermion/.style={line width=0.8pt},
  centered arrow/.style={postaction={decorate,decoration={markings,mark=at position 0.58 with {\arrow{Stealth[length=4.5pt,width=3.5pt]}}}}},
  early arrow/.style={postaction={decorate,decoration={markings,mark=at position 0.30 with {\arrow{Stealth[length=4.5pt,width=3.5pt]}}}}},
  late arrow/.style={postaction={decorate,decoration={markings,mark=at position 0.70 with {\arrow{Stealth[length=4.5pt,width=3.5pt]}}}}}
}

\begin{tikzpicture}[baseline={(current bounding box.center)}]

  \begin{scope}[xshift=0.00cm,scale=0.82,transform shape]
    \coordinate (T) at (0,2.80); \coordinate (B) at (0,0);
    \def\r{0.32}
    \coordinate (LU) at (-0.72,1.72); \coordinate (LL) at (-0.72,1.08);
    \coordinate (RU) at (0.72,1.72);  \coordinate (RL) at (0.72,1.08);
    \draw[photon] (T) .. controls (-0.58,2.60) and (-0.78,2.08) .. (LU);
    \draw[fermion,centered arrow] (LU) arc[start angle=90,end angle=-90,radius=\r];
    \draw[fermion,centered arrow] (LL) arc[start angle=-90,end angle=-270,radius=\r];
    \draw[photon] (LL) .. controls (-0.78,0.72) and (-0.58,0.20) .. (B);
    \draw[photon] (T) .. controls (0.58,2.60) and (0.78,2.08) .. (RU);
    \draw[fermion,centered arrow] (RU) arc[start angle=90,end angle=-90,radius=\r];
    \draw[fermion,centered arrow] (RL) arc[start angle=-90,end angle=-270,radius=\r];
    \draw[photon] (RL) .. controls (0.78,0.72) and (0.58,0.20) .. (B);
    \foreach \p in {LU,LL,RU,RL}{\node[interaction vertex] at (\p) {$\times$};}
    \node[operator insertion] at (T) {}; \node[operator insertion] at (B) {};
  \end{scope}

  \begin{scope}[xshift=3.15cm,scale=0.82,transform shape]
    \coordinate (top insertion) at (0,2.80);
    \coordinate (bottom insertion) at (0,0);
    \def\fermionRadius{0.28}

    \coordinate (upper left vertex) at (-0.28,1.72);
    \coordinate (upper right vertex) at (0.28,1.72);
    \draw[photon]
      (top insertion)
      .. controls (-0.88,2.80) and (-0.88,1.72) ..
      (upper left vertex);
    \draw[photon]
      (top insertion)
      .. controls (0.88,2.80) and (0.88,1.72) ..
      (upper right vertex);

    \draw[fermion,centered arrow]
      (upper left vertex)
      arc[start angle=180,end angle=0,radius=\fermionRadius];
    \draw[fermion,centered arrow]
      (upper right vertex)
      arc[start angle=0,end angle=-180,radius=\fermionRadius];

    \coordinate (lower left vertex) at (-0.28,1.08);
    \coordinate (lower right vertex) at (0.28,1.08);
    \draw[photon]
      (bottom insertion)
      .. controls (0.88,0.00) and (0.88,1.08) ..
      (lower right vertex);
    \draw[photon]
      (bottom insertion)
      .. controls (-0.88,0.00) and (-0.88,1.08) ..
      (lower left vertex);

    \draw[fermion,centered arrow]
      (lower left vertex)
      arc[start angle=180,end angle=360,radius=\fermionRadius];
    \draw[fermion,centered arrow]
      (lower right vertex)
      arc[start angle=0,end angle=180,radius=\fermionRadius];

    \node[interaction vertex] at (upper left vertex) {$\times$};
    \node[interaction vertex] at (upper right vertex) {$\times$};
    \node[interaction vertex] at (lower left vertex) {$\times$};
    \node[interaction vertex] at (lower right vertex) {$\times$};
    \node[operator insertion] at (top insertion) {};
    \node[operator insertion] at (bottom insertion) {};
  \end{scope}

  \begin{scope}[xshift=6.30cm,scale=0.82,transform shape]
    \coordinate (T) at (0,2.80); \coordinate (B) at (0,0);
    \coordinate (UL) at (-0.62,1.98); \coordinate (UR) at (0.62,1.98);
    \coordinate (LL) at (-0.62,0.82); \coordinate (LR) at (0.62,0.82);
    \draw[photon] (T)..controls(-.4,2.7)and(-.57,2.3)..(UL); \draw[photon] (T)..controls(.4,2.7)and(.57,2.3)..(UR);
    \draw[photon] (LL)..controls(-.57,.48)and(-.38,.1)..(B); \draw[photon] (LR)..controls(.57,.48)and(.38,.1)..(B);
    \draw[fermion,centered arrow](UL)--(LL); \draw[fermion,centered arrow](LL)--(LR); \draw[fermion,centered arrow](LR)--(UR); \draw[fermion,centered arrow](UR)--(UL);
    \foreach \p in {UL,UR,LL,LR}{\node[interaction vertex] at (\p) {$\times$};} \node[operator insertion]at(T){};\node[operator insertion]at(B){};
  \end{scope}

  \begin{scope}[xshift=9.45cm,scale=0.82,transform shape]
    \coordinate (T) at (0,2.80); \coordinate (B) at (0,0);
    \coordinate (UL) at (-0.62,1.98); \coordinate (UR) at (0.62,1.98); \coordinate (LL) at (-0.62,0.82); \coordinate (LR) at (0.62,0.82);
    \draw[photon](T)..controls(-.4,2.7)and(-.57,2.3)..(UL);\draw[photon](T)..controls(.4,2.7)and(.57,2.3)..(UR);\draw[photon](LL)..controls(-.57,.48)and(-.38,.1)..(B);\draw[photon](LR)..controls(.57,.48)and(.38,.1)..(B);
    \draw[fermion,centered arrow](UL)--(UR);\draw[fermion,early arrow](UR)--(LL);\draw[fermion,centered arrow](LL)--(LR);\draw[fermion,late arrow](LR)--(UL);
    \foreach \p in {UL,UR,LL,LR}{\node[interaction vertex] at (\p) {$\times$};}\node[operator insertion]at(T){};\node[operator insertion]at(B){};
  \end{scope}

  \begin{scope}[xshift=12.60cm,scale=0.82,transform shape]
    \coordinate (T) at (0,2.80); \coordinate (B) at (0,0);
    \coordinate (UL) at (-0.62,1.98); \coordinate (UR) at (0.62,1.98); \coordinate (LL) at (-0.62,0.82); \coordinate (LR) at (0.62,0.82);
    \draw[photon](T)..controls(-.4,2.7)and(-.57,2.3)..(UL);\draw[photon](T)..controls(.4,2.7)and(.57,2.3)..(UR);\draw[photon](LL)..controls(-.57,.48)and(-.38,.1)..(B);\draw[photon](LR)..controls(.57,.48)and(.38,.1)..(B);
    \draw[fermion,centered arrow](UL)--(LL);\draw[fermion,early arrow](LL)--(UR);\draw[fermion,centered arrow](UR)--(LR);\draw[fermion,late arrow](LR)--(UL);
    \foreach \p in {UL,UR,LL,LR}{\node[interaction vertex] at (\p) {$\times$};}\node[operator insertion]at(T){};\node[operator insertion]at(B){};
  \end{scope}
\end{tikzpicture}

%% file: Figures/Schwinger_F2FF.tikz
\tikzset{
  photon/.style={
    line width=0.8pt,
    decorate,
    decoration={snake,amplitude=0.9pt,segment length=5.2pt}
  },
  operator insertion/.style={circle,fill=black,inner sep=1.15pt},
  interaction vertex/.style={text=red,inner sep=0pt,font=\normalsize},
  fermion/.style={line width=0.8pt},
  centered arrow/.style={
    postaction={
      decorate,
      decoration={
        markings,
        mark=at position 0.58 with {
          \arrow{Stealth[length=4.5pt,width=3.5pt]}
        }
      }
    }
  }
}

\begin{tikzpicture}[baseline={(current bounding box.center)}]

  \begin{scope}[xshift=2.35cm,yshift=3.20cm,scale=0.86,transform shape]
    \coordinate (top) at (0,2.20);
    \coordinate (left) at (-1.35,0);
    \coordinate (right) at (1.35,0);

    \def\bubbleRadius{0.36}
    \coordinate (left upper vertex) at (-0.487,1.407);
    \coordinate (left lower vertex) at (-0.863,0.793);
    \coordinate (right upper vertex) at (0.487,1.407);
    \coordinate (right lower vertex) at (0.863,0.793);

    \draw[photon] (top) -- (left upper vertex);
    \draw[photon] (left lower vertex) -- (left);
    \draw[photon] (top) -- (right upper vertex);
    \draw[photon] (right lower vertex) -- (right);

    \draw[fermion,centered arrow]
      (left upper vertex)
      arc[start angle=58.47,end angle=238.47,radius=\bubbleRadius];
    \draw[fermion,centered arrow]
      (left lower vertex)
      arc[start angle=238.47,end angle=418.47,radius=\bubbleRadius];
    \draw[fermion,centered arrow]
      (right upper vertex)
      arc[start angle=121.53,end angle=301.53,radius=\bubbleRadius];
    \draw[fermion,centered arrow]
      (right lower vertex)
      arc[start angle=301.53,end angle=481.53,radius=\bubbleRadius];

    \node[interaction vertex] at (left upper vertex) {$\times$};
    \node[interaction vertex] at (left lower vertex) {$\times$};
    \node[interaction vertex] at (right upper vertex) {$\times$};
    \node[interaction vertex] at (right lower vertex) {$\times$};

    \node[operator insertion] at (top) {};
    \node[operator insertion] at (left) {};
    \node[operator insertion] at (right) {};
  \end{scope}

  \begin{scope}[xshift=7.95cm,yshift=3.20cm,scale=0.86,transform shape]
    \coordinate (top) at (0,2.20);
    \coordinate (left) at (-1.35,0);
    \coordinate (right) at (1.35,0);

    \def\lowerBubbleRadius{0.32}
    \coordinate (lower left vertex) at (-0.32,0);
    \coordinate (lower right vertex) at (0.32,0);

    \draw[photon] (left) -- (lower left vertex);
    \draw[photon] (lower right vertex) -- (right);
    \draw[fermion,centered arrow]
      (lower left vertex)
      arc[start angle=180,end angle=0,radius=\lowerBubbleRadius];
    \draw[fermion,centered arrow]
      (lower right vertex)
      arc[start angle=0,end angle=-180,radius=\lowerBubbleRadius];

    \node[interaction vertex] at (lower left vertex) {$\times$};
    \node[interaction vertex] at (lower right vertex) {$\times$};

    \def\upperPhotonRadius{0.64004}
    \def\upperFermionRadius{0.31}
    \coordinate (upper photon center) at (0,1.55996);
    \coordinate (upper left vertex) at (-0.31,1.00);
    \coordinate (upper right vertex) at (0.31,1.00);

    \draw[photon]
      (upper left vertex)
      arc[start angle=241.02,end angle=-61.02,radius=\upperPhotonRadius];

    \draw[fermion,centered arrow]
      (upper left vertex)
      arc[start angle=180,end angle=0,radius=\upperFermionRadius];
    \draw[fermion,centered arrow]
      (upper right vertex)
      arc[start angle=0,end angle=-180,radius=\upperFermionRadius];

    \node[interaction vertex] at (upper left vertex) {$\times$};
    \node[interaction vertex] at (upper right vertex) {$\times$};

    \node[operator insertion] at (top) {};
    \node[operator insertion] at (left) {};
    \node[operator insertion] at (right) {};
  \end{scope}

\end{tikzpicture}